\documentclass[rmp,longbibliography,aps,twocolumn]{revtex4-1}
\usepackage{bm,dcolumn,amsmath,graphicx,natbib,url,textcase}

\usepackage{epsfig}
\usepackage{epstopdf}
\usepackage{euscript}
\usepackage{amsfonts}
\usepackage{float}
\usepackage{multirow}
\usepackage{eufrak}

\DeclareMathAlphabet{\mathpzc}{OT1}{pzc}{m}{it}

\def\prn#1{{\left(#1\right)}}
\def\cbrk#1{{\left\{#1\right\}}}
\def\sbrk#1{{\left[#1\right]}}
\def\abrk#1{{\langle#1\rangle}}
\def\ket#1{{|#1\rangle}}
\def\bra#1{{\langle#1|}}

\def\dbyd#1#2{{\frac{d #1}{d #2}}}

\def\cg(#1,#2)(#3,#4)(#5,#6){\bra{#1,#2,#3,#4}#5,#6\rangle}

\def\ts#1{{_{\mbox{\scriptsize #1}}}}

\def\threej(#1,#2)(#3,#4)(#5,#6){\begin{pmatrix}#1&#3&#5\\#2&#4&#6\end{pmatrix}}
\def\sixj(#1,#2,#3)(#4,#5,#6){\begin{Bmatrix}#1&#2&#3\\#4&#5&#6\end{Bmatrix}}
\def\ninej(#1,#2,#3)(#4,#5,#6)(#7,#8,#9){\begin{Bmatrix}#1&#2&#3\\#4&#5&#6\\#7&#8&#9\end{Bmatrix}}

\def\sA{{\ensuremath{\EuScript A}}}

\def\sJ{{\ensuremath{\EuScript J}}}
\def\sR{{\ensuremath{\EuScript R}}}

\def\sV{{\ensuremath{\EuScript V}}}
\def\sL{{\ensuremath{\EuScript L}}}

\def\sX{{\ensuremath{\EuScript X}}}

\def\mr{\mathrm}
\def\mb{\mathbf}
\def\bs{\boldsymbol}
\def\mc{\mathcal}

\newcommand{\Esca}{\mathcal{E}}
\newcommand{\Evec}{\hbox{\boldmath{$\mathcal{E}$}}}

\newcommand{\Bvec}{\hbox{\boldmath{$\mathcal{B}$}}}
\newcommand{\SMtvec}{\hbox{\boldmath{$\mathcal{S}$}}}
\newcommand{\SMt}{\mathcal{S}}
\begin{document}
\title{Search for New Physics with Atoms and Molecules}
\author{M.S. Safronova$^{1,2}$, D. Budker$^{3,4,5}$,  D. DeMille$^{6}$, Derek F. Jackson Kimball$^{7}$,  A. Derevianko$^8$ and 
Charles W. Clark$^{2}$\\ \vspace{0.1in}
$^1$University of Delaware, Newark, Delaware, USA,\\
$^2$Joint Quantum Institute, National Institute of Standards and Technology  and the University of Maryland, College Park, Maryland, USA,\\
$^3${H}elmholtz Institute, Johannes Gutenberg University, Mainz, Germany, \\
$^4$University of California, Berkeley, California, USA,\\
$^5$Nuclear Science Division, Lawrence Berkeley National Laboratory, Berkeley, California, USA\\
$^6$Yale University, New Haven, Connecticut, USA,\\
$^7$California State University, East Bay, Hayward, California, USA,\\
$^8$University of Nevada, Reno, Nevada, USA}

    \begin{abstract}
    This article reviews recent developments in tests of fundamental physics using  atoms and molecules, including the subjects of
    parity violation, searches for permanent electric dipole moments,
    tests of the {$CPT$} theorem and {L}orentz symmetry, searches for spatiotemporal variation of fundamental constants, tests of quantum electrodynamics, tests of general relativity and the equivalence principle, searches for dark matter, dark energy
 and extra forces, and tests of the spin-statistics theorem.
 Key results are presented in the context of potential new physics and  in the broader context  of similar investigations in other fields.  Ongoing and future experiments of the next decade are discussed.
    \end{abstract}
 \maketitle

\tableofcontents

\section{Introduction}

\subsection{Recent advances in AMO physics}
 The past two decades have been a transformational era for atomic, molecular and optical (AMO) physics, due to extraordinary accomplishments in the control of matter and light.  Experimental breakthroughs, including laser cooling and trapping of atoms, attainment of Bose-Einstein condensation, optical frequency combs and quantum control - subject of Nobel Prizes in Physics in 1997 \cite{Phi98,Chu98,Coh98}, 2001
\cite{CorWie02,Ket02}, 2005 \cite{Han06,Hal06,Gla06}, and 2012 \cite{Win13,Har13}, respectively - have led to widespread availability of ultracold (temperature $T < 1~ \mu$K)  ions, atoms and molecules, subject to precise interrogation and control.
Revolutionary developments on several fronts have been made possible by these advances, aided by improvements in precision time and frequency metrology, measurement techniques such as atomic magnetometry and interferometry, and first-principles atomic and molecular theory.  These advances brought forth a plethora of new AMO applications, including
novel  tests of the fundamental laws of physics.
\subsection{Problems with the Standard Model}
The Standard Model (SM) of particle physics \cite{PDG} has been exceptionally successful in predicting and describing numerous phenomena and has been extensively tested by a multitude of different approaches spanning most fields of physics. Despite its great success, the SM has major problems. Indeed, it is inconsistent with the very existence of our Universe: the Standard Model cannot account for the observed imbalance of matter  and antimatter \cite{Dine2003}. In addition, all attempts to combine gravity with the fundamental interactions described by the SM have been unsuccessful.

A long-standing mystery dating back to the 1930s \cite{Zwi33,Zwi09}
is the apparent existence of   ``dark matter'' that is observed only via its gravitational interactions. This is confirmed by numerous studies of astronomical objects, which show that
 the particles of the SM make up only $\approx$16~\% of the total matter present in our Universe.
Decades of investigation have not identified the nature of dark matter \cite{Ber13}. We do now know what most of it is not---any of the particles of the SM.

Studies of the Type I supernovae which were originally aimed at measuring the deceleration rate of the Universe arrived at a completely unexpected result: the expansion of the  Universe
 is now accelerating \cite{Rei12,Sch12,Per12}. This seems to be possible only  if our Universe contains a kind of ``dark energy'' which effectively acts as repulsive gravity. While we do not know what dark matter is, we know even less what such dark energy could be---while  vacuum energy is a handy potential candidate, the discrepancy between the sum of known contributions to vacuum energy in the universe and the cosmologically
observed value is 55 orders of magnitude \cite{Sol13}.
According to the 2015 results of the Planck Mission study of cosmic microwave background radiation \cite{Pla15}, our present Universe  is 69~\% dark energy, 26~\% dark matter, and 5~\% ordinary (Standard Model) matter.

In summary, we are at an extraordinary point in time for physics discovery. We have found all of the particles of the SM and have tested it extensively, but we do not know what makes 95~\% of the Universe, nor how ordinary matter survived against annihilation with antimatter in the aftermath of the Big Bang.  This provides strong motivation to search for new particles (and/or the associated fields)  beyond those described in the SM.

\subsection{Search for new physics with precision measurements}
While one can search for new particles  directly with large-scale collider experiments at the TeV energy scale, such as those carried out at the Large Hadron Collider (LHC) at CERN,  new physics may also be observed  via low-energy precision measurements.
%\subsubsection{An early example: atomic parity violation}
An early example of the use of  AMO physics in this paradigm, beginning in the 1970s, was the deployment of highly sensitive laser-based techniques  to observe parity violation in optical transitions in atoms.  This parity violation occurs due to exchange of $Z$ bosons between electrons and nuclei in atoms, and quantitative measurements of the strength of the effect can be used to test the predictions of theoretical models of the electroweak interaction \cite{Khr91a}.  These investigations quickly led to the realization that the accuracy of first-principles theory of atomic structure needed radical improvement in order to interpret the experimental results.  This was particularly true for heavy atoms like cesium (Cs, $Z=55$), which required the development of novel theoretical methodologies enabled by modern computing architecture \cite{PorBelDer09}.
 Improved computational resources and development of high-precision methodologies have led to essential progress in related theoretical investigations, enabling improved analyses of precision experiments, development of new experimental proposals, and  improved theoretical predictions for yet unmeasured quantities.
As a result,  analyses of the Cs atomic parity violation (APV) experiment \cite{WooBenCho97} provided  the most accurate to-date tests of the low-energy electroweak sector of the SM and constraints on a variety of scenarios for physics beyond the SM \cite{PorBelDer09,DzuBerFla12}. Combined with the results of high-energy collider experiments, Cs APV studies confirmed the energy dependence of the
electroweak force over an energy range spanning four orders of magnitude \cite{PorBelDer09}. Further details are given in Sec.~\ref{Sec:APV-intro}.

Similarly, for several decades AMO experiments have been employed to search for violation of time-reversal ($T$) symmetry, as manifested by an electric dipole moment (EDM) along the angular momentum axis of a quantized system. $T$-violation is required to generate a cosmological matter-antimatter asymmetry, and sources beyond those in the SM are required to explain the magnitude of the observed imbalance \cite{Dine2003}.  Extensions to the SM frequently introduce new sources of $T$-violation that are associated with new particles  \cite{Barr1993}.  In theories where these new particles have mass at the TeV scale---or, sometimes, well above it---EDMs are typically predicted with size near the limits set by current AMO experiments \cite{Pospelov2005,Engel2013}.  Hence these EDM experiments probe commonly-predicted physics at similar or higher energy scales than those accessible with the LHC.  New experiments based on large enhancements of the observable EDM effects (in experiments using polar molecules or deformed nuclei)  hold the promise to increase the energy reach for probing new $T$-violating physics by an order of magnitude or even more in the near future. Further details are given in Sec.~\ref{Sec:EDM}.

AMO experiments also probe even higher energy scales. A number of theories aiming to unify gravity with other fundamental interactions suggest violations of cornerstones of modern physics such as  Lorentz symmetry and combined charge-conjugation ($C$), parity ($P$), and time reversal ($CPT$) invariance \cite{ColKos98,KosRus11} and imply spatiotemporal variation of fundamental constants \cite{Uza11}.
 Whereas the energy scale of such  physics is much higher than that attainable at present by particle accelerators, Lorentz violation may nevertheless be detectable via precision measurements at low energies.

The unprecedented accuracy of AMO precision measurements coupled with accurate theory predictions  facilitated  significant expansion of AMO fundamental physics studies.
As a result, AMO physics now addresses questions in fields from which it was once quite remote, such as nuclear, particle and gravitational physics and  cosmology.

For example, a number of AMO technologies  such as high-precision magnetometery \cite{Pus13,Bud14}, atom interferometry \cite{Ham15}, atomic clocks \cite{DerPos14}, and ultra
 high-intensity lasers  \cite{DiPMulHat12} are aimed at the search for axions and other
 dark matter and dark energy candidates.
The principles of a new technique for detecting
transient signals of exotic origin using a global
network of synchronized optical magnetometers were
demonstrated by \citet{Pus13}. The network may probe
stable topological defects (e.g., domain walls) of axion-like fields.

A recent Cs matter-wave interferometry experiment   constrained a wide class of dynamical dark energy theories \cite{Ham15}. The  exceptional sensitivity of matter-wave  interferometers operated with quantum gases has generated new ideas for probing the fundamental concepts of quantum mechanics, tests of general relativity, and gravitational wave detection \cite{MunAhlKru13,Bie15,HogKas16}. The first quantum test of the universality of free fall  with matter waves of two different atomic species was reported by \citet{SchHarAlb14}.

The accuracy of atomic clocks has  improved by a factor of 1000 in the past 10 years, to a fractional frequency uncertainty of two parts in $10^{18}$ \cite{NicCamHut15,UshTakDas15} which corresponds to  a temporal uncertainty of one second in the lifetime of our Universe. As a result, atomic clocks are now used to search for possible time variations
of the dimensionless fine-structure constant, $\alpha$, and proton-electron mass ratio,
$m_\mr{p}/m_\mr{e}$ \cite{RosHumSch08,GodNisJon14,HunLipTam14}.

A demonstration of  the potential of quantum-information techniques in
the search for physics beyond the SM was provided by \citet{PruRamPor15}. Using a  pair of trapped calcium (Ca, $Z$ = 20) ions
in a decoherence-free subspace, they improved by a factor of 100 the bounds on a number of
 Lorentz-symmetry violating  parameters of the Standard Model Extension (SME)  for electrons.

\subsection{Scope of this review}
The examples above show the diversity of recent AMO searches for new physics.
Here, we review this subject as a whole rather than limit the treatment to a few specific topics, since this field is based on a commonality of approaches that is likely to have even wider applicability in the future, given the growth that we have witnessed recently.
 %The main subjects covered in this review are: parity violation; searches for permanent electric dipole moments; tests of the CPT theorem and Lorentz % symmetry; searches for variation of fundamental constants; tests of quantum electrodynamics; tests of general relativity and the equivalence % principle;  searches for dark matter, dark energy
 %and extra forces; and tests of the spin-statistics theorem.

Another  active area of AMO physics is the simulation of condensed-matter systems using ultracold atoms in optical potentials. This field  has aspects of searches for new physics associated with novel quantum phases, non-Abelian gauge potentials, atomtronics and the like.  Our review will not deal with such topics since they are already addressed by other reviews \cite{2007AdPhy..56..243L,2008RvMP...80..885B,2012NatPh...8..267B,GJOS2014,Ueda2014,WS2013,2013RvMP...85.1191S,GeoAshNor14}, and constitute a vast subject in their own right. We will also exclude detailed consideration of quantum mechanics tests with AMO systems which  have been recently reviewed as well \cite{BasLocSat13,HorGerHas12,AspKipMar14}.

Since the field of AMO tests of fundamental physics is a vast subject spanning decades of research, we limit this review to recent developments and proposals.
For each topic, we begin with an introduction to its specific relevance to physics beyond the SM. We present recent key results in the context of potential new physics and summarize ongoing and future experiments of the next decade.

\section{Search for variation of fundamental constants}
\label{Sec:FC}
\subsection{Fundamental constants: an introduction}
\label{FC1}
First, we have to define what we mean by ``fundamental constants.'' Opening a physics textbook on various physics fields would produce different
lists of measured quantities of specific importance to a given field.
In this review, we follow the definition of \citet{Uza15}: a fundamental constant is \textit{“any parameter not
determined by the theories in which it appears”}. This definition has the following implications:
\begin{itemize}
\item the number of fundamental constants depends on a particular theory and
\item the fundamental constants are not predicted by any theory and thus their values must be determined through measurements.
\end{itemize}

Present physics is described by general relativity (GR) and the Standard Model (SM) of particle physics
that combines quantum chromodynamics (QCD) with the electroweak theory \cite{PDG}. The minimal SM has 19 parameters, with
somewhat different sets of these parameters given
 in the literature \cite{Hog00,Uza13,Sco06}. Following the summary of \citet{Sco06}, the list contains 6 quark masses,
 3 lepton masses, 3 quark mixing angles ($\theta_{12}$, $\theta_{23}$, $\theta_{13}$) and phase $\delta$,  3 electroweak parameters (fine-structure constant $\alpha$, Fermi coupling constant $G_\textrm{F}$,
 and mass of the $Z$ boson $M_Z$), Higgs mass, strong combined charge-conjugation and parity ($CP$) violating phase  and the QCD coupling constant.
The incorporation of the neutrino masses leads to additional parameters.

 To reproduce known physics, the SM  parameters must be supplemented by the
Newtonian constant of gravitation $G$ of GR, the speed of light in vacuum $c$, and the Planck constant $h$.
We note that this list of fundamental constants lacks  any description of either dark matter or dark energy and contains no
cosmological information about the Universe. The Standard Cosmological Model adds 12 more parameters, listed by \citet{Sco06}
which include the Hubble constant, baryon, cold dark matter and dark energy densities, and others.
Further understanding of these phenomena may increase the required number of fundamental constants, while
developing a unified theory might reduce them.

Measurements of fundamental constants and numerous other derived quantities, some of which can be predicted from current
theories with varying levels of accuracy, is a vast area of research. We refer the
reader to the publications of Committee on Data for Science and Technology (CODATA), \cite{CODATA2014} and Particle Data
Group \cite{PDG} for measurement techniques, analysis of data and current recommended values. The data are continuously
revised and improved, with critical assessment of various types of experiments carried out prior to new CODATA and PDG publications.

It should be kept in mind that there is {\em no single experiment} that determines the CODATA recommended value of a given
fundamental constant.  There is a complex web of deep and sometimes subtle  connections between fundamental constants -
for example, between the fine structure constant and the molar Planck constant $N_{\mathrm A}h$ \cite{CODATA2014} - and
the CODATA recommended values are determined by a least-squares adjustment that keeps inconsistencies within limits.

An example of this interdependence, that is  highlighted by \citet{CODATA2014} and is of particular relevance to atomic, molecular, and optical (AMO) physics,
is the determination of the
fine-structure constant
\begin{equation}
\alpha=\frac{1}{4\pi \epsilon_0}\frac{e^2}{\hbar c},
\label{FC-2}
\end{equation}
which characterizes the strength of the electromagnetic interaction, see Sec.~\ref{Sec:QED}. Here, $e$ is the elementary charge,
$\hbar=h/2\pi$ is the reduced Planck constant, and
$\epsilon_0$ is the electric constant.
A recent  overview of the determinations of fundamental constants from low-energy measurements is given by \citet{Kar13}.

We note the values of the coupling constants of the SM depend on the energy at which they are measured (so-called ``running'' of
the coupling constants discussed using the example of  sin$^2\theta_W$ in Sec.~\ref{Sec:APV-intro}).
The fine-structure constant $\alpha$ is defined in the limit of zero momentum transfer.

\subsection{Units of measurement vs. fundamental constants}
\label{Units}
 Experimental measurements can be reduced to comparing two physical systems, one of which defines the unit of measurement.
  For example, the International System of Units (SI)  unit of time is defined as:
``The second is the duration of 9\,192\,631\,770 periods of the radiation corresponding to the transition between the two
hyperfine levels of the ground state of the cesium 133 atom'' \cite{BIPM}.
This definition refers to a Cs atom at rest at a temperature of 0~K.
Therefore, absolute values of all other frequencies are determined relative to this Cs frequency and no absolute
frequency measurement can be performed with smaller fractional frequency uncertainty than that of the best Cs frequency standard,
which is presently
on the order of $10^{-16}$ \cite{GueRosLau10,SzyEonMar10,GueAbgRov12a,HeaDonLev14,LevCalCal14}.
Note that one can still make \textit{relative} comparison of two frequencies to much better precision than the Cs
standard provides \cite{UshTakDas15}. To make absolute frequency measurements accurate to, for example, $10^{-18}$ of a second,
we would need to change the
definition of the second from the Cs microwave frequency transition to another physical system. Such system must allow for the
construction of the frequency standard with $10^{-18}$ uncertainty in a consistently reproducible way, accompanied by a global
technology infrastructure for frequency comparison \cite{PolOatGill13,LudBoyYe15}.

 Changing the values of the constants in such a way that all dimensionless combinations are unchanged will simply change the units.
 For example, in atomic units the values
of  $e$, the electron mass $m_{\rm e}$, and the reduced
Planck constant $\hbar$ have the numerical value 1, and the electric constant $\epsilon_0$ has the numerical value $1/(4\pi)$. However,
the value of the dimensionless fine-structure constant $\alpha$ is still the same as in SI units as given by Eq.~(\ref{FC-1}).

%The Planck units of mass, length, and time, can be constructed from products of  $G$, $c$, and $h$
%leaving 26 dimensionless fundamental constants that determine the relative strengths of fundamental interactions, hierarchy of masses
% for quark and leptons, etc. \cite{Uza15}.

Dimensionless fundamental constants play a special role in discussions of spatiotemporal variations of physical laws. Their values
are, by construction, independent of the choice of units of measurement, which are arbitrary conventions that have changed in
the past and may change in the future.

For example, it is difficult to see how one could measure unambiguously a time variation in the speed of light, $c$. This may be
 viewed from the perspective of 1982. Then, the second and the meter were defined independently: the second as it is today, a
 defined multiple of the period of the ground hyperfine transition of $^{133}$Cs, and the meter as a defined multiple of the
 wavelength of the 2p$_{10}-$5d$_5$ spectral line of an isotope of krypton (Kr, $Z = 36$), $^{86}$Kr. On a simple observational basis, if a change in the 1982 value
 of $c$ was well established on the basis of multiple independent observations, it seems impossible to disentangle that effect
 from changes in either, or both, of the Cs frequency or the Kr wavelength.

The focus of modern studies of variation of fundamental constants is thus on  dimensionless constants, and as concerns AMO physics,
 particularly on $\alpha$ and the proton/electron mass ratio, $m_{\mathrm{p}}/m_{\mathrm{e}}$.

\subsection{Theories with varying fundamental constants}
\label{FC-theory}
While the 2014 CODATA value of fine-structure constant $\alpha$ has a remarkably small $2.3\times10^{-10}$ uncertainty, it remains an open question whether the value of $\alpha$ is variable across space and time. In the
 SM, all fundamental constants are invariable.
 The dimensionless constants become dynamical (i.e. varying) in a number of theories beyond the SM and GR.   Detailed review and references to theories with varying fundamental constants are given by \citet{Uza11} so we only
 give a brief summary here.

 Higher-dimensional theories, in particular string theories, naturally lead to varying fundamental constants.
 String theories predict the existence of a scalar field, the dilaton, that couples directly to matter \cite{TayVen88}.
 The 4-dimensional coupling constants are determined in terms of a string scale and various dynamical fields. As a result, the coupling constants  naturally become varying, evaluated as the expectation values of these dynamical fields.
The variation of the gauge couplings and of the gravitational constant may also arise from the variation of the size of the extra dimensions.

Many other theories beyond the SM and GR have been proposed in which fundamental constants become dynamic fields.
These include: discrete quantum gravity \cite{GamPul03}; loop quantum gravity \cite{TavYun08}; chameleon models \cite{KhoWel04};
 dark energy models with a non-minimal coupling of a quintessence field \cite{AveMarNun06} and others. As a result, studies of
 the variation of fundamental constant may
provide some information on potential origin of dark energy.  Analysis of experiments on the variation of fundamental constants
 also depends on the nature of the particular model. For example, a chameleon field is expected to be more massive in high-density
  regions on Earth than in low-density regions of the solar system \cite{KhoWel04}. Since the constants would be dependent on the local
   value of the chameleon field, the values of the constants become dependent on their (mass density) environment.

 While one can construct models in which only one or a few constants vary, in most realistic current models, all constants vary if
 one  does \cite{Uza15}. In  unified theories of fundamental interactions the variations of
 fundamental constants are correlated. However,  including such correlations in the analysis of experiments
 leads to dependence of the results on the particular model.

It has been pointed out that searching for variation of fundamental constants
 is a test of the local position invariance hypothesis and thus of the equivalence principle [see \citet{Uza11} and \citet{Uza15}
 and references therein].

 Searches for variation of fundamental constants are conducted in a number of systems  including atomic clocks,
astrophysical studies of quasar spectra and observation of the HI 21 cm line, the {O}klo natural nuclear reactor, meteorite dating,
stellar physics, cosmic microwave background (CMB), and Big Bang nucleosynthesis (BBN). A detailed review of these  topics is given
by \citet{Uza11}. We limit our coverage to recent results, ongoing experiments, and proposals relevant to AMO physics.

 Laboratory tests for the variation of fundamental constants, such as carried out with atomic clocks, are only sensitive to  present-day variation, while  other searches are probing whether $\alpha$ and other constants were different in the past compared to what they are now, with different look-back times. We discuss this further in Sec.~\ref{quasar}. The analysis of CMB and BBN in terms of constraining
the variation of fundamental consistent is also dependent on the cosmological model.

In this section, we consider the ``slow-drift'' model of variation of fundamental constants, as well as
coupling of fundamental constants to a changing gravitational potential, and
testing for a dependence of fundamental constants on the  mass density of the environment.
Searches for oscillatory and transient variation of fundamental constants, and their
relevance to the nature of dark matter and dark energy, are discussed in Sec.~\ref{Sec:LightDarkMatter}.

\subsection{Tests of fundamental constant variations with atomic clocks}
\label{FC-D}
The most precise tests of modern-epoch variation of fundamental constants  are carried out using atomic clocks.
From the standpoint of metrology and other precision experiments, testing the variation of  fundamental constants is necessary to ensure that
the experiments are reproducible at the level of their uncertainties.  This became particularly important
 due to exceptional improvement of AMO precision metrology in recent years.   If  $\alpha$ or $\mu=m_\mr{p}/m_\mr{e}$
are space-time dependent, so are atomic and molecular spectra. Therefore, the variation of the fundamental constants makes the clock tick rate  dependent on location,
time, or type of the clock - since
 frequencies of Cs or Sr depend differently on fundamental constants. We have arrived at
 a level of precision such that new physics might show up unexpectedly as an irreducible systematic error!
 An important
question for AMO theory is predicting   the best
systems for dedicated experiments where the variation of fundamental
constants is strongly enhanced.

We start with the discussion of  the dependence of atomic spectra on the dimensionless constants of interest.
The possibility of using atomic spectroscopy to detect variations in the fine-structure constant $\alpha$ is suggested in Dirac's
theory of the hydrogen atom. The energies of $E_{n,j}$ of a Dirac electron bound to an infinite-mass point nucleus are given by
\begin{multline}
\label{DiracHydrogenEnergy}
E_{n,j} = m_{\mathrm{e}}c^2  \\
\times \left[ 1 +
\frac{\left( Z \alpha\right)^2}
{\left[ n - j - \frac{1}{2} +
\sqrt{ \left(j + \frac{1}{2} \right)^2 -\left( Z \alpha \right)^2} \right]^2}\right]^{-1/2} ,
\end{multline}
%alternative
%\begin{equation}
%\label{DiracHydrogenEnergys}
%\frac{E_{nj}}{m_\mathrm{e} c^2} =
%\left[ 1 +
%\frac{\left( Z \alpha\right)^2}
%{\left[ n - j - \frac{1}{2} +
%\sqrt{ \left(j + \frac{1}{2} \right)^2 -\left( Z \alpha \right)^2} \right]^2}\right]^{-1/2}
%\end{equation}
where $Ze$ is the charge of the nucleus, with $e$ the elementary charge, $n$ is the principal quantum number, and $j$ the electronic
angular momentum in units of $\hbar$ \cite{Greiner2000,Joh07}. With reference to the discussion of Sec.~\ref{FC1}, note that the
Rydberg constant is given by

\begin{equation}
\label{Rydbergconstant}
R_\infty = \frac{1}{hc} \cdot \frac{\alpha^2}{2} m_{\mathrm{e}}c^2.
\end{equation}

The only fundamental constants present in Eq.~(\ref{DiracHydrogenEnergy}) are $\alpha^2$ and the rest mass energy of the electron, $m_\mathrm{e}c^2$.
 Expansion of $E_{n,j}$ in powers of $\alpha^2$ shows that for electronic  states with
different values of the principal quantum number $n$, the energy splitting scales with $\alpha ^2$, whereas the splitting scales
 with $\alpha ^4$ for  states with the same $n$ but different $j$.
  Thus, ratios of the wavelengths of these two types of transitions are sensitive to variations in $\alpha$. The dependence of the atomic spectra of more complicated atoms on fundamental constants is discussed below.

\subsubsection{Dependence of hyperfine and electronic transitions on dimensionless constants}
Interaction of atomic electrons with the magnetic and electric multipole fields of the nucleus
leads to a splitting of atomic energy levels referred to as hyperfine structure.
For example,  the  nuclear angular momentum of $^{133}$Cs is $I=7/2$ and the
ground state electronic configuration consists of a closed Xe-like core (xenon, $Z = 54$) with  an unpaired single valence  electron  with $j=1/2$.
 Therefore,  Cs [Xe]$6s$ ground state
splits into two hyperfine levels, with $F=3$ and $F=4$, where the total angular momentum  $\mb{F}=\mb{I}+\mb{J}$.
The  frequency of electromagnetic radiation associated with transitions between these  levels is
conventionally expressed as

\begin{equation}
\nu_\textrm{hfs} \sim c R_{\infty} A_\textrm{hfs} \times g_i \times \frac{m_\mr{e}}{m_\mr{p}} \times \alpha^2 F_\textrm{hfs}(\alpha),
\label{eq2-FC}
\end{equation}
where $R_{\infty}$ is given by Eq.~(\ref{Rydbergconstant}), $A_\textrm{hfs}$ is the numerical quantity depending on a particular atom,
 and
$F_\textrm{hfs}(\alpha)$ is a relativistic correction specific to each hyperfine transition.
The dimensionless $g_i=\mu_i/\mu_\textrm{N}$   is the   g-factor associated with the nuclear magnetic moment $\mu_i$, where
$\mu_\textrm{N}=e\hbar/2m_\mr{p}$ is the nuclear magneton.

The Cs hyperfine  $F=3 - F=4$ transition frequency
$\nu_{\textrm{Cs}}$ of $\approx 9$~GHz defines the second, and all \textit{absolute}
frequency measurements are actually measurements of $\nu/ \nu_{\textrm{Cs}}$ frequency ratios. Atomic clocks based on hyperfine
transitions are referred to in the literature as ``microwave clocks'', specifying the relevant  region of the electromagnetic spectrum \cite{PolOatGill13,LudBoyYe15}.

The transition frequency between electronic energy levels in an atom can be expressed as
\begin{equation}
\nu \sim c R_{\infty} A F(\alpha),
\label{eq3-FC}
\end{equation}
where $A$ is the numerical factor depending on an atom and $F(\alpha)$
depends upon the particular transition. Atomic clocks based on electronic
transitions with frequencies from $\approx 0.4\times 10^{15}$~Hz  to  $\approx 1.1\times 10^{15}$~Hz  are referred to in the literature as ``optical clocks''.

\subsubsection{Theoretical determination of the sensitivity of atomic transitions to variations of $\alpha$}

The coefficient $F(\alpha)$ in Eq.~(\ref{eq3-FC}) is obtained by calculating the $\alpha$-dependence of the energies of the two atomic
 levels involved in the transition.
The dependence of electronic energy level $E$ on $\alpha$ is usually parameterized
 by the  coefficient $q$~\cite{DzuFlaWeb99,DzuFlaWeb99a}
\begin{equation}
E(\alpha) = E_0 +q \left[\left(\frac{\alpha}{\alpha_0}\right)^2-1\right],
\label{eq:wx}
\end{equation}
which can be determined rather accurately [generally to 1~\% - 10~\%] from atomic-structure computations.
Here, $\alpha_0$ is the current value of $\alpha$ \cite{CODATA2014}, the measurement of  which  was discussed in Sec.~\ref{FC1},
 and $E_0$ is the energy corresponding to this  value of $\alpha_0$ .
The coefficient $q$ depends weakly on electron correlations, so it can be calculated more accurately than the actual energy of a level.

The  coefficient $q$ of an atomic state is  computed by varying the numerical value of $\alpha$ in the computation of the
respective energy  level \cite{DzuFla09}. Generally,
 three energy level calculations are performed  which differ only  by the values of $\alpha$.  The first calculation uses
  the current $\textrm{CODATA}$ value of $\alpha^2$ ~\cite{CODATA2014}.  Two
 other calculations are performed with
  $\alpha^2$ varied by a small but non-negligible amount, commonly selected at $\delta=0.01$.   Then,  the value of $q$ is derived from a
  numerical derivative
 \begin{equation}
 q=\frac{E(\delta)-E(-\delta)}{2\delta},
 \end{equation}
 where $E(\pm \delta)$ are results of the energy calculations. The additional calculation
  (with the CODATA value of $\alpha$)
 is used to verify that the change in the energy is close to linear.

The parameter $q$ links variation of the transition energy $E$, and hence the atomic frequency $\nu=E/h$, to the
variation of $\alpha$
\begin{equation}
\frac{\delta E}{E_0} = \frac{2q}{E_0}\frac{\delta
  \alpha}{\alpha_0} \equiv K  \frac{\delta \alpha}{\alpha_0},
\label{eq:K}
\end{equation}
where
\begin{equation}
K=\frac{2q}{E_0}
\label{eq4-FC}
\end{equation}
 is a dimensionless sensitivity factor.

In $\alpha$-variation tests with atomic clocks, the ratio of two
clock frequencies is monitored, and the
sensitivity to the variation of $\alpha$ is then given by the difference in their
respective $K$ values for each clock transition, i.e. $\Delta K=|K_2-K_1|$. The larger the value of $K$, the more sensitive is a particular
atomic energy level to the variation of $\alpha$.

A note of caution has to be added here: while small $E_0$ may lead to large $K$ following Eq.~(\ref{eq4-FC}), it may also lead to technical difficulties in  measuring the relevant frequency with the extremely high accuracy that is required for tests of variation of fundamental constants. Small energy $E_0$ corresponds to transitions in the infrared, with wavelength that may exceed 3000~nm. Accurate theory predictions are particulary difficult for such transitions, as small $E_0$ is the result of strong cancellations of upper and lower energies, leading to difficulties in locating weak clock transitions. Moreover, the clock instability is inversely proportional to the transition frequency, so lower frequency leads to higher ultimate instability, which is particulary problematic with single ion clocks. Therefore, the actual transition frequency and other experimental considerations have to be taken into account when designing  dedicated experiments. This issue is further discussed in Sec.~\ref{FC-F}.

\begin{table}
\caption{\label{tab1-FC}
Sensitivity factors $K$ to the variation of the fine-structure constant $\alpha$ for clock transitions \cite{DzuFla09}.
 $K$ is defined by Eq.~(\ref{eq4-FC}).
All transitions except Rb and Cs are optical frequency standards.}
\begin{ruledtabular}
\begin{tabular}{llc}
\multicolumn{1}{l}{Atom}&
 \multicolumn{1}{l}{Transition}&
 \multicolumn{1}{c}{$K$}\\
\hline
  $^{87}$Rb             & ground hyperfine &                             0.34     \\
  $^{133}$Cs             & ground hyperfine&                               0.83         \\
Al$^+$          &$3s^2 \, ^1S_0 - 3s3p \, ^3P_0$                      &0.008  \\
Ca$^+$          &$4s \, ^2S_{1/2}- 3d \, ^2D_{5/2}$                   &0.15 \\
 Sr             &$5s^2 \, ^1S_0 - 5s5p \, ^3P_0$                      & 0.06  \\
Sr$^+$          &$5s \, ^2S_{1/2}- 4d \, ^2D_{5/2}$                   & 0.43\\
 Yb             &$6s^2 \, ^1S_0 - 6s6p \, ^3P_0$                      &  0.31      \\
  Hg$^+$        &$6s \, ^2S_{1/2}- 5d \, ^2D_{5/2}$                   & -2.94\\
 Yb$^+$\, $E2$  &$4f^{14}6s \, ^2S_{1/2}- 4f^{14} 5d \, ^2D_{5/2}$    & 1.03\\
 Yb$^+$\, $E3$  &$4f^{14}6s \, ^2S_{1/2}- 4f^{13} 6s^2 \, ^2F_{7/2}$  &-5.95 \\
\end{tabular}
\end{ruledtabular}
\end{table}

\subsubsection{Microwave vs. optical clock-comparison experiments}

At the lowest level of the analysis that requires only atomic structure calculations,
 measuring the ratios $R=\nu_1/ \nu_2$ of two clocks over time may set limits on variation of $\alpha$, the proton-to-electron mass ratio
 $\mu=m_\mr{p}/m_\mr{e}$, and nuclear $g$ factors, specifically $g_\textrm{Cs}$ and $g_\textrm{Rb}$ as these correspond to two  microwave clocks with
  the smallest uncertainties.  We summarize the dependence of clock-frequency ratios on the dimensionless constants as follows:
 \begin{itemize}
 \item The ratio of two microwave clock frequencies  depends on $\alpha$ and $g$-factors of the corresponding nuclei according
 to Eq.~(\ref{eq2-FC}). For example, the ratio of Cs to Rb (rubidium, $Z = 37$) clock frequencies is proportional to
  \begin{equation}
 \frac{\nu_{\textrm{Cs}}}{\nu_{\textrm{Rb}}}\propto \frac{g_{\textrm{Cs}}}{g_{\textrm{Rb}}} \times
  \alpha^{K_{\textrm{Cs}}-K_{\textrm{Rb}}}= \frac{g_{\textrm{Cs}}}{g_{\textrm{Rb}}}  \alpha^{0.49},
  \end{equation}
  where the $K$ factors defined by Eq.~(\ref{eq4-FC}) are given in Table~\ref{tab1-FC}.
  \item The ratio of frequencies of any two optical clocks depends only upon $\alpha$, according to Eq.~(\ref{eq3-FC}).
  \item The ratio of   optical  to  microwave clock frequencies depends on $\alpha$, $\mu=m_\mr{p}/m_\mr{e}$ ratio, and the $g$-factor of
  the atomic nucleus of the microwave clock.
 \end{itemize}
Reducing the potential variation of $g$-factors to more fundamental quantities, such as $X_q=m_q/\Lambda_{\textrm{QCD}}$, and calculation of the
corresponding dimensionless sensitivity factors $\kappa_{\textrm{Cs}}$ and $\kappa_{\textrm{Rb}}$,
requires nuclear structure calculations which are dependent on a particular model \cite{FlaTed06,DinDunDzu09,Kim15}. Here, $m_q$ is the
average light-quark mass and  $\Lambda_{\textrm{QCD}}$ is the  QCD energy scale.
\begin{figure}[t]
            \includegraphics[scale=0.5]{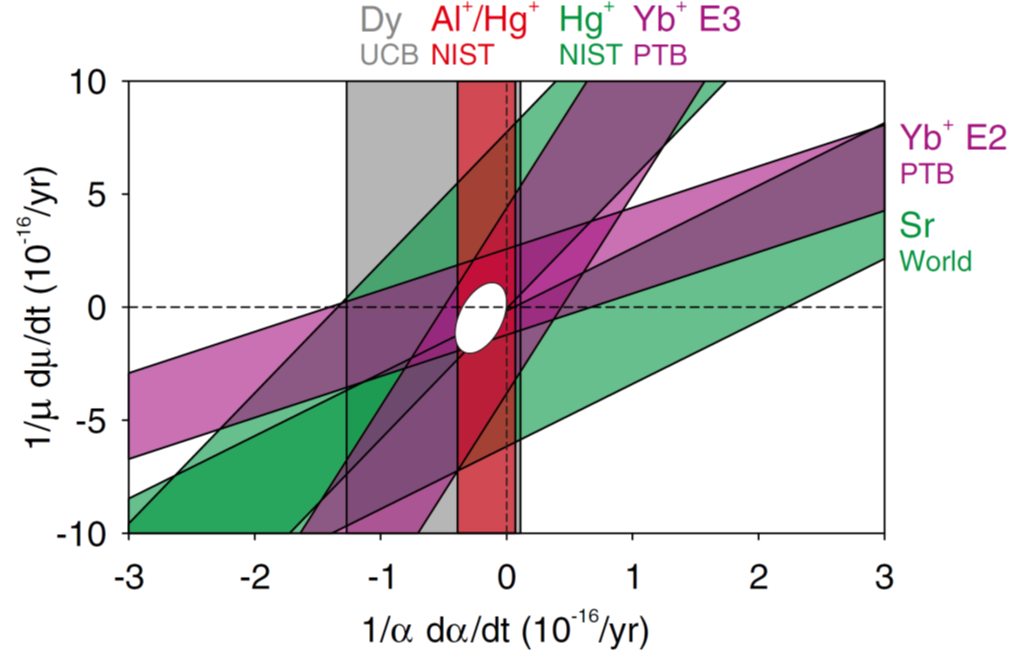}
            \caption{(color online). Constraints on temporal variations of $\alpha$
and $\mu$ from comparisons of atomic transition frequencies. Filled stripes mark the one standard deviation $\sigma$
uncertainty regions of individual measurements and the central
blank region is bounded by the standard uncertainty ellipse
resulting from the combination of all data. From \citet{HunLipTam14}.}
\label{fig1-FC}
\end{figure}

\subsection{Current limits on $\alpha$ and $\mu$ variations from atomic clocks and Dy spectroscopy}
\label{FC-clocks}

At present, the best constraints on temporal variations of $\alpha$
and $\mu$ from comparisons of atomic transition frequencies are due to combination of several
experiments tracking ratios of different clock transitions  \cite{HunLipTam14,GodNisJon14}.
The analysis of current $\alpha$
and $\mu$ clock constraints of \citet{HunLipTam14} is illustrated  in  Fig.~\ref{fig1-FC}.
 Filled stripes mark one-standard-deviation
uncertainty regions of individual measurements and the central
blank region is bounded by the standard uncertainty ellipse
resulting from the combination of all data.
The results of the experiments measuring the stability of the
 frequency ratios  $R=\nu/\nu_{\textrm{Cs}}$ of optical Hg$^+$  \cite{ForAshBer07}, Yb$^+$  quadrupole (E2) \cite{TamHunLip14},
  Yb$^+$ octupole (E3) \cite{HunLipTam14}, and Sr \cite{TagLorCoq13} clocks  to the Cs microwave clock plotted in Fig.~\ref{fig1-FC} were
parameterized by
\begin{equation}
\frac{1}{R} \frac{dR}{dt} = (K-K_{\textrm{Cs}}-2) \frac{1}{\alpha} \frac{d\alpha}{dt} + \frac{1}{\mu} \frac{d\mu}{dt} - \kappa_{\textrm{Cs}}
 \frac{1}{X_q}\frac{dX_q}{dt},
\label{eq5-FC}
\end{equation}
\noindent where the coefficients $K$ for the optical clocks and Cs are listed in Table~\ref{tab1-FC}. We note that the extra  ``2''
  in the parenthesis of the first term
appears due to the presence of a factor of $\alpha^2$ in the  hyperfine frequency expression given by Eq.~(\ref{eq2-FC}).

The contribution due to the third term in Eq.~(\ref{eq5-FC}) was taken to be zero in the analysis of \citet{GodNisJon14}.
\citet{HunLipTam14} accounted for this term by  using the result
\begin{equation}
\kappa_{\textrm{Cs}}  \frac{1}{X_q}\frac{dX_q}{dt} = 0.14(9) \times 10^{-16}/\textrm{year}
\end{equation}
inferred from the comparison of $^{87}$Rb and $^{133}$Cs clocks over 14 years reported by \citet{GueAbgRov12}.

Figure~\ref{fig1-FC} also includes constraints on temporal variation of $\alpha$  from comparisons of transition frequencies
 of Al$^+$ (aluminum, $Z = 13$) and Hg$^+$ (mercury, $Z = 80$) optical clocks \cite{RosHumSch08}
and from the measurement of Dy transition frequencies \cite{LeeWebCin13}.
 The  Al$^+$/Hg$^+$ optical clock comparison \cite{RosHumSch08} currently provides the most accurate single
test of only $\alpha$-variation, setting the limit
\begin{equation}
\frac{\dot{\alpha}}{\alpha}=(-1.6\pm2.3) \times 10^{-17} \, \textrm{yr}^{-1}.
\label{eq6-FC}
\end{equation}

The Dy limit
on $\alpha$-variation comes from spectroscopy of radio-frequency transitions between nearly degenerate, opposite-parity
excited states rather than from an
atomic clock comparison. These states are
sensitive to variation of $\alpha$ due to large relativistic corrections of opposite sign
for the opposite-parity levels. The near degeneracy reduces the relative precision needed to place strict
constraints on $\alpha$-variation. We note that filled stripes representing results of both Al$^+$/Hg$^+$  and dysprosium (Dy, $Z = 66$) experiments in Fig.~\ref{fig1-FC}
are vertical, since they are sensitive only to variation of $\alpha$ and not $m_\mr{p}/m_\mr{e}$.

We emphasize that Yb$^+$ (ytterbium, $Z=70$) has two ultranarrow optical clock transitions at 467~nm and 436~nm:
electric octupole ($E3$) $4f^{14}6s \, \, ^2S_{1/2}- 4f^{13} 6s^2 \, ^2F_{7/2}$
and electric quadrupole  $4f^{14}6s \, \, ^2S_{1/2}- 4f^{14} 5d \,\, ^2D_{5/2}$. This is the only case among the clocks presently under
 development for which there is more than one  clock transition.

\begin{figure}[tbp]
            \includegraphics[scale=0.35]{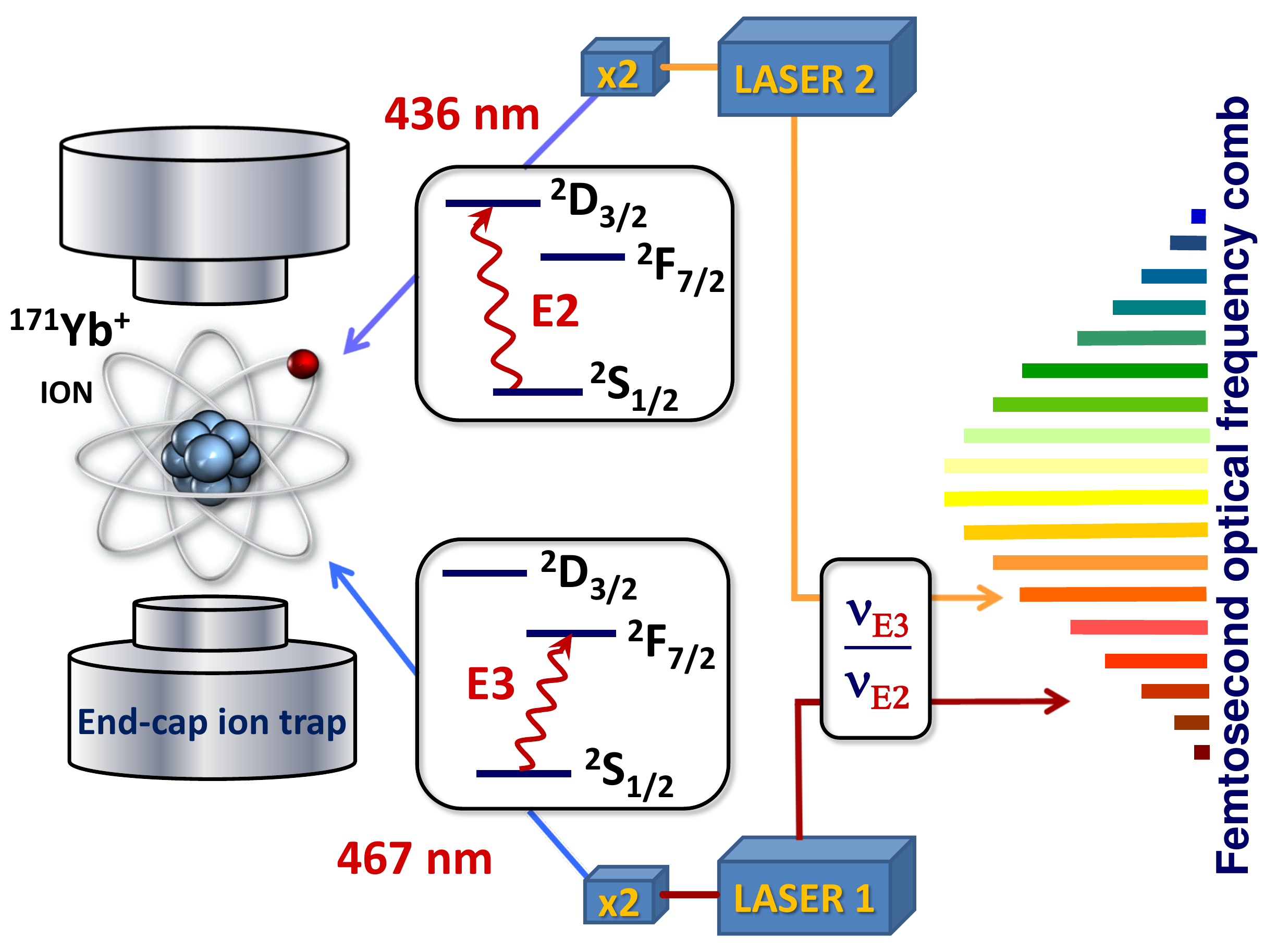}
            \caption{Schematic experimental arrangement for measuring the $E2$ and $E3$ clock frequencies of a single $^{171}$Yb$^+$ ion.
            The $E3/E2$ frequency ratio was determined by stabilizing one laser to the $E3$ transition and the other laser to the E2 transition
            and measuring the ratio between the laser frequencies with an optical frequency comb. (For experimental reasons, the researchers
             used infrared lasers that have to be frequency doubled to excite the $E2$ and $E3$ optical transitions. Adapted from \citet{Saf14}.}
\label{fig2-FC}
\end{figure}

The frequency ratio of those two transitions in Yb$^+$ was measured directly for the first time by \citet{GodNisJon14},
without reference to the Cs primary standard, and using the same single ion of $^{171}$Yb.
This measurement is illustrated in Fig.~\ref{fig2-FC}.  The $E3/E2$ frequency ratio was determined by stabilizing one laser to the
 $E3$ transition and the other laser to the $E2$ transition and measuring the ratio between the laser frequencies with an optical
 frequency comb. Both lasers were
simultaneously stabilized to their respective transitions in
the same ion ensuring  experimental simplicity and
common-mode rejection of certain systematic effects such
as the gravitational redshift and relativistic time dilation.
Such direct measurements of the ratio of the two optical frequencies are  free from the additional uncertainties introduced
by the primary Cs frequency standard.

Combining this measurement with constraints from previous experiments, \citet{GodNisJon14} set the following limits to the present day variation of
$\alpha$ and $\mu$:
\begin{eqnarray}
\frac{\dot{\alpha}}{\alpha}&=&(-0.7 \pm 2.1) \times 10^{-17}\,  \textrm{yr}^{-1}\\
\frac{\dot{\mu}}{\mu}&=&(0.2 \pm 1.1) \times 10^{-16} \, \textrm{yr}^{-1},
\end{eqnarray}
which are similar to limits set by the analysis of \citet{HunLipTam14}.

\subsection{Prospects for the improvement of atomic clock constraints on fundamental-constant variations}
\label{FC-F}
The limits on the variation of the fundamental constants from comparison of two clock frequencies are determined
 by (1) uncertainties of  both clocks, (2) sensitivity factors of each  clock to the variation of different constants,
 and (3) the time interval over which the ratios are repeatedly measured.
 Therefore,
 strategies to improve the limits set by atomic clocks on the variation of fundamental constants may arise from the improvement
 of any of the three factors: building clocks with lower uncertainties, building conceptually different clocks with higher
 sensitivities to variation of fundamental constants, and making measurements over longer time intervals.

For example, the  Al$^+$/Hg$^+$ clock constraint on $\alpha$-variation  reported in 2008 \cite{RosHumSch08} was obtained from
 repeated measurements during one year.
 Even with the same accuracy for both
 Al$^+$ and Hg$^+$ clocks,
 repeating the frequency-ratio measurements now  would improve the 2008 limit (\ref{eq6-FC}) by almost of factor of 10, since almost a decade
 has passed since the first measurements. For clock-ratio experiments that have already accumulated more than a decade of data, such as the Cs/Rb
 ratio  \cite{GueAbgRov12}, only moderate  improvements can be achieved in the next decade without the reduction of clock uncertainties.
 We start with a discussion of the prospects for further improvements in searches for variation of fundamental constants with
 current clocks and then explore new clock proposals.
  \subsubsection{Improvements of current clocks}

\begin{figure}[tbp]
            \includegraphics[scale=0.1]{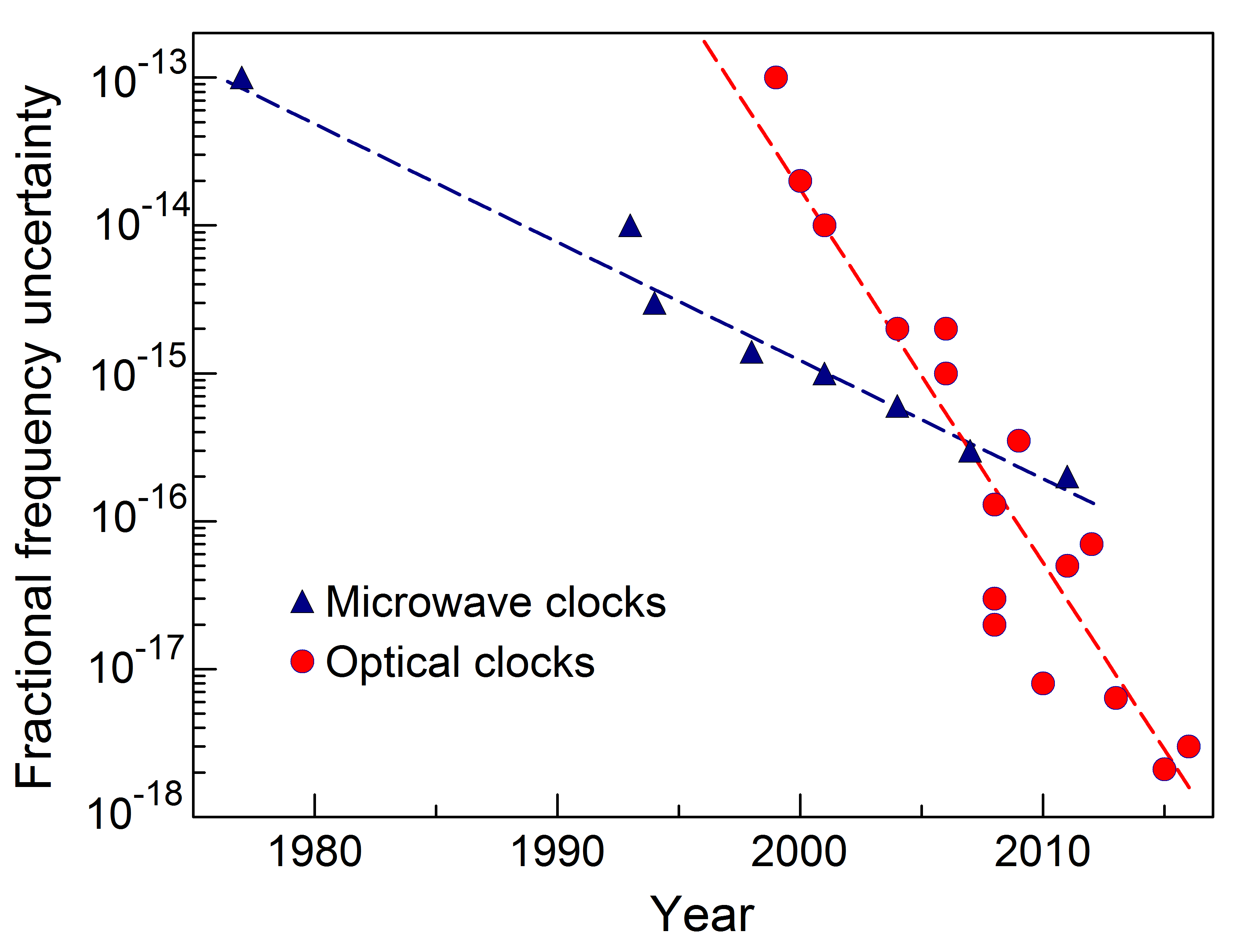}
            \caption{Evolution of fractional frequency uncertainties of atomic frequency standards based
on microwave (Cs clocks) and optical transitions. Data points are from \citet{PolOatGill13,NicCamHut15,HunSanLip16}.}
\label{fig3-FC}
\end{figure}

Figure~\ref{fig3-FC} illustrates the  evolution of fractional frequency uncertainties of atomic frequency standards based
on microwave and optical transitions. All microwave data in this figure come from Cs clocks. The figure is adapted from
\citet{PolOatGill13} with addition of  recent data up to 2016.

 The present-day  state-of-the-art Cs microwave clocks are approaching uncertainties of $10^{-16}$ \cite{HeaDonLev14,LevCalCal14},
 which is near their practical limitations.
 This is a remarkable achievement considering that the
Cs atomic clock transition  has an intrinsic quality factor $Q$, defined as the ratio of the absolute frequency of the transition to its natural
the linewidth, of $Q\approx 10^{10}$. The $Q$ factors of optical atomic clocks are five orders of magnitude higher than those of  microwave clocks,
giving optical clocks a tremendous advantage in terms of frequency stability \cite{PolOatGill13,LudBoyYe15}.
Recent progress in the accuracy of the optical clocks has been extraordinary, with
the world's best optical lattice atomic clocks
  approaching fractional frequency uncertainties  of
10$^{-18}$~\cite{NicCamHut15,UshTakDas15}. The smallest
uncertainty attained to date is $2\times10^{-18}$ in a strontium (Sr, $Z=38$) optical lattice clock \cite{NicCamHut15}. In 2016, a
systematic uncertainty of $3 \times 10^{-18}$  was reported in a single-ion atomic clock based on the electric-octupole transition
  in Yb$^+$ \cite{HunSanLip16}.
 As a result, the most rapid improvement in this field is expected to come from optical to optical clock comparison,
 with optical to microwave comparison being limited by the ultimate accuracy of microwave clocks.

Results of experiments measuring the stability of two optical
clock-frequency ratios  $R=\nu_2/\nu_{1}$ are parameterised by a simpler version of Eq.~(\ref{eq5-FC}):
\begin{equation}
\frac{\dot{R}}{R} = (K_2-K_1) \frac{\dot{\alpha}}{\alpha},
\label{eq81-FC}
\end{equation}
where $K_1$ and $K_2$ are $K$ sensitivity coefficients for clocks 1 and 2.
 Therefore, the sensitivity of optical clock frequency ratios to $\alpha$-variation is described by the difference in the corresponding
$K$ values, i.e. $\Delta K=|K_2-K_1|$.
 The  $K$ factors are small ($0.008-1.0$, see Table~\ref{tab1-FC})  for most
 clocks currently in development: Mg, Al$^+$, Ca$^+$, Sr$^+$,
 Sr, Yb, Yb$^+$ quadrupole transition, and Hg.
  The $K$ factors for Hg$^+$ and Yb$^+$ octupole clock transitions
are $-3$ and $-6$, making them the best candidates for one member of
a clock-comparison pair, with the  other member taken from the
previous group. Recently reported drastic reductions in the fractional frequency uncertainty of the Yb$^+$
octupole clock \cite{HunSanLip16} are expected to lead to a more accurate test of $\alpha$-variation, with the second clock being, perhaps, Sr.
Future prospects for development of optical atomic clocks are discussed in recent reviews \cite{PolOatGill13,LudBoyYe15},
which envisage further decreases in
atomic clock uncertainties  during the next decade.
Comparison of different clocks frequencies beyond 10$^{-18}$ accuracy will become  more challenging
 due to the sensitivity to the environment,
including temperature and gravitational
potential ~\cite{LudBoyYe15}. For example, a clock on the surface of
the Earth that is higher by just 1~cm than
another identical clock runs faster by $\delta \nu/ \nu_0\approx
10^{-18}$ ~\cite{LudBoyYe15}.  The blackbody radiation (BBR) shift has a leading temperature dependence of $T^4$, making clock frequencies sensitive to the temperature fluctuations. The BBR shift for a given temperature also varies significantly, by orders of magnitude, between different clock transitions.   The strategies for reducing BBR shifts in current clocks are discussed by \citet{LudBoyYe15}
and references therein.
Comparisons of clocks based on two transitions in a single ion, such as Yb$^+$
 quadrupole to octupole clock comparisons
illustrated by Fig.~\ref{fig2-FC} or with two ions held in the same trap, may be used to  reduce the environmental sensitivities of
the clock ratios.

\subsubsection{Prospects for optical clocks with highly charged ions}
\label{FC-new}
Another pathway toward improved tests of $\alpha$-variation with atomic clocks  is the development of frequency standards
based on new systems, which have higher $K$ sensitivities,  while still enabling highly accurate measurements of the frequency ratios.
Put simply,  it is much easier to measure large effects, so the search for high-sensitivity systems
is a major ongoing effort of  AMO theory.

This brings us to a question: what are the requirements for such new systems? If we would like to build a  clock with accuracy at the present state-of-the-art level, we need,
at the very least, a
system with a transition in a laser accessible range with very high $Q$, at least $Q\approx10^{15}$.
The high-$Q$ requirement means that the upper state of the transition is metastable, i.e. long-lived.  There are a number of other
 considerations to ensure small
frequency uncertainties by minimizing  systematic effects. The system also has to be amenable to cooling and trapping.
If we want to use our new clock to search for $\alpha$-variation, the clock transition has to be between states of different
electronic configurations, i.e. not between fine- or hyperfine-structure levels since the $K$ factors for such states are similar.

 These  requirements were formulated in
 the criteria for good clock transitions   proposed by \citet{DzuFlaKat15}:
\begin{itemize}
\item The transition  is in the near-optical region ($230 \ \rm{nm} < \lambda < 2000 \ \rm{nm}$
  or $5000 \ \rm{cm}^{-1} < 1/\lambda < 43000 \ \rm{cm}^{-1}$) as such transitions are accessible with available laser systems.
\item The lifetime of the clock state is  between 100 and $\approx10^4$ seconds as this enables high $Q$.
\item There are other relatively strong optical transitions in the same atomic system with
 a  lifetime of the upper level on the order of  $\tau \lesssim 1$ ms, which may be useful for laser cooling or optical pumping/probing.

\item The clock transition is not sensitive to perturbations caused by blackbody radiation, gradients of external electric fields, etc.
\end{itemize}
The first requirement seems to limit the potential systems to neutral atoms or singly charged ions, all of which have been considered
as potential clock systems. Examination of the National Institute of Standards and Technology (NIST) atomic spectra database \cite{NIST} establishes that the energies of the  relevant ion transitions involving
the ground states tend to be  outside of the laser-accessible range with the degree of ionization
exceeding two. Remarkably, selected highly charged ions with degrees  of ionization ranging from 9 to 18, actually have potential clock
 transitions in the optical range between different electronic configurations, as was discovered by \citet{BerDzuFla10}.
 This phenomenon arises from the rearrangement of the
order of electronic configurations: as more electrons are  removed, the order of levels becomes more hydrogenic, for example,
restoring the Coulomb ordering where the $4f$ shell becomes occupied prior to the  $5s$ shell. For example, the ground
state of cadmium (Cd, $Z=48$) is [Kr]$4d^{10}5s^2$. Proceeding along the Cd isoelectronic sequence, the ground state remains in this configuration
 up to Nd$^{12+}$, but the ground-state configuration of Sm$^{14+}$ (samarium, $Z=62$) becomes [Kr]$4d^{10}4f^2$ \cite{SafDzuFla14b}.
As a result, different electronic configurations move close together for two or three ions in an  isoelectronic sequence when the
order of levels is rearranged. An example is the Nd$^{12+}$  (neodymium, $Z=60$), Pm$^{13+}$ (promethium, $Z=61$), Sm$^{14+}$ (samarium, $Z=62$) part of the Cd-like sequence. This  provides an unexpected
gift of optical transitions for metrology applications.  Extensive theoretical effort during the past five years resulted in the
 identification of many such candidate systems in highly charged ions (HCIs), predictions of their properties,  and assessments
 of their potential for tests of  $\alpha$-variation
\cite{BerDzuFla11,BerDzuFla11a,BerDzuFla12,BerDzuFla12a,DerDzuFla12,DzuDerFla12,DzuDerFla12,DzuDerFla13E,KozDzuFla13,DzuFlaKat15,YudTaiDer14}.
The most accurate calculations were done using a state-of-the-art
hybrid approach that combines coupled-cluster and configuration
interaction methods \cite{SafDzuFla14,SafDzuFla14a,SafDzuFla14b,DzuSafSaf15}.
   Proposals for $\alpha$-variation searches in HCIs were reviewed by \citet{BerFlaOng13} and \citet{OngBerFla14}.

A particular attraction of the HCIs for constructing highly accurate clocks is the suppression of the   clock-frequency shifts
due to external electric fields which can lead to systematic errors, due to the contraction of  the electron cloud with increasing ionization stage.
Stronger relativistic effects resulting from  localization of the electron cloud  also provide enhanced sensitivity to $\alpha$-variation.
Assessments of systematic
effects in optical clocks based on HCIs concluded that an uncertainty of
10$^{-19}$ is achievable
\cite{DerDzuFla12,DzuDerFla12,DzuDerFla13E,DzuFlaKat15}.

Up to this point we have not discussed the technical feasibility of using HCIs to build clocks. Until very recently, the realm of  HCI
 research had little overlap with field of  ultracold precision metrology. In 2015, a breakthrough experiment \cite{SchVerSch15}
 demonstrated  sympathetic cooling of Ar$^{13+}$ (argon, $Z=18$) ions with laser-cooled Be$^+$ (beryllium, $Z=4$) ions in a cryogenic Paul trap.
This result  removes a major obstacle for HCI investigations with high-precision laser spectroscopy,
paving the way toward future experiments with cold and precisely controlled HCIs. Experimental work toward this goal is underway,
 starting with the
identification of the HCI spectra of interest to $\alpha$-variation studies \cite{WinCreBek15}.
Optical transitions in HCIs and their applications will be reviewed in detail in a separate Reviews of Modern Physics article.

Hyperfine transitions of hydrogen-like HCIs, such as  $^{207}$Pb$^{81+}$ (lead, $Z=82$), have also been proposed for tests of
 fundamental constant variation \cite{Sch07}. Due to high degree of ionization, the ground-state hyperfine transition wavelength
  is in the infrared, with a $Q$-factor of about $10^{14}$. The importance of such HCI transitions is their sensitivity to the
  variation of $\mu$ and $g$-factors, and $Q$ factors that are  much larger than those of Cs and Rb hyperfine transitions.

\subsubsection{A candidate nuclear clock}
With atoms and ions of the periodic table now considered, we turn our attention to  nuclei. Can we build clocks based on transitions
between different states of a nucleus? A great attraction of such an idea is the suppression of the field-induced frequency shifts since the nucleus
is highly isolated from the environment due to the electron cloud and
interacts only via the relatively small nuclear moments \cite{YamKolKas15}. There is a vast catalog of nuclear energy levels \cite{FirShi98}, but their transition frequencies are
higher by factors of $10^4-10^6$ than those accessible by modern laser technologies. Only one sufficiently long-lived nuclear transition,
 between the ground state of $^{229}$Th (thorium, $Z=90$)
and a low-lying  isomer (i.e. long-lived excited nuclear state), has a suitable wavelength, predicted to be 160(10)~nm \cite{BecBecBei07,BecWuBei09}.
This transition was proposed for application in a nuclear clock \cite{PeiTam03,CamRadKuz12}, but a decade of searches did not result in its detection
\cite{PelOkh15,YamKolKas15,JeeSchSul15}. Finally, in 2016, the existence of the isomer was confirmed \cite{WenSeiLaa16}, although there remains a significant uncertainty in its energy, motivating continued searches.  The measurement of the internal-conversion decay half-life of neutral $^{229\rm{m}}$Th was reported by \citet{SeiWenThi17}.
The hyperfine structure of
$^{229\rm{m}}$Th$^{2+}$ was investigated by \citet{ThiOkhGlo17} using the laser spectroscopy, yielding values of the magnetic dipole
and electric quadrupole moments as well as the nuclear charge radius.

 \citet{Fla06a} estimated that the relative effects of the variation of $\alpha$  and $m_q/\Lambda_{\textrm{QCD}}$  in this $^{229}$Th
 nuclear transition are
 enhanced by $5-6$  orders of magnitude using the Walecka model of nuclear forces and other assumptions. Other nuclear calculations
 predicted no enhancement \cite{HayFriMol08}, but their uncertainly was also very large, with a $4\times10^3$ enhancement  factor
 still being within their uncertainty limit, [see \citet{BerDzuFla09} for a discussion].
 With nuclear calculations  currently being unable to determine the sensitivity factor,  an alternative method for extracting
sensitivity to $\alpha$ variation using laboratory measurements
of the change in nuclear mean-square charge radius and electric-quadrupole moment between
the isomer and the ground-state nucleus
was proposed by \citet{BerDzuFla09}. The first experimental results were reported
by \citet{ThiOkhGlo17} but the precision of the electric-quadrupole moment  was not sufficient to extract the  sensitivity of the nuclear clock to $\alpha$-variation.

\subsection{Laboratory searches for variation of fundamental constants with molecules}
\label{FC-mol}

Molecular spectroscopy provides further possibilities for testing the stability of fundamental constants owing to rich spectra with many different transition types.
The proton-to-electron mass ratio $\mu$  defines the scales of electronic, vibrational, and rotational
intervals in molecular spectra:
\begin{equation}
E_\mathrm{el} : E_\mathrm{vib} : E_\mathrm{rot} \sim 1 : \mu^{-1/2} :\mu^{-1}.
\label{eq10-FC}
\end{equation}
 Purely vibrational and rotational transitions will
have the $K_{\mu} = -1/2$ and $K_{\mu} = -1$ sensitivity  factors to variation of $\mu$, respectively.
Moreover, molecules have fine
and hyperfine structures, $\Lambda$-doubling, hindered rotation, etc., which  adds a variety of dependences
on the fundamental constants \cite{ChiFlaKoz09}.

The first experimental comparison of a molecular clock to an
atomic clock \cite{SheButCha08} was obtained by comparing the frequency of a rovibrational
transition in SF$_6$ with the  hyperfine transition of the Cs clock.
The measured rovibrational transition frequency  in SF$_6$ depends only on $\mu$ (Rydberg constant cancels out when comparing to any other transition) :
$$
\nu_{\mathrm{SF_6}}=A \left(\frac{m_\mr{e}}{m_\mr{p}}\right)^{1/2} R_{\infty},
$$
where $A$ is a numerical factor. The resulting constraint
on the fractional temporal variation of the proton-to-electron mass ratio
was reported as $\dot{\mu}/\mu=(-3.8\pm 5.6)\times 10^{-14}$~yr$^{-1}$.
While this limit is less stringent than that set by optical clocks \cite{HunLipTam14,GodNisJon14}, this study offers a clean separation
 of  $\mu$-variation from $\alpha$-variation.

Proposals for future tests of variation of fundamental constants with ultracold molecules were reviewed by \citet{ChiFlaKoz09} and
we provide only a brief summary here. These proposals are based on the enhanced sensitivities to $\alpha$, $\mu$, and
$m_q/\Lambda_{\mathrm{QCD}}$ for accidentally closely spaced levels.
We note that there is a difference between proposals with enhanced relative sensitivities and those with enhanced absolute sensitivities. The relative-sensitivity proposals, for example \cite{Fla06}, are practical for cases where the frequency uncertainty scales with the frequency, such as Doppler shifts in astrophysical measurements. Most of the laboratory measurement are limited by absolute frequency uncertainties, so transitions with large overall shifts may be better candidates, for example \citet{DemSaiSag08,ZelKotYe08,KajGopAbe14,HanCarLan16}. In special cases, there are both absolute and relative enhancements to $\mu$-variation \cite{DemSaiSag08,HanCarLan16}.

The relative effect of  $\alpha$-variation  in microwave transitions between very close and narrow rotational-hyperfine levels may
 be enhanced by 2-3 orders of magnitude in diatomic molecules with unpaired electrons like LaS, LaO, LuS, LuO, YbF, and similar
  molecular ions due to accidental degeneracies of  hyperfine and rotational structures \cite{Fla06}.
Degeneracies between  the fine and vibrational structures within the electronic ground
states of diatomic molecules, such as Cl$^+_2$, CuS, IrC, SiBr, and HfF$^+$, lead to enhanced sensitivities to the variation of
 both $\alpha$ and $\mu$
 \cite{FlaKoz07,BelBorSch10}.
  Strong enhancements of $\alpha$- and $\mu$-variation effects in
dihalogens and hydrogen halides,  HBr$^+$, HI$^+$, Br$_2^+$, I$_2^+$, IBr$^+$, ICl$^+$, and IF$^+$, were reported by \citet{PavBorFla15}.
The calculation of \citet{FlaStaKoz13} demonstrated enhanced sensitivity to the variation $\alpha$ and $m_q/\Lambda_{\mathrm{QCD}}$
 in opposite-parity closely spaced levels of the $^{207}$Pb$^{19}$F molecule due to a near cancellation of the omega-type doubling and
 magnetic hyperfine-interaction-energy shifts.

Experiments with  cold diatomic molecules Cs$_{2}$ \cite{DemSaiSag08} and  Sr$_{2}$ \cite{ZelKotYe08} have also been proposed.
 \citet{DemSaiSag08} predicted that the splitting between
pairs of Cs$_2$ nearly-degenerate
vibrational levels, where each state is associated with a different electronic potential, could be measured precisely enough
to sense a fractional change of $\delta\mu/\mu \lesssim10^{-17}$.
Detailed spectroscopy of the Cs$_2$  $a^3\Sigma_u^+$ state was performed by \citet{SaiSagTie12}, who further discussed the
prospects for $\mu$-variation measurements. Coherent control of molecular quantum states, which is a prerequisite for a
``molecular lattice clock'', was
achieved for  Sr$_2$ \cite{MucGMcDIwa15}.
Searches for $\mu$-variation might  be made using vibrational transitions in diatomic alkali-alkaline-earth molecules and
alkaline-earth hydride molecular ions
\cite{KajGopAbe14}.

Several high-sensitivity transitions with narrow linewidths were identified in the deeply
bound O$_2^+$ molecular ion \cite{HanCarLan16}. The authors suggested the experimentally feasible routes toward the $\mu$-variation measurements in this system.
Another method to search for the $\mu$-variation using vibrational transitions in O$_2^+$ with high accuracy was proposed by \citet{Kaj17,Kaj17a}.
\citet{KajGopAbe14a} proposed the search for $\mu$-variation using vibrational transitions in N$_2^+$.

The leading systematic effects for realization of optical clocks with
rovibrational levels of the molecular  ions
H$_2^+$ and HD$^+$ were   assessed by \citet{SchBakKor14} and \citet{Kar14},
who also discussed their  potential sensitivity to $\mu$-variation.
The principle issues limiting the accuracy of such clocks involved effects due to light shifts, though it is possible these could be suppressed with appropriate pulse sequences~\cite{YudTaiOat10,HunLipTam14}.
 Ramsey-type spectroscopy in a beam of metastable
 CO molecules was reported by \citet{NijUbaBet14} for further tests of variation of $\mu$.

\citet{SanDiSRic14} discussed the
design of an experiment aimed to constrain the fractional temporal variation of $\mu$ at a level of $10^{-15}$/yr using
 spectroscopic frequency measurement on a beam of cold CF$_3$H molecules. Progress toward precision spectroscopic
  measurement with ultra-cold deuterated ammonia, ND$_3$, for future laboratory tests for variation of $\mu$ was reported
  in a paper by \citet{QuiWallHoe14} and references therein. Prospects for high-resolution microwave spectroscopy of methanol,
   CH$_3$OH, and CD$_3$OH molecules in a Stark-deflected molecular beam
were discussed by \citet{JanKleMen13}, but the precision must be significantly enhanced for laboratory tests. A current goal of methanol
studies is to improve precision to reference the astrophysical searches of $\mu$-variation described in Sec.~\ref{FC-mu}.

An alternative proposal to test variation of fundamental constants with atoms and molecules involves precise measurements of the
scattering lengths in Bose-Einstein condensate and Feshbach molecular experiments \cite{ChiFla06,GacCot14}.
A measurement of the scattering length accurate to $10^{-6}$ performed near narrow Feshbach resonances in two consecutive years was estimated to
probe the variation of $\mu$ at the level of $10^{-15}$~yr$^{-1}$ - $10^{-18}$~yr$^{-1}$ depending on the choice of atomic species \cite{ChiFlaKoz09}.

Recent advances in cooling and control of molecules \cite{Kobayashi2015,Park2017,ChoChrHum17,GerTonWil14,WolWanHei16,Kozyryev2017b,Norrgard2016,Truppe2017,Prehn2016,Cheng2016,WuGanKol17,Hutzler2012}  promise future  progress in laboratory tests of variation of fundamental constants with molecules.

\subsection{Limits on variation of $\alpha$ and $\mu$ from quasar absorption spectra}
\label{quasar}

The discussion of Secs.~\ref{FC-D} - \ref{FC-mol} concerns with a question: \textit{Do fundamental constants vary now? }
The dependence of atomic and molecular spectra  on fundamental constants may also be used to probe for their variation  in a distant past, as far back as
$\approx 10$ billion years ago, the scale given by the age of the Universe. The basic idea is the same: to compare the spectra from two different times, but to increase the time separation $\delta t$
from $\delta t=(1-15)$~years of the laboratory tests to $\delta t=(3-13)$~billions of years. In practice, we need a particularly bright, distant astrophysical light source, such as a quasar, to serve as a backlight of
  high-redshift gas clouds in which  atomic or molecular absorption spectra can be observed. Emission spectra are also used in
  some studies. The sensitivities of those spectra to the variations
  of $\alpha$ and $\mu$ are defined and calculated in the same way as for the terrestrial experiments.

Due to the expansion of the Universe, all wavelengths of light $\lambda$ from the Universe's past  are redshifted. A cosmological redshift $z$ is
 defined as the ratio
\begin{equation}
z=\frac{\lambda_{\mathrm{lab}}-\lambda}{\lambda},
\end{equation}
where $\lambda$ is the wavelength of the absorbed/emitted light and $\lambda_{\mathrm{lab}}$ is the wavelength of the light which is observed on Earth.
A redshift of $z=1$  means that a 500 nm~absorption wavelength is observed on Earth as 1000~nm instead. This corresponds to a
``look back'' time of $\approx8$ billion years \cite{Pil13}.

To separate the redshift, one needs to compare ``ancient'' and present  wavelengths of  at least two spectral lines that have
 different sensitivities to the constants of interest.

\citet{Uza11} provided a detailed review of atomic and molecular quasar absorption studies, so we will provide only
 key points and more recent results here.

\subsubsection{Limits on variation of $\alpha$ from quasar absorption studies of atomic spectra}
\label{FC-alpha}
Quasar absorption studies of $\alpha$-variation use alkali-doublet \cite{MurWebFla01},
 many-multiplet \cite{WebFlaChu99},
 and single-ion differential $\alpha$-measurement
\cite{LevCenMol06}
 methods.  The alkali-doublet method uses the $ns-np_{1/2}$, $ns-np_{3/2}$ fine-structure intervals of alkali-metal atoms as a probe of
 $\alpha$-variation. The many-multiplet method is a  generalization of this approach  which uses many atomic transitions with
 different dependences on $\alpha$, and yields more accurate results than the alkali-doublet method. The single-ion differential
  $\alpha$-measurement (SIDAM) method uses different transitions of one ionic species in an individual exposure, in an attempt
  to reduce some of the systematics of the many-multiplet method. It is mainly used with Fe$^+$ (iron, $Z=26$) which
has several transitions with both positive and negative $K$, allowing one to compare lines within a single spectrum.
Most of the quasar-absorption studies with atoms are based on strong UV lines redshifted into the visible spectrum range. Unfortunately,
these transitions depend weakly  on $\alpha$
for most atoms visible from these sources, since the atoms are relatively light, $Z\leq 30$, which generally leads to smaller values of $K$.  For example, the maximum $\Delta K$ difference for any two lines of Fe$^+$,
is $\Delta K=0.11$, with an estimated 30~\% uncertainty
\cite{PorKosTup07}. Another difficulty of the many-multiplet method is ensuring that one compares transition lines from the
same object, i.e. at the same redshift $z$. The advantage of SIDAM is using lines of the same element, eliminating or simplifying this issue.
Another significant systematic arises from the assumption of the
isotopic-abundance ratios for each atom or ion used for the analysis
in the distant past, in particular, the $^{25,26}$Mg/$^{24}$Mg (magnesium, $Z=12$) ratios,
and their possible deviations from the  terrestrial values. This issue, discussed by \citet{KozKorBer04}, is further complicated by the lack of isotope-shift measurements for a number for
transitions that are used in the quasar absorption studies \cite{BerDzuFla11d}.

A large-scale many-multiplet  analysis of  the Keck telescope
high-resolution Echelle-spectrometer (HIRES) data from 143 absorption
systems at  $0.2 < z < 4.2$, indicated a variation of $\alpha$
\cite{MurFlaWeb04}:
\begin{equation}
\frac{\Delta \alpha}{\alpha} = \frac{\alpha_{\mathrm{obs}}-\alpha_{\mathrm{lab}}}{\alpha_{\mathrm{lab}}}  =(-0.57 \pm 0.11) \times 10^{-5},
\label{eq8-FC}
\end{equation}
where $\alpha_{\mathrm{obs}}$ corresponds to a value of $\alpha$ in the distant past, between 2 and 12.4 gigayears  here,
and  the $\alpha_{\mathrm{lab}}$ is the current terrestrial value.

  However, the analysis of data from 23 absorption systems taken by the Very Large Telescope (VLT) ultraviolet and visual Echelle spectrograph (UVES)
  yielded a null result
  \begin{equation}
\frac{\Delta \alpha}{\alpha} = (-0.06 \pm 0.06) \times 10^{-5}, \hspace{1cm}
\end{equation}
for $0.4 < z < 2.3$
\cite{SriChaPet04,ChaSriPet04}.
This analysis was disputed by \citet{MurWebFla07,MurWebFla08,MurWebFla08a}, who obtained different  results from an analysis of the
 same data; this was followed by the reply of \citet{SriChaPet07}.

An intriguing solution to this discrepancy  was suggested by \citet{WebKinMur11}: since Keck and VLT data come from different
hemispheres, both results can be made consistent by introducing a dipole spatial
variation of $\alpha$; this topic is discussed further in Sec.~\ref{spatial}.
The
VLT data were  reanalysed in the more recent work by \citet{WilWebKin15}.
Considering both statistical and
systematic error contributions, \citet{WilWebKin15} obtained $\delta \alpha/\alpha = (0.22 \pm 0.23)\times 10^{-5}$,
consistent with the dipole spatial variation limits introduced by \citet{WebKinMur11}.

 An impact of instrumental systematic errors on $\alpha$-variation results obtained from atomic quasar-absorption data was recently studied by   \citet{WhiMur15}
  using 20 years of archival spectra from  VLT and Keck spectrographs. \citet{WhiMur15} concluded
that systematic errors in their wavelength scales were substantial and capable of significantly weakening the
evidence for variations of $\alpha$ from quasar absorption lines. However, they  still can not entirely explain the Keck/HIRES result (\ref{eq8-FC}).

    To summarize, atomic quasar-absorption data remains a subject of open controversy which requires further study and
    future deployment of high-resolution ultrastable
spectrographs like ESPRESSO (for the
VLT) and ELT-HIRES \cite{Mar15} for improved astrophysical measurements.
   Laser frequency-comb techniques for precise astronomical spectroscopy were described by \citet{MurLocLi12}.

 \subsubsection{Limits on variation of $\mu$ from quasar absorption studies of molecular spectra}
 \label{FC-mu}
Molecular  spectra provide clean  constraints on $\mu$-variation since rotational and vibration transitions have different
$\mu$-dependences given by
Eq.~(\ref{eq10-FC}).
 There are two main considerations when selecting molecules for astrophysical studies of $\mu$-variation:
\begin{itemize}
\item How sensitive are the molecular transitions to variation of $\mu$? This is quantified  with a dimensionless sensitivity
factor $K_\mu$, analogous to the $K$
factor for sensitivity to $\alpha$-variation.
\item How abundant is this molecule in the Universe? A high sensitivity factor would be good for laboratory tests of Sec.~\ref{FC-mol},
but useless for astrophysical studies if it is impossible to observe the corresponding transitions.
\end{itemize}
It is particularly advantageous if a molecule has several transitions with different $K_{\mu}$, preferably of opposite sign.
Then, transitions in the same molecule can be used for the astrophysical search for $\mu$-variation, reducing important systematic effects.

Until recently, the most accurate astrophysical limits on the variation of $\mu$
came from H$_2$ studies,  recently reviewed by \citet{UbaBagSal16}, since
H$_2$ is the most abundant molecule observed, with 56 absorption systems known at the present time.
A combined H$_2$ result from 10 systems with $2.0<z<4.2$ sets the limit on the variation of $\mu$ at
\begin{equation}
\left| \frac{\Delta \mu}{\mu}\right| = \left| \frac{\mu_{\mathrm{obs}}-\mu_{\mathrm{lab}}}{\mu_{\mathrm{lab}}} \right| \le 5\times 10^{-6}\, (3\sigma),
\end{equation}
where $\mu_{\mathrm{obs}}$ corresponds to the value of $\mu$ in the distant past, from 10 to 12.4 gigayears in this study,
and  $\mu_{\mathrm{lab}}$ is the current terrestrial value \cite{UbaBagSal16}.
These molecular-hydrogen studies use
the
 UV  transitions in  Lyman
and Werner bands that are redshifted into the visible spectrum.
The $B^1\Sigma^+_u - X^1 \Sigma^+_g$
Lyman and $C^1\Pi^+_u - X^1 \Sigma^+_g$
Werner band lines are strong dipole-allowed absorption lines of the H$2$
molecule with $\lambda$ = 910~\AA~ -- 1140~\AA.
 All of these transitions have weak dependence on $\mu$, with a maximum
 sensitivity coefficient $\Delta K_\mu \approx 0.06$ \cite{KozLev13}.

Improved limits on the variation of $\mu$ are obtained by going
from optical to microwave frequencies, where a number of molecular transitions are available with values of $K_\mu$ greater by factors of 100-1000.
The dependence of microwave and submillimeter molecular transitions  on fundamental constants was reviewed by \citet{KozLev13}.
The following molecules were considered: CH, OH, NH$^+$, C$_3$H, H$_3$O$^+$, NH$_3$ (ammonia), H$_2$O$_2$ (hydrogen peroxide),
CH$_3$OH (methanol),  and CH$_3$NH$_2$ (methylamine).
Nine diatomic and 16 polyatomic molecular candidates for $\mu$-variation studies were reviewed by \citet{JanBetUba14}.

In 2011, a number of polyatomic molecules, including methanol and  methylamine were observed for the first time at high redshift, $z=0.89$,
 corresponding to look-back time of $7.5\times 10^9$  years.
 The $K_\mu$ coefficient for ammonia is -4.5 \cite{KozLev13},
 which represents a two orders of magnitude enhancement in comparison with H$_2$.
  However, all of ammonia lines exhibit the same sensitivity, so  comparison with other systems
  is required. Two absorption systems are known with NH$_3$ lines, at $z=0.69$ and $z=0.89$.
  Studies of $\mu$-variation in these systems resulted in a $2\sigma$ limit  $|\Delta \mu/\mu|<1.8 \times 10^{-6}$ \cite{MurFlaMul08}
  and $3\sigma$ limit of $|\Delta \mu/\mu|<1.4 \times 10^{-6}$ \cite{HenMenMur09}, respectively.
A joint three-component fit to the NH$_3$, CS, and H$_2$CO lines
yielded $|\Delta \mu/\mu|<3.6 \times 10^{-6}$, for $z=0.69$ \cite{Kan11}.

The sensitivity coefficients in methanol transitions range from 17 to -43, potentially allowing for the maximum enhancement of
 $|\Delta K_\mu| \approx 60$ \cite{KozLev13}\footnote{We caution the reader that here $\mu=m_\mr{p}/m_\mr{e}$ but $\mu=m_\mr{e}/m_\mr{p}$ is frequently
 used in the literature, leading to opposite signs of the $K_\mu$ coefficients in different sources.}.  \citet{BagJanHen13}
  set the most stringent limits of past variation of $\mu$, $|\Delta \mu/\mu|<1 \times 10^{-7}$ at (1$\sigma$),  using four methanol
  lines at $z=0.89$. An extended study of $\mu$-variation based
on 17 measurements of ten different absorption lines of
methanol was carried out by \citet{BagDapJan13a},
allowing for
a quantitative analysis of previously
unaddressed underlying systematic effects  yielding
$\Delta\mu/\mu=(-1.0\pm0.8{_\mathrm{stat}} \pm 1.0_{\mathrm{sys}})\times 10^{-7}$.
 Assuming a linear variation of $\mu$ with time, this limit translates into
$
{\dot{\mu}}/{\mu}< 2 \times 10^{-17} \textrm{yr}^{-1}
$
which is  more constraining than the atomic clock limit \cite{HunLipTam14,GodNisJon14} associated with the same
linear model of fundamental constant variation.
 We note that there is no theoretical basis to assume the linear variation of fundamental constants. We make such comparison only as an illustration of the accuracies reached by the astrophysical and laboratory studies.

In 2015, one of the four methanol lines observed at $z=0.89$ and used in the analysis of this absorption system, was noted to have a
 different line profile: the  line full widths at half-maximum was larger,
 at $4.3\sigma$ significance, suggesting that the sightline in this
transition traces different absorbing gas from that detected in the
other three lines \cite{KanUbaMen15}. Therefore, it was recommended to exclude this line from the analysis, resulting in a 2$\sigma$  constraint of $|\Delta \mu/\mu|<4 \times 10^{-7}$.

Using combinations of atomic and molecular lines allows one to probe variation of various combinations of fundamental constants.
A comparison of the atomic hydrogen 21 cm hyperfine ground-state transition with atomic UV spectral lines
 \cite{TzaWebMur05,TzaMurWeb07,RahSriGup12} or OH molecular transitions \cite{CheKan03} constrains
combinations of $\alpha$, $\mu$, and the proton $g$-factor.

Comparing the 21 cm line to molecular rotational  transitions in CO, HCO$^+$, and HCN eliminates the dependence
on $\mu$, which is $1/\mu$ for both types of transition
 \cite{MurWebFla01a}.

 The combination $F=\alpha^2 \mu$ was probed with a C$^+$ and CO transitions
\cite{LevReiKoz08,LevComBoo12},
 thus eliminating the dependence on $g_p$ and yielding a constraint $\Delta F/F< 2 \times 10^{-5}$ at $z=5.2$.

In summary, currently the best astrophysical constraint on the $\mu$-variation for high redshifts,
up to $z=4.2$, (12.4~Gyr), come from the H$_2$ data \cite{UbaBagSal16}, while the strictest constraints for lower redshifts,
$z=0.89$, are obtained from the methanol data \cite{BagDapJan13a,KanUbaMen15}.
Further improvement may come from observation of ammonia, methanol and other more complicated molecules with high sensitivities to $\mu$-variation
at higher redshifts, increased
sensitivity and spectral resolution of astronomical
observations and
increased
precision of the laboratory measurements for the most sensitive molecular
transitions \cite{KozLev13}.

\subsection{Spatial variation of fundamental constants}
\label{spatial}
As discussed in Sec.~\ref{FC-theory},
if the fundamental constants  depend on some
dynamical scalar field $\phi$ they become dynamical.
A coupling of such scalar field $\phi$ to electromagnetic fields induces a coupling to matter
which may depend on the local
matter density. Such density-dependent couplings may lead to a spatial variation of fundamental constants:
 fundamental constants will be different  in the regions of high density of matter in comparison to regions of low density.
 However, such spatial variation at
 the cosmological scales is expected to be much smaller in most theories than
a temporal variation unless under extreme densities, such as in the vicinity of a neutron star \cite{Uza11}.
Therefore, the \citet{WebKinMur11} hypothesis of a dipole spatial variation of $\alpha$ introduced to explain the discrepancy
 between Keck and VLT
 data discussed in  Sec.~\ref{FC-alpha} was quite extraordinary.

The spatial variation  idea arises from the geographical positions of Keck and VLT telescopes in northern (Hawaii) and
 southern (Chile) hemispheres, respectively, separated by 45$^{\circ}$ in latitude. These two telescopes, on average,
 observe different directions in the sky and Keck and VLT $\alpha$-variation  results can be made consistent by introducing a spatial
variation of $\alpha$. The result of \citet{WebKinMur11} would mean that $\alpha$ was larger in
the past in one direction and smaller in the past in the
opposite direction according to
\begin{equation}
\frac{\Delta \alpha}{\alpha}=\frac{\alpha(\mathbf{r})-\alpha_0}{\alpha_0}=(1.10\pm 0.25) \times 10^{-6}\, r \,cos \psi \,\mathrm{Gly^{-1}},
\label{FC-eq15}
\end{equation}
where $(\alpha(\mathbf{r})-\alpha_0)/\alpha_0$ is a  variation
of $\alpha$ at a particular place $\mathbf{r}$ in the Universe relative
to $\alpha_0$ on Earth at $\mathbf{r}=0$. The function $r\, cos \psi$ describes the
geometry of the spatial variation, where  $\psi$ is the angle between
the direction of the measurement and the axis of the dipolar variation.
 The distance function $r$ is the light-travel
distance $r =ct$ measured in giga-lightyears.
The Keck/VLT data were further analyzed in terms of spatial variation of $\alpha$ by \citet{BerFlaKim11,BerKavFla12} and \citet{KinWebMur12}.

A subsequent analysis of the Keck and VLT systematic instrumental errors by \citet{WhiMur15} weakened but not completely eliminated such a scenario.
The extraordinary claim of the spatial variation of $\alpha$ will require future extraordinary evidence obtained with next-generation
 ultra-stable high-resolution spectrographs and a higher level of control of systematic errors.

Nevertheless, the subject of the spatial variation of fundamental constants is an interesting subject and various scenarios for new physics could exist
that may be tested with astrophysics and laboratory studies.
Regardless of validity of the \citet{WebKinMur11} result, we invite the reader to use it as an example to consider the following question:
\textit{What type of new physics can induce a spatial cosmological variation of fundamental constants and how can we test for it?}
\citet{BerFla12}, \citet{BerKavFla12}, and
\citet{OliPelUza11,OliPelPet12} considered three scenarios, described below.   \\

\noindent I.  Fundamental constants may fluctuate on a cosmological scale involving regions not in causal contact
 due to super-Hubble quantum fluctuations of a light field during inflation; further constraints from CMB are described by \citet{SigKurKam03}.\\

\noindent II. A background value of $\phi$ depends on position and time, i.e. there is a non-zero spatial gradient of the field $\nabla \phi \neq 0$.
It was pointed out by \citet{OliPelUza11} that the generalization of the Copernican principle that assumes a homogeneous Universe at large scales
 is not fully satisfied in such a model and its theoretical foundation is unclear.
Such a model will result in a dipole variation of the fundamental constants
in the general form of Eq.~(\ref{FC-eq15})
with the axis of the dipole being in  the direction of the gradient $\nabla \phi$. The spatial variation of fundamental constant $X$
is described by
\begin{equation}
\frac{\delta X}{X}=k_X {\delta \phi},
\end{equation}
where $k_X$ is a dimensionless factor quantifying the spatial variation of the fundamental constant $X$.
Assuming that the same field is responsible for the variation of all fundamental constants, the direction of the dipole is the same for all
fundamental constants.

\citet{BerFla12} proposed that such a dipole variation can be tested using
atomic clocks, quasar atomic and molecular  spectra, the {O}klo natural nuclear reactor, meteorite dating, and cosmic microwave background.
The Earth  is moving along with the Sun with
respect to the rest frame of the CMB and this motion has a component
along the direction of the $\phi$ gradient. This model results  in a small spatial variation  as well as annual modulation
of fundamental constants with Earth motion around the Sun. The result of \citet{WebKinMur11} roughly translates
into a $\dot{\alpha}/\alpha\approx 10^{-19}$~y$^{-1}$ variation
with a $\Delta \alpha/\alpha\approx 10^{-20}$ annual modulation.
Therefore, atomic clocks with high sensitivities to $\alpha$-variation described in Sec.~\ref{FC-new} are particulary desirable for such tests.
Present CMB constraint on the  dipolar
modulation  of $\alpha$ (corresponding to a gradient
across the observable Universe) from 2015 Planck data is $(-2.7 \pm 3.7)\times 10^{-2}$ at the 68~\% confidence level \cite{Pla15}.\\

\noindent III. \citet{OliPelUza11} theorized that such spatial variations of $\alpha$ may be
a signature of a domain wall produced in the spontaneous symmetry breaking in the early Universe, involving
a  scalar field coupled to electromagnetism. In this scenario,  there is no spatial gradient but a discontinuity
in the values of fundamental constants at the domain wall (or walls) in our
Hubble volume. The fundamental constants on either side of the wall differ, and the quasar absorption
spectra may not be actually testing deviation of  $\alpha$ from the current Earth value, but probe locations of the domain walls in
 our Hubble volume.  Attempts to fit Keck/VLT quasar absorption data into the one or two-wall models faced difficulties \cite{OliPelPet12,BerKavFla12}.

Atomic clocks are only sensitive to such a scenario of spatial $\alpha$-variation  if the Earth actually
passes thorough a domain wall at the present time. Precision magnetometery and atomic clock experiments aimed at detection of domain walls are discussed in  Sec.~\ref{Sec:LightDarkMatter}.

In a  different type of experiment, \citet{WieNevSch16}
used an optical
resonator fabricated from crystalline silicon at 1.5 K continuously for over one year, repeatedly
comparing its resonance frequency with the frequency of a GPS-monitored hydrogen maser.  The resonator frequency is determined by the physical length and the speed
of light and \citet{WieNevSch16} measure it with respect to the atomic unit of time, ruling out , to first order, a
hypothetical differential effect of the Universe’s expansion on rulers and atomic clocks. \citet{WieNevSch16} also constrain a
hypothetical violation of the principle of local position invariance for resonator-based clocks and derived
bounds for the strength of space-time fluctuations.

Analysis of  H$_2$ molecular spectra in terms of  spatial dependence of $\mu$ on cosmological scales  is presented by \citet{UbaBagSal16}.
Spatial variation of fundamental constants may also manifest itself at local scales (Milky Way and the Solar system).
Two types of tests are being pursued with atoms and molecules described below.

\subsubsection{Search for coupling of fundamental constants to a changing gravitational potential}

First, a spatial change in fundamental constants may be
induced by light scalar fields that change
linearly with changes in the local gravitational potential. This may be tested by searching for a dependence of fundamental constants
on a  varying gravitational potential.

Variations in fundamental constant $X$ with the change in the gravitational potential are modeled as
\begin{equation}
\frac{\Delta X}{X}=k_X \frac{\Delta U(t)}{c^2},
\label{FC-eq16}
\end{equation}
where $\Delta U(t)$ is the variation in the gravitational
potential. The goal of experiments is to measure or set constraints on  the quantities  $k_X$ which quantify the
couplings of various fundamental constants to the changing gravitational potential.
Due to the eccentricity of the Earth's orbit around the Sun,
the gravitational potential has a seasonal 3~\% variation
and a corresponding modulation of the constants may be studied with atomic clocks and other precisions instruments.

\begin{figure}[t]
            \includegraphics[scale=0.57]{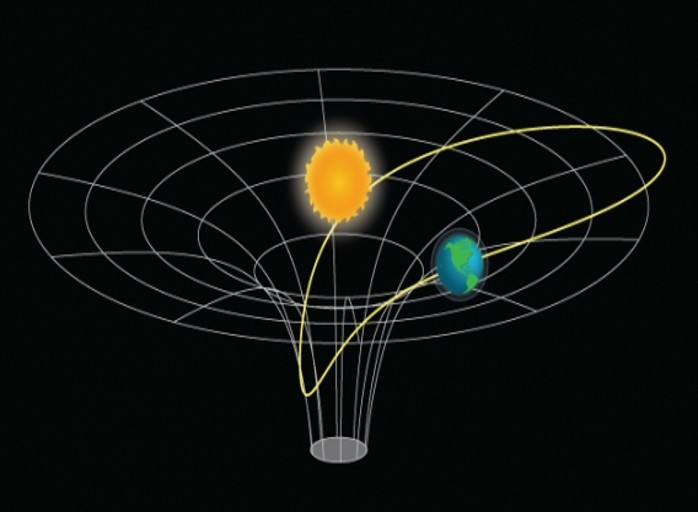}
            \caption{(color online). Earth orbiting  the Sun (mass $m_{\odot}$) in
gravitational potential;  the orbit
eccentricity is exaggerated. Picture credit: Jun Ye's group and Greg Kuebler, JILA. }
\label{fig4-FC}
\end{figure}

The idea for such experiment is illustrated in Fig.~\ref{fig4-FC}.
\citet{BlaLudCam08} searched for such  change in fundamental constants by monitoring the ratio of Sr and Cs clock frequencies. They
combined their result
with  Hg$^+$/Cs \cite{ForAshBer07} and H-maser/Cs \cite{AshHeaJef07} clock experiments to set constraints on the
couplings of fundamental constants $\alpha$, $1/\mu=m_\mr{e}/m_\mr{p}$ (designated by the subscript $\mu$ in this section), and $X_q = m_q/\Lambda_{\mathrm{QCD}}$ (designated by the subscript $q$) to the changing gravitational potential
 defined by Eq.~(\ref{FC-eq16}):
 \begin{eqnarray*}
k_{\alpha} &=& (2.5 \pm 3.1)\times 10^{-6},\\
k_{\mu} &=& (-1.3 \pm 1.7)\times 10^{-5},\\
k_q &=& (-1.9 \pm 2.7)\times 10^{-5},
\end{eqnarray*}
We note that decoupling of $k_q$ is not straightforward and is dependent on the nuclear model \cite{FlaTed06,Kim15}.
Only the dependence of the H-maser frequency on the light quark mass was taken into account, but not of the Cs clock.

A limit on the $k_{\alpha}$ was obtained with a measurement of frequencies
 in atomic dysprosium, which is only sensitive to $\alpha$-variation \cite{FerCinLap07}.

\citet{GueAbgRov12} reported a  limit  on a fractional variation
of the Rb/Cs frequency ratio with gravitational potential at the level of
$(0.11\pm 1.04)\times 10^{-6}$. A global fit to  available clock data yielded constraints  similar to the analysis of \citet{BlaLudCam08}.
\citet{TobStaMcF13} constrained  a fractional variation
of the Cs/H and Rb/H frequency ratios with gravitational potential at the level of
$3.6(4.8)\times 10^{-6}$ and $6.3(10)\times 10^{-6}$, respectively, over 8 years of measurements.

\citet{PeiCraHan13} reported  limit on fractional variation
of the Rb/Cs, Rb/H, and Cs/H frequency ratios with the gravitational potential at the level of
$(-1.6\pm 1.3)\times 10^{-6}$, $(-2.7\pm 4.9)\times 10^{-6}$, and $(-0.7\pm 1.1)\times 10^{-6}$, respectively, over 1.5 years of measurements.
\citet{PeiCraHan13} performed a global fit of constraints which included clock data from \citet{ForAshBer07,AshHeaJef07,BlaLudCam08} and \citet{GueAbgRov12}
which gave improved values:
\begin{eqnarray*}
k_{\alpha} &=& (1.7 \pm 7.5)\times 10^{-7},\\
k_{\mu} &=& (-2.5 \pm 5.4)\times 10^{-6},\\
k_q &=& (3.8 \pm 4.9)\times 10^{-6}.
\end{eqnarray*}

In 2013, a new Dy frequency measurement set an improved limit on the $k_{\alpha}$ \cite{LeeWebCin13}
$$
k_{\alpha} = (-5.5 \pm 5.2)\times 10^{-7}.
$$

Clock experiments intended for the ACES space mission on the
International Space Station (ISS) \cite{CacDimSan07} could improve upon the precision
of absolute redshift measurements. However, this does not help differential
measurements since the annual modulation of the gravitational potential due to the Sun is the same for clock on Earth and ISS, and the orbit around the Earth is circular.

Further improvements may come from further optical-clock tests and proposed space
missions that would put clocks in a highly eccentric earth
orbit \cite{SchTinGil09} or a solar-system-escape trajectory \cite{WolBorCla09}.
The use of  optical clocks based on the Yb$^+$ octupole transition \cite{HunSanLip16} as well as new clock schemes with
high sensitivity to $\alpha$-variation described in Sec.~\ref{FC-new} may significantly improve the constraints.

\citet{BerFlaOng13a} proposed
a new test of  $\alpha$-variation in a strong gravitational
field using metal lines in the spectra of white dwarf stars. A goal of such studies
 is to probe $\alpha$-variation
with  gravitational potential mediated by a light scalar field at a much stronger, by five orders of magnitude, gravitational potential than probed by clock experiments.
Laboratory measurements of relevant metal lines, such as the Fe$^{4+}$ and Ni$^{4+}$ (nickel, $Z=28$) spectra, are needed to improve on the results of \citet{BerFlaOng13a}.
Limits on a gravitational field dependence of $\mu$  from H$_2$ spectra in white dwarfs was reported in \citet{BagSalPre14}.

\subsubsection{Search for chameleons: testing the dependence of fundamental constants on the mass density of the environment}

Second, a spatial variation of
fundamental constants may  result from a shift in the expectation
value of $\phi$ between dense and rarefied environments \cite{Oli08}, when coupling of field to matter depends on its density, via,
for example chameleon mechanism. Such tests also probe  local position invariance.

In  chameleon models, the scalar fields which are dark energy  candidates
are ultra-light
  in a cosmic vacuum but possess
an effectively large mass  when  coupled to normal matter \cite{Oli08,Joy15} as discussed in Sec.~\ref{Sec:LightDarkMatter}, hence the ``\textit{chameleon}'' name.
Chameleon dark matter models and relevant experimental tests have been recently reviewed by \citet{Joy15}.
 The chameleon mechanism can potentially significantly affect quasar absorption spectra used to search for  variation of fundamental constants
  as well as comparison of laboratory and astrophysics limits.

 Here, we describe testing a class of chameleon models with
 scalar-field couplings to matter that are
  much stronger than the gravitational coupling \cite{Oli08}. In such a scenario, fundamental constants depend on the local  matter density
$\rho$  and one expects $\delta \alpha/\alpha \neq 0$ and
 $\delta \mu/\mu \neq 0$  for \textit{all}
interstellar clouds, when compared to terrestrial laboratory values. This change is due to differences in densities of the interstellar clouds and Earth environments,
 $\rho_{\oplus}/ \rho_{\mathrm{cloud}} > 10^{10}$ \cite{LevMolKoz11}. The large matter density on Earth  results in a screening of the cosmological
  chameleon field for
 terrestrial frequency measurements.

  Molecular studies with CO, CH, ammonia (NH$_3$) and methanol
CH$_3$OH have provided the most accurate limits on matter-density couplings of fundamental constants  because of high
 sensitivity of some molecular absorption spectra in our galaxy to $\mu$-variation.
Variation of the quantity $F=\alpha^2 \mu$ with matter density  was probed using a combination of  C$^+$ and CO transitions
\cite{LevMolRei10}, yielding a constraint $|\Delta F/F|< 3.7 \times 10^{-7}$.
The best-quality radio astronomical data for methanol  lines were used to constrain the variability
of $\bar{\mu}=1/\mu$ in the Milky Way at the level of $|\Delta \bar{\mu}/\bar{\mu}|< 2.8 \times 10^{-8}$ \cite{LevKozRei11}. This result can be further
improved with better laboratory spectroscopy of the CH$_3$OH microwave lines.

In 2010, \citet{LevMolLap10,LevLapHen10} reported a
 surprising non-zero $\Delta \bar{\mu}/\bar{\mu}= (26 \pm 1_{\mathrm{stat}} \pm 3_{\mathrm{sys}}) \times 10^{-9}$ result for $\mu$-variation
using the ammonia method. This approach  involves observations of the NH$_3$ inversion lines complemented by rotational lines of
other molecules in the Milky Way and comparing  these frequencies with terrestrial values.
In 2010-2013, \citet{LevReiHen13} carried out additional observations  in the Milky Way
 to test for hidden errors  and found a systematic error in the radial velocities of earlier studies.
A revised value of $\Delta \bar{\mu}/\bar{\mu}  < 2 \times 10^{-8}$ at the $3\sigma$ confidence level obtained using the ammonia method was reported in
 \citet{LevReiHen13}, resolving the discrepancy.

A spectroscopic method for pulsed beams of cold molecules was developed by \citet{TruHenTok13} and applied to measure the frequencies of
microwave transitions in CH. Comparing new CH values  and OH  laboratory results \cite{HudLewSaw06}
with those measured from Milky Way sources of CH and OH, \citet{TruHenTok13} constrained the variation of $\alpha$ and $\mu$
between the high- and low-density environments of the Earth and the interstellar medium at the levels of
$\Delta \alpha/\alpha  =( 0.3\pm1.1) \times 10^{-7}$  and $\Delta \bar{\mu}/\bar{\mu} =(-0.7\pm2.2)  \times 10^{-7}$.
Sensitivities for relevant transitions were calculated by \citet{Koz09}.

%\bibliography{master}

%\end{document}
% Typos corrected - Marianna Feb. 5

\section{Precision tests of Quantum Electrodynamics}
\label{Sec:QED}
%{\bf Notes: \\}

\subsection{Introduction}
\label{Sec:QED-intro}

In this section, we give an overview of low-energy precision tests of Quantum Electrodynamics (QED) with tools of atomic physics.
Historically, QED is the first relativistic quantum field theory, which laid  the foundation of the modern formalism of the Standard Model. It is arguably the most stringently tested sector of the Standard Model.

We focus on recent results and existing inconsistencies. The reader is referred to textbooks [e.g.,~\citet{BjoDre64,PeskinSchroeder1995QFTbook}] for a general introduction to QED and recent reviews by \citet{Eides2001,DraYan08,Bei10}; and \citet{Karsh2005} for a more detailed exposure.

Precision tests of QED are carried out by comparing experimental results with theoretical predictions.
For example, QED predictions depend on the value the electromagnetic fine-structure constant $\alpha$.
QED is then validated to the extent that the deduced values of $\alpha$ from different methods, one of which incorporates QED calculations,  agree with each other, as described in Sec.~\ref{el}.
The comparisons are  affected by the uncertainties in the values of fundamental constants [such as masses, Rydberg constant, etc.] and by the uncertainties in the strong-force [hadronic] contributions beyond QED.

In general, one distinguishes between free-particle properties, such as the anomalous magnetic moment of the electron, and bound-state QED [Lamb shift being the most prominent example].

Bound-state QED can be tested in a number of  systems and we highlight the advantages of various approaches below. Such tests are expected to be more accurate in
light systems such as H, D, $^3$He$^+$, He, positronium (Ps) and muonium (Mu), where the contribution of inter-electron interaction is either absent  or can be evaluated to high accuracy. QED tests with these systems were reviewed  by \citet{Karsh2005}. The relative importance of inter-electron interactions is also reduced in
highly charged ions [HCIs]. However, in HCIs  the nuclear structure uncertainty is the limiting factor and
QED calculation in heavy ions require a development of non-perturbative methods.
Interesting intermediate cases are few-electron systems, where the electron-electron correlations must be  taken into account on par with the nominally field-theoretic (QED) contributions.
High precision QED atomic calculations for Li and Be$^+$ were carried out by \citet{YanNorDra08,YanNorDra09} and resulting energies were found to be in good agreement with experiment, with the exception of the Be$^+$ ionization potential. QED corrections to the $2p$ fine structure in
Li were calculated by \citet{PucPac14}, yielding the splitting value with $6\times 10^{-6}$ uncertainty, in agrement with recent high-precision experiment \cite{BroWuPor13,BroWuPor13a}.
Precision test of many-body QED was reported by \citet{NorGepKri15}  using  the Be$^+$ $2p$ fine-structure
doublets measured in short-lived isotopes.
 Simple molecules, H$_2$, HD, and D$_2$, and H$^{2+}$, HD$^+$ molecular ions~\cite{SalDicIva11,DicNiuSal13,BieKarHil16} offer additional QED tests.

 Below we highlight a few recent examples of precision QED tests and review the recent progress in QED tests with HCIs.

\subsection{Anomalous magnetic moment of the electron}
\label{el}

At present, the most accurate contributions to the determination of $\alpha$  come from comparison of theory and experiment for the electron
magnetic-moment anomaly $a_\mathrm{e}$ \cite{CODATA2010,CODATA2014}. This quantity is defined as follows.
The magnetic moment of the electron is
\begin{equation}
\bm{\mu}_e=g_\mathrm{e} \,\frac{e}{2m_\mathrm{e}} \bm{s},
\end{equation}
where $g_\mathrm{e}$ is the (dimensionless)  electron g-factor, $m_\mathrm{e}$ is its mass, and $\bm{s}$ is
its spin.
The
magnetic-moment anomaly $a_\mathrm{e}$ is defined by
$$
|g_\mathrm{e}|=2(1+a_\mathrm{e}).
$$
The solution of the Dirac equation for a free electron  gives $g_\mathrm{e}=-2$ and thus $a_\mathrm{e} = 0$. In the Standard Model,
 $a_e\neq0$: it is given by
$$
a_\mathrm{e(th)}=a_\mathrm{e}(\textrm{QED})+a_\mathrm{e}(\textrm{weak})+a_\mathrm{e}(\textrm{hadronic}),
$$
where the three terms account respectively for
the purely quantum electrodynamic, predominantly electroweak,
and predominantly hadronic
contributions [using the notation of \citet{CODATA2014}].
The $a_\mathrm{e}(\textrm{QED})$ contribution depends on $\alpha$ and can be expressed as a powers series of $\alpha/\pi$ whose
coefficients are calculated from QED.
The dependence of $a_\mathrm{e}$ on $\alpha$ for the other two contributions is negligible.

The most accurate measurement of $a_\mathrm{e}$ was carried out with a single electron that was suspended for months at a time
in a cylindrical Penning trap \cite{HanFogGab08}.
The ratio of electron spin-flip  frequency to the
 cyclotron frequency in the trap determines
 $a_\mathrm{e}$. The resulting value of $\alpha$ extracted by combining the 2008 measurement \cite{HanFogGab08} and theoretical
 result for $a_\mathrm{e}$ is
\begin{equation}
1/\alpha=137.035\,999\,084(51),
\label{FC-1}
\end{equation}
which has a relative uncertainty of $3.7\times10^{-10}$. The theoretical uncertainty contribution is $2.8\times10^{-10}$.

Alternatively, the value of $\alpha$ can be obtained from the expression \cite{WicHenSar02}
\begin{equation}
\alpha^2=\frac{2R_{\infty}}{c} \frac{M}{m_e} \frac{h}{M},
\label{alpha1}
\end{equation}
where $M$ is the mass of any atom.
The relative standard uncertainties
of the $R_{\infty}$ and $M/m_{\rm{e}}$ are about $6\times 10^{-12}$ and a few times $10^{-10}$ or less \cite{CODATA2014}.
 Therefore, a precision measurement
of the ratio $h/M$ for a particular atom can provide a value of $\alpha$  with a precision competitive
to that of the determination of $\alpha$ from $a_\mathrm{e}$ described above.

\citet{WicHenSar02} used atom interferometry to measure the recoil velocity,
$
v_r=\hbar k/M
$
of a $^{133}$Cs atom when it absorbs a photon with momentum $\hbar k$. The resulting value of $h/M(^{133}\textrm{Cs})$ yielded a value of $\alpha$ with a relative uncertainty of $7\times 10^{-9}$.

\citet{CadMirCla08} used Bloch oscillations\footnote{The atoms in an optical lattice created by
two counterpropagating laser beams whose frequency difference
is swept linearly undergo a succession
of Raman transitions which correspond to the absorption of
one photon from a beam and a stimulated emission of a
photon to the other beam. The internal state is unchanged
while the atomic velocity increases by $2v_r$ per oscillation \cite{CadMirCla08}.}
of $^{87}$Rb atoms in an optical lattice to  impart to the atoms up to
1600 recoil momenta and a Ramsey-Bord\'{e} interferometer to precisely measure the induced atomic velocity variation.
\citet{BouClaGue11} improved this method further and measured the ratio of the Planck constant to the mass of the $^{87}\textrm{Rb}$ atom to obtain
a value of $\alpha$, accurate to $6.6\times10^{-10}$:
\begin{equation}
1/\alpha=137.035\,999\,037(91),
\label{FC-1A}
\end{equation}
improving the  result of \citet{CadMirCla08} by a factor of seven.
\citet{TerTin14} discussed the potential for further improvement of $h/M$ measurements, demonstrating that it may be possible to attain the level of precision needed to test the anomaly for the magnetic moment of the muon \cite{Bennett2006}.

The agreement of two determinations of $\alpha$, from the measurements of $a_\mathrm{e}$ and of $h/M\left(^{87}\textrm{Rb}\right)$,
validates the theoretical QED calculation of $a_\mathrm{e}$ \cite{PhysRevD.77.053012}.  This, in turn, provides the most accurate test
 of quantum electrodynamics and the SM to date. We emphasize that $a_\mathrm{e}$ is calculated
 in terms of the fundamental constant $\alpha$, but $\alpha$ is a SM parameter   as it cannot be calculated from the first principles.

\subsection{Quantum electrodynamics tests with polyelectrons}

In 1946, Wheeler denoted as ``polyelectrons'' all bound complexes consisting of only electrons and positrons \cite{Wheeler1946}. Although all such complexes are likely unstable with respect to electron-positron annihilation into gamma rays, there are some that are stable with respect to dissociation into simpler complexes, and thus may live sufficiently long to have physical and even chemical significance. Positronium (Ps), the atom consisting of one electron and one positron, is the simplest example: it has the same discrete spectrum as the hydrogen atom in nonrelativistic quantum mechanics, up to a multiplicative factor of
$\left(m_\mathrm{p} + m_\mathrm{e}\right)/2 m_\mathrm{p}$. \citet{Wheeler1946} used a simple variational calculation to show that that Ps$^-$ should also be stable, and \citet{PhysRev.71.493} determined that Ps$_2$ should be stable. These three species were subsequently found experimentally. Reviews of developments in this field up to 2012 were given by \citet{Karsh2005} and \citet{Namba2012}, and of more recent work by \citet{Nagashima2014} and \citet{Mills2014}.

As purely leptonic systems, polyelectrons offer a testbed for precision comparison of QED theory with experiment, particularly for bound-state systems.  We review recent results and future prospects below. There is no experimental evidence for more complex polyelectrons. \citet{FroWar08} suggest that Ps$_2^-$ and Ps$_3$ should be stable, but \citet{Varga2014} and \citet{Bubin2013}, respectively, consider these two species to be unstable.

\subsubsection{Positronium}
Positronium (Ps), the atom consisting of one electron and one positron, was first identified in the laboratory by \citet{Deutsch1951}.
It is a system in which bound-state QED has been studied with precision.  Most recently, the structure of the lowest triplet term of Ps was measured by optical spectroscopy \cite{Cassidy2012}, and the  transition between the triplet and singlet terms of the Ps ground state has been observed directly \cite{Yamazaki2012,Miyazaki2015}. This  energy splitting is a benchmark for first-principles QED calculations of two-particle systems.  It has been calculated by QED theory up to order $\alpha^6$ to an accuracy of 1 ppm. The result differs by $4 \sigma$ from the experimental determination, which presently has an uncertainty of around 3 ppm \cite{Cassidy2012}. Improvements in precision are required to resolve this discrepancy. [A more recent experiment does claim to have a result closer to theory \cite{Ishida2014, Ishida2015}].  There are suggestions about beyond SM physics mechanisms to which positronium might be particularly susceptible \cite{Lamm2017}, and there are substantial efforts to extend QED theory to order $\alpha^7$ in order to sharpen the comparison with experiment\cite{Adkins2015,Eides2017}.

Another noteworthy recent development is the advent of Ps Rydberg spectroscopy, in which states have been resolved with principal quantum numbers $n$ as large as 50 \cite{PhysRevLett.108.043401,PhysRevA.90.012503,PhysRevLett.114.173001,PhysRevLett.115.183401}. Such states may be of fundamental interest for testing QED, as some QED corrections appear at lower orders of $\alpha$ than they do for the ground state \cite{Lamm2017}. Ps Rydberg states can also have longer lifetimes than the $n = 1$ ground state, since the electron-positron annihilation rate is proportional to the probability density at the point of contact, which scales as $n^{-3}$ for $s$ states and can be further reduced by using states with higher values of $l$. This could be advantageous for precision spectroscopic study, or for use of Rydberg Ps states as reservoirs for positrons used to produce the antihydrogen required for the studies described in Sec.~\ref{Sec:CPT}. If Ps could be placed in highly ``circular'' Rydberg states, it could be a candidate for studies of the Einstein equivalence principle for a mixed matter-antimatter system, either via free--fall measurements or gravitational quantum state spectroscopy \cite{Dufour2015}.

\subsubsection{Positronium anion, Ps$^-$}

In the laboratory, the positron that ends up in Ps is typically born at energies around 0.5~MeV  by beta decay of a radionuclide such as $^{22}$Na. The positron is moderated down to $ \approx 10 \,$ meV energies by passage through matter, at which stage it can be controlled by conventional electron-optical techniques for use in scattering experiments, or produce Ps by
electron capture from solids \cite{Charlton2000,Mills2014}.  The positronium anion, Ps$^-$, was first obtained in the laboratory by \citet{Mills1980}.  An experimental breakthrough in 2008 made
it possible to generate  Ps$^-$ with efficiencies above 1~\%, i.e. for every 100 positrons entering the moderator, one Ps$^-$ emerges
\cite{Nagashima2014}. This development transformed the study of Ps$^-$, for example, making possible the observation of a shape resonance in its photodetachment \cite{Michishio2016}. It
also provides a means for producing energy-tunable beams of Ps, by applying standard acceleration procedures to Ps$^-$ and then neutralizing it by photodetachment.  There is a
considerable body of theory on the structure of Ps$^-$, including treatment of QED corrections \cite{Drake2002,Frolov2005}. In time, the postronium anion may become a benchmark
for testing QED in three-particle systems.

\subsubsection{Diatomic positronium, Ps$_2$}

The Ps$_2$ molecule was also observed, both in its ground state \cite{Ps22007} and in an $L = 1$ bound excited state that was predicted by
\citet{PhysRevLett.80.1876} and \citet{PhysRevA.58.1918} and subsequently observed by  \citet{PhysRevLett.108.133402}. The wavelength of the ground - excited state transition in Ps$_2$ was
predicted to be 250.9179(11) nm. The observed wavelength reported by
\citet{PhysRevLett.108.133402} is 250.979(6) nm. At present, the reason for the difference  is not understood in detail.  The Ps$_2$ is thought to be located in a porous silica matrix, which
has been found to produce shifts in Ps transition wavelengths that are comparable to the difference between the theoretical and observed values for Ps$_2$.

 \subsection{Tests of QED in highly charged ions}
 QED tests in highly charged ions were recently reviewed by \citet{Bei10,VolGlaPlu13,StuWerBla13,ShaGlaPlu15,StuVogKoh17} and we focus on key results and
 new developments. The spectroscopic properties involved in the HCI QED tests are atomic transition energies, hyperfine splittings and $g$ factors.

\subsubsection{Energies}

The ground-state Lamb shift in H-like uranium (U, $Z=92)$ was measured by \citet{GumStoBan05} with 1~\% uncertainty, $460.2\pm 4.6$ eV.  The theory prediction is 463.99(39)~eV, with
QED contributing 265.2~eV [of which -1.26(33)~eV is due to 2nd order QED] \cite{YerIndSha03} and finite nuclear size contributing 198.54(19)~eV \cite{KozAndSha08}. Combining theory and experiment
provided a test of QED at the 2~\% level.

 The $2S-2P_{1/2}$ transition energy in Li-like U$^{89+}$, 280.645(15)\,eV, was measured with much higher, 0.005~\%
 precision, in agreement with theoretical value of 280.71(10)~eV \cite{KozAndSha08}. Li-like uranium study tested second-order (in $\alpha$) QED effects to 6~\% \cite{VolGlaPlu13}.  Theoretical accuracy of HCI QED tests is
limited by the nuclear polarization correction.

The experimental accuracy is much higher for lighter ions. The $1s2p ~^1P_1 - 1s^2$~$^1S_0$
resonance  line in He-like Ar$^{16+}$ was measured with a relative  uncertainty of $2\times10^{-6}$ for a test of two-electron and two-photon QED radiative corrections \cite{BruBraKub07}. The experimental value was in perfect agreement with theoretical prediction \cite{ArtShaYer05}, as well as with a later 1.5~ppm measurement of 3139.581(5)~eV \cite{KubBraBru12}.
However,
a measurement of the He-like Ti (titanium, $Z=22$) resonant line, 4749.85(7) eV, by \citet{ChaKinGil12}
differed by 3$\sigma$ from the theoretical prediction \cite{ArtShaYer05}. \citet{ChaKinGil12} noted that there appeared to be
an evidence for a
$Z$-dependent divergence between experiment and calculation in He-like isoelectronic sequence with $Z>20$. This analysis was disputed
by  \citet{Epp13}, in particular the omission of the \citet{KubBraBru12} value from the fit. This issue was addressed by \citet{ChaKinGil13} and further discussed by \citet{Gil14}, indicating need for further experimental and theoretical work.
Measurement of the resonant line in He-like Fe$^{24+}$ \cite{KubMokMac14} was found to be in agreement with theory \cite{ArtShaYer05}
and  inconsistent with  a claim of systematic divergence between theory and experiment \cite{ChaKinGil12} at a
3$\sigma$ level.
The other Fe$^{24+}$ spectroscopy data \cite{RudBerEpp13} are also consistent with QED theory values.

 The energy of the $1s2s ~^3S_1 - 1s^2$~$^1S_0$ magnetic dipole transition in helium-like Argon was measured to 2.5~ppm accuracy by \citet{AmaSchGue12}, differing
 by $1.6~\sigma$ from the theoretical prediction of \citet{ArtShaYer05}. Even higher precision of 0.6~ppm was achieved for $1s^{2} 2s^{2} 2p$~$^2P_{3/2}$ - $^2P_{1/2}$ transition in boron-like
Ar$^{13+}$ ions, 441.25568(26) nm \cite{MacKlaBre11}. The theory value \cite{ArtShaTup07} is in agrement with the experiment, but two orders of magnitude less accurate. Since nonrelativistic energies of $p_{1/2}$  and $p_{3/2}$ states are the same, this transition energy is determined by the relativistic and QED effects, making it an excellent candidate for precision QED tests. Experimental accuracy can be significantly increased by recent demonstration of sympathetic cooling of HCIs \cite{SchVerSch15}, and theory accuracy urgently needs to improve.

A high-precision nonperturbative
(in $Z\alpha$) calculation of the nuclear-recoil effect on the Lamb shift of light hydrogenic atoms
was carried out by \citet{YerSha15}. This   resolved
 previously-reported disagreements
between the numerical all-order and analytical
$Z\alpha$-expansion approaches,  which were caused by unusually large higher-order
terms omitted in the $Z\alpha$-expansion calculations.
 This work eliminated the second-largest
theoretical uncertainty in the $1S$ and $2S$ Lamb shift of H.

\subsubsection{Hyperfine splittings}
\citet{KlaBorEng94} reported the first direct observation of a hyperfine splitting in the
optical regime and measured the  wavelength of the M1 transition between the hyperfine levels of the ground state of hydrogenlike $^{209}$Bi$^{82+}$.
A number of measurements of the hyperfine splitting in H-like HCIs with about $1\times 10^{-4}$ uncertainty [for  example, measurements in $^{203}$Tl$^{80+}$ and $^{205}$Tl$^{80+}$ (thallium, $Z=81$) by  \citet{BeiUttWon01}] motivated corresponding theoretical efforts. Since the theoretical uncertainty  is dominated by the correction due to nuclear magnetization distribution [the Bohr-Weisskopf (BW) effect], \citet{ShaArtYer01} proposed
to consider a specific difference of the ground-state hyperfine splitting in Li-like ion, $\Delta E(2s)$, and H-like ion, $\Delta E(1s)$, for the same nucleus:
\begin{equation}
\Delta^{\prime}E=\Delta E(2s)-\xi \Delta E(1s).
\label{gg}
\end{equation}
The parameter $\xi$ introduced to cancel the Bohr-Weisskopf effect can be calculated with high precision. In 2012, the theoretical accuracy of the specific difference between the hyperfine splitting values of H- and Li-like Bi (bismuth, $Z=83$) ions was significantly improved [to a relative uncertainty of $\approx 10^{-4}$] due to a new evaluation~\cite{VolGlaAnd12} of the two-photon exchange corrections to the hyperfine structure   in Li-like ion. Measurements of the H-like and Li-like Bi hyperfine splittings at the $10^{-6}$ level will allow probing the many-body QED effects at a few percent level \cite{VolGlaPlu13}.

Hyperfine splitting of the $2s$ and $2p_{1/2}$ levels in Li-like and Be-like ions of $^{141}$Pr were measured by \citet{BeiTraBro14}
using high-resolution spectroscopy of the $2s - 2p_{1/2}$ transition in the extreme ultraviolet region (EUV).
This work demonstrated  that EUV spectroscopy
can be used to measure the hyperfine structure in
high-$Z$ ions with a few valence electrons at the meV level.

\citet{UllAndBra17} measured the specific difference given by the Eq.~(\ref{gg}) between the hyperfine splittings in
hydrogen-like $^{209}$Bi$^{82+}$ and lithium-like $^{209}$Bi$^{80+}$  with more than an order of magnitude improvement in precision.
The parameter $\xi=0.16886$ was chosen from
theory to cancel the BW correction for $^{209}$Bi \cite{ShaArtYer01}.
 While it was expected that the specific difference is largely
insensitive to nuclear structure, the experimental result -61.012(5)(21)~meV  differs by 7$\sigma$  with the theoretical prediction -61.320(4)(5)~meV \cite{VolGlaAnd12}. For the experimental value, the parenthesis list the statistical and systematic
uncertainties. The first and second uncertainties in the theory value arise from uncalculated higher-order terms and the uncertainty of the complete cancellation of all nuclear effects.
This is the largest deviation reported in strong-field QED up to now.

\subsubsection{QED tests for $g$ factors}

The experimental determination of the electron $g$ factor in Penning traps was reviewed by \citet{StuWerBla13,StuVogKoh17} while
the theory of bound-electron $g$ factors in HCIs was  reviewed by \citet{ShaGlaPlu15}. Here, we highlight the most recent results.

The  $g$ factor of the electron bound in H-like $^{28}$Si$^{13+}$ was measured to 10 significant
digits by \citet{StuWagSch11}
using a single ion confined in a cylindrical Penning trap.  The experimental $g$ factor is  determined via the relation
\begin{equation}
g = 2 \frac{\nu_L}{\nu_c} \frac{q}{|e|} \frac{m_e}{M},
\label{qed1}
\end{equation}
where $\nu_c$  is the cyclotron frequency, $\nu_L$  is the Larmor precession frequency, $M$ is the ion mass, and $q$ is the ion charge.
The experimentally determined quantity is the frequency ratio
$\Gamma = {\nu_L}/{\nu_c}$.
The details of the setup, the experimental procedure and the data evaluation were given by \citet{SchStuWag12}. An improved result was reported by \citet{StuWagKre13}, where the electron $g$ factor  in $^{28}$Si$^{13+}$ was measured with a $4\times 10^{-11}$ fractional uncertainty using a phase-detection method to determine the cyclotron  frequency.
This measurement presented a challenge for theory as the
 theoretical uncertainty, mostly determined by uncalculated two-loop QED
corrections of order $\alpha^2(\alpha Z)^5$ and higher \cite{PacCzaJen05}, became a factor of
 two larger than the experimental one.
This uncertainty can be reduced by combining theoretical and experimental
 values for two different H-like ions as demonstrated by \citet{StuKohZat14}.

Following Eq.~(\ref{qed1}),
the combination of the precision $g$-factor measurement and the state-of-the-art QED calculation, may be used to
determine the electron mass.
In 2014, a very precise measurement of the magnetic moment of a
single electron in H-like $^{12}$C$^{5+}$, combined with QED theory and previous measurement of the electron $g$ factor with $^{28}$Si$^{13+}$ \cite{StuWagKre13}, were used to extract a new value of the electron mass \cite{StuKohZat14}, improving its accuracy by a factor of 13.
Carbon ions were used since the $^{12}$C atom defines the atomic mass unit, and the mass of the ion is
known to high precision.
The measurement details, including a
comprehensive discussion of the systematic shifts and their uncertainties are presented by \citet{KohStuKra15}.
\citet{ZatSikKar17}
reevaluated the extraction of the electron mass taking into account recently
calculated additional QED contributions \cite{YerHar13,CzaSza16}. The
resulting value for the electron mass,
\begin{equation}
m_e = 0.000\,548\,579\,909\,065(16)\, \rm{u},
\end{equation}
is shifted by $0.3~\sigma$ with respect to earlier
evaluations of the same experimental data  \cite{KohStuKra15} due to the
inclusion of light-by-light scattering terms of the order of
$\alpha^2(Z\alpha)^4$ calculated by \citet{CzaSza16}.
The theoretical uncertainty of the $g$ factor is now an order of magnitude less than that of the uncertainty in the
measurement of the frequency ratio $\Gamma$ for $^{12}$C$^{5+}$.
\citet{ZatSikKar17} discussed the prospects for improved determination of $m_\textrm{e}$, $M(^4\textrm{He})$ and $\alpha$.

A recent comparison of the cyclotron frequencies of the
protons and  $^{12}$C$^{6+}$ ions yielded a value of the proton mass in atomic mass units
with a precision of 32 parts per trillion \cite{HeiKohRau17}. The resulting value is more precise than the current
CODATA recommeded value by a factor of 3, but  disagrees with it by  about 3 standard deviations \cite{CODATA2014}.

The most stringent bound-state QED test of the ground state $g$ factor for a three-electron systems was carried out for Li-like $^{28}$Si$^{11+}$ (silicon, $Z=14$) by \citet{WagStuKoh13,VolGlaSha14}. The $g$-factor  measurement carried out in a Penning trap,
2.000\,889\,889\,9(21) was found to be in excellent agreement with the theoretical value.
 The theory precision was further improved by \citet{VolGlaSha14} due to rigorous QED evaluation of the two-photon exchange corrections to the $g$ factor, yielding 2.000\,889\,892(8).
A comparison of this new theoretical value with the  experimental result~\cite{WagStuKoh13}  provides tests of relativistic interelectronic
interaction at the  $10^{-5}$ level of precision, the one-electron bound-state QED   in the presence
of a magnetic field at the 0.7~\% level, and the screened bound-state QED at the 3~\% level.

By comparing the $g$ factors of two
isotopes, it  is also possible to cancel most of the bound-state QED contributions and  probe nuclear
effects. \citet{KohBlaBlo16} presented calculations and experiments on the isotope dependence of the
Zeeman effect
by studying $g$ factors of Li-like  $^{40}$Ca$^{17+}$
and $^{48}$Ca$^{17+}$ ions.

 For heavy ions, a specific difference scheme similar to Eq.~(\ref{gg}) can be employed to largely cancel the nuclear effects in the $g$-factor HCI calculations \cite{ShaGlaSha02}.
In 2014, \citet{VolPlu14} performed a systematic study of the nuclear polarization effects in one-electron and few-electron heavy ions, which
included the calculation of the nuclear polarization corrections to the
binding energies, the hyperfine splitting, and the bound-electron $g$ factor
in the zeroth and first orders in $1/Z$.  Strong cancellation of nuclear polarization effects determining
the ultimate accuracy of the QED tests was observed in all cases for the specific differences described above.
 A possibility for a determination of $\alpha$ in bound-electron $g$ factor experiments via the study of a specific difference of the $g$ factors of B-like and H-like ions, for the same isotope with zero nuclear spin in the Pb region, was discussed by \citet{ShaGlaOre06}.

Further improvement of the experimental accuracy of $\Gamma$ is expected for any ion from
the currently commissioned ALPHATRAP, Penning-trap setup
at the Max Planck Institute for Nuclear Physics \cite{StuWerBla13,StuVogKoh17}.
ALPHATRAP receives ions from an external electron beam ion trap that can produce  charge
states of up to Pb$^{81+}$ for QED tests  and a determination of $\alpha$.
Another experiment currently under commissioning, ARTEMIS \cite{LinWieGla13}, located at the
HITRAP \cite{KluBeiBla07} facility at GSI, Darmstadt will employ a spectroscopic technique for ions with non-zero
nuclear spin and is designed to work with ions up to the highest charge states.

\subsection{Proton radius puzzle}
The proton radius puzzle, as it is known colloquially, has perplexed the physics community for over half a decade
\cite{Jentschura2011,PohGilMil13,Carlson2015,PohlJPSJ,Hill2017}. The highly-precise root mean square (r.m.s.) charge radius $r_p=0.84087(39)$ fm extracted from the $2S-2P$ Lamb shift in muonic hydrogen \cite{PohAntNez10,AntNezSch13} is in significant disagreement with the result $r_p=0.8758(77)$ fm deduced from spectroscopy with ordinary hydrogen \cite{CODATA2010}. The latter value is also supported by electron scattering experiments, further exacerbating the problem \cite{CODATA2010,BerAchAye10,ArrSic15}. This outstanding discrepancy has prompted speculations that the discrepancy may hint at physics beyond the Standard Model [see e.g., \citet{LiuMcKeenMiller2016,Dahia2016,Onofrio2013}]. Resolution may be more mundane, such as missing systematic corrections both in theory and experiment or  incorrect value of the Rydberg constant~\cite{PohGilMil13}.
In fact, \citet{CzaSza16} have recently pointed out that light-by-light scattering diagrams have been erroneously neglected in the computations of the Lamb shift; these authors estimate that such contributions
decreases the theoretical prediction for the $1S-2S$ splitting in hydrogen
by an amount 28 times larger than the experimental error~\cite{Parthey2011}.
  Interestingly, the same  muonic-hydrogen collaboration (CREMA) has reported~\cite{Pohl2016} the value of deuteron radius
that shows a similar discrepancy with results of  deuterium spectroscopy.

Two spectroscopic experiments on hydrogen were reported as this paper was being prepared for publication.
\citet{BeyMaiMat17} measured the $2S-4P$ transition frequency in H
using a cryogenic beam. The extracted value of the proton radius, $r_p = 0.8335(95)$~fm, is 3.3 combined standard deviations smaller than the previous ``H world data'', which is the consensus of previous experiment on spectroscopy of atomic hydrogen. However, this radius is in good agreement with that inferred from the spectroscopy of muonic hydrogen.
On the other hand, \citet{FleGalTho18} measured  the $1S-3S$ two-photon transition frequency of hydrogen,
realized with 205~nm continuous-wave laser excitation of a room-temperature atomic beam. They extracted the value
$r_p = 0.877(13)$~fm, which is in good agreement with the current CODATA value. Thus,
new hydrogen spectroscopy experiments that were intended to unravel this mystery, have only deepened it further.

\subsection{Conclusion}
Finally, we would like to emphasize that the detailed understanding of atomic structure and, in particular, QED contributions,
is crucial for a number of precision tests of physics beyond the SM. QED contributions are needed for determining fundamental constants. While the stark discrepancies in proton and deuteron radii determinations from various methods have spurred re-examinations of both theory and experiment, the puzzle still remains unresolved.
A number of technological advances, such as high-precision Ramsey-comb spectroscopy at deep ultraviolet wavelengths \cite{AltGalDre16} and HCI trapping~\cite{SchVerSch15} are anticipated to extend the QED tests to new regimes. % APD Jan 11, 2018 Referee suggested edits
\section{Atomic parity violation}
\label{Sec:APV}

\subsection{Introduction}
\label{Sec:APV-intro}

In this section, we give an overview of parity violation in atomic and molecular physics. This field is generally referred to as  parity non-conservation (PNC) or atomic parity violation (APV) in the literature.  The field of parity violation started  with
the seminal paper by~\citet{LeeYan56} and discovery of
PNC in  nuclear $\beta$-decay~\cite{WuAmbHay57}, followed by the Nobel Prize in physics awarded to  Yang and Lee in 1957. Soon
after this discovery, \citet{Zel59} contemplated a possibility of
observing PNC  in atoms. He
concluded that the effect was too small to be of experimental
significance. However,  %M.-A. Bouchiat and C. Bouchiat~
\citet{BouBou74} realized that  PNC is amplified in {\em heavy}
atoms. They showed that the relevant PNC amplitude scales steeply
with the nuclear charge $Z$, roughly as $Z^3$. PNC amplitudes  in heavy atoms, such as Cs,   are enhanced by over those in hydrogen by a factor of $10^5-10^6$. This crucial observation  enabled experiments on APV.
In atomic physics, the first $P$-violating signal was observed in
1978 by \citet{BarZol78} in Bi, followed closely by a measurement of P-violation in Tl~\cite{ConBucChu79,BucksbaumComminsHunter81}.
%(Conti et al.,1979; Bucksbaum et al., 1981).
In the same year, parity violation was reported in  inelastic electron-deuterium scattering \cite{PreAtwCot78}. Direct observations of the charged $W^\pm$ boson and neutral $Z$ boson (responsible for APV) were not made until 1983 when they were observed at CERN's
proton-antiproton collider~\cite{Arnison1983,Arnison1983a}.

Over the following decades, AMO experiments were refined, with
PNC effects observed in several atoms.
% (see, e.g., Refs.\cite{ContiBucksbaumChu79,EdwPhiBai95,FortsonPb1993,VetMeeMaj95,Stacey1991}).
The most accurate measurement to date was performed
in Cs \cite{WooBenCho97} and the most recent reported measurement is in  ytterbium (Yb, $Z$ = 70) \cite{TsiDouFam09}. New experiments on atomic and molecular PNC are underway or in planning stages as described in Sec.~\ref{Sec:APV:NewExpt}.
There are a number of extensive review articles of
APV \cite{Bouchiat1997,Budker1999,GinFla04,DerPor07,Roberts2014} as well as  a monograph \cite{Khr91a}.
Basics  of electroweak theory can be found in a number of textbooks, e.g.,~\citet{commins1983weak} which has a discussion on APV.

%This Section is organized as follows. After giving a general introduction ...

The parity transformation $P$, or spatial inversion, is equivalent to mirror reflection and rotation by $180^{\circ}$. The eigenvalues of $P$ are $\pm$~1, referred to as even and odd, respectively.  Under this transformation,
 all position vectors $\mathbf{r}$ change sign: $\mathbf{r} \rightarrow -\mathbf{r}$, while spin and orbital angular momenta remain unaffected. Electric and magnetic fields are transformed as $\Evec \rightarrow - \Evec$ and $\Bvec \rightarrow \Bvec$.

 The QED Lagrangian governing AMO physics commutes with $P$, leading to distinct selection rules in atomic physics.  For the electronic  configuration $n_1\ell_1 \ldots n_{N_e} \ell_{N_e}$ of $N_e$-electron atom, the parity eigenvalue is given by $\Pi=(-1)^{\sum_{i=1}^{N_e} \ell_i}$. The parity of a given configuration is determined by the parity of the open electron shell, e.g. the parity of the [Hg]$6p^3$~$^4\mathrm{S}^\circ_{3/2}$ ground state of Bi is odd. The conventional spectroscopic notation of  electronic terms arising from a given electronic configuration includes the label ``$\circ$'' for odd parity states.

 \citet{lap:24} discovered parity conservation in atoms from analysis of the iron spectrum, and formulated a rule:
\textit{electric dipole transitions between states of like parity are strictly forbidden}. To see this,   consider the electric-dipole (E1) transition amplitude
${\bf{T}}_{fi}^{({\rm{E1}})} = \left\langle {{\Psi _f}} \right|{\bf{D}}\left| {{\Psi _i}} \right\rangle$, where the atomic states ${{\Psi _{i,f}}}$ are parity eigenstates and  $\bf{D}$ is  the electric dipole moment operator. Inserting the identity $1=P^\dagger P$,

\begin{eqnarray*}
{\bf{T}}_{fi}^{({\rm{E1}})}& =& \left\langle {{\Psi _f}} \right| P^\dagger P {\bf{D}}\left| {{\Psi _i}} \right\rangle\\
& = &\left\langle P{{\Psi _f}}  \right|  P ( {\bf{D}}\left| {{\Psi _i}} \right\rangle) = - \Pi_f \Pi_i {\bf{T}}_{fi}^{({\rm{E1}})}.
\end{eqnarray*}
Now if  the  two states have  the same parities,  ${\bf{T}}_{fi}^{({\rm{E1}})} = -{\bf{T}}_{fi}^{({\rm{E1}})}$ and thereby ${\bf{T}}_{fi}^{({\rm{E1}})}=0$.  If parity is not conserved, the eigenstates of the full atomic Hamiltonian no longer possess a well defined parity. In other words, PNC leads to [usually small] mixing of opposite-parity states leading to non-vanishing values of ${\bf{T}}_{fi}^{({\rm{E1}})}$.
The theory and experiments described below show how  Laporte's rule is violated in atoms and molecules. %, as a consequence of $Z$-boson exchange between  atomic electrons and the nucleus.

\begin{figure}[t]
\begin{center}
\includegraphics[width=0.9\columnwidth]{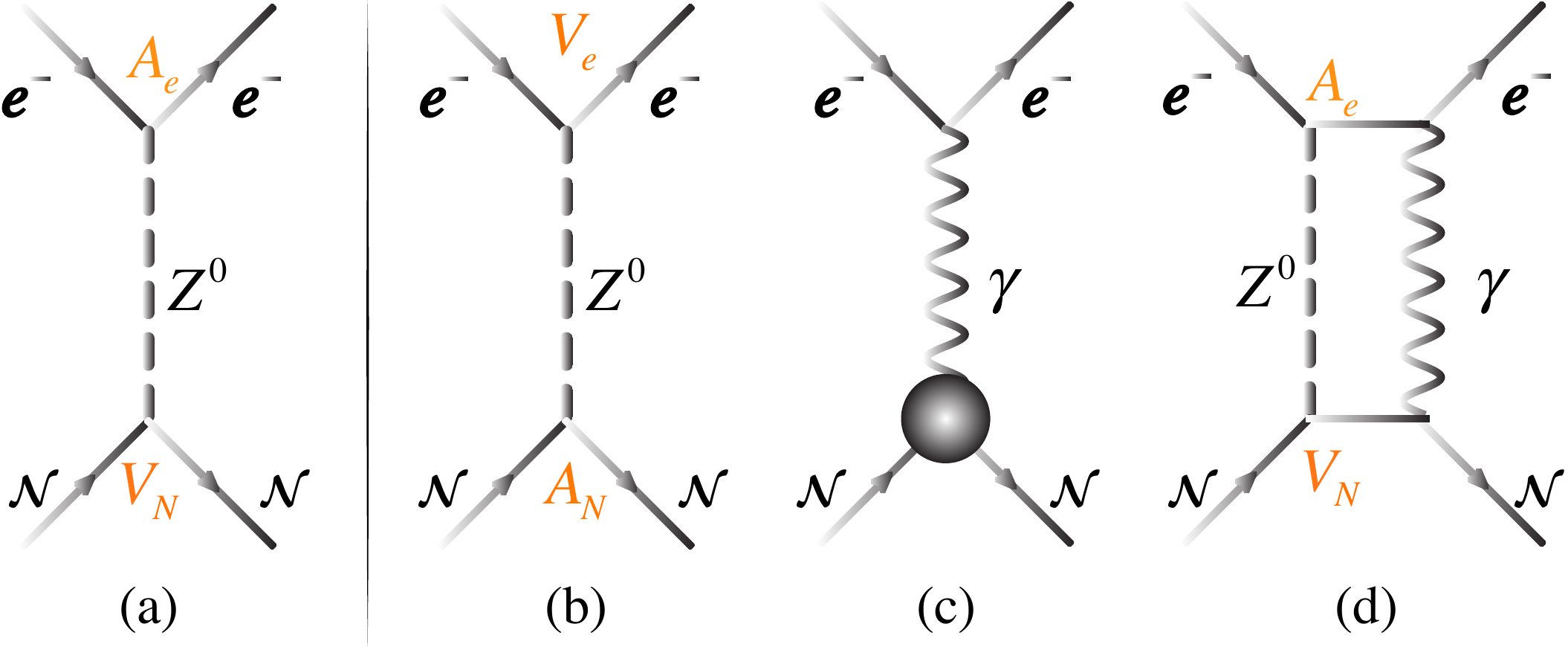}
\caption{ \label{Fig:APV-diagrams} (Color online)
Major diagrams contributing to the parity violation in atoms. $\mathcal{N}$ and $e^-$  label nucleons and atomic electrons. $A_{e,N}$ and $V_{e,N}$ denote axial-vector  and  vector currents.
(a) $Z$-boson exchange  between electron axial-vector and nucleon vector currents ($A_n V_e$);
(b) $Z$-boson exchange  between nucleon  axial-vector and electron  vector currents ($V_n A_e$);
(c) Electromagnetic interaction of atomic electrons with the nuclear anapole moment (shown as a blob);
(d) Combined effect of the $A_n V_e$  diagram (a) and hyperfine interaction.
The vertical line separates nuclear spin-independent (a) and spin-dependent (b)--(d) diagrams.
}
\end{center}
\end{figure}

Microscopically, APV is caused by the weak interaction mediated by the exchange of a $Z$ boson. Since the range of this interaction is $\sim \hbar/ (m_\textrm{Z} c)\approx 2 \times 10^{-3}\, \mathrm{fm}$ [$m_\textrm{Z} \approx 91 \, \mathrm{GeV}/c^2$ is the mass of the $Z$ boson], it is essentially a contact interaction on the scale of atomic distances.
% Weak interaction being short-ranged is essentially contact in APV since a typical momentum transfer $q$ in electron interaction with a nucleus of  size $r_p \sim \mathrm{fm}$, $q \sim \hbar/r_p \approx 0.2 \, \mathrm{GeV}/c$, is much smaller than $m_Z c$, where  $m_Z \approx 91 \, \mathrm{GeV}$ is the Z-boson mass~\cite{PDG-2014}.
The relevant contact contribution to the SM Hamiltonian density reads \cite{Mar95}
\begin{equation}
{H_\mathrm{PV}} = \frac{{{G_\mathrm{F}}}}{{\sqrt 2 }}\sum_{q} {\left( {{C_{q}^{(1)}}\,\bar e{\gamma _\mu }{\gamma _5}e\,\bar q{\gamma ^\mu }q + {C_{q}^{(2)}}\,\bar e{\gamma _\mu }e\,\bar q{\gamma ^\mu }{\gamma _5}q} \right)} \,,
\end{equation}
where the Fermi constant $$G_\mathrm{F} \approx  1.17 \times 10^{-5} (\hbar c)^3 \, \mathrm{GeV}^{-2} = 2.22 \times10^{-14}  \,\mathrm{a.u.}$$  determines the overall strength of the weak interaction,
the summation is over quark flavors, $q = \{ u,d,s,...\}$, $e$ and  $q$ are field operators for electrons and quarks respectively,
$\gamma _\mu$ are Dirac matrices, and $\gamma _5$ is the Dirac matrix associated with pseudoscalars.

The  coupling of the electron axial-vector currents to the quark vector currents is parametrized
by the constants $C_{q}^{(1)}$;
the constants
${C_{q}^{(2)}}$ describe the coupling of the electron vector currents to quark axial-vector currents. These interactions and constants could be further combined into couplings to protons and neutrons of atomic nuclei~\cite{MarcianoSanda1978}, e.g.,
\begin{eqnarray*}
C_{p}^{(1)}&=&2C_{u}^{(1)}+C_{d}^{(1)},\\ C_{n}^{(1)}&=&C_{u}^{(1)}+2C_{d}^{(1)},
\end{eqnarray*}
reflecting the quark composition of nucleons. Explicitly in terms of the Weinberg angle $\theta_\textrm{W}$: \begin{eqnarray*}
C_p^{(1)}& = &\frac{1}{2}\left( {1 - 4\,{\sin }^2}{\theta_\mathrm{W}}\right),\\
C_n^{(1)} &=&  - \frac{1}{2},\\ C_p^{(2)} &=&  - C_n^{(2)} = {g_A}C_p^{(1)},
\end{eqnarray*}
where $g_A \approx 1.26 $  is the scale factor accounting for the partially conserved axial vector current and $\sin^2 \theta_\mathrm{W}  = 0.23126(5)$~\cite{PDG}. Since $\sin^2\theta _\mathrm{W} \approx 1/4$, the $C_n^{(1)}$ contribution dominates ${H_\mathrm{PV}}$ except for the $^1$H atom.

The main diagrams contributing to PNC processes in atoms are shown in Fig.~\ref{Fig:APV-diagrams}.
The
$H_\mathrm{PV}$ terms discussed above are illustrated by
diagrams (a) and (b). In addition,  there is also a contribution from the nuclear anapole moment (c) and a combined effect of $Z$-boson exchange and hyperfine interaction (d).
The effective weak Hamiltonian arising from diagram (a) does not depend on the nuclear spin, while that from the set of diagrams (b)--(d) does. We will consider the former in Sec.~\ref{Sec:NSI} and the latter in  Sec.~\ref{Sec:NSD}.

%The absence of charged leptons in $\nu + N$  scattering experiments carried out in years 1973-1975 was consistent with
%the existence of neutral currents that interact through $Z$ exchange,
% which one of the original predictions of the standard model.
%Experiments to detect parity violating effects of $Z$ boson exchange on atoms
%were discussed Bouchiat and Bouchiat \cite{BouBou74,BouBou74a,BouBou74b,
%BouBou74c,BouBou74d}. The first positive evidence
%for the parity violating effects of $Z$ boson exchange on atoms were found
%reported in experiments on bismuth atoms reported by Barkov and Zolaterov In the same year and effects of $Z$ exchange in inelastic electron-deuterium scattering were reported by Prescott et al \cite{PreAtwCot78}. Direct observations of the charged $W_\pm$ boson and neutral $Z$ boson were not made until 1983 when they were observed at CERN's
%proton-–antiproton collider.
%T
%
%
%It should be noted that
%\vspace{1ex}
%
%The material in this section will be based on Refs....

\subsection{Nuclear-spin independent effects}
\label{Sec:NSI}
\subsubsection{Overview}
\label{NSI-1}
The dominant contribution to parity violation in atoms arises from the electron
axial-vector -- nucleon-vector term in $H_\text{PV}$, Fig.~\ref{Fig:APV-diagrams}(a).
If we treat
 the nucleon motion non-relativistically,  average over the nucleon distribution, and neglect the difference between proton and neutron distributions, we reduce the  corresponding part of
$H_\mathrm{PV}$ to  an effective weak Hamiltonian in the electron sector
\begin{equation}
{H_\mathrm{W}} = {Q_\mathrm{W}}\, \frac{{{G_\mathrm{F}}}}{{\sqrt 8 }}{\gamma _5}\,{\rho}\left( r \right) \,,
\label{Eq:HW}
\end{equation}
where $\rho\left( r \right)$ is the nuclear density and $Q_\mathrm{W}$ is a nuclear weak charge.  The non-relativistic limit of
the operator ${\gamma _5}\,{\rho}\left( r \right)$ is
$$
\frac{1}{2c}
 \left[ 2\rho(r) (\bs{\sigma} \cdot \mb{p})  -i (\bs{\sigma} \cdot \nabla \rho)   \right] \,,
 $$
 where $\mb{p}$ is the linear momentum operator and $\bs{\sigma}$ are electron Pauli matrices.

The nuclear weak charge $Q_\mathrm{W}$ entering the effective weak Hamiltonian is
\begin{equation}
{Q_\mathrm{W}} \equiv {{ 2 Z }}\,C_p^{(1)} + 2N\,C_n^{(1)}, \nonumber
\end{equation}
where $Z$ and $N$ are the numbers of protons and neutrons in the nucleus. Electrons predominantly couple to neutrons and ${Q_\mathrm{W}}\approx -N$. This is a ``tree-level'' [or the lowest-order] value of   ${Q_\mathrm{W}}$; more accurate values include SM radiative corrections~\cite{Mar95}, which are typically a few percent and can be computed to high accuracy. A major theme in APV is a comparison of the extracted  ${Q_\mathrm{W}}$ with its SM calculated value: a difference between the two values can indicate physics beyond the SM.

$H_\mathrm{W}$ is a pseudo-scalar operator with  its matrix element accumulated inside the nucleus. Its largest  matrix element is between $s_{1/2}$ and $p_{1/2}$ atomic orbitals.  Since  parity is no longer a good quantum number, yet the total angular momentum $J$ is conserved, atomic states of nominal parity acquire admixtures of state of opposite parity  with the same $J$. The relative size of the admixture is governed by the ratio of the matrix element of $H_\mathrm{W}$ [typically $\propto Z^3$] to the energy splitting  between the nearby states of opposite parity [typically $\approx 1$ a.u.].
%$\Delta E_\mathrm{opp. parity}$
%The figure of merit $Z^3/\Delta E_\mathrm{opp. parity}$ for several  recent and ongoing APV experiments is plotted in Fig.~\ref{XXX}. {\bf Need to make Fig.}

Since $G_\mathrm{F} \approx 10^{-14}\, \mathrm{a.u.}$, matrix elements of $H_\mathrm{W}$ are exceptionally small [$\approx 10^{-11}$ a.u. for Cs] compared to  the typical $1\, \mathrm{a.u.}$ transition amplitudes in atomic physics. To amplify the PNC signal, all experiments rely on an interference technique, where the $H_\mathrm{W}$-induced amplitude $T_\mathrm{PV}$ is amplified by beating it against an allowed amplitude $T_0$. Indeed, if the total transition amplitude is $T_\mathrm{tot}=T_0 + T_\mathrm{PV}$, then the transition probability (amplitude squared) acquires an interference term  $T_0^* T_\mathrm{PV}  + c.c.$ and the experiments extract $T_\mathrm{PV}$ by measuring this interference term.

%By reversing  externally applied fields , the sign of the interference term can be flipped and it can be extracted from  the $|T_0|^2$ background. %(i.e., looking at the mirror image of the probed atom)

The first APV signal was observed by the Novosibirsk group in 1978 using the ``optical rotation'' technique in Bi \citet{BarZol78}.  This technique  is based on the interference between the APV and the allowed magnetic dipole (M1) transition amplitudes. Parity violation leads to optical activity, i.e.,  atoms interacting differently with left- and right-circularly polarized light. Thereby the polarization vector of linearly polarized light is rotated as the light passes through an atomic vapor.  The measured quantity, the rotation angle, is  proportional to the ratio of APV and M1 amplitudes. APV was measured in optical-rotation experiments  with $^{209}$Bi, $^{209}$Pb, and $^{205}$Tl~\cite{Phipp1996,WarThoSta93,FortsonPb1993,Stacey1991,EdwPhiBai95,VetMeeMaj95}.
%\subsubsection{Single-isotope APV}

An alternative to the optical rotation scheme is the Stark interference technique~\cite{BouBou75}, which we illustrate below using a $^{133}$Cs experiment~\cite{WooBenCho97} as an example.
This technique was used in Cs~\cite{WooBenCho97,Lintz2007}, Tl \cite{ConBucChu79} and Yb \cite{TsiDouFam09} APV experiments. Additional interference techniques are described in Sec.~\ref{Sec:APV:NewExpt}.

\subsubsection{Parity violation in cesium}
The measurement of APV in $^{133}$Cs   ~\cite{WooBenCho97}  is the most accurate to date, and supplemented with sophisticated atomic theory, it probes  the  SM low-energy electroweak sector with exquisite precision. %The same measurement also provided the first evidence  for the nuclear anapole moment which we discuss in Sec.~\ref{Sec:NAM}.

An alkali-metal atom with 55 electrons; Cs has a single valence electron outside a tightly-bound  Xe-like core: its ground electronic level is designated [Xe]$6s$~$^2S_{1/2}$, sometimes called $6S_{1/2}$.   We focus on the optical transition between a $6S_{1/2}$ ground state  and an excited state of the same parity, $7S_{1/2}$. This  transition is E1-forbidden due to the parity selection rule: $\left\langle {6{S_{1/2}}} \right|D\left| {7{S_{1/2}}} \right\rangle =0$. The weak interaction leads to an admixture of states of opposite parity: $P_{1/2}$ states mix with the $S_{1/2}$ states, leading to a small E1 transition amplitude\footnote{It is conventional to define $E_{PV}$ parity violating amplitude as the transition matrix element $\left\langle {{\Psi _f}}\right|{\bf{D}}\left| {{\Psi _i}}\right\rangle$  between the states with the
with maximum values of the magnetic quantum numbers $m$.}, $E_{PV}$, of magnitude
$E_{PV}  \approx 10^{-11}\, \mathrm{a.u.}$

%$E_{PV} = \left\langle {\overline {6{S_{1/2}}} } \right|D\left| {\overline {7{S_{1/2}}} } \right\rangle \sim 10^{-11}\, \mathrm{a.u.}$, where %$\overline {6{S_{1/2}}}$ designates a state that is predominately $6{S_{1/2}}$ but has an admixture of $P$ states.

The Stark interference technique mentioned in Sec.~\ref{NSI-1} is used to amplify the parity-violating signal.
Application of an external electric field $\Evec$ induces an additional admixture of $P$ states.
This provides a strong E1 pathway with a transition amplitude $\beta \mathcal{E}$, where $\beta$ is the vector transition polarizability. The optical excitation rate  for the $6S_{1/2} - 7S_{1/2}$ transition is proportional to the square of the transition amplitude, $\beta^2 \mathcal{E}^2 + ( \beta \mathcal{E} E_{PV}+c.c)$, where the term quadratic in $E_{PV}$ is negligible. By changing direction of the electric field, the excitation rate can be modulated, and the PNC amplitude $E_{PV}$ can be isolated.

The nuclear spin of $^{133}$Cs is $I=7/2$, so each of the $S_{1/2}$ electronic states is split into $F=3$ and $F=4$  hyperfine components. Measuring the transition amplitudes between the different hyperfine states enables one to separate nuclear spin-dependent and spin-independent effects.  Multiple reversals of the electric field, magnetic substates and laser polarization are used to further isolate the APV effect.
The measured quantity is the ratio $R_\mathrm{Stark}=\mathrm{Im}(E_{\rm{PV}})/\beta$ for $F=3 \rightarrow F=4$ and $F=4 \rightarrow F=3$  transitions between hyperfine states.

A first measurement of  $R_\mathrm{Stark}$, accurate
to 10~\%,  was performed by the Paris group~\cite{BouGuePot84},  who ultimately reached an accuracy of 2.6~\%~\cite{Lintz2007}. A  series of measurements by the JILA group culminated in a determination of $R\ts{Stark}$ with an accuracy of  0.35~\%~\cite{WooBenCho97}.  The JILA measurements also resolved the difference between $R_\mathrm{Stark} (6S_{F=3} \rightarrow 7S_{F=4})$ and $R_\mathrm{Stark} (6S_ {F=4} \rightarrow 7S_{F=3})$, providing the first signature of a nuclear anapole moment. This is discussed further in Sec.~\ref{Sec:NSD}.

The nuclear-spin-independent parity-violating amplitude  is extracted from the measured  $R_\mathrm{Stark}$ and $\beta$~\cite{BenWie99}:
$$\mathrm{Im}(E_{\mathrm{PV}} )= -0.8374(31)_\mathrm{exp}(21)_\mathrm{th} \times 10^{-11}\mathrm{a.u.} $$ Extraction of the weak charge $Q_\mathrm{W}$ requires calculations of an atomic structure factor $k_{\rm{PV}}$, defined as
\begin{equation}
 E_{\rm{PV}} = k_{\rm{PV}} Q_\mathrm{W} \,.
 \label{e1-APV}
\end{equation}
Reaching theoretical accuracy in $k_{\rm{PV}}$ equal to or better than the experimental accuracy of 0.35~\%
has been a challenging task. In fact, theoretical calculations of $k_{\rm{PV}}$ and extraction of the weak charge from the
 Cs APV experiment ~\cite{WooBenCho97} has been a subject of
controversy and lively activity over the past 15 years.
%In 1997, the  0.35\%-accurate measurement of APV in Cs has been
%carried out~\cite{WooBenCho97}.
At the time of the 1997 APV measurement, the
accuracy of the theoretical  calculations~\cite{DzuFlaSus89,BluJohSap90}
was estimated to be 1~\%.
New atomic lifetime and polarizability data reported by 1999  improved the agreement of theory and experiment
and the theoretical uncertainty was reduced to 0.4~\% \cite{BenWie99}.
The resulting
value of $Q_\mathrm{W}$ differed by $2.5\,\sigma$ from  the prediction of the SM \cite{BenWie99}.
That discrepancy  prompted
substantial interest in the particle physics community
\cite{Ram99,CasCurDom99,Ros00,Ros02}. At the same time, the
reduced theoretical uncertainty raised the question of whether some ``small''
sub-1~\%  atomic-structure effects could be the reason for the
discrepancy. Several groups  contributed to understanding such small corrections~\cite{Der00,DerPor02,DzuHarJoh01,Sushkov2001,KozPorTup01,JohBedSof01,KucFla02,SapPacVei03,MilSusTer02,MilSusTer03,ShaPacTup05} [reviewed by \citet{DerPor07}].
The dominant corrections were found to be due to the Breit interaction, radiative QED effects, and the neutron skin correction, which is the
difference between the well-known proton nuclear distribution and the relatively poorly known neutron distribution that dominates $H_\mathrm{W}$. This issue is described in Sec.~\ref{Sec:APV:skin}. In 2005, these corrections
 essentially reconciled   APV in Cs with the SM, with theoretical uncertainty standing at 0.5~\%, still larger than the experimental error bar.

With the small corrections sorted out, major theoretical effort turned to more accurate calculation of the dominant
 many-body Coulomb correlation contribution to the structure factor $k_{\rm{PV}}$ ~\cite{PorBelDer09,PorBelDer10}.
State-of-the-art calculations
 built upon the {\em ab initio} relativistic coupled-cluster scheme of \citet{BluJohSap90} and included a large class of higher-order many-body effects. All relevant atomic properties were reproduced at a level better than 0.3~\%, leading to an overall 0.27~\% theoretical uncertainty in the structure factor $k_{\rm{PV}}$  \cite{PorBelDer09,PorBelDer10}. The final value of the extracted $Q_\mathrm{W}$ was in essential agreement with the SM value.
 
 Recent reevaluation of some sub-leading correlation contributions to $k_{\rm{PV}}$ by \citet{DzuBerFla12} [contributions of the core and highly-excited states] raised the theoretical uncertainty back to 0.5~\%, slightly shifting the value of $k_{\rm{PV}}$ from that of \citet{PorBelDer09,PorBelDer10}, but maintaining agreement with the SM. The difference in the core contribution 
 [$0.0038 \times 10^{-11} i(-Q_\textrm{W}/N)$\,a.u.] came from the inclusion of the core polarization [i.e. the change in the self-consistent
Hartree-Fock potential due to an  electric dipole
field of the external photon and the weak interaction of
atomic electrons with the nucleus] and  Brueckner type correlations which describe the correlation
interaction of the external electron with the atomic core \cite{DzuBerFla12}.
 One of us, A.D., thinks that the correction to the 
 contribution of highly-excited states \citet{DzuBerFla12}  may have come  from the use of many-body intermediate states by \citet{DzuBerFla12}  that is inconsistent with the one employed by~\citet{PorBelDer09}, as the summation over intermediate states while evaluating $k_{\rm{PV}}$ must be carried out over a complete  set and thereby the results of \citet{DzuBerFla12}  require revision. This matter remains unresolved at  present but new methods are being developed to address it.
The ever-increasing power of computation is anticipated to bring further improvements in the atomic-structure analysis.

\subsubsection{Implications for particle physics and the dark sector}
\begin{figure}[t]
\begin{center}
\includegraphics[width=0.95\columnwidth]{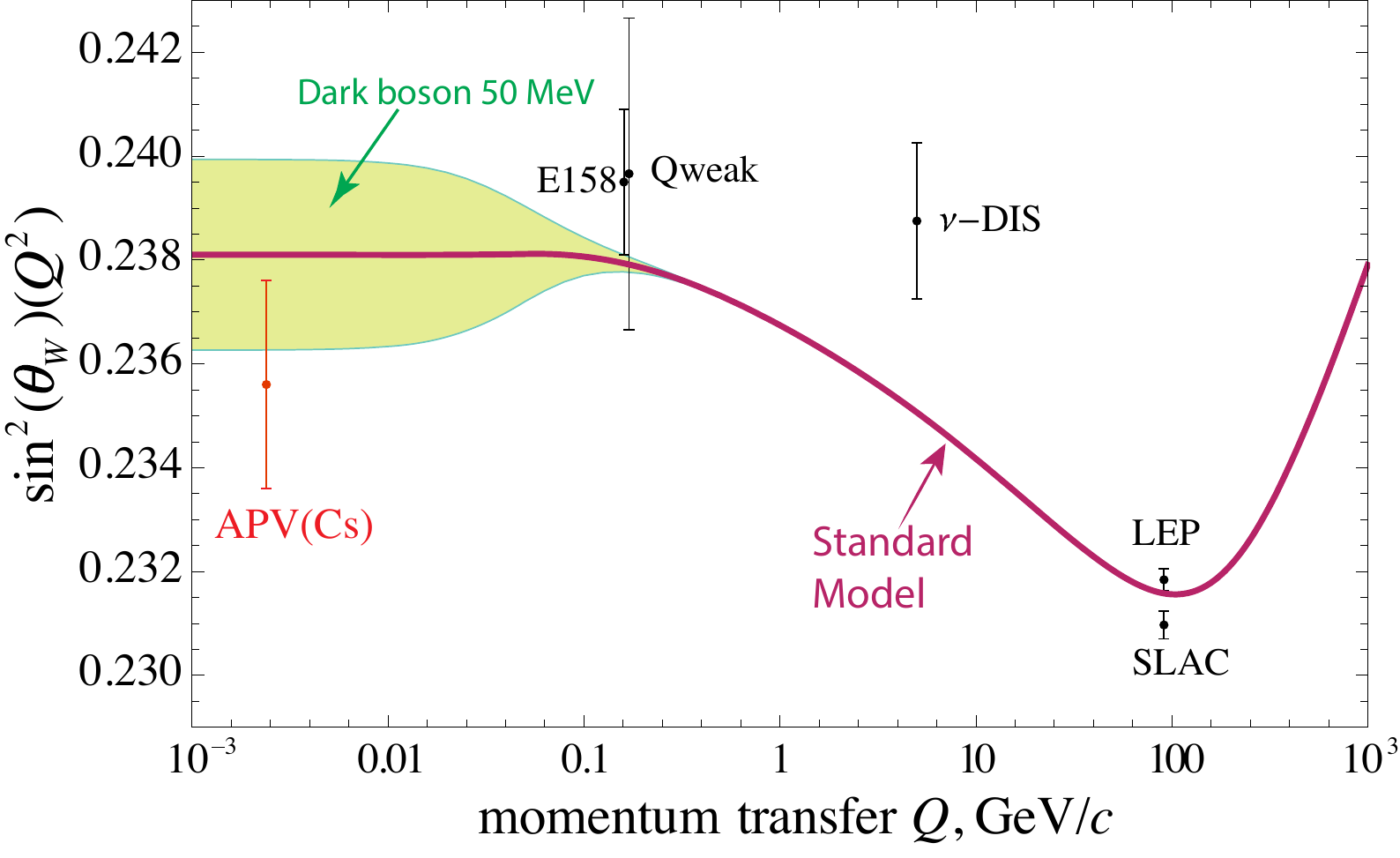}
\caption{ \label{Fig:running} (Color online)
Running of the weak mixing angle  with momentum transfer $Q$. The solid red curve is the SM prediction.
The Cs APV result is supplemented with data from particle physics experiments: E158, M{\o}ller [electron-electron] scattering; Q-weak, PNC electron-proton scattering, $\nu$-DIS, deep inelastic scattering; LEP and SLAC results.   Area colored in yellow comes from one of the ``new physics'' scenarios~\cite{Davoudiasl2014}: a dark boson of mass 50 MeV. The colored area is limited by constraints on model parameters that would explain
the discrepancy~\cite{Bennett2006} between the muon's experimental anomalous magnetic moment and the SM prediction.
Adopted from \citet{Davoudiasl2014}.
}
\end{center}
\end{figure}

Atomic parity violation yields the most accurate up-to-date probe of the low-energy electroweak sector of the SM, playing
 a unique role  complementary to that of
high-energy physics experiments. Figure~\ref{Fig:running} illustrates the energy dependence (or ``running'') of the electroweak interaction
and places APV in the context of other precision electroweak measurements. The solid curve  is the SM prediction for the dependence of $\sin^2 \theta_\mathrm{W}$ on the four-momentum transfer $Q$. At low $Q$, it describes the evolution primarily through quark loops with small leptonic corrections; the minimum at 100~GeV/c  occurs when the W$^{+}$W$^{-}$ loop starts contributing substantially at $Q \sim 2 m_\mathrm{W}$, $m_\mathrm{W}$ being the mass of W bosons.
%While the Cs APV result comes from the latest paper~, it is a subject to revision.
The Cs APV result is placed at $Q=2.4\, \mathrm{MeV/c}$~\cite{BouPik83}, which is roughly $\hbar/(a_0/Z)$, where $a_0$ is the Bohr radius.
This relates the momentum to the
radius of the innermost electron shell of the Cs atom.
Together with the results of high-energy collider experiments, APV demonstrates the validity of the
running
of the electroweak interaction over an energy range spanning five orders of magnitude.
An alternative and more detailed plot in a different  %($\overline{\mathrm{MS}}$)
renormalization scheme can be found in~\citet{PDG}; this Particle Data Group review also provides further discussion of relevant particle physics experiments.

The transitions  measured
 in APV studies are typically on a 1 eV energy scale, yet the exquisite accuracy of
those measurements and calculations  probes minute contributions of the sea of virtual  particles
at a much higher mass scale, including candidates beyond the SM. For example, APV is uniquely
sensitive to extra  $Z$ [$Z^\prime$] bosons  predicted in grand
unified theories, technicolor models, SUSY, and string theories~\cite{Langacker2009}.  Limits on their masses set by APV are at the TeV scale~\cite{PorBelDer09}, and these were only recently improved upon by direct searches  at the Large Hadron Collider~\cite{PDG}.
Such $Z^\prime$ bosons can also mediate new spin-dependent interactions, see Sec.~\ref{Sec:ExoticSpin:TheoreticalMotivation:hidden-photons-Z-bosons}.

Low-energy precision measurements are also  uniquely sensitive to possible ``dark forces'' which are motivated by the intriguing possibility of a ``dark sector'' extension to the SM \cite{And12}. Dark sector is understood broadly as new physics constituents and forces that couple to SM fields weakly or do not couple at all, so that the current experiments are blind to their existence.  Dark matter may be a small part of the dark sector,  or many dark sectors could exist, each with their own forces and constituent particles. Dark matter may be accompanied by heretofore unknown gauge bosons (dark force carriers) which can couple dark matter particles and ordinary particles with exceptionally weak couplings. Modern colliders can be blind to such new forces, even though the mass  of the dark force carriers can be quite small. This is because the cross-sections of  relevant processes for ordinary matter are so small that the dark force events are  statistically insignificant and are discarded in high-energy experiments.

Light-mass, weakly-coupled dark-sector particles that interact with ordinary matter have been proposed as explanations of astronomical anomalies \cite{Fay04,Ark09} as well as discrepancies between the calculated and measured magnetic moment of the muon  \cite{Fay07,Pos09}.
%Such interactions would be inevitably below the weak force scale, ergo, the dark sector has so far escaped detection.
There are several proposed inroads into the detection of weakly-coupled particles and their associated dark forces~ \cite{Essig:2013lka}. One such example is the dark photon \cite{Hol86}, discussed in Sec.~\ref{Sec:LightDarkMatter}, that is hypothesized to be a massive particle which couples to electromagnetic currents just like the photon does. In addition, dark $Z$ bosons have been proposed \cite{Davoudiasl2014} that couple to the weak neutral currents, i.e., their interactions are parity violating. In a sense, dark photons are massive photons, while dark $Z$ bosons are light versions of $Z$ bosons. In Fig.~\ref{Fig:running}, the yellow-colored area represents the limits on dark $Z$ bosons  in the model by~\citet{Davoudiasl2014}; the unique sensitivity and discovery potential of APV are apparent.  We also point the reader to \citet{BouPik83}, who considered APV mediated by a light gauge boson.
%{\bf Perhaps move to EDM section: Dark force contributions to atomic EDMs have been considered by~\citet{GhaDer15}.}

%Additionally, the latest results from APV  raise limits on the masses of extra $Z$ bosons~\cite{ErlLan00,Riz06,Lan08} ($Z^\prime$) previously derived from direct searches at the Tevatron collider. The raised bound on the $Z^\prime$ masses carves out a lower-energy part of the discovery reach of the Large Hadron Collider.

%{\bf This paragraph and above should refer to DM section which should explain what DM virial distribution/ energy density are.}
Another novel possibility for probing the dark sector  with APV experiments is associated with the search for axions and axion-like particles~\cite{Stadnik2014,RobStaDzu14a,RobStaDzu14}.
 The axion [see also  Secs.\ref{Sec:ExoticSpin} and \ref{Sec:LightDarkMatter}] is a pseudo-scalar particle introduced by~\citet{Peccei1977a} to solve the strong $CP$ problem, which is the ``unnatural'' smallness of the $\bar{\theta}_{\rm QCD}$ parameter in the QCD Lagrangian that quantifies the amount of $CP$-violation \cite{Weinberg1976}, see Sec.~\ref{Sec:EDM} for more detail. Axions are also  viable dark matter candidates \cite{Ber13}. The relevant pseudo-scalar coupling is $$\mathcal{L'} =
 i \zeta_1 m_\mr{e}\, \phi  \,  \bar e \gamma_5  e  +
\zeta_2 (\partial_\mu\phi)\, \bar e  \gamma^\mu\gamma_5 e,$$ where  $\zeta_1$, $\zeta_2$  are the coupling strengths. The  spin-0 bosonic field
\begin{equation}
\phi(r,t)=\phi_0 \cos( \omega_\phi t - \mathbf{k}\cdot \mathbf{r}+...)
\label{l99}
\end{equation}
 has an amplitude $\phi_0$ related to the DM energy density and it oscillates at the particle Compton frequency.
 $\omega_\phi = m_\phi c^2/\hbar$ for a particle of mass $m_\phi$ and $...$ in Eq.~(\ref{l99}) stands for an unknown phase.
 The $k$-vectors follow the virial distribution of DM velocities. This interaction induces small oscillations in the APV amplitude at the Compton frequency. A power spectral density of the measured time series of APV amplitude would exhibit a characteristic peak at the Compton frequency with a characteristic strongly asymmetric profile derived by~\citet{Derevianko2016a}. Such proposals are complementary to searches for axion-induced P-conserving M1 transitions~\cite{Sikivie2014}.

%+  \zeta_2 m_\mr{e}\, \phi  \,  \bar e \gamma_0 \gamma_5  e

\subsubsection{Isotopic chains and neutron skin}
\label{Sec:APV:skin}
All APV studies to date have been conducted with a single isotope and required the theoretical calculation of a
$k_{\rm{PV}}$ factor in Eq.~(\ref{e1-APV}).  Considering challenges faced by such calculations, an alternative approach  was proposed
by~\citet{DzuFlaKhr86}. The  idea was to form a ratio
$\mathcal{R}$ of the PNC amplitudes for two isotopes of the same
element. Since the factor $k_{\rm PV}$ remains the same, it cancels out in the ratio.
However, \citet{ForPanWil90} pointed out a conceptual limitation to this approach -- an enhanced sensitivity of
possible constraints on ``new physics'' to uncertainties in the
{\em neutron} distributions. This problem is usually
referred to as that of  ``neutron skin.''
The neutron skin is defined  as the difference between the root-mean-square radii $R_n$ and $R_p$  of neutron and proton distributions. While nuclear charge densities (i.e., proton distributions) have been accurately measured
with electron scattering,  and the mean-square charge radii are well determined from isotope-shift measurements, neutron distributions, while expected to largely follow the proton distributions, are poorly known \cite{BroDerFla09}.

%The value of neutron skin is sensitive to the parameters of the  model-dependent nuclear-structure
%calculations.

Even in the interpretation of the most accurate to date single-isotope measurement in Cs~\cite{WooBenCho97},  neutron skin was a point of concern, as the induced uncertainty was comparable to the experimental uncertainty in the APV amplitude~\cite{PolWel99,VreLalRin00}.
%From  {\em nonrelativistic}
%nuclear-structure calculations  \citet{PolWel99}
%concluded for $^{133}$Cs $\Delta R_{np}/R_{p} = 0.016$ or 0.022 depending on the
%model of nuclear forces.
%The calculations \cite{VreLalRin00,PanDas00} of nuclear distributions were
%{\em relativistic} and the corrections as twice as large
%$\Delta R_{np}/R_{p} = 0.043-0.053$ were found.
%bab The question was settled in Ref.~\cite{DerPor02},
The question was addressed  in \citet{DerPor02},
where  empirical antiprotonic-atom data fit for the neutron skin was used~\cite{TrzJasLub01},
and the associated uncertainty in the neutron skin contribution to $k_{\rm{PV}}$ was substantially reduced. An analysis for multiple
isotopes~\cite{BroDerFla09} shows that in Fr and Ra$^{+}$, the present uncertainty in neutron skin would limit extraction of weak charge to 0.2~\% accuracy.

%A closer examination of Eq.~(\ref{Eq:HW}), focusing on the difference between the proton and neutron distributions, leads to the substitution of $Q_{W}$ with
%$
%\bar{Q}_{W}=                                                    % (3)
%-N \, q_n + Z \, q_p \ (1 - 4 \,{\rm sin}^2 \theta_W)
%\label{QW}
%$
%The quantities $q_n$ and $q_p$ depend on the neutron and proton distributions
%convoluted with atomic wavefunctions:
%$
%q_n = 1+f_n\left( R_n/R_p\right) \, .
%$
%In the ``sharp edge'' model of nuclear density distribution,
%%------------------------------------------------------------------
%\begin{eqnarray}
%f_n\left( \frac{R_n}{R_p}\right) \approx
%- \frac{3}{70} \left( \alpha Z \right)^2
%     \left[1+5 \left( \frac{R_n}{R_p} \right)^{2} \right] \,.        % (11)
%\label{Eq:qn}
%\end{eqnarray}

The question of determining neutron skin is of interest in its own right,
for example, as it relates to the equation of state for neutron stars.
The $^{208}$Pb   Radius   Experiment   (PREX)  at  Jefferson Lab  (JLAB) \cite{PREX-2012} uses PNC asymmetry in elastic scattering
of electrons from $^{208}$Pb with the goal of  measuring $R_{n} $ to
 1~\% accuracy. \citet{BroDerFla09} examined the question of whether  neutron skin can be probed with APV.
 The neutron skin correction is  about 0.2~\% for Cs APV and 0.6~\% for  Fr and Ra$^+$.  Yb, francium (Fr, $Z$=87), and radium (Ra, $Z=88$) have a number of isotopes
 available for APV experiment and highly accurate measurements of APV in two isotopes may, in principle, be used to extract
 the neutron skin data.
  In isotopic chain experiments the
largest effect is attained for a pair of isotopes comprised of the lightest (neutron-depleted) and the heaviest (neutron-rich) isotopes of the chain. For Yb the accuracy in the ratio determination should be smaller than 0.2~\% [0.3~\% for Fr and Ra$^+$] just to detect the effects of having different rms neutron radii for two isotopes. This may prove more  challenging than the single isotope approach, unless common systematics in measuring APV amplitudes in individual isotopes cancel out in the ratio. APV measurements on Yb~\cite{Leefer2014} also benefit from a 100-fold enhancement in $E_{\rm{PV}}$ compared to Cs; the enhancement is due to the presence of closely spaced opposite parity levels \cite{DeMille1995}.

While the single isotope measurements are sensitive to new physics associated with electron-neutron couplings, the isotopic ratios predominantly probe electron-proton (e-p) couplings~\cite{Ram99}. Bounds on the e-p new physics  can also  be directly established
from PNC electron scattering off protons in the Q-weak experiment at JLAB~\cite{Qweak2013}.
While it was previously argued~\cite{ForPanWil90,Ram99,DerPor02} that APV ratios, due to neutron skin uncertainties, are not competitive to such direct experiments,
 \citet{BroDerFla09}  showed that the induced neutron skin uncertainties for isotopes are highly correlated and tend to strongly cancel
 while forming $\mathcal{R}$. This observation makes APV isotopic ratio experiments a competitive tool in probing new physics e-p couplings,  provided the experiments can reach the required level of accuracy.

%, i.e., the uncertainty brought in by the difference in the nuclear proton and %neutron distributions.
%The problem was
%Here we show that the neutron skins in different isotopes are correlated; this leads to a substantial cancelation in the neutron skin induced uncertainties
%in the PNC ratios. The use of modern experimental data and nuclear calculations
%makes the isotopic ratio method a competitive tool in search for new physics beyond the standard model.
%
%
%
%At the same time this question is yet to be settled
%for on-going PNC experiments with unstable analogs of Cs: Fr~\cite{GomOroSpr06} and Ra$^{+}$ \cite{WanVerWil08}.
%%bab Below we provide the necessary calculation of these effects.
%In this letter we use results of recent advances in our theoretical
%and experimental understanding of
%neutron skins to address these important questions.

\subsection{Nuclear-spin-dependent effects and the  nuclear anapole moment}
\label{Sec:NSD}
\subsubsection{Overview}
The three nuclear-spin-dependent diagrams, Fig.~\ref{Fig:APV-diagrams}(b-d),  can be reduced to the
effective interaction in the electron sector
\begin{equation}
{H_{{\rm{NSD}}}} = \,\frac{{{G_\mathrm{F}}}}{{\sqrt 2 }}\left( {{\eta _{{\rm{axial}}}} + {\eta_{{\rm{NAM}}}} + {\eta_{{\rm{hf}}}}} \right)\, \left(\bs{\alpha } \cdot \,{\mb{I}} \right) \,{\rho} \left( r \right) \,,
\label{Eq:APV:HNSD}
\end{equation}
where $\bm{\alpha}$ is the velocity operator ($\alpha_i=\gamma_0\gamma^i$) for atomic electrons, $\rho$ is the nuclear density, and $I$ is the nuclear spin.  This contribution is only present for $I\ne 0$ isotopes and open-shell atoms.
The dimensionless parameters $\eta$ primarily come from nuclear physics.
In the ideal nuclear shell-model limit these coefficients are associated with the properties of the valence nucleon $\mathcal{N}$:  $\mathcal{N}=p$ or $\mathcal{N}=n$ depending on a specific nucleus. The non-relativistic reduction of the operator $\left(\bs{\alpha } \cdot \,{\mb{I}} \right) \,{\rho} \left( r \right)$ in ${H_{{\rm{NSD}}}}$  reads
%$(\bm{I} \cdot \nabla \rho{r}) $
$$
 \frac{1}{2c} \left[ 2\rho(r) (\mb{I}\cdot\mb{p}) - i (\mb{I} \cdot \nabla \rho) +\bs{\sigma} \cdot (\nabla{\rho} \times \mb{I}), \right] \,.
$$
The coefficient $\eta _\mathrm{axial}$ is associated with the $Z$ exchange interaction from nucleon axial-vector ($A_n V_e$) currents, Fig.~\ref{Fig:APV-diagrams}(b), and its nuclear shell-model value is~\cite{Flambaum1980}
\begin{equation}
\eta _\mathrm{axial} =  - C_\mathcal{N}^{(2)}\frac{{{\kappa_\mathcal{N}} - 1/2}}{{I(I + 1)}} \,,
\end{equation}
where the weak-interaction constants $C_{n,p}^{(2)}$ were introduced in Sec.~\ref{Sec:APV-intro} and
 $$\kappa_\mathcal{N} = (I+1/2) (-1)^{I+\ell_\mathcal{N} +1/2}$$ is the relativistic angular quantum number for the unpaired nucleon in a state with orbital angular momentum $\ell_\mathcal{N}$. Notice that this contribution is substantially suppressed compared to the $V_n A_e$ diagram~\ref{Fig:APV-diagrams}(a)
because  $$|C_\mathcal{N}^{(2)}/C_n^{(1)} |= g_A (1- 4 \sin^2 \theta_\mathrm{W})  \approx 0.1$$  and only the unpaired nucleon contributes to Fig.~\ref{Fig:APV-diagrams}(b)
whereas all nucleons coherently contribute to Fig.~\ref{Fig:APV-diagrams}(a).
%As to the numerical values of $\eta _\mathrm{axial}$, for $^{133}$Cs with $1g_{7/2}$ valence proton, the above formula results in $\eta _\mathrm{axial} = 0.010$ consistent with the result  $\eta _\mathrm{axial} = 0.014$ of a more sophisticated nuclear shell-model estimate~\cite{HaxLiuRam01}.

The $\eta_\mathrm{NAM}$ coefficient parameterizes the nuclear anapole moment (NAM) contribution to atomic parity violation. It is illustrated in  Fig.~\ref{Fig:APV-diagrams}(c) and discussed in Sec.~\ref{Sec:NAM}.
Parity violation in the nucleus leads to toroidal currents
that in turn generate a parity-odd, time-reversal-even (P-odd, T-even) moment, known as the nuclear anapole moment,  that couples electromagnetically to atomic electrons.
The nuclear shell model expression for the anapole moment ~\cite{FlaKhrSus84},
\begin{equation}
\eta _\mathrm{NAM} = 1.15 \times {10^{ - 3}}\frac{{{\kappa _\mathcal{N}}}}{{I(I + 1)}}{\mu _\mathcal{N}}\,{g_\mathcal{N}}{A^{2/3}}, \label{Eq:APV:eta-NAM}
\end{equation}
depends on the atomic number $A$, the magnetic moment $\mu _\mathcal{N}$ of the unpaired nucleon expressed in units of the  nuclear magneton, and the weak coupling constant $g_\mathcal{N}$.
Their values are $\mu_p \approx 2.8$, $\mu_n \approx -1.9$, $g_p\approx 5$, and $g_n \approx -1$.

The combined action of the hyperfine interaction and the spin-independent $Z$-exchange interaction from nucleon vector ($V_n A_e$) currents
leads to
the third nuclear-spin dependent parity violating effect,  Fig.~\ref{Fig:APV-diagrams}(d). This contribution is quantified by a parameter $\eta_\mathrm{hf}$.
An analytical approximation for $\eta_\mathrm{hf}$ was derived by ~\citet{FlaKhr85} and  values of
$\eta_\mathrm{hf}$ were  determined for various cases of  experimental  interest by \citet{Bouchiat1991-anapole} and \citet{Johnson2003}.
%Atomic many-body calculations \cite{Johnson2003} lead to the value of $\eta_\mathrm{hf}=  4.9 \times 10^{-3}$ for $^{133}$Cs that is 40\% smaller than from~\citet{BouPik91} but in agreement with \citet{FlaKhr85}.
\citet{Johnson2003}  also tabulated the values of  $\eta_\mathrm{hf}$ for microwave transitions between ground-state hyperfine levels in atoms of potential experimental interest.

Recently, \citet{Flambaum2016} pointed out a novel nuclear spin-dependent effect: the quadrupole moment of the neutron distribution leads to a tensor weak interaction that mixes opposite parity states in atoms  with total angular momentum difference $\le 2$. This effect should be carefully investigated in future work to see if it influences determination of the anapole moments from APV measurements. The effect is of interest on its own as a probe of the neutron distributions in nuclei~\cite{FlaDzuHar2017-APV-NQM}.
%, and should be measured in atoms as well as in molecules, where it is systematically enhanced due to the proximity of levels of opposite parity.
The atom or molecule should contain a nucleus with $I>1/2$, and there is an enhancement for heavy and deformed nuclei.

An outstanding question is the relative importance of the  nuclear spin-dependent contributions. The $\eta_\mathrm{hf}$ coefficient can be carefully evaluated and it is usually suppressed compared to $\eta_\mathrm{NAM}$  and  $\eta _\mathrm{axial}$.
Generically, because of the $A^{2/3}$ scaling, the anapole contribution dominates for heavier nuclei.
%The anapole contribution, because of the $A^{3/2}$ scaling, dominates for heavy nuclei.
%Because of the $A^{2/3}$ scaling, the anapole contribution
%dominates for odd-proton nuclei with $A \gtrsim 20$ and below we examine the anapole moments in detail.
For lighter   nuclei, the axial contribution is more important and  APV experiments can be a sensitive probe of $C_{n,p}^{(2)}$ electroweak parameters, providing a window on the $A_n V_e$ interactions that are typically studied with deep inelastic scattering~\cite{PVDIS2014}.  The boundary between the axial- and anapole-dominated regimes depends on  quantum numbers of the valence and type of the valence nucleon~\cite{DeMCahMur08}.
Values of  $C_{n,p}^{(2)}$ can set constraints on exotic new physics such as  leptophobic $Z^\prime$ bosons~\cite{Buckley2012}, while NAMs probe hadronic PNC.

%\begin{center}
%\begin{tabular}{cc}
% $^{133}$Cs: & $\kappa_\text{hf} =  0.0078$ \\
% $^{205}$Tl: & $\kappa_\text{hf} =  0.044$
%\end{tabular}
%\end{center}
%\vspace{2ex}

\subsubsection{Nuclear anapole moments as a probe of hadronic parity violation}
\label{Sec:NAM}
The traditional multipolar expansion of electromagnetic  potentials generated by a finite distribution of
currents and charges leads to  the identification of magnetic (MJ) and  electric (EJ)  multipolar moments~\cite{Jac99}.
Non-vanishing nuclear multipolar moments (charge E0, magnetic-dipole M1, electric-quadrupole E2, \ldots)  respect parity and  time reversal, i.e. they are
P-even and T-even,  and describe multipolar fields outside the finite distribution.
Weak interactions inside the nucleus lead to additional P-odd moments~\cite{Gray2010}; the leading moment
is referred to  as the anapole moment.  Zel'dovich and Vaks were the first to point out the possibility
of such a moment~\cite{Zeldovich1958-anapole}.

\begin{figure}[t]
\begin{center}
\includegraphics[width=0.8\columnwidth]{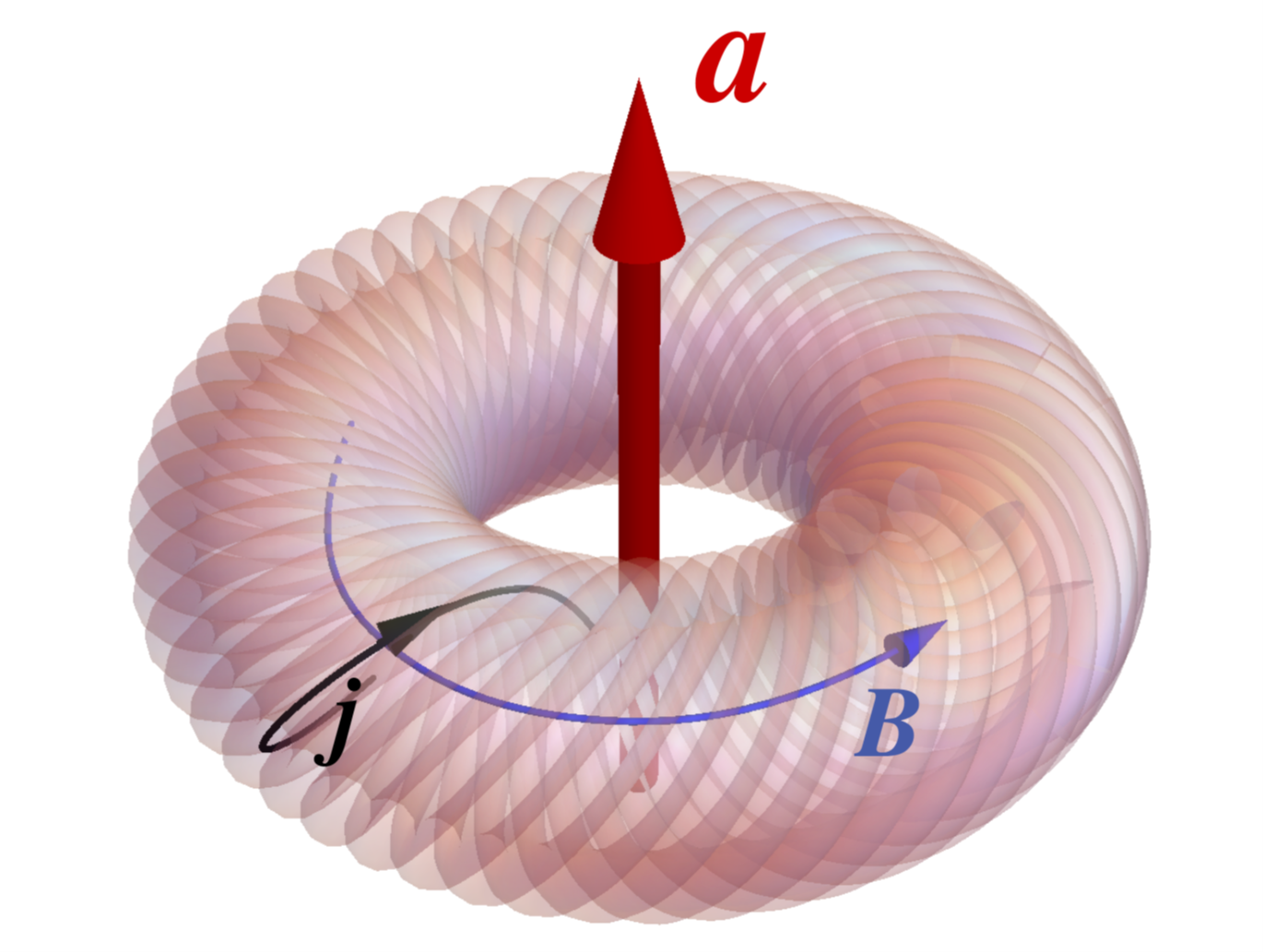}
\caption{ \label{Fig:anapole-tokamak} (Color online)
 The toroidal component of current density $\bm{j}$ produces anapole moment $\bm{a}$, with magnetic field  $\bm{B}$ that is entirely confined inside the ``doughnut''.
The azimuthal component of current density generates magnetic dipole moment aligned with $\bm{a}$, with its associated conventional dipolar magnetic field not shown.}
\end{center}
\end{figure}

The anapole moment  $\bm{a}$  of a current density distribution $\bm{j}(r)$ is defined as
\begin{equation}
 \bm{a} =   -\pi \int d^3r\, r^2\, \bm{j}(\bm{r}) , \label{Eq:APV:anapole-def}
\end{equation}
with magnetic vector potential $\bm{A}= \bm{a} \delta(\bm{r})$, leading to the electromagnetic coupling of electrons to the nuclear anapole moment,  $(\bm{\alpha} \cdot \bm{A})$. %It can be re-expressed in terms of magnetic fields $ \bm{a} = c/2 \int d^3r\, (\bm{r} \times \bm{B})$.
%The B-field outside of the distribution must vanish.
A classical
analog of the anapole moment is a Tokamak-like configuration shown in Fig.~\ref{Fig:anapole-tokamak}. The inner and outer parts of the toroidal
currents are weighted differently by $r^2$ in Eq.~(\ref{Eq:APV:anapole-def}), leading to a nonvanishing value of the anapole moment.
%The magnetic field is non-zero only inside the ``doughnut'', and namely  it is the  inner region of the doughnut that
%gives rise to the anapole moment.
Microscopically, a nuclear anapole moment  can be   related to a chiral distribution of nuclear magnetization caused by parity-violating nuclear forces~\cite{Bouchiat1991-anapole}. Due to the Wigner-Eckart theorem, the NAM (just as the nuclear magnetic moment)
is proportional to the nuclear spin $I$ so that
$$
 \bm{a} = \frac{G_\mathrm{F}}{{|e|} \sqrt{2}}\ \eta_\mathrm{NAM} \mb{I},
$$
defining the constant $\eta_\mathrm{NAM}$ in Eq.~(\ref{Eq:APV:HNSD}).
Atomic electrons interact with NAM only inside the nucleus, as is apparent from the classical analog, since the magnetic field is entirely
confined inside the ``doughnut''.  Another important observation is that the NAM  is proportional to the area of the toroidal winding,
i.e., $\propto (\mathrm{nuclear\,\, radius})^2 \propto A^{2/3}$, where $A$ is the atomic number, illustrating the trend in Eq.~(\ref{Eq:APV:eta-NAM}).

Microscopically, the  nuclear anapole arises due to  nucleon-nucleon
interaction, mediated by meson exchange, where one of the nucleon-meson
vertexes is strong and another is weak and P-violating. Thus, determination of anapole
moments from atomic parity violation provides an important window into
hadronic PNC~\cite{HaxWie01}. The innards of the anapole bubble in Fig.~\ref{Fig:APV-diagrams}(c) are shown in Fig.~7 of the review by~\citet{HaxWie01}.
The nuclear-physics approach is to characterize weak meson-nucleon
couplings in terms of parameters of Desplanques, Donoghue and Holstein (DDH) ~\cite{DDH1980}, who deduced SM estimates of their values.
These six hadronic PNC parameters are $f_\pi, h_\rho^{0,1,2}, h_\omega^{0,1}$,
where the subscript ($\pi,\rho,\omega$) indicates meson type
and the superscript stands for isoscalar (0), isovector (1), or isotensor (2).
We refer the reader to~\citet{HaxWie01}
for a detailed review of nuclear structure calculations of
NAMs within the DDH parameterization. The effective field theory parameterizations of hadronic PNC, an alternative to DDH, are also discussed~\cite{RamseyMusolf2006}, although NAM analysis in this framework remains to be carried out.
It should be pointed out that a more recent review~\cite{Haxton2013} omits the Cs result. These authors explain the omission by the fact that  the accuracy of the constraints on the nucleon-nucleon PNC interaction derived from the NAM experiments is  somewhat difficult to assess due to complex nuclear polarizability issues.

\begin{figure}[t]
\begin{center}
\includegraphics[width=0.9\columnwidth]{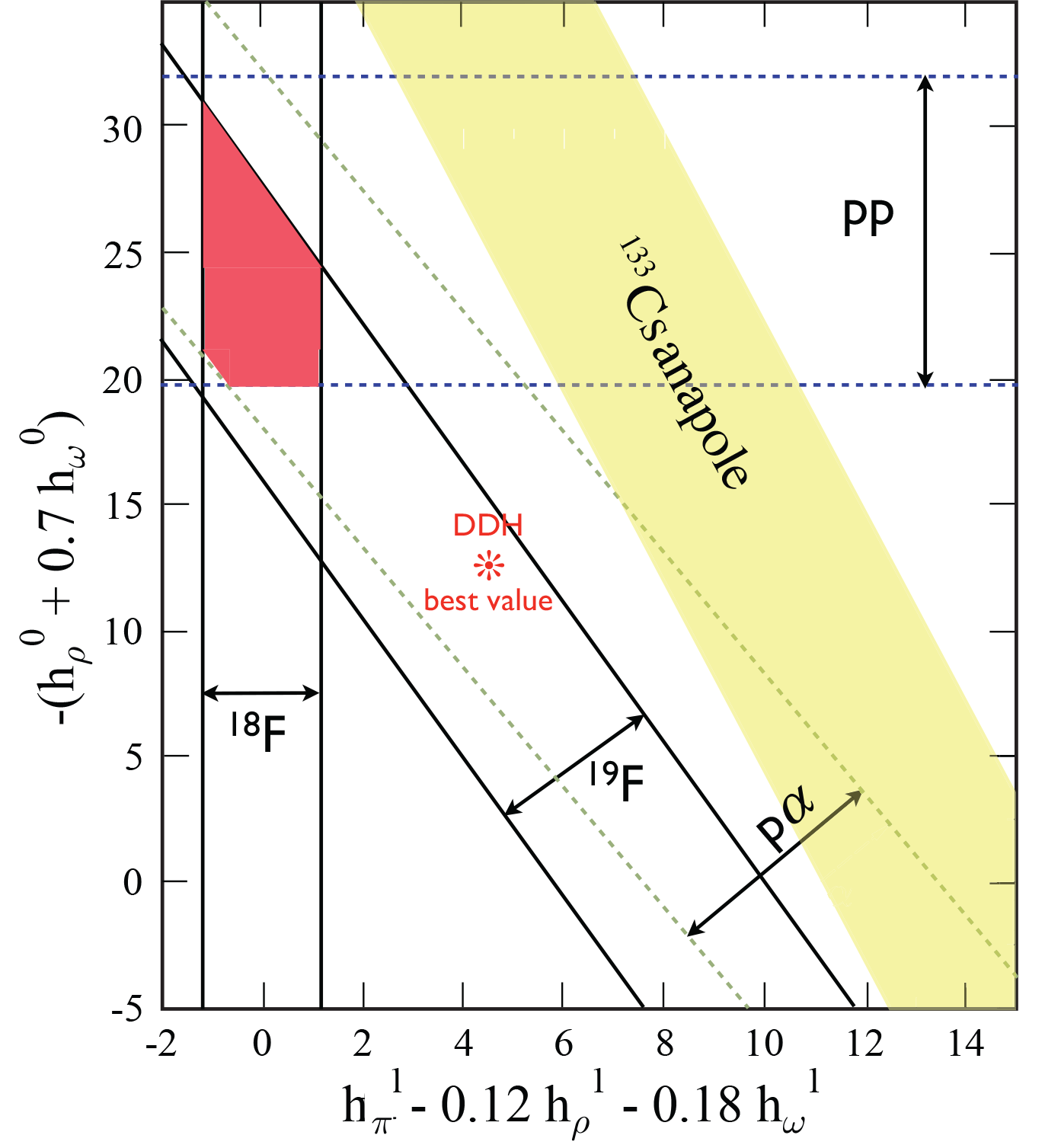}
\caption{ \label{Fig:anapole-DDH} (Color online)
Constraints on combinations of parity violating meson couplings ($\times 10^7$) derived from Cs anapole moment (yellow band) and nuclear experiments. Bands have a width of one standard deviation. Best value predicted by the DDH analysis is also shown. This figure combines Cs NAM band  from~\citet{HaxWie01} with more recent nuclear-physics constraints figure from \citet{Haxton2013}.
}
\end{center}
\end{figure}

The derived bounds~\cite{HaxWie01,Haxton2013} on PNC meson couplings are shown in Fig.~\ref{Fig:anapole-DDH}.  The   $^{133}$Cs APV result  is shown in addition to
constraints from scattering of polarized protons on
unpolarized proton and $^{4}$He targets and  emission of circularly polarized photons
from $^{18}$F and $^{19}$F nuclei. The area colored red lies at the intersection of nuclear experimental bands. There is some tension with the Cs anapole moment result, although the Cs result is consistent with ``reasonable ranges'' of the DDH parameters.
\citet{HaxWie01} point out that additional APV experiments with unpaired-neutron nuclei would produce a band perpendicular to the Cs band (the $^{133}$Cs anapole moment is primarily  due to a valence proton). This provides strong motivation for the ongoing experiments to measure nuclear-spin-dependent APV effects in nuclei with unpaired neutrons such as $^{171}$Yb~\cite{Leefer2014}, $^{212}$Fr~\cite{Aubin2013},
and $^{137}$Ba~\cite{DeMCahMur08}.

\subsection{New and ongoing APV experiments}
\label{Sec:APV:NewExpt}
We limit our discussion to APV experiments that are now being actively pursued. We refer the reader to the earlier reviews \cite{Bouchiat1997,Budker1999,GinFla04,Roberts2014} for a discussion of various proposals.

Experimental efforts to improve the accuracy of PNC measurements in Cs are underway at Purdue University~\cite{Antypas2014}.
This group is exploring a new two-pathway coherent control technique. Here,
two optical excitations, starting from the same initial state ($6S_{1/2}$) and leading to the same final ($7S_{1/2}$)
state, are driven by two different mutually-coherent fields.  One of the lasers is resonant with the $6S_{1/2}-7S_{1/2}$ transition and the other operates at half the resonant frequency driving an allowed two-photon E1 amplitude. The absorption rate contains an interference term between the two-photon amplitude and a sum of Stark-induced and PNC amplitudes, and it depends on the relative phase of the applied laser fields.  By experimentally varying the relative phase one would observe oscillating modulation of the transition rate. As a demonstration, the Purdue group has measured several atomic properties of Cs~\cite{AntEll11,AntEll2013,AntEll2013-radialmel}.

Francium and the Ra$^{+}$ ion have an electronic atomic structure similar to Cs, but larger nuclear charge $Z$ and thereby larger PNC amplitude due to the $Z^3$ enhancement.  Both atoms are amenable to the application of the same theoretical techniques as  Cs~\cite{Rupsi2009,Wansbeek2008,Dzuba2001-s2d-apv,Dzuba1995-Fr,Sahoo2010,SafJoh00a}   and potentially offer improved probes of the low-energy electroweak sector. The experimental challenge with these systems lies in their radioactivity which requires special experimental  facilities. A Fr experiment is in preparatory stages at the TRIUMF facility in Vancouver~\cite{Aubin2013}, while  Ra$^{+}$ ion is investigated in Groningen~\cite{GomOroSpr06,NunezPortela2013}. Ra$^{+}$ is an ion and requires application of novel experimental techniques \cite{For93}.

Since the accuracy of atomic calculations for multivalent systems is unlikely to  approach that achieved for atoms with a single valence electron [Cs, Fr, Ra$^+$], the strategy for ongoing experiments in Yb is to pursue isotopic ratios, as discussed in Sec.~\ref{Sec:APV:skin}. One of the most immediate goals of Yb APV experiments~\cite{Leefer2014} is verification of the isotopic dependence of the weak charge, with the Yb experiment (recently moved from Berkeley to Mainz) currently taking data. Experiments with Dy, where there are nearly-degenerate states of opposite parity, have not yet detected APV~\cite{NguBudDeM97}; however, this is expected in the new generation of the apparatus~\cite{Leefer2014}.

While  Cs is the only experiment to date that has measured NAM~\cite{WooBenCho97}, there are several  proposals on NAM detection in atomic and molecular experiments. \citet{Bou07} discusses a NAM-induced linear dc Stark shift
of the individual substates of an alkali atom in its ground state, dressed by a circularly polarized laser field.
\citet{Choi2016} propose an application of the two-pathway coherent control technique for direct measurement of the anapole moment using the
ground-state  hyperfine splitting of Cs.  Measurements in a chain of Fr isotopes~\cite{GomAubSpr07,Aubin2013} are being actively explored, with future plans for APV measurements using  both $7S_{1/2}-8S_{1/2}$ and $7S_{1/2}$ hyperfine transitions.
\citet{DeMCahMur08}  outline a  Stark-interference
technique to measure spin-dependent APV effects to determine the mixing between opposite-parity rotational/hyperfine
levels of ground-state molecules. By using a magnetic field to tune these levels to near-degeneracy, the usual PNC-induced mixing is dramatically amplified~\cite{Kozlov1991}.
This method can in principle give a large enhancement in sensitivity relative to traditional experiments with atoms.
The technique is applicable to nuclei over a wide range of atomic numbers in diatomic species that
are theoretically tractable. Both NAMs and $C_{n,p}^{(2)}$ electroweak parameters, discussed in Sec.~\ref{Sec:NSD}, can be probed. Such experiments are underway at Yale~\citep{Cahn2014,Altuntas2017,Altuntas2018}.

While PNC interactions do not normally cause first-order energy shifts because they mix states of opposite parity, such energy shifts do occur in chiral systems. This fact has been recognized since 1970s~\cite{Let75}, and searches for minute PNC energy shifts between states of chiral enantiomers (molecules that are mirror images of one another) via high-resolution spectroscopy have been ongoing ever since then [see, for example, \citet{Tokunaga2013} and references therein]. So far there have been no  conclusive observations of a parity violating effect in chiral molecules.
\citet{EilBlaBou17} proposed a new experiment to search for PNC
in chemical shifts of chiral molecules using nuclear
magnetic resonance (NMR) spectroscopy. A proof-of-principle experiment with $^{13}$C-containing molecules was presented, with molecules containing heavier nuclei with enhanced PNC effects to be used next.
Precision measurements of this kind may be
useful for studying nuclear PNC and testing exotic physics models that predict the presence
of parity-violating cosmic fields \cite{RobStaDzu14a,RobStaDzu14}.

\section{Time-reversal violation: electric dipole moments and related phenomena} \label{Sec:EDM}
\subsection{Introduction} \label{Sec:EDM:Introduction}
 %\citep{Phi98,Chu98,Coh98}
In this section, we review phenomena related to simultaneous time-reversal- ($T$-) and parity- ($P$-) violation in atomic and molecular physics.  As we will describe, recent searches for $T$-, $P$-violating (T,PV) effects in these systems are probing energy scales well above 1 TeV in particle theory models widely considered as natural extensions to the SM. Clear prospects for future improvements make it likely that work in this area will remain at the forefront of particle physics for some time.  This topic has been reviewed frequently, with emphasis on different aspects of the related physics: see, for example, \citet{Barr1993, Khriplovich1997, GinFla04, Pospelov2005, Commins2009, Chupp2010, Fukuyama2012, Engel2013, Jungmann2013}, and \citet{Yamanaka2017}. Here we focus on the connection between underlying physics and observable signals in atomic and molecular systems, and the resulting impact on particle physics.

\begin{figure}
\centering
\includegraphics[width=70mm]{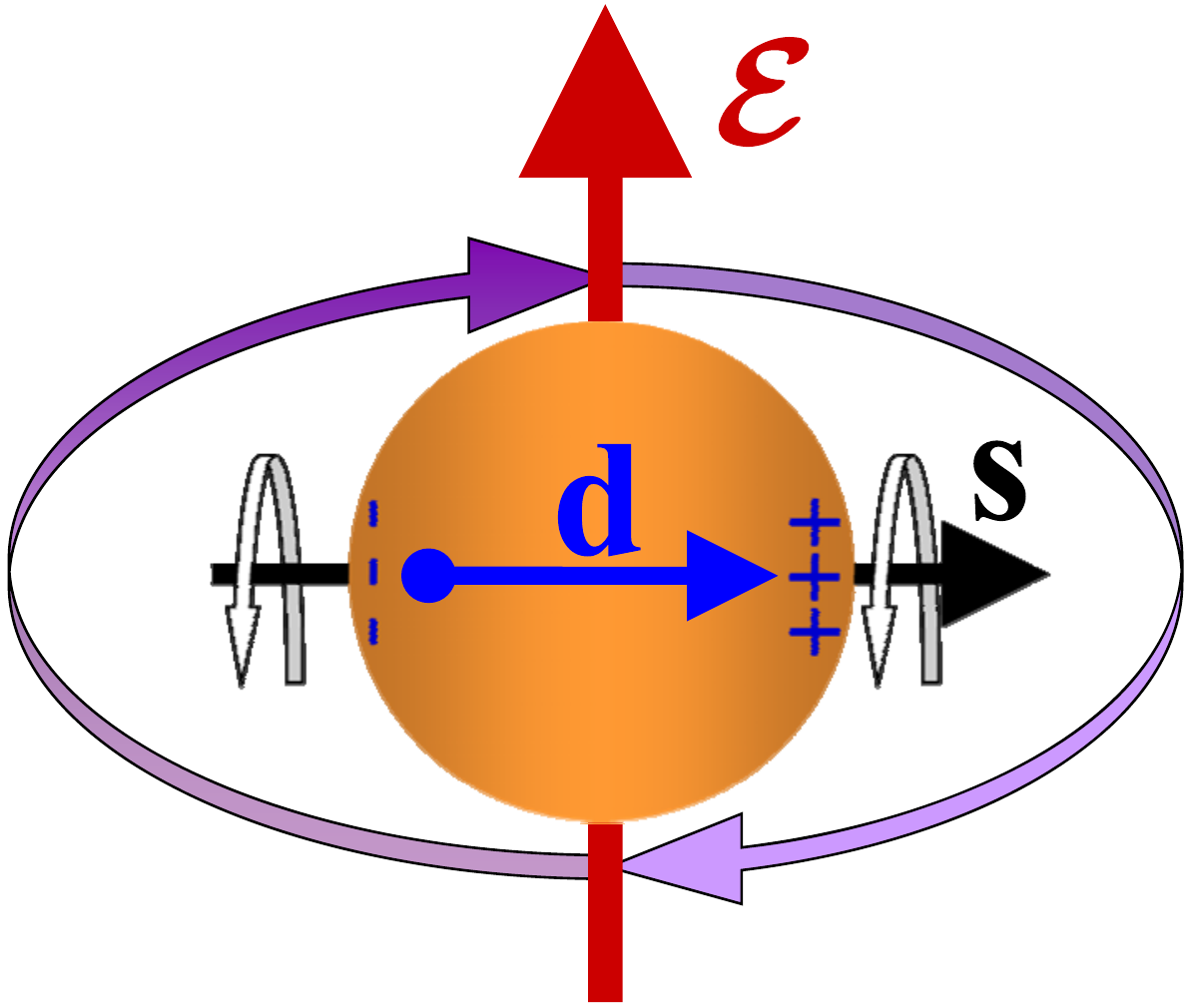}
\caption{Basic concept of EDM measurements.  When a particle has an EDM $\mathbf{d}$  along its spin axis $\mathbf{s}$, an electric field $\Evec$ causes $\mathbf{s}$ to precess about $\Evec$.}\label{Fig_EDM_basic_edm}
\end{figure}

A relevant example of a T,PV effect is when a particle has an electric dipole moment (EDM), $\mathbf{d}$, along its spin $\mathbf{s}$,
i.e., $\mathbf{d} = d \,\mathbf{s}/s$ (Fig. \ref{Fig_EDM_basic_edm}).
The idea that elementary particles might possess a permanent
electric dipole moment (EDM) in addition to a
magnetic dipole moment was proposed by \citet{Pur50}.
This leads to an interaction with an electric field $\Evec$ described by the Hamiltonian $$H_{\rm EDM} = - \mathbf{d}\cdot\Evec \propto d \,\mathbf{s} \cdot \Evec.$$  $H_{\rm EDM} $ is odd under both $T$ and $P$: $\mathbf{s}$ changes sign under $T$ but $\Evec$ does not, while under $P$, $\Evec$ changes sign but the axial vector $\mathbf{s}$ does not.
Most T,PV effects in atomic and molecular systems result in an EDM or some closely related quantity (for example, an interaction between a  spin and the internuclear axis in a molecule).\footnote{In this section, expressions related to electomagnetism will use the cgs system of units so that electric and magnetic fields, as well as electric and magnetic dipole moments, have the same dimensions.  For many expressions we use mixed units that are standard for the field, such as electric fields in units V/cm and electric dipole moments in units $e\cdot\mr{cm}$.}

%We will discuss the microscopic physics that can lead to this Hamiltonian, as well as its observable consequences, later in this section.

In relativistic quantum field theories, the combined symmetry $CPT$ (where $C$ is charge conjugation) is always conserved \citep{Streater2000}.  Moreover, $CPT$ conservation has been experimentally confirmed to extraordinary precision (see Sec.~\ref{Sec:EDM}).  Hence, in typical theoretical extensions to the SM, it is assumed that $T$-violation is equivalent to $CP$-violation (CPV), and for the remainder of this section we do so as well. Based on very general considerations in quantum field theory, at low energies the largest effects of CPV are expected to appear as T,PV interactions rather than $T$-violating but $P$-conserving (TV) signals \citep{Khriplovich1997}. In fact, limits on T,PV effects in combination with established principles of field theory rule out TV effects \citep{Conti1992} far below the level of any conceived experiment to detect them \citet[see]{Kozlov1989, Hopkinson2002}. Hence for the remainder of this section we use the terms CPV and T,PV interchangeably.

The time-reversal operator $T$ is anti-unitary: it can be represented as the product of a unitary operator and the complex conjugation operator \citep{Sakurai2011}.  Hence while quantum wave equations with real-valued potentials are $T$-invariant (i.e., if some wavefunction $\Psi(t)$ is a solution, then $\Psi^*(-t)$ is also a solution), $T$-violating effects arise for complex-valued potentials.  Hence, $T$- (and $CP$-) violation is associated with the irreducible presence of complex numbers in the underlying theory.  The strength of CPV interactions is proportional to $\sin{\phi_{\rm CP}}$, where $\phi_{\rm CP}$ is the phase of such a complex number \citep{Fortson2003}. It is known that CPV occurs in nature, from observations of CPV in decays of $K^0$  and $B^0$ mesons \citep{PDG}.  These observations are all consistent with a single source of CPV in the SM: a complex phase in the
Cabibbo -- Kobayashi -- Maskawa (CKM) matrix that describes the mixing between quark flavors to form mass eigenstates.  The measured value of this phase is large: $\delta_{\rm CKM} \sim 1$ rad.  However, the linkage with flavor mixing causes the observable effects of CPV to be systematically small in the SM \citep{Khriplovich1982, McKellar1987, Bernreuther1991, Pospelov2014, Yamanaka2016}.  In particular, EDMs within the SM are exceptionally small, despite the large value of $\delta_{\rm CKM}$.  By contrast, theories that extend the SM  naturally can, and frequently do, include new CPV phases that contribute to EDMs and related phenomena in a more direct way, with no obvious mechanisms for suppression \citep{Barr1993}.
This makes EDMs a nearly background-free signal for detecting new physics associated with CPV \citep{Pospelov2005, Engel2013}.
% even in flavor-changing processes such as $K^0$ decays

Moreover, there \textit{must} be new sources of CPV in nature.  This conclusion arises from the observation of a baryon asymmetry---a cosmological imbalance between matter and antimatter \citep{Dine2003}.   Since matter-antimatter annihilations in the aftermath of the Big Bang produce photons, this asymmetry is typically parameterized by the cosmological baryon-to-photon ratio $\eta \sim 10^{-10}$. Sakharov showed that, among other conditions, CPV is necessary to generate this asymmetry \citep{Sakharov1967}.  While the SM in principle incorporates all the Sakharov conditions, the size of CPV effects in the SM is far too small to account for the observed value of $\eta$  \citep{Gavela1994}.
By contrast, theoretical models containing new particles with masses near the electroweak scale $M_Z \sim\!100$ GeV, together with new CPV phases, could explain the experimentally observed value of $\eta$.  In these scenarios---known as electroweak baryogenesis--- CPV (or, equivalently, T,PV) signals typically are predicted to appear at a level near current experimental sensitivities \citep{Engel2013}.
%%%something about hierarchy problem...?

\subsection{Observable effects in atoms and molecules}  \label{Sec:EDM:ObservableEffects}
Atomic and molecular experiments searching for T,PV can be broadly classified into two categories, based on whether the system is  paramagnetic---unpaired electron spins---or diamagnetic---closed electron shells, but nonzero nuclear spin) \citep{Khriplovich1997, Barr1993}.  Paramagnetic systems are most sensitive to effects that depend explicitly on electron spin: the electron EDM (eEDM) and one type of semileptonic (electron-nucleus) interaction.  Diamagnetic systems are most sensitive to effects that depend explicitly on nuclear spin: purely hadronic T,PV interactions, EDMs of nucleons, and other types of semileptonic interactions.

%Though the intrinsic sensitivity of diamagnetic systems to electron-spin dependent effects is deeply suppressed, the extraordinary precision reached in experiments with diamagnetic atoms makes them often competetive in setting limits on such effects.
%Here, we discuss how these microscopic physics mechanisms translate to observable effects in atoms and molecules.

To understand how T,PV effects are related to EDMs in atoms and molecules, consider a toy system consisting of two states with opposite parity eigenvalues $\Pi = \pm 1$, each with angular momentum $j$, split in energy by $\Delta$, and in particular a pair of substates $|j, m, \Pi\rangle$ with the same projection $m$. These states can be mixed by a \mbox{$T$-,$P$-odd} Hamiltonian $H_{\rm TP}$; since this is a rank-0 tensor, its matrix elements $\delta_{\rm TP}$ are independent of $m$.  The levels can also be mixed by the Stark Hamiltonian
 $$H_{\rm St} = -\mathbf{D}\cdot\Esca \hat{z},$$ with electric dipole matrix element $$\left\langle j,m,+| D_z | j,m,-\right\rangle \equiv D{\rm sgn}(m),$$ where the $m$-dependence follows from the Wigner-Eckart theorem.   This system is described by the Hamiltonian
\begin{equation}
H_{\rm toy} = \left( \begin{array}{cc} -\Delta/2 & -D\Esca{\rm sgn}(m) + \delta_{\rm TP}\\
                                       -D\Esca{\rm sgn}(m) + \delta_{\rm TP} & \Delta/2   \end{array} \right)
\end{equation}
In addition to the usual Stark shifts there are T,PV  shifts, given to 1$^{\rm st}$ order in $\delta_{\rm TP}$  by
$$\Delta E_{\rm TP}^\pm = \mp \mathcal{P} \delta_{\rm TP} {\rm sgn}(m).$$
 Here, the dimensionless quantity $$\mathcal{P} \equiv \frac{D\Esca}{\sqrt{(\Delta/2)^2 + D^2\Esca^2}}$$ (with values $0 \le \mathcal{P}<1$) quantifies the polarization of the system. The superscript $\pm$ refers to the upper/lower state of the system.

This has simple behavior in the limiting cases where $D\Esca \ll \Delta$ or $\gg \Delta$.
Consider the weak-field limit, where $\mathcal{P} \ll 1$; then $$\Delta E_{\rm TP}^\pm \approx \mp (2D\Esca/\Delta) \delta_{\rm TP} {\rm sgn}(m).$$  This is exactly the shift that would be found for a system with permanent dipole moment $\mathbf{d}^\pm = \pm d\mathbf{j}/j$, where $$d = 2D\delta_{\rm TP}/\Delta.$$  Hence, in this weakly-polarized regime it is sensible to say that $H_{\rm TP}$ induces a dipole moment (of opposite sign in the upper/lower states of the system), and the energy shifts of interest are proportional to the strength of the applied field, $\Esca$.  Next, consider the strong-field case, where $1-\mathcal{P} \ll 1$.
%Here, in the absence of $H_{\rm TP}$ the eigenstates of the system are symmetric or antisymmetric mixtures of the parity eigenstates, with energies given by $E^\pm \approx \pm D\Esca$ and Stark-induced dipole moments $\langle \mathbf{D} \rangle^\pm = \mp D\hat{z}$, independent of $m$.
In this regime, $\Delta E_{\rm TP}^\pm \approx \mp \delta_{\rm TP} {\rm sgn}(m)$.  Here the T,PV energy shifts are independent of $\Esca$, and it is no longer sensible to speak of a T,PV dipole moment of the system.  The shifts are also maximal in this regime.

It is infeasible to reach the strong-field regime in the ground  states of atoms: for typical splittings between opposite-parity levels $\Delta \sim E_{\rm h}$  (the atomic unit of energy, $e^2/a_0$)  and dipole matrix elements $D \sim ea_0$, the required field strength $\Esca \gtrsim \Esca_{\rm at}$, where $\Esca_{\rm at} = e/a_0^2 \sim 5\times 10^9$ V/cm is the atomic unit of electric field, is far too large to apply in the lab.  However, in polar molecules there are  levels of opposite parity with much smaller energy splittings but similar dipole matrix elements, making it far easier to polarize these systems. Such pairs of levels---associated with rotational structure (where $\Delta \sim [m_\mr{e}/m_\mr{p}]E_{\rm h}$) or $\Omega$-doublet structure (where $\Delta \sim [m_\mr{e}/m_\mr{p}]^n E_{\rm h}$, with $n=1$ or $2$ depending on the type of electronic state)---make it routine to reach the regime of nearly full polarization in these systems \citep{Sandars1967, Sushkov1978}. The increase in observable T,PV energy shifts, relative to the case of atoms in lab-scale $\Esca$-fields, is typically 3-5 orders of magnitude.   Hence, experiments with molecules play an important role in this field \citep{Kozlov1995}.

%%%%%%%%%%%%

\subsection {Underlying physical mechanisms for T,PV}  \label{Sec:EDM:Underlying}

\subsubsection{Semileptonic interactions} \label{Sec:EDM:Underlying:Semileptonic}

Semileptonic interactions (SLIs) arise in several particle-theory models. They can be described as a 4-fermion interaction, related to the exchange of a heavy force-carrying boson between electrons and the nucleus. Effects due to exchange of lighter force-carriers are discussed in Sec.~\ref{Sec:ExoticSpin:Parametrization}. A few distinct forms of interaction give nonzero effects \citep{Khriplovich1997}.  The first is the coupling of a scalar current from nucleons $n$ in the nucleus to a pseudoscalar electron current, described by the relativistic Lagrangian density
$$\mathcal{L}_{\rm SP} \propto \sum_{n} \bar{\psi}_e i \gamma^5\psi_e \bar{\psi}_n \psi_n.$$
  This yields a relativistic Hamiltonian for the interaction of a single electron with a pointlike nucleus, $$H^{\rm rel}_{\rm SP} = i frac{G_F}{\sqrt{2}} \frac{1}{2m_e c} Q_{\rm SP} \delta^3(\mathbf{r}) \gamma^0\gamma^5.$$
    Here $Q_{\rm SP}$ is the effective charge of the nucleus for the scalar-pseudoscalar interaction, analogous to the weak charge $Q_W$ for the $PV$ weak interaction.  This is frequently written in the form $Q_{\rm SP} = AC_{\rm S}$, where $A$ is the mass number and $C_S$ is the average effective charge per nucleon.  In the nonrelativistic $\mathrm{(n.r.)}$  limit, this Hamiltonian takes the form
$$
H^{\rm nr}_{\rm SP} = i \frac{G_F}{\sqrt{2}}\frac{1}{2m_e c\hbar} AC_{\rm S}
 \{ \mathbf{s}\cdot \mathbf{p}, \delta^3(\mathbf{r}) \},
$$
where $\{ \}$ denotes the anticommutator. This has the same form as the $P$-odd (but $T$-even) Hamiltonian arising from $Z^0$-boson exchange, aside from the factor of $i$.
Due to the contact nature of the interaction, $H^{\rm SP}$ mixes only $s_{1/2}$ and $p_{1/2}$ orbitals in atoms, with typical matrix element
\begin{eqnarray}
\delta_{\rm SP} &=& \langle s_{1/2} | H_{\rm SP} | p_{1/2} \rangle \nonumber \\
&\sim& C_{\rm S} AZ^2 G_F \frac{\hbar}{m_e ca_0^4} \sim 10^{-16} \times C_{\rm S} AZ^2 E_{\rm h}. \nonumber
\end{eqnarray}
The explicit dependence of $H^{\rm nr}_{\rm SP}$ on $\mathbf{s}$ shows that, to lowest order, the effect of $H_{\rm SP}$ is nonzero only for paramagnetic systems; in diamagnetic systems, hyperfine-induced mixing leads to a nonzero effect at higher order \citep{Flambaum1985b}.

Other forms of SLIs lead to Lagrangian densities with the form of a pseudoscalar nucleon-scalar electron current or a tensor-tensor interaction,
%, interactions that include derivatives of the fermion fields (i.e. momenta).
which give rise to Hamiltonians that depend on the nuclear spin $\mathbf{I}$ in the system.  However, the effects of these interactions are usually strongly suppressed, either in the underlying particle theory models
%(e.g. the $TT$ interaction, which is a dimension-eight operator)
or at the atomic/nuclear level.
%(e.g. the $PS$ interaction, which vanishes in the limit of an infinitely heavy nucleon).
We refer the reader to \citet{Khriplovich1997} for more details.

%%%%%%%%%Read and fix as needed!!!!!

\subsubsection{EDMs of constituent particles: Schiff's theorem} \label{Sec:EDM:Underlying:EDMConstituents}
We next turn to the question of how EDMs of constituent particles in an atom or molecule---electrons or nuclei---can lead to observable T,PV.  The answer is subtle.  \citet{Sch63} showed that under reasonable first-order assumptions---i.e., non-relativistic point particles moving in a purely electrostatic potential---there is no energy shift when an $\Esca$-field is applied to a neutral system built from such consitutents.  The proof is simple.  The total electric field $\Evec^{\rm tot}$ experienced by the particle of interest---which is the sum of an externally applied field $\Evec$ and the internal field $\Evec^{\rm int}$ due to other particles in the system---can be expressed as $\Evec^{\rm tot} = -\mathbf{\nabla}\Phi$, where $\Phi$ is an electrostatic potential.  The Hamiltonian for the particle of charge $q$ and mass $m$, neglecting the EDM, is
$$H_0 = p^2/(2m) +  q\Phi.$$
  Since $\mathbf{p} = -i\hbar\mathbf{\nabla}$,
$\Evec^{\rm tot} \propto \left[ p, H_0 \right]$.  Thus, for any eigenstate $|\psi\rangle$ of $H_0$, the expectation value of the total $\Esca$-field vanishes: $\langle \Evec^{\rm tot} \rangle = 0$.  Hence, the energy shift due to the constituent particle's EDM $\mathbf{d}$ also vanishes: $\langle H_{\rm EDM} \rangle = -\mathbf{d}\cdot \langle \Evec^{\rm tot} \rangle = 0$.  The physical meaning of this result, known as Schiff's theorem, is that other parts of the system rearrange so as to completely screen the external $\Esca$-field felt by the charged particle; otherwise, it would undergo a net acceleration. Mechanisms for evading Schiff's theorem are thus central to experiments searching for constituent particle EDMs in atoms and molecules.

\subsubsection{Electron EDM} \label{Sec:EDM:Underlying:eEDM}
First, we consider the eEDM in a paramagnetic atom.  Remarkably, the relativistic motion of the bound electron can lead to energy shifts orders of magnitude larger than the shift for a free electron, $\Delta E_{\rm TP} = -\mathbf{d}_e \cdot \Evec$. This enhancement, first recognized by \citet{Sandars1965}, makes atomic and molecular experiments particularly sensitive to the eEDM.  We discuss the underlying mechanism here.

The relativistic Lagrangian density associated with the interaction between the eEDM, $d_e$, and an electromagnetic field, described by the field tensor $F^{\mu \nu}$, is
\begin{equation}
\mathcal{L}_{\mathrm{eEDM}} = -i\frac{d_e}{2} \overline{\Psi} \sigma^{\mu\nu} \gamma^{5} \Psi F_{\mu\nu},
 \label{eq:eEDM_L_rel}
\end{equation}
where $\Psi$ is the Dirac bispinor for the electron and $\sigma^{\mu \nu}=\frac{i}{2}\left( \gamma^\mu \gamma^\nu - \gamma^\nu \gamma^\mu \right)$. This yields the single-electron relativistic Hamiltonian $H_{\mathrm{eEDM}}^{\rm rel}$:
\begin{equation}
H_{\mathrm{eEDM}}^{\rm rel} = -d_e \gamma^{0} \hbox{\boldmath{$\displaystyle\Sigma$}} \cdot \hbox{\boldmath{$\mathcal{E}$}},
%\mathbf{\Sigma} \cdot \Evec^{\rm tot}
\label{eq:eEDM_H_rel}
\end{equation}
where $\hbox{\boldmath{$\displaystyle\Sigma$}}$ is a Dirac spin operator.
From  Schiff's theorem, on application of an external field $\Evec$ the $\mathrm{n.r.}$ version of this Hamiltonian (still expressed in terms of bispinors), $-d_e \mathbf{\Sigma} \cdot \Evec^{\rm tot}$, will yield a vanishing energy shift.  Hence, we may subtract this term away to find an \textit{effective} Hamiltonian that will account for any observable energy shift due to the eEDM:
 \begin{equation}
H_{\mathrm{eEDM}}^{\rm rel, eff} = -d_e (\gamma^{0}-1) \hbox{\boldmath{$\displaystyle\Sigma$}} \cdot \hbox{\boldmath{$\mathcal{E}$}}^{\rm tot}. \label{eq:eEDM_H_rel_eff}
\end{equation}
In the $\mathrm{n.r.}$ limit, this takes the form
 \begin{equation}
H_{\mathrm{eEDM}}^{\rm nr, eff} = 4\frac{d_e}{m_\mr{e}^2 c^2 \hbar^3} \left[ (\mathbf{s}\cdot\mathbf{p}) (\mathbf{s}\cdot \hbox{\boldmath{$\mathcal{E}$}}^{\rm tot}) (\mathbf{s}\cdot\mathbf{p}) \right].  \label{eq:eEDM_H_nr_eff}
\end{equation}
The matrix elements of $H_{\mathrm{eEDM}}^{\rm nr, eff}$ between atomic $s$ and $p$ orbitals are
$$\delta_{\rm eEDM} \sim  d_e  (Z^3\alpha^2) \Esca_{\rm at}$$ \citep{Sandars1966, Khriplovich1997}.  On application of a polarizing external field $\Evec$, this gives rise to energy shifts $$\Delta E_{\rm eEDM}^\pm = \mp \mathcal{P} \delta_{\rm eEDM}.$$ For a fully polarized system, we can write $$\Delta E_{\rm eEDM}^\pm = -d_e \Esca^{\rm eff},$$ where $$\Esca^{\rm eff} \sim \pm (Z^3\alpha^2) \Esca_{\rm at}.$$  This effective electric field can be orders of magnitude larger than the applied field $\Esca$: for $Z \approx 90$, typically $\Esca^{\rm eff} \sim 100$ GV/cm.  For a weakly-polarized system, $\Delta E_{\rm eEDM}^\pm = \mp 2D \Esca \delta_{\rm eEDM}/\Delta$ can be written as
$$\Delta E_{\rm eEDM}^\pm = \mp d_e F_e(Z) \Esca,$$ where the quantity $F_e(Z)$ is referred to as the eEDM enhancement factor for atoms: it describes the factor by which, in the limit of weak polarization, $\Delta E_{\rm eEDM}$ exceeds the shift for a free electron. With $D \sim ea_0$ and  $\Delta \sim E_{\rm h}$, $F_e(Z) \sim 2Z^3 \alpha^2$, with the typical values $F_e \approx 100-600$ for $Z \approx 55-80$.

The evasion of Schiff's theorem here is remarkable, since even in the relativistic case the expectation value of $\Evec^{\rm tot}$ vanishes. The nonzero effect can be understood heuristically as arising from the relativistic length contraction of the eEDM, acting in concert with the spatial variation of the Coulomb field $\Evec^{\rm int}$ \citep{Commins2007}.  Since neither the length-contracted dipole moment $\mathbf{d}_e^{\rm rel}$ nor the electric field $\Evec^{\rm tot} = \Evec + \Evec^{\rm int}$ are constants over the atomic volume, it makes sense that $\langle -\mathbf{d}_e^{\rm rel}\cdot \Evec^{\rm tot} \rangle \neq  0 = \langle \mathbf{d}_e^{\rm rel}\rangle \cdot \langle \Evec^{\rm tot} \rangle$.

\subsubsection{Hadronic T,PV: nuclear Schiff moment and related effects} \label{Sec:EDM:Underlying:SchiffMoment}
Much like how $T$-,$P$-odd SLIs and/or the eEDM can induce an atomic EDM, the presence of a proton or neutron EDM, or of $T$-,$P$-odd hadronic interactions, can mix nuclear states to induce a nuclear EDM. However, within an atom the motion of a nucleus is deeply nonrelativistic.  Hence, according to Schiff's theorem, any nuclear EDM is very effectively screened from external fields and leads to negligible energy shifts.  Nevertheless, the same $T$-,$P$-odd hadronic effects can induce changes in the nuclear charge and current distributions corresponding to electromagnetic moments other than an EDM.  These modified distributions, unlike a nuclear EDM, can give rise to T,PV energy shifts in an electrically-polarized atom or molecule.

The primary mechanism for these shifts is associated with the finite size of the nucleus.  Penetration of valence electrons into the finite nuclear volume allows them to interact with a local (intra-nuclear) $\Esca$-field different from that of a point dipole, which would be completely screened.  The charge distribution that leads to the lowest-order observable T,PV energy shift is known as the Schiff moment (SMt), $\SMtvec = \SMt\mathbf{I}/I$ \citep{Sch63}.
$$\SMtvec \equiv \frac{Ze}{10} \left[ \int\rho_Z(\mathbf{r})\mathbf{r}r^2d^3\mathbf{r} - \frac{5}{3}\int\rho_Z(\mathbf{r})\mathbf{r}d^3\mathbf{r} \int \rho_Z(\mathbf{r})r^2d^3\mathbf{r} \right],$$
 where $\rho_Z$ is the nuclear charge density normalized as $\int \rho_Z(\mathbf{r})d^3\mathbf{r} = 1$.
Physically, $\SMt$ corresponds to the charge distribution that gives a constant electric field $\Evec_\SMt \parallel \mathbf{I}$ within the volume of the nucleus \citep{Flambaum2002}; it has dimensions of [charge$\cdot$volume].

This yields a term $H_\mr{S}$ in the $\mathrm{n.r.}$ atomic/molecular Hamiltonian that, for a spherical nucleus of radius $R_N$, has the form
\begin{equation}
H_\mr{S} = -15e (\SMt/R_N^5) \mathbf{r}\cdot \mathbf{I}/I~(r\! <\! R_N).
\end{equation}
This interaction gives first-order effects in both diamagnetic and paramagnetic systems. The associated T,PV atomic/molecular matrix elements have typical size $\delta_\mr{S} \equiv \langle p |H_\mr{S} | s\rangle \sim Z^2 \SMt/(ea_0^3) E_{\rm h}$ \citep{Khriplovich1997}.

A nuclear SMt can be induced by a variety of microscopic physics effects.  An example is when the nucleus contains a valence nucleon $n$ with dipole moment $d_{n}$.  In a nuclear shell model where $n$ moves around a uniform spherical core of radius $R_N = R_0 A^{1/3}$ (where $A$ is the nuclear mass number and $R_0 \approx 1.2$ fm is the characteristic nuclear size), the SMt has magnitude $\SMt \sim 0.1 d_{n} A^{2/3} R_0^2$.  In the weak polarization limit, $H_S$ induces an atomic/molecular EDM $d_a$.  Since $\SMt \propto d_{n}$, the quantity $F_{n} = d_a/d_n$ is analogous to the eEDM enhancement factor $F_e$ for the eEDM.  However, here there is instead a suppression: $F_{n} \sim Z^2 A^{2/3} R_0^2/a_0^2$, with typical numerical value $F_{n} \approx 10^{-3}$ for $Z=80$ \citep{Khriplovich1997}.

In most theoretical models, $T$-,$P$-odd intranuclear interactions, rather than the nucleon EDM, give dominant contributions to the nuclear SMt \citep{Sushkov1984}.  For example, in many theories quarks acquire a chromo-EDM (cEDM), $\tilde{d}_q$, which is the strong-interaction analogue of the ordinary EDM.  The color field resulting from the cEDM induces a $T$-,$P$-odd strong interaction---typically described as an effective T,PV nucleon-nucleon interaction---between a valence nucleon and the remainder of the nucleus.  This mechanism generally leads to a larger nuclear SMt than that from the ordinary nucleon EDM \citep{Fischler1992}, by a factor of $\sim\! 40$, for the same size of $\tilde{d}_q$ and $d_n$ \citep{Khriplovich1997}.\footnote{The color field within the nucleon from $\tilde{d}_q$ induces $d_n \sim e \tilde{d}_q$.} Hence these experiments are particlularly sensitive to new physics at high energy scales that is related to quark cEDMs \citep{Pospelov2005, Engel2013}.

In addition, there is the possibility in quantum chromodynamics (QCD) of an irreducible CPV interaction (see, for example, the reviews by   \citet{Pec08}, \citet{Kim10R}, and \citealt{Sikivie2012}), described by the Lagrangian density
$$\mathcal{L}_\theta = -\bar{\theta}_{\rm QCD} (\alpha_s/8 \pi) (\epsilon^{\mu\nu\kappa\lambda}/2) G^{\rm a}_{\mu\nu}G^{\rm a}_{\kappa\lambda}.$$  Here, $\alpha_s$ is the strong-interaction analogue of $\alpha = e^2/(\hbar c)$ in electromagnetism; $\epsilon_{\mu\nu\kappa\lambda}$  is the completely antisymmetric tensor; $ G^{\rm a}$ is the gluon field tensor; and $\bar{\theta}_{\rm QCD}$ is a dimensionless constant parameterizing the strength of this term relative to the ordinary strong interaction.  $\mathcal{L}_\theta$ also leads to an effective T,PV nucleon-nucleon interaction \citep{Crewther1979, Pospelov2005, Engel2013}.  Typical calculations in spherical nuclei of the relation between $\bar{\theta}_{\rm QCD}$ and $\SMt$ yield $\SMt \sim 10^{-3} \bar{\theta}_{\rm QCD} eR_0^3$ \citep{Khriplovich1997}.  Searches for nuclear SMts (and the bare neutron's EDM \citep{Pendlebury2015, Bak06}) set a strong bound $\bar{\theta}_{\rm QCD} \lesssim 10^{-10}$, while naively one expects dimensionless fundamental parameters to have values of order unity.  The hypothetical particle known as the axion was first devised as a mechanism to solve this so-called ``strong $CP$ problem''. (Axions are discussed further in Secs.~\ref{Sec:ExoticSpin:EarlyWork:axions-ALPs} and Sec.~\ref{Sec:LightDarkMatter}.)

% For example, consider again a simple nuclear shell model, in which a valence nucleon $v = p,n$ interacts with core nucleons $c$ via the $T$-,$P$-odd contact potential
%$V_{TP}^{\rm nucl} = (G_F/\sqrt{2}) \sum_{c}\xi_{vc} \mathbf{s}_{v} \mathbf{\nabla} \delta^3(\mathbf{r}_v-\mathbf{r}_c)/m_v$. Here $\xi_{vc}$ is a dimensionless parameter that %characterizes the %strength of this interaction relative to the usual weak interaction.  This induces a SMt of typical size $S \sim e\xi A^{2/3} \times 10^{-8} %R_0^3$.  In most particle physics models, this or similar T,PV hadronic interactions give the dominant contribution to the SMt.

While the SMt makes the dominant contribution to T,PV energy shifts in diamagnetic systems, in paramagnetic systems with nuclear spin $I \ge 1$, another mechanism typically leads to larger nuclear spin-dependent T,PV effects \citep{Sushkov1984, Khriplovich1997}.  Here, the underlying hadronic T,PV  physics leads to a current distribution in the nucleus that corresponds to a magnetic quadrupole moment (MQM).  This MQM couples to the gradient of the magnetic field produced by the electron and mixes $s_{1/2}$ and $p_{3/2}$ atomic orbitals.  The nuclear spin-dependent T,PV energy shifts associated with the MQM can exceed those due to the nuclear SMt by a factor of $\sim\!10-100$.

\subsection{State-of-the-art experiments} \label{Sec:EDM:StateoftheArt}
\subsubsection{General remarks} \label{Sec:EDM:StateoftheArt:General}

All recent atomic and molecular experiments that have set stringent limits on T,PV effects rely on the same basic measurement principle.  A $T$-,$P$-odd Hamiltonian $H_{\rm TP}$, together with an applied electric field $\Evec = \Esca \hat{z}$, results in an energy shift $\Delta E_{\rm TP}$ of the state $|j,m_j\rangle$, given by  $\Delta E_{\rm TP} = \mathcal{P} \delta_{\rm TP} {\rm sgn}(m_j)$.  Since $\delta_{\rm TP}$ grows rapidly with $Z$, all experiments use heavy atoms.  To measure $\Delta E_{\rm TP}$, an equal superposition of states with $\pm m_j$ is prepared and allowed to freely evolve for time $\tau$.  This state is typically prepared with high efficiency by optical pumping, sometimes in combination with radiofrequency spin-flips.  The energy splitting leads to a relative phase accumulated between the states, $\phi = 2\Delta E_{\rm TP}\tau/\hbar$. The superposition state corresponds to an orientation or alignment of $\mathbf{j}$ in the $x-y$ plane, and the phase evolution is equivalent to a precession of $\mathbf{j}$ about $\Evec$ by angle $\phi$. For a single particle, $\Delta E_{\rm TP}$ can be measured with minimum uncertainty $\hbar/(4\tau)$; hence with $N$ uncorrelated particles, the T,PV energy shift can be measured with uncertainty $\delta ( \Delta E_{\rm TP} ) = \hbar/(4\tau\sqrt{N})$.

Experiments of this type contend with certain common issues related to the fact that $m_j$-dependent energy shifts easily can be caused not only by T,PV effects, but also by magnetic fields $\hbox{\boldmath{$\mathcal{B}$}}$ due to their interaction with the magnetic moment $\mathbf{\mu} \propto \mathbf{j}$.  Random $\mathcal{B}$-field fluctuations can degrade the signal-to-noise ratio.  Nearly all experiments minimize this effect by reversing $\Evec$, in order to reverse the system polarization $\mathcal{P}$, as frequently as possible. It is also common to perform measurements on side-by-side regions with opposing $\Esca$-fields; common-mode magnetic field shifts cancel in the difference between energy shifts in these regions.
In addition, $\hbox{\boldmath{$\mathcal{B}$}}$-fields correlated with $\Evec$ can lead to systematic errors that mimic $\Delta E_{\rm TP}$. These can arise due to leakage currents associated with the $\Esca$-field and due to motional effects (since a particle moving in an electric field $\Evec$ with velocity $\mathbf{v}$ experiences a magnetic field $\hbox{\boldmath{$\mathcal{B}$}}_{\rm mot} = \Evec \times \mathbf{v}/c$).  Frequently, experiments replicate the measurements on an EDM-insensitive system [e.g., a lighter species as in \citep{Regan2002}] or on a state of the same system with opposite sign of $\mathcal{P}$  [e.g., the excited state of the pair in the toy model of section \ref{Sec:EDM:ObservableEffects}, as first used in \citep{DeMille2001, Eckel2013}] and hence also $\Delta E_{\rm TP}$. These ``co-magnetometers'' act as a useful probe for systematic errors.

\subsubsection{Experiments on paramagnetic systems} \label{Sec:EDM:StateoftheArt:Paramagnetic}
\begin{figure}
\centering
\includegraphics[width=85.8mm]{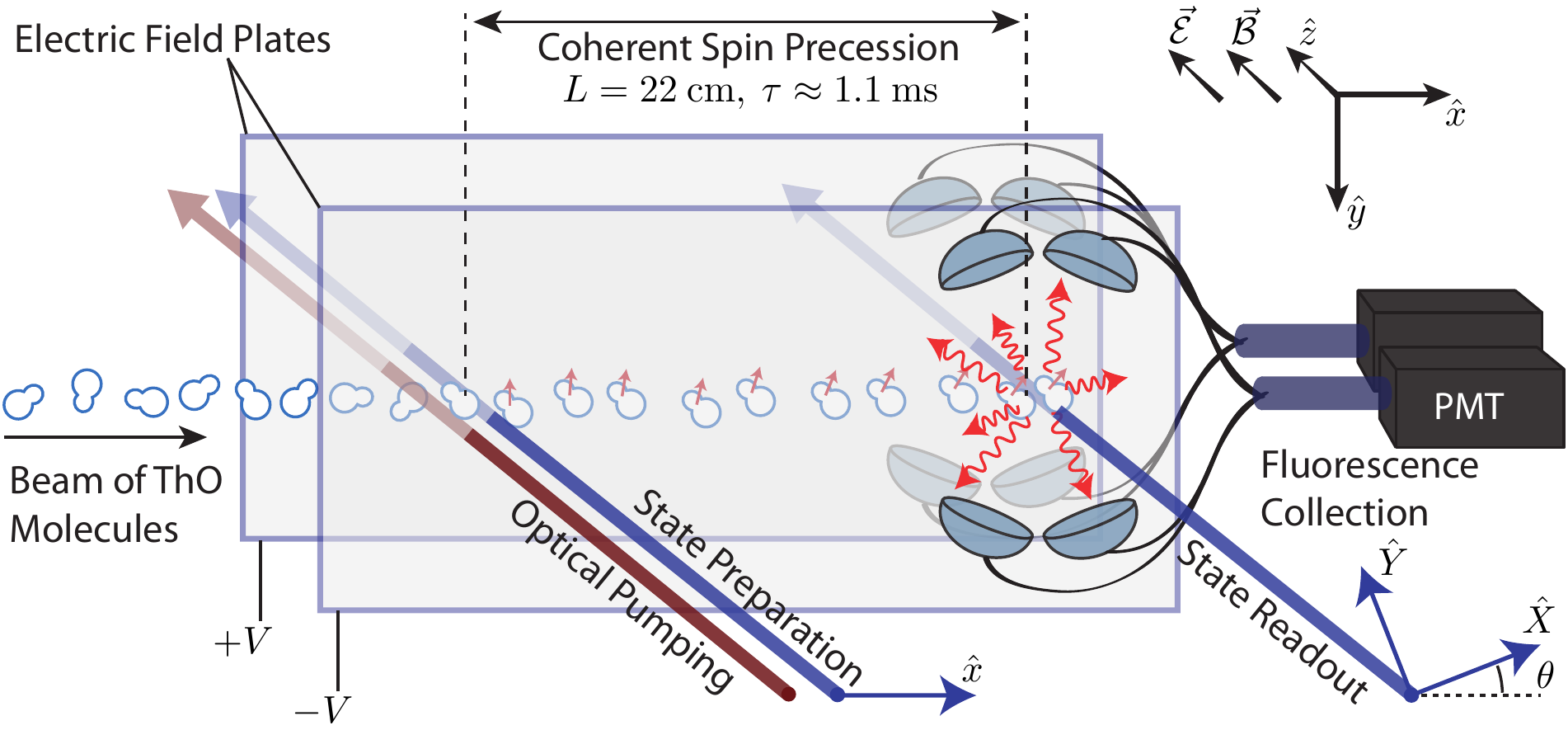}
\caption{Schematic of the ACME eEDM experiment. The figure shows only the magnetically-shielded region where spin precession takes place.  Reproduced from \citet{Baron2014}.}\label{Fig_EDM_ACME}
\end{figure}

The ACME collaboration (Yale/Harvard) recently completed the most sensitive experiment using a paramagnetic system \citep{Baron2014, Baron2017}.  In ACME, ThO molecules are prepared in a metastable triplet state with two valence electrons.  In this state (labeled $H ^3\Delta_1$), one electron is in a $\sigma_{1/2}$ orbital---roughly, a linear combination of $s_{1/2}$ and $p_{1/2}$ atomic Th orbitals---and provides excellent sensitivity to T,PV effects. The second electron, in a $\delta_{3/2}$ orbital, nearly cancels the magnetic moment of the first electron, and also gives rise to high polarizability due to a small $\Omega$-doublet splitting \citep{Meyer2006, Meyer2008, Vutha2010}.  In the experiment (see Fig.\ \ref{Fig_EDM_ACME}), a beam of ThO molecules is produced with a cryogenic source that yields, relative to conventional molecular beam sources, a low forward velocity, low internal temperature, and high flux.  The sequence of events experienced by molecules in the beam proceeds as follows.  First, a set of `'rotational cooling'' lasers optically pumps ground-state ThO molecules to accumulate population in a single rotational level.  Next, they enter a magnetically-shielded interaction region where an electric field $\Evec$ with magnitude $\Esca \sim 100$ V/cm is applied to achieve polarization $\mathcal{P} \cong 1$.  Once in this region, a laser pumps population from the enhanced rotational level into the $H$ state.   Next, another laser is used to spin-align the $H$-state molecules in a direction perpendicular to $\Evec$, after which they fly freely for a distance of $\approx 20$ cm. In the slow molecular beam, this corresponds to spin evolution time $\tau \approx 1$ ms, comparable to the metastable level's lifetime, $\tau_H \approx 2$ ms.  A magnetic field
$\hbox{\boldmath{$\mathcal{B}$}}$ is applied parallel to $\Evec$ to provide a bias (typically $\pi/4$ rad) to the spin precession. After the free-flight region, the final direction of the spin alignment axis is detected by the relative strength of laser-induced fluorescence when molecules are excited by a laser beam with alternating orthogonal polarizations. To suppress a wide range of systematic errors, the measurement is performed in both the positively- and negatively-polarized states of the $\Omega$-doublet and at different magnitudes of the applied field $\Esca$. With a rate $dN/dt \sim 5\times 10^4$/s of detected molecules and $\sim\! 2$ weeks of data, ACME was sensitive to an energy shift $\Delta E_{\rm TP}/h < 2$ mHz.  Given the calculated sensitivity of the ThO H state to the eEDM, from $\Esca^{\rm eff} \approx 80$ GV/cm \citep{Skripnikov2016, Denis2016, Skripnikov2013, Meyer2006}, and to the pseudoscalar electron-scalar nucleon SLI
\citep{Skripnikov2016, Denis2016},
this corresponds to limits $d_e < 9 \times 10^{-29} e\cdot$cm or $C_{\rm S} < 6\times 10^{-9}$ (both at 90\% $\mathrm{c.l.}$) \citep{Baron2017} (in each case assuming only one of the two terms is nonzero).

Very recently, results from a new experiment at JILA were reported \citep{Cairncross2017}.  Here, HfF$^+$ molecular ions in a metastable $^3\Delta_1$ state are exposed to a rotating $\Evec$-field ($\Esca \sim 20$ V/cm) that serves both to fully polarize the $\Omega$-doublet levels and to trap the ions \citep{Loh2013, Gresh2016, Leanhardt2011}.  A small, static quadrupolar magnetic field is applied; since molecules orbit a finite distance from the center of the trap, this lab-frame field gradient causes them to experience a rotating $\hbox{\boldmath{$\mathcal{B}$}}$-field, parallel to the rotating field $\Evec$. All state preparation and readout operations are performed in synchrony with the rotating fields. The spin precession frequency is measured relative to this rotating frame, with a Ramsey measurement sequence of two $\pi/2$ pulses to prepare a superposition of $\pm m$ states and then to transfer information on the final direction of the spin to the populations of these states.  Metastable state population and spin polarization along $\Evec$ is achieved with a series of laser pulses.  The $\pi/2$ pulses are induced by a rotation-induced 3$^{rd}$-order coupling between the states, amplified by briefly reducing $\Esca$. The population in one $m$ state is detemined by a series of laser pulses that photodissociates molecules only in that state, and detection of resulting Hf$^+$ ions. A remarkably long spin coherence time $\tau \approx 700$ ms is achieved. However, ion-ion Coulomb interactions in the trap limit the useful molecular density, leading to a low counting rate $dN/dt \sim 10$/s.  With $\sim\! 2$ weeks of data, this experiment was sensitive to an energy shift $\Delta E_{\rm TP}/h < 0.8$ mHz. With the calculated value $\Esca^{\rm eff} \approx 23$ GV/cm for the HfF$^+$ $^3\Delta_1$ state \citep{Skripnikov2017, Fleig2017, Fleig2013, Petrov2007, Meyer2006}, this corresponds to $d_e < 13 \times 10^{-29} e\cdot$cm (90\% $\mathrm{c.l.}$), only a factor of $1.4$ less stringent than the ACME result.  This experiment can also be interpreted as a limit on $C_{\rm S}$; from the calculated sensitivity of HfF$^+$ \citep{Skripnikov2017, Fleig2017}, we infer $C_{\rm S} < 14 \times 10^{-9}$, about $2.3$ times less sensitive than ACME.  Due to the different relative sensitivity to $d_e$ and $C_{\rm S}$ in ThO and HfF$^+$, a combination of the two can be used to set joint limits on both quantities \citep{Khriplovich1997, Jung2013, Chupp2014}.  Earlier experiments, one using a beam of YbF molecules \citep{Hudson2011} and the other using side-by-side beams of both Tl and sodium (Na, $Z$=11) atoms \citep{Regan2002}, each set limits about $10\times$ less stringent than those of ACME.

%In principle a combination of these experiments can be used to set formally correct limits on both $d_e$ and $C_S$ simultaneously rather than on a linear combination of them as (strictly speaking) follows from any single experiment.  However, in this case the weights in the linear combinations differ by little between the most sensitive experiments, and the formally correct limits on individual parameters are a factor of 30-100$\times$ less stringent than found if one or the other is assumed to vanish.  However, saturating these formal bounds would require a remarkable level of cancellation between the contributions of the two parameters, which has no \textit{a priori} rationale for occurring.

\subsubsection{Experiments on diamagnetic systems} \label{Sec:EDM:StateoftheArt:Diamagnetic}
\begin{figure}
\centering
\includegraphics[width=85.8mm]{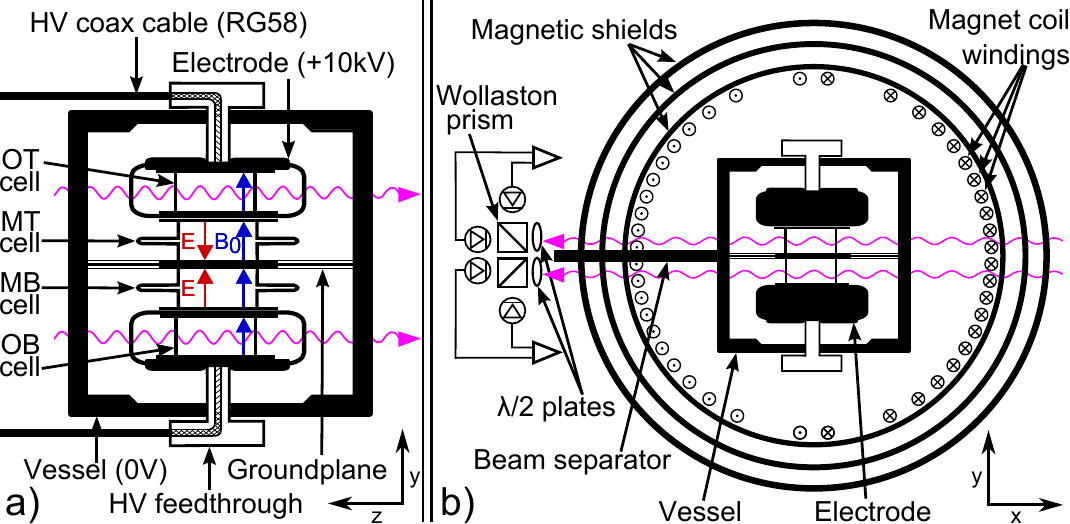}
\caption{Schematic of the $^{199}$Hg EDM experiment. Purple arrows show probe laser beams, polarization-analyzed by the combination of Wollaston prisms and photodiodes. (a)  Section through the y-z plane of the vessel containing all four vapor cells, showing probe beams through the outer cells (where $\Esca = 0$).  (b) Section through the x-y plane showing probe beams through the inner cells. Reproduced with permission from \citep{Graner2016}.}\label{Fig_EDM_199Hg}
\end{figure}

By far the most sensitive experiment using a diamagnetic system is the long-running Hg EDM search at the University of Washington \citep{Graner2016, Swallows2013}.  Here, $^{199}$Hg atoms (with a $^1S_0$ closed-shell ground state) are contained at high density in vapor cells (see Fig.\ \ref{Fig_EDM_199Hg}.  Their nuclear spins ($I =1/2$) are polarized by optical  pumping with a resonant laser beam, whose intensity is modulated at the precession frequency of the atoms in the nominally uniform and static applied $\hbox{\boldmath{$\mathcal{B}$}}$-field.  A stack of four nominally identical cells is used; the inner cells have strong, equal and opposite $\Evec$ fields along the $\hbox{\boldmath{$\mathcal{B}$}}$-field axis, while the outer cells have $\Esca = 0$.  This configuration makes it possible to cancel fluctuations not only in the average value of $\mathcal{B}$, but also in its first-order gradient.  At the applied field  $\Esca \approx 10$ kV/cm, the atomic Hg reaches a polarization $\mathcal{P} \sim 3\times 10^{-5}$.  The cells are filled with $\sim 0.5$ atm of CO buffer gas to slow diffusion of the Hg atoms to the walls, which are coated with paraffin to suppress spin relaxation.  After initial polarization, the spins freely precess over $\tau = 170$ s, after which nearly all remain polarized.   The final spin direction is probed by monitoring the angle by which the linear polarization of a near-resonant probe laser beam is rotated as it passes through the atomic vapor.  Decades of development led to cells with extremely low leakage currents ($<40$ fA). The slow diffusion ensures small motional field effects.  The primary systematic errors were associated with nm-scale voltage-induced movements of the vapor cells together with uncontrolled $\mathcal{B}$-field gradients.  With $N \sim 10^{14}$ atoms detected in each measurement cycle and $\sim\! 250$ days of data, the experiment was sensitive to an energy shift $\Delta E_{\rm TP}/h < 20$ pHz.  From the calculated sensitivity of the atomic EDM to the nuclear SMt $\SMt_{\rm Hg}$, this set a limit $\SMt_{\rm Hg} < 3 \times 10^{-13}~e\cdot$fm$^3$ (95\% c.l.). This can be interpreted in terms of underlying mechanisms that give rise to $\SMt$.  For example, this yields a limit on the neutron EDM, $d_n < 1.6 \times 10^{-26} e$ cm, that is more stringent than the best limit from direct measurements with free neutrons by a factor of $\sim 2$.  Similarly, the $^{199}$Hg experiment sets the best limits on quark cEDMs ($\tilde{d}_u - \tilde{d}_d < 6\times 10^{-27}$ cm) and on the observable QCD $\theta$-parameter, $\bar{\theta}_{\rm QCD} < 1.5\times 10^{-10}$ as well as on hadronic $T$-,$P$-odd couplings, pseudoscalar-scalar and tensor-tensor SLI couplings, and the proton EDM $d_p$.  Remarkably, despite having no sensitivity to the
scalar-pseudoscalar SLI at lowest order, the limit on $C_S$ from $^{199}$Hg is only $\sim\! 2\times$ less strict than that from ACME.

Other experiments with diamagnetic systems also have set limits on nuclear SMts and nuclear spin-dependent SLI couplings.  The most sensitive include searches for an EDM of $^{129}$Xe \citep{Rosenberry2001} and $^{225}$Ra atoms \citep{Bishof2016}, and for a T,PV energy shift in $^{205}$TlF molecules \citep{Cho1989, Cho1991}.  All of these are several orders of magnitude less sensitive to the underlying physics than is the $^{199}$Hg experiment.  Nevertheless, their sensitivity to different linear combinations of the large set of parameters needed to describe T,PV in these systems makes them useful for providing global constraints \citep{Chupp2014}.

\subsubsection{Role of low-energy theory} \label{Sec:EDM:StateoftheArt:LowEnergyTheory}
Interpreting the results of atomic and molecular EDM experiments in terms of underlying physical parameters requires knowledge of electronic wavefunctions.  Calculations of EDM sensitivity (i.e. ratio of atomic EDM to eEDM or SLI coupling strength) for paramagnetic atoms are similar to those needed to interpret APV experiments, and have accuracy $\lesssim 5\%$ for single valence-electron atoms such as Tl, Cs, and Fr [see, for example, \citet{Dzuba2009}].  Remarkably, calculations for paramagnetic molecules with one valence electron (YbF, BaF) or even two (ThO, HfF$^+$) now have accuracy of $10\%$ or better [see, for example, \citet{Abe2014}, \citet{Denis2016}, and \citet{Skripnikov2016}].  Calculations of electronic structure for diamagnetic systems---both atoms (Hg, Ra, Xe) and molecules (TlF)---give the ratio between observable energy shift and nuclear SMt with accuracy $\lesssim 20\%$ [see, for example, \citet{Dzuba2002}].  For the null results from all current EDM experiments, these small uncertainties have negligible impact on the limits that can be set on underlying physics.

By contrast, theoretical uncertainties associated with strongly interacting particles are not negligible for interpretation of underlying hadronic T,PV parameters.  There are difficulties with the relations both between quark- and nucleon-level parameters (e.g.,  what value of the proton EDM $d_p$ results from a given value of $\bar{\theta}_{\rm QCD}$ or the up-quark chromo-EDM $\tilde{d}_u$) and between nucleon- and nucleus-level parameters (e.g., what value of a nuclear SMt arises from $d_p$ or from a given strength of effective nucleon-nucleon T,PV interaction).  In the former case, the uncertainties are estimated to be at the level of $\sim\! 100\%$; in the latter, they can be as large as
$\sim\! 500\%$, i.e.\ even the sign of the relation is not reliably known \citep{Pospelov2005, Engel2013}.  These uncertainties are typically not folded into quoted limits on fundamental parameters from diamagnetic system EDM experiments; if properly included, the corresponding limits would typically be weaker by factors of a few.

\subsection{Impact on particle physics} \label{Sec:EDM:Impact}

To discuss the impact of these experiments, it is useful to begin with crude estimates for the size of the underlying effects in models with new T,PV physics at a high energy scale \citep{Pospelov2005, Commins2009}.  First, consider effects associated with the EDMs (and cEDMs) of the light fundamental fermions that make up atoms: the electron and the up and down quarks.  The non-renormalizable EDM Lagrangian $\mathcal{L}_{\mathrm{EDM}}$ describes the effect of radiative corrections (Feynman loop diagrams) in the underlying theory.  If the assocated diagram for a fermion with mass $m_f$ has $n_\ell$ loops that contain heavy new particles with mass up to $m_X$, a typical size of the associated EDM will be $$d \sim \mu_f \sin\phi_{\rm CP} \left( g^2/ 2 \pi \right)^{n_\ell} m_f^2/m_X^2,$$ where $\mu_f = e\hbar/(2m_f c)$ is the magnetic moment for a Dirac fermion and $g$  is a dimensionless coupling strength (e.g., $g^2 = \alpha$ for electromagnetic interactions). The factor $1/m_X^2$ is associated with the propagator of the heavy particle in the loop.

In the SM, electron and quark EDMs appear only at four- and three-loop level, respectively \citep{Khriplovich1997}.  There is a strong additional suppression of EDMs in the SM due to a near-cancellation in the sum over all contributing amplitudes \citep{Shabalin1978, Nanopoulos1979, Hoogeveen1990}.  This  mechanism, which arises from the explicit linkage of flavor mixing and CPV via the CKM matrix, makes the SM predictions for EDMs extraordinarily small---for example, some 5-10 orders of magnitude below current limits for $d_n$ and $d_e$, respectively.  By contrast, for an uncancelled 1-loop diagram ($n_\ell = 1$) and with $\sin\phi_{\rm CP} \sim 1$, the current limit on the eEDM corresponds to $m_X \gtrsim 10$ TeV; bounds from $^{199}$Hg on the quark chromo-EDMs probe a similar scale \citep{Barr1993, Pospelov2005, Engel2013}.

%Consider first the $SP$ semileptonic interaction: if this is associated with the exchange of a heavy boson $X$ that couples with $CP$-odd phase
%$\phi_{\rm CP}$, the limit on $C_S$ indicates that the mass $m_X$ must satisfy $m_X \gtrsim \sqrt{m_Z / (C_S \sin\phi_{\rm CP})} $, which for
%$\sin\phi_{\rm CP} \sim 1$ means $m_X \gtrsim 1000$ TeV.  Although this bound is smaller in most realistic theoretical models, this gives a rough sense
% of the extraordinary scale of energies probed with this type of atomic/molecular experiment.

Detailed calculations of the size of the relevant T,PV parameters have been made in a wide range of theoretical models.
Among the most widely explored are models that incorporate Supersymmetry (SUSY) that is broken near the electroweak scale, i.e.\ which predict superpartner particles with mass $M_{\rm SUSY} \sim M_Z \sim 0.1$ TeV. Weak-scale SUSY naturally includes many attractive features \citep{Kane2002}: it stabilizes the Higgs mass against radiative corrections, at around its observed value; includes candidate particles for dark matter; modifies the energy-dependent running of strong, weak, and electromagnetic couplings so that they converge at a sensible scale for grand unification; and includes new CPV phases $\delta_{\rm SUSY}$ that could produce the cosmic baryon asymmetry.

%\begin{figure}
%\centering
%\includegraphics[width=70mm]{RMP_EDM_Fig_2_vertical}
%\caption{Some Feynman diagrams.}\label{Fig_EDM_Feynman_diagrams}
%\end{figure}

\begin{figure*}
\centering
\includegraphics[width=150mm]{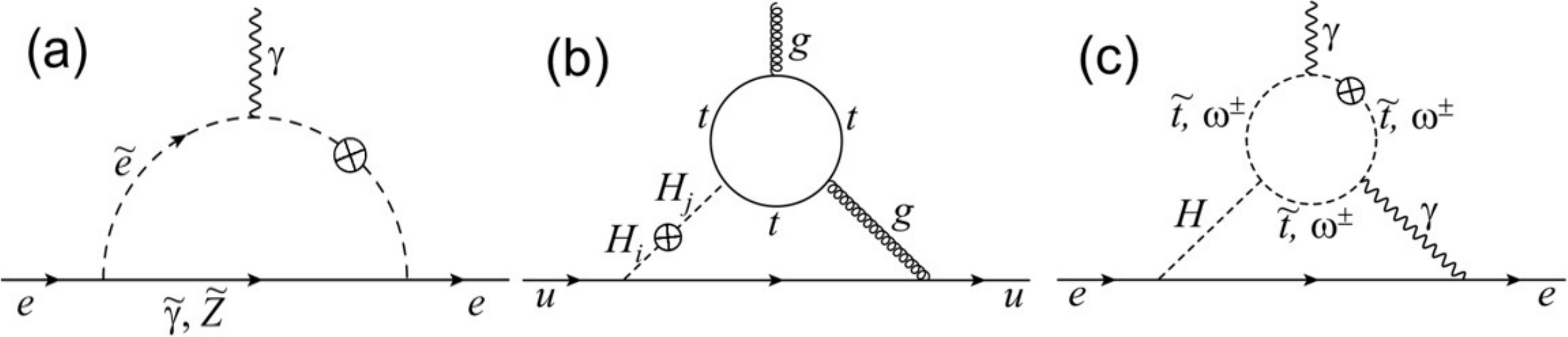}
\caption{Example Feynman diagrams leading to particle EDMs. Crosses represent CPV phases and tildes indicate SUSY partners of SM fields. (a) One-loop diagram leading to an eEDM in a SUSY model. The vertical photon represents the coupling of the EDM to an electric field. The CPV phase arises from the mechanism that leads to breaking of SUSY at low energies.  (b) Two-loop diagram leading to an up quark cEDM in a multi-Higgs doublet model.  $H_{i,j}$ represent different Higgs fields, and the CPV phase arises from mixing between them.  The dominant diagram includes $t$ quarks since their large mass indicates a strong coupling to the Higgs field. (c) Dominant two-loop diagrams leading to an eEDM in SUSY models where partners of fermions are heavy.  Here $\omega^\pm$ are SUSY partners of $W^\pm$ bosons.}\label{Fig_EDM_Feynman_diagrams}
\end{figure*}

The simplest weak-scale SUSY models include one-loop diagrams that lead to EDMs much larger than the experimental limits, unless $\delta_{\rm SUSY} M_{\rm SUSY}^{-2} \lesssim$ (10 TeV)$^{-2}$ \citep{Barr1993, Pospelov2005, Engel2013, Feng2013}.  Improved EDM sensitivity by 1-2 orders of magnitude will either yield a discovery or conclusively rule out SUSY models, such as these, that are compatible with electroweak baryogenesis \citep{Balazs2016, Cirigliano2010, Huber2007}.  There is growing interest in models where only a few of the new particles have  $M_{\rm SUSY} \sim M_Z$
%(e.g., in ``split SUSY'' only the so-called gauginos and higginos, the fermionic partners to the gauge and Higgs bosons;
%and in ``natural SUSY'' only the higgsinos, gluino, and the bosonic partners to the top and bottom quarks,
%the so-called stop and sbottom squarks)
while all other SUSY partners have much higher mass \citep{Arkani2005}.  Here, the primary contribution to EDMs usually comes from two-loop diagrams.  Even in these scenarios the eEDM and quark cEDM limits correspond to lower bounds of $\sim\! 2-4$ TeV on the masses of the lighter SUSY particles, if $\delta_{SUSY} \sim 1$ \citep{Giudice2006, Nakai2016}.  This is well beyond the direct reach of the Large Hadron Collider (LHC) for these types of particles \citep{PDG}.   Some typical Feynman diagrams leading to particle EDMs are shown in Fig.\ \ref{Fig_EDM_Feynman_diagrams}.

SUSY is a well-motivated and thoroughly investigated extension to the SM.  However, in nearly every model that predicts new physics near the electroweak scale, new CPV phases appear and T,PV signals in atomic/molecular experiments should arise at values near current experimental bounds.  For example, there may be additional scalar fields in nature, analogous to the Higgs boson.  In such multi-Higgs models, the relative phase between the fields, $\phi_h$, can lead to CPV \citep{Weinberg1976}.  Exchange of the Higgs bosons between electrons and nucleons can lead to T,PV SLIs \citep{Barr1992}, and two-loop diagrams including the Higgs can lead to fermion EDMs and cEDMs \citep{Barr1990}.  The relative importance of the SLI and EDM contributions to T,PV signals depends on the details of parameters in the theory.  Broadly speaking, however, in these models the $^{199}$Hg and ACME experiments set limits of $$\sin\phi_hM_{h^\prime}^{-2} \lesssim (5~\mr{TeV}/c^2)^{-2},$$ where $M_{h^\prime}$ is the mass of a new Higgs particle  \citep{Barr1993, Pospelov2005,Engel2013}. Again, this substantially exceeds direct LHC bounds on $M_{h^\prime}$, if $\phi_{h} \sim 1$.

A few authors have also begun to explore the implications of EDM limits on the possible existence of new particles with mass below the electroweak scale, but with very weak couplings to ordinary matter---``dark sector'' particles and their associated dark forces, discussed in Secs.~\ref{Sec:APV} and \ref{Sec:ExoticSpin:TheoreticalMotivation:hidden-photons-Z-bosons}.  It has been argued \citep{LeDall2015} that within a broad class of models where new particles appear \textit{only} at low mass scales, EDM limits provide less stringent constraints than other types of experiments (aside from limits on $\bar{\theta}_{\rm QCD}$, which is not associated with a high mass scale and can appear in such a scenario).  However, more generally (e.g., where new particles are present at both low and high mass), EDMs can provide very strict limits on T,PV couplings to dark force carriers with mass below the electroweak scale.  The first example of such an analysis (for tensor SLIs) was carried out in \citet{Gha15}.

\subsection{Future directions}  \label{Sec:EDM:Future}
Here we briefly review ongoing or planned T,PV experiments known to us.  A few themes are common in new experimental approaches.  For example, because of the importance of obtaining long spin precession times, techniques of laser cooling (to obtain lower velocities) and trapping (for the longest hold times) are beginning to be used.  Several new and ongoing experiments are also exploiting the high polarizability of polar molecules for enhanced sensitivity.  A few groups plan to employ both these concepts, leveraging the recent initial demonstrations of laser cooling and trapping of polar molecules.  These methods require optical cycling behavior, which in itself enables both efficient detection (when each molecule emits many fluorescent photons) and cooling of internal states, such as rotation, via optical pumping.

\subsubsection{Paramagnetic systems} \label{Sec:EDM:Future:Paramagnetic}
Improvements in all the recent experiments using paramagnetic molecules are underway. The ACME collaboration plans several upgrades to improve statistics.  In the ongoing second generation of ACME, improved efficiency of state preparation and detection is anticipated to improve the sensitivity to $d_e$ by $\sim\!\! 20\times$ \citep{Panda2016}.  A third generation with increased molecular beam flux from a new source and electrostatic focusing, enhanced detection efficiency via optical cycling, and longer integration time could yield another $\sim\!\! 30$-$50\times$ improvement, corresponding to sensitivity at the level $d_e \sim10^{-31} e\cdot$cm \citep{Vutha2010}.
Simultaneously, the YbF experiment at Imperial College is also improving beam flux and velocity by use of a cryogenic beam and rotational cooling, plus optical cycling fluorescence for efficient detection \citep{Rabey2016}.  YbF reaches $\mathcal{P} \approx 60\%$ at $\Esca = 10$ kV/cm, yielding $\mathcal{P}\Esca_{\rm eff} \approx 15$ GV/cm, $\approx\! 5\times$ smaller than $\Esca_{\rm eff}$ in ThO \citep{Kara2012}.  Future plans call for a dramatic increase in interaction time by use of a laser-cooled molecular fountain similar to that used for atomic Cs clocks \citep{Tarbutt2013}. \citet{Kozyryev2017} recently proposed using certain types of polyatomic paramagnetic molecules that could be laser cooled, and also have a favorable energy level structure for eEDM measurements (similar to the $\Omega$-doublet states used in ACME). Meanwhile, the JILA trapped molecular ion experiment plans a new trap electrode geometry to allow use of much larger ion clouds.  A $\sim\!\!10\times$ improved sensitivity to $d_e$ is projected.  In the longer term, use of the heavier species ThF$^+$ will improve $\mathcal{E}_{\rm eff}$ by a factor of $1.5\times$, and possibly also enable longer spin coherence time since here $^3\Delta_1$ is the ground state \citep{Gresh2016,Cairncross2017}.  \citet{Kawall2011} also proposed to perform similar experiments in a storage ring to further increase the trapping volume.

%In the meantime, the JILA uses HfF$^+$ molecular ions in a metastable $^3\Delta_1$ electronic state (similar to the one in ThO used for ACME), and has demonstrated very long spin coherence times  ($\sim\! 0.5$ s) in an EDM-sensitive state for ions trapped by a rotating electric field\citep{Loh2013, Gresh2016, Leanhardt2011}.  At present, the statistics are low due to a limit on the density of ions in the trap (Coulomb interactions between colliding ions lead to systematic errors), and $\mathcal{E}_{\rm eff}$ is $\sim\! 4\times$ smaller than in ThO.  Significantly improved state preparation and detection efficiencies are anticipated, and use of the heavier species ThF$^+$ will improve $\mathcal{E}_{\rm eff}$ and possibly also coherence time (since here $^3\Delta_1$ is the ground state) \citep{Gresh2016}.  It has also been proposed to perform similar experiments in a storage ring to increase the trapped volume \citep{Kawall2011}.

Several other efforts are also under development.  A group at Pennsylvania State University is using laser cooled and optically trapped Cs atoms, with co-trapped Rb as a co-magnetometer \citep{Weiss2003}.  Here long coherence times, $\sim\! 3$ s, are anticipated, along with large counting rates to compensate for the low $Z$ and low $\mathcal{P}$ in Cs atoms, which yield an effective field $F_e\mathcal{E} \sim 10^7$ V/cm with $F_e$(Cs)$= 120$ and $\mathcal{E} \sim10^5$ V/cm.  A group at Tohoku University is planning a similar experiment using Fr atoms \citep{Inoue2015}, with $F_e = 900$.  Finally, a group at the University of Groningen is constructing an apparatus to electrically decelerate \citep{Mathavan2016} a beam of BaF molecules (with $\mathcal{P}\Esca_{\rm eff} \approx 5$ GV/cm), then apply transverse laser cooling to obtain a very bright and slow molecular beam \citep{Hoekstra2015}.

\subsubsection{Diamagnetic systems} \label{Sec:EDM:Future:Diamagnetic}
A new generation of the $^{199}$Hg experiment is planned, with various technical improvements to increase sensitivity by $2$-$3\times$ \citep{Heckel2016}. A newer effort at Argonne National Lab uses laser-cooled and optically trapped $^{225}$Ra atoms  \citep{Bishof2016}.  Here there is a large enhancement in the SMt of $^{225}$Ra, for the same size of underlying parameters such as quark cEDMs.  An octupole deformation of the $^{225}$Ra nucleus leads to closely-spaced opposite-parity nuclear energy levels, analogous to $\Omega$-doublet levels in polar molecules, and a similar enhancement of the induced SMt due to T,PV effects in the nucleus \citep{Auerbach1996}.  However, the relatively short half-life of $^{225}$Ra ($\sim\! 15$ days) complicates the experimental protocol.  Calculations of the SMt for given values of the microscopic T,PV parameters \citep{Ban2010} generally indicate a 100-1000$\times$ larger SMt in $^{225}$Ra as compared to $^{199}$Hg; the atomic structure of Ra gives another $3\times$ enhancement \citep{Dzuba2002}.  In addition, it may be possible to make the uncertainty in the relation between fundamental parameters and nuclear SMt smaller in octupole-deformed nuclei such as $^{225}$Ra than in more spherical nuclei such as $^{199}$Hg. The Argonne group recently reported a limit on the EDM of atomic $^{225}$Ra, $d_{\rm Ra} < 1.4\times 10^{-23}~e\cdot$cm (95\% c.l.), corresponding to a limit on the SMt of $S_{\rm Ra} < 2 \times 10^{-7} e\cdot$fm$^3$ \citep{Bishof2016}.  While this is $\sim\! 1000 \times$ less sensitive than the $^{199}$Hg experiment at present, dramatic improvements are anticipated in trapped atom number, detection efficiency, and $\Esca$-field strength \citep{Bishof2016}.  Another effort to take advantage of an enhanced SMt due to octupole deformation is underforway at TRIUMF using $^{233}$Rn, which can be collected in a vapor cell after production at a radioactive beam facility \citep{Tardiff2014}.  Groups at Munich \citep{Kuchler2016}, Mainz \citep{Zimmer2017}, and Tokyo \citep{Sato2015} are preparing new measurements of the $^{129}$Xe EDM.  All will use vapor cells, where extraordinarily long spin coherence times can be achieved; all also use $^3$He as a comagnetometer.  However, Xe ($Z=54$) has lower intrinsic sensitivity than $^{199}$Hg; moreover, the inaccessability of optical transitions in Xe forces the use of direct magnetic field sensing of the nuclear spins, with lower signal/noise than is routinely achieved with laser-based detection methods.

Finally, groups at Yale, Columbia, and University of Massachusetts are constructing a new experiment (CeNTREX) to measure the SMt of $^{205}$Tl.  CeNTREX will use a cryogenic beam of TlF molecules, with rotational cooling and electrostatic focusing for a large useful flux, plus optical cycling for efficient detection of laser-induced fluorescence \citep{Norrgard2017}.   The $^{19}$F nucleus will be used as a co-magnetometer.  With near-unit polarization, the sensitivity of TlF to the SMt is $\sim\! 10^4\times$ larger in TlF than in the $^{199}$Hg experiment \citep{Dzuba2002}.  This helps to overcome a small spin coherence time of $\sim\!15$ ms in the TlF cryogenic beam.  Future generations will employ transverse laser cooling for improved beam flux and, eventually, optically trapped molecules for long spin coherence time.  New experiments also have been proposed to search for a nuclear MQM \citep{Flambaum2014}.  Here two enhancement mechanisms can be employed. First, using a nucleus with large quadrupole deformation enhances the MQM by a factor of  $\sim\! 10-20$ relative to spherical nuclei \citep{Flambaum1994}.  Second, using molecules in a $^3\Delta_1$ state gives the unpaired electron spin needed to couple to the nuclear MQM, high polarization $\mathcal{P}$, and suppressed magnetic moment relative to typical paramagnetic systems, just as in the ThO eEDM experiment. For the same underlying T,PV parameters, the energy shifts in such a system could be $\sim\! 10^7\times$ larger than in the  $^{199}$Hg experiment.

%\bibliography{EDM_v3}
%
%
%\end{document}

\section{Tests of the $CPT$ theorem, matter-antimatter comparisons}
\label{Sec:CPT}

Current physical laws are believed to be invariant under the $CPT$ transformation (the $CPT$ theorem), i.e. combined transformations of charge conjugation, spatial inversion and time reversal. Within conventional field theory, the $CPT$ symmetry is closely related to Lorentz invariance; however, in more general frameworks such as string theory, there is a possibility in principle to violate one symmetry without violating the other \cite{Gre02}. This topic has been a subject of recent research and lively debates \cite{Dolgov2012,Tureanu2013,Kostelecky2015}.

Since weak interactions are not invariant under charge conjugation and also violate $CP$, a prudent question is whether violation of these symmetries may result in a difference of properties between particles and antiparticles. As it turns out, within the framework of conventional field theory, $CPT$ invariance ensures the equality of masses and total lifetimes between particles and antiparticles \cite{Luders1957} and the same is true for the magnitude of the magnetic moments \cite{BluKosRus97}.

Comparison of particle and antiparticle properties, therefore, provides tests of the $CPT$ theorem and detection of any discrepancies will be an unambiguous signal of new physics, motivating such experiments which have seen significant progress in recent years.
Tests of the $CPT$ theorem were recently reviewed  by
\citet{Yamazaki2013}, \citet{GabHooDor14}, and \citet{Kellerbauer2015}. Here,
we present only a brief account of recent results and progress toward future $CPT$ tests with antiprotons and antihydrogen.

The ALPHA experiment at CERN demonstrated trapping of antihydrogen ($\overline{\textrm{H}}$) atoms for 1000 s in 2011
\cite{Alpha11}. With the goals of performing spectroscopy
 of the $1S-2S$  and hyperfine transitions for a comparison with
their values in hydrogen,
 the ALPHA team carried out a proof-of-principle experiment using resonant microwave
radiation to flip the spin of the positron in magnetically trapped antihydrogen atoms \cite{Alpha12}.
The spin flip caused trapped $\overline{\textrm{H}}$  to be ejected from the trap and
detected via annihilation.
 While this experiment  was not aimed
at precision frequency measurement, it bounded the resonance within 100~MHz
of the hydrogen hyperfine frequency, corresponding to a
relative precision of about $4\times 10^{-3}$ \cite{Alpha12}.
The ATRAP collaboration reported
accumulation of $4 \times 10^9$ electron-cooled positrons in a Penning trap for production and storage of antihydrogen atoms for future tests of $CPT$ and
antimatter gravity \cite{Fitzakerley2016}.

In 2014, the ASACUSA experiment succeeded for the first time in producing a beam of antihydrogen atoms;
 detection of 80 antihydrogen atoms 2.7 metres downstream of their production was reported \cite{ASACUSA14}. This result represents a milestone towards precision measurements of the ground-state hyperfine splitting of antihydrogen using beam spectroscopy.

An experimental limit on the charge $Qe$ of antihydrogen, in which $e$ is the elementary charge,
was reported by ALPHA collaboration \cite{Amo14}. In 2016, they further improved this bound to
$|Q|<0.71$ parts per billion (one standard deviation) \cite{Alpha16}.
Assuming charge superposition and using the best measured value of the antiproton charge \cite{PDG}, this measurement
placed a new limit on the positron charge anomaly, i.e. the relative difference between the positron and elementary charge, of about one part per billion (ppb).

In December of 2016, ALPHA reported a long awaited breakthrough result \cite{ALPHA1S2S2017}: they have further improved the efficiency of antihydrogen production (trapping about 14 antiatoms per trial), and employed two-photon laser excitation with 243 nm light to drive the $1S-2S$ transition. The initial measurements of the transition frequency indicated that it is equal to its hydrogen counterpart at the level of $2\times 10^{-10}$ with further significant improvements anticipated in the near future.
In 2017, the results of a microwave spectroscopy experiment which probed the response of antihydrogen over a controlled range of frequencies were reported \cite{AhmAlvBak17} providing a direct, magnetic-field-independent measurement of the hyperfine splitting of $1420.4 \pm 0.5$ MHz, consistent with expectations from atomic hydrogen at the level of four parts in $10^4.$

ATRAP collaboration \cite{DisMarMar13} measured the antiproton magnetic moment with a 4.4 parts per million (ppm) uncertainty with a single particle.

The Baryon Antibaryon Symmetry Experiment (BASE) experiment aims at precise comparisons of the fundamental properties of antiprotons and protons for tests of $CPT$ \cite{BASE15a}.
 The BASE collaboration observed the first spin flips with a single trapped proton \cite{UlmRodBla11} and  performed a direct measurement of the magnetic moment of a single trapped proton with a precision of 3.3 ppb, which is the most precise measurement
of $g_p$ to date \cite{MooUlmBla14}.
\citet{NagSmoSel17} measured the magnetic moment of a
single trapped antiproton in a single Penning trap with a
superimposed magnetic bottle achieving fractional precision
is at 0.8 ppm at the 95\% confidence level improving the
fractional precision  by a factor
of 6. To avoid the broadening of the resonance lines
due to the magnetic bottle, a two-trap method was developed separating the high-precision frequency measurements to a homogeneous precision trap. The spin-state analysis is performed in a trap with a superimposed magnetic inhomogeneity. Further extension of this method to an advanced cryogenic multi-Penning trap system enabled a parts-per-billion measurement of the antiproton
magnetic moment  \cite{SmoSelBor17}.
 This experiment used a particle with an effective temperature of 300~K for magnetic field measurements and a cold particle at 0.12~K for spin transition spectroscopy.
\citet{SmoSelBor17} improved the precision of the $\mu_{\bar{p}}$ measurement by a
factor of $\approx 350$, reporting the value
$\mu_{\bar{p}}  = 2.792\,847\,3441(42)\mu_N$ (at the 68\% confidence level).
 A measurement of the proton magnetic moment at the 0.3 ppb level was reported by \citet{SchMooBoh17}; the resulting value  is in agreement with the currently accepted CODATA value \cite{CODATA2014} but is
an order of magnitude more precise. A comparison of the $\mu_{\bar{p}}$ with the new proton value $\mu_p  = 2.792 847 344 62 (82) \mu_N$ constrains some CPT-violating effects.

 The BASE collaboration proposes to use quantum-logic technologies \cite{HeiWin90,DubNiePas13} to trap and probe (anti)protons by  coupling the (anti)proton to an atomic ``qubit'' ion trapped in its vicinity via Coulomb interaction. This coupling will be used for both ground-state cooling of single (anti)protons and for the state readout.
Such sympathetically cooling the (anti)proton will reduce
particle preparation times by more than two orders
of magnitude, potentially enabling the
proton and antiproton magnetic moment measurements at the
parts per trillion level \cite{SchMooBoh17}.

The BASE collaboration also performed a comparison of the  charge-to-mass ratio for the antiproton $(q/m)_{\overline{\mr{p}}}$
to that for the proton $(q/m)_{\mr{p}}$ using
high-precision cyclotron frequency comparisons of a single
antiproton and a negatively charged hydrogen ion (H$^-$) carried
out in a Penning-trap system \cite{BASE15}. This experiment established a limit
\begin{equation}
\frac{(q/m)_{\overline{\mr{p}}}}{(q/m)_{\mr{p}}} -1= 1(69)\times 10^{-12}
\label{eq1-CPT}
\end{equation}
and gave a bound on sidereal variations in the measured ratio of $< 720$
parts per trillion  \cite{BASE15}.

Three-body metastable antiprotonic helium $\bar{p}$He$^+$ consists of
an $\alpha$-particle, an electron and an antiproton, $\bar{p}$. When He  captures a slow $\bar{p}$ in an atomic collision, $\bar{p}$He$^+$ is often formed in a high Rydberg state of  $\bar{p}$ orbiting He$^+$. Such states are
 amenable to precision laser spectroscopy in order to determine the antiproton-to-electron mass ratio and to test the equality between the magnitudes of antiproton and proton charges and masses. Two-photon spectroscopy of $\bar{p}$He$^+$ performed by \citet{HorSotBar11} resulted in the determination of the
 antiproton-to-electron mass ratio $m_{\bar{\mathrm{p}}}/m_{\mathrm{e}}  = 1836.1526736(23)$. Recently, \citet{Hori2016} employed buffer-gas cooling and performed single-photon spectroscopy of $\bar{p}$He$^+$.
  Combining the experimental results and the high-precision calculations of the relevant transition
frequencies performed by \citet{KorHilKar14,KorHilKar14a,Kor14,KorHilKar15},
 yielded a more precise value of 1836.1526734(15) \cite{Hori2016}, which agrees with the CODATA recommended value of $m_\mathrm{p}/m_\mathrm{e}$  at a level of 0.8 ppb. Laser spectroscopy of pionic helium atoms to determine the charged-pion mass was proposed by \citet{HorSotKor14}.

The experimental efforts on matter-antimatter comparisons
aimed at testing whether antimatter is affected by gravity
in the same way as matter are described in Sec.~\ref{antigrav}.

Due to the deep intrinsic connection between $CPT$ and other symmetries such as Lorentz invariance, testing $CPT$ does not always require antimatter \cite{Pospelov2004}. A recent review of ``magnetometry'' experiments in this area was given by  \citet{Kimball2013Tests}; see also Sec.~\ref{Sec:LV} of this review.

\section{Review of laboratory searches for exotic spin-dependent interactions}
\label{Sec:ExoticSpin}

\subsection{Early work}
\label{Sec:ExoticSpin:EarlyWork}

Ever since the discovery of intrinsic spin [see the historical review by \citet{Com12}], a central question in physics has been the role of spin in interactions between elementary particles.  Leptons and quarks, the fundamental fermions, are spin-$1/2$ particles which in principle can possess only two possible multipole moments: monopole moments (such as mass and charge) and dipole moments (such as the magnetic moment). A particle's dipole moment is necessarily proportional to its spin based on the Wigner-Eckart theorem. In fact, the inception of the idea of spin was based on the observation of the anomalous Zeeman effect, a consequence of the interaction of the electron's magnetic dipole moment with an external magnetic field. It is natural to ask what other sorts of spin-dependent interactions might exist between fermions apart from the magnetic dipole interaction.

\subsubsection{Torsion in gravity}
\label{Sec:ExoticSpin:EarlyWork:torsion}

There were a number of hypothetical dipole interactions postulated and searched for soon after the discovery of intrinsic spin. An early theoretical question was how to incorporate the concept of spin into the framework of general relativity. The fact that intrinsic spin possessed all the usual properties of angular momentum but yet could not be understood as arising from the physical rotation of an object posed a deep question for attempts to extend our understanding of gravity to the quantum level. There were indeed general relativistic interactions, such as frame-dragging \cite{Len18,Tho85}, between macroscopic rotating bodies possessing angular momentum. But it was unclear if analogous effects would exist for particles with spin since general relativity, being a geometrical theory, did not directly include the possibility of intrinsic spin.  At the macroscopic scale, mass-energy adds up due to its monopole character and leads to observable gravitational effects. On the other hand, spin, due to its dipole character, tends to average out for astrophysical bodies such as stars and planets. Thus any gravitational effects related to spin would tend to be difficult to detect through astronomical observations, which are the principal vehicles for tests of general relativity to this day. Nonetheless, soon after the invention of general relativity by \citet{Ein16}, and even before the discovery of electron spin, Cartan proposed an extension of general relativity that opened the possibility of incorporating spin through its effect on the torsion of spacetime \cite{Car22,Car23,Car24,Car25}. Significant later work \cite{costa1942dynamique,Wey47relativistic,stueckelberg1948possible,papapetrou1949,weyl1950remark,costa1964translational} strengthened the theoretical connection between intrinsic spin and spacetime torsion. Torsion quantifies the twisting of a coordinate system as it is transported along a curve. In Einstein's general relativity, mass-energy generates curvature of spacetime but the torsion is zero, and so vectors curve along geodesics via parallel transport but do not twist. In Cartan's extension, spin generates nonzero torsion, and so frames transported along geodesics curve due to the effect of mass-energy and twist due to the effect of spin [see, for example, the review by \citet{Heh76}]. The consequence is that gravitational dipole interactions are possible within the framework of Einstein-Cartan theory. From another point-of-view, assuming there is a way to parameterize gravity in terms of a quantum field theory, in addition to the spin-2 graviton (the hypothetical quantum of the gravitational field associated with Einstein's general relativity), there might exist spin-0 and spin-1 gravitons associated with the torsion field.

\subsubsection{Electric dipole moments}

In 1950, \citet{Pur50} proposed another significant idea: elementary particles might possess a permanent electric dipole moment (EDM) in addition to a magnetic dipole moment. As discussed in detail in Sec.~\ref{Sec:EDM} of this review, this hypothetical electric dipole coupling has stimulated intensive experimental and theoretical interest ever since.

\subsubsection{Axions and axion-like-particles (ALPs)}
\label{Sec:ExoticSpin:EarlyWork:axions-ALPs}

The above ideas involved new dipole couplings to known fields, gravitational and electric. It was later realized that another possibility existed: there could be heretofore undiscovered fields generating dipole couplings between fermions. Among the earliest and most influential of these proposals was the suggestion that a light spin-0 boson, the axion \cite{Wei78,Wil78,Din81,Shi80,Kim79}, could possess a coupling to dipoles that might be detectable in laboratory experiments \cite{Moo84}. As Moody and Wilczek note, a spin-0 field $\varphi$ can couple to fermions in only two possible ways: through a scalar vertex or through a pseudoscalar vertex. In the nonrelativistic limit (small fermion velocity and momentum transfer), a fermion coupling to $\varphi$ via a scalar vertex acts as a monopole and a fermion coupling to $\varphi$ via a pseudoscalar vertex acts as a dipole. This can be understood from the fact that in the particle's center of mass frame, there are only two vectors from which to form a scalar/pseudoscalar quantity: the spin $\mb{s}$ and the momentum $\mb{p}$ (since the field $\varphi$ is a scalar), so the either the vertex does not involve $\mb{s}$ (monopole coupling) or if it does, it depends on $\mb{s}\cdot\mb{p}$, which is a $P$-odd, pseudoscalar term. Hence it is the pseudoscalar coupling of $\varphi$ that is the source of new dipole interactions.

The axion emerged from an elegant solution to the strong-$CP$ problem (see Sec.~\ref{Sec:EDM:Underlying:SchiffMoment}). The strong-$CP$ problem is that the observable $CP$-violating  phase that can appear in the QCD Lagrangian, $\bar{\theta}_{\rm QCD}$, is known from EDM limits to be extemely small: $\bar{\theta}_{\rm QCD} \lesssim 10^{-10}$. This presents a so-called fine-tuning problem, since na\"ively one would expect $\bar{\theta}_{\rm QCD} \approx 1$. The solution to the strong-$CP$ problem proposed by \citet{Peccei1977a,Pec77b} was that $\bar{\theta}_{\rm QCD}$ does not possess a constant value, but rather evolves dynamically. In this model, $\bar{\theta}_{\rm QCD}$ is replaced in the Lagrangian by a term representing a dynamical field, and the quantum of this field is known as the axion (or, more specifically, the QCD axion). The underlying physics of the Peccei-Quinn solution to the strong-$CP$ problem is closely related to the physics behind the Higgs mechanism endowing particles with mass in the Standard Model: there exists a global continuous symmetry in QCD that is spontaneously broken, and a result of the spontaneous symmetry breaking is the appearance of a new ``pseudo-Nambu-Goldstone'' boson (in this case the axion). It turns out that the mass of the axion is very small [upper limits on the axion mass based on astrophysical observations are $\lesssim 10~{\rm meV}$ \cite{Raf99}], thus producing long-range dipole forces that can be searched for in laboratory experiments \cite{Moo84}. The idea of axions spurred theorists to consider other possibilities for light bosons that could mediate dipole interactions between fermions, such as familons \cite{Wil82,Gel83}, majorons \cite{Gel81,Chi81}, arions \cite{Ans82}, and new spin-0 or spin-1 gravitons \cite{Sch79,Nev80,Nev82,Car94}. Familons are pseudo-Nambu-Goldstone bosons arising from spontaneous breaking of flavor symmetry; majorons were developed to understand neutrino masses and are constrained by searches for neutrinoless double-$\beta$ decay; and arions are the bosons corresponding to a spontaneous breaking of the chiral lepton symmetry.

\subsubsection{Early experiments}
\label{Sec:ExoticSpin:Intro:early-expts}

On the experimental front, early work searching for new dipole interactions focused on EDMs of neutrons, nuclei, and electrons (discussed in Sec.~\ref{Sec:EDM} of this review). Later, some attention turned to the role of spin in gravity. \citet{Mor62} proposed a test of the equivalence principle for a spin-polarized body and \citet{Lei64} pointed out that a gravitational monopole-dipole interaction would violate $P$ and $T$ (time-reversal) symmetries. If a gravitational monopole-dipole interaction existed, the energy of a particle would depend upon the orientation of its spin relative to the local gravitational field of the Earth. Since no such dependence had been experimentally observed, Leitner and Okubo were able to derive corresponding constraints on monopole-dipole couplings based on the absence of gravitationally induced splitting of Zeeman sublevels in measurements of the ground state hyperfine structure of hydrogen. A later experiment searching for such a gravitational dipole moment (GDM) of the proton by \citet{Vel68} in fact found a nonzero value for the proton GDM, but this result was later proved erroneous by \citet{You69} and \citet{Vas69}. \citet{Win72} searched for a nuclear GDM with orders of magnitude greater sensitivity than previous experiments by using a deuterium maser. \citet{Ram79} established the first precise constraints on exotic dipole-dipole interactions between protons by comparing the measured magnetic dipole interaction between protons in molecular hydrogen with theoretical calculations.

\subsection{Theoretical motivation}
\label{Sec:ExoticSpin:TheoreticalMotivation}

Speculation concerning the possibility of a spin-gravity coupling manifesting as a GDM of elementary fermions \cite{Mor62,Kob63,Lei64,Har76,Per78} or a torsion field \cite{Nev80} stood as a principal theoretical impetus encouraging experimental searches for exotic spin-dependent interactions for some time until the appearance of the idea of spin-dependent potentials generated by light spin-0 particles such as the axion \cite{Moo84} and arion \cite{Ans82}. The theoretical motivation to search for axions was significantly boosted when it was realized that axions could be the dark matter permeating the universe [see, for example, Sec.~\ref{Sec:LightDarkMatter} and also the reviews by \citet{Duf09}, \citet{Raf99}, and \citet{Gra15review}].

More recently, the ideas underpinning the concept of the axion have been extended to a diverse array of problems opening new frontiers of research. The numerous light pseudoscalar bosons proposed to address a panoply of theoretical problems in modern physics are known collectively as axion-like particles (ALPs).

\subsubsection{Axion-like-particles (ALPs) in string theory}

ALPs generically arise in string theory as excitations of quantum fields that extend into compactified spacetime dimensions beyond the ordinary four \cite{Svr06,Bai87}. It has been further proposed by \citet{Arv10} that, in fact, because of the topological complexity of the extra-dimensional manifolds of string theory, if string theory is correct and there are indeed spacetime dimensions beyond the known four, there should be many ultralight ALPs, possibly populating each decade of mass down to the Hubble scale of $10^{-33}$~eV, a so-called Axiverse.

\subsubsection{The hierarchy problem}

Another intriguing hypothesis where axions and ALPs appear is a novel proposed solution to the electroweak hierarchy problem \cite{Gra15}. The electroweak hierarchy problem is essentially the question of why the Higgs boson mass is so much lighter than the Planck mass, for one would expect that quantum corrections would cause the effective mass to be closer to the Planck scale. Phrased another way, it is surprising that the electroweak interaction should be so much stronger than gravity. Attempts to solve the hierarchy problem include, for example, supersymmetry \cite{Dim81} and large (sub-mm) extra dimensions \cite{Ark98,Ran99}. \citet{Gra15} propose that instead the hierarchy problem is solved by a dynamic relaxation of the effective Higgs mass from the Planck scale to the electroweak scale in the early universe that is driven by inflation and a coupling of the Higgs boson to a spin-0 particle dubbed the relaxion, which could be the QCD axion or an ALP. Inflation in the early universe causes the relaxion field to evolve in time, and because of the coupling between the relaxion and the Higgs, the effective Higgs mass evolves as well. The coupling between the relaxion and the Higgs generates a periodic potential for the relaxion once the Higgs' vacuum expectation value becomes nonzero. When the periodic potential barriers become large enough, the evolution of the relaxion stalls and the effective mass of the Higgs settles at its observed value. A key idea of this scenario is that the electroweak symmetry breaking scale is a special point in the evolution of the Higgs mass, and that is why the Higgs mass eventually settles at the observed value relatively close to the electroweak scale and far from the Planck scale.

\subsubsection{Dark energy}

A further theoretical motivation for ALPs comes from attempts to explain the observed accelerating expansion of the universe, attributed to a so-called dark energy permeating the universe \cite{Pee03}. \citet{Ark04} have proposed an infrared (i.e., at very low energy scales corresponding to the large distances over which the accelerating expansion of the universe is observed) modification of gravity that posits dark energy is a ghost condensate, a constant-velocity scalar field permeating the universe. The ghost condensate acts as a fluid filling the universe which turns out to behave identically to a cosmological constant by possessing a negative kinetic energy term, and thus matches astrophysical observations. The direct coupling of the ghost condensate to matter leads to both apparent Lorentz-violating effects and new long-range spin-dependent interactions \cite{Ark04,Ark05}. Along these same lines, \citet{Fla09} point out that if dark energy is a cosmological scalar/pseudoscalar field (which could be considered to be a spin-0 component of gravity) there would be a spin-gravity coupling. This implies that fermions would possess GDMs (as discussed in Sec.~\ref{Sec:ExoticSpin:Intro:early-expts}), and also predicts spatial and temporal variations of particle masses and couplings.

In general, it should be noted that most other such theories proposing that cosmic acceleration is due to the dynamical evolution of a scalar field (termed quintessence), by virtue of possessing a conventional kinetic energy term, require a certain level of fine-tuning at least at the level of invoking a nonzero cosmological constant, see for example the review by \citet{Joy15}. For example, in many quintessence models there must exist a screening mechanism of some kind in order to avoid existing astrophysical and laboratory constraints from tests of gravity (see also Sec.~\ref{Sec:5thForces}).

\subsubsection{Unparticles}
\label{Sec:ExoticSpin:TheoreticalMotivation:unparticles}

Yet another theoretical idea that motivates searches for spin-dependent interactions is the unparticle \cite{Geo07}. It is possible in the context of quantum field theory that interactions may be scale invariant \cite{Wil70,Ban82}. A scale-invariant interaction's strength is independent of the energy of the interacting particles. This is not the case for Standard Model fields: in quantum electrodynamics, for example, the strength of the electromagnetic interaction is energy-dependent because of the appearance of virtual particles (i.e., higher-order processes). In fact, unlike Standard Model fields, quantum excitations of scale-invariant interactions cannot be described in terms of particles (like the photon): rather they are objects known as unparticles that are unconstrained by any dispersion relation and without definite mass.  The coupling of unparticles to fermions results in long-range spin-spin interactions that depend on a nonintegral power of distance between the fermions \cite{Lia07} that can be searched for in laboratory experiments.

\subsubsection{Paraphotons, dark photons, hidden photons, and new $Z^\prime$ bosons}
\label{Sec:ExoticSpin:TheoreticalMotivation:hidden-photons-Z-bosons}

An entirely different source of new spin-dependent interactions are exotic spin-1 bosons. There are twelve known gauge bosons in the standard model: the photon, the W$^\pm$ and $Z$ bosons, and the eight gluons. Generally speaking, a massless spin-1 boson accompanies any new unbroken $U(1)$ gauge symmetry [such symmetries arise quite naturally, for example, in string theory \cite{Cve96} and other standard model extensions; $U(1)$ refers to the unitary group of degree 1, the collection of all complex numbers with absolute value 1 under multiplication]. Massless spin-1 bosons are referred to as paraphotons $\gamma'$ \cite{Hol86} in analogy with photons, the quanta arising from the $U(1)$ gauge symmetry of electromagnetism. If paraphotons couple directly to standard model particles, in order to generate fermion masses and avoid gauge anomalies (quantum corrections that break the gauge symmetry and lead to theoretical inconsistencies), the gauge symmetry corresponding to the paraphoton must be $U(1)_{B-L}$ \cite{App03}, where $B-L$ refers to difference between the baryon ($B$) and the lepton ($L$) number: in other words, the ``charge'' of standard model particles with respect to $\gamma'$ is given by $B-L$ (so, for example, a proton has $B-L = 1$ and an electron has $B-L = -1$). However, if the paraphoton coupling to standard model particles is indirect, i.e., through higher-order processes [so that all standard model particles have zero charge under the new $U(1)$ symmetry], this restriction on the possible charge is removed and the coupling of quarks and leptons to $\gamma'$ can take on a range of possible values \cite{Dob05}. Such couplings generate long-range spin-dependent interactions \cite{Dob06}. A closely related hypothesis is that of the dark photon, which would communicate a ``dark'' electromagnetic interaction between dark matter particles, and could be detectable via mixing with photons \cite{Ack09}. Of course, it is also possible that exotic spin-1 bosons possess non-zero mass, as does the $Z$ boson in the standard model. A non-zero mass for such a hypothetical $Z^\prime$ boson could arise from the breaking of a new $U(1)$ gauge symmetry. There is a plethora of theoretical models predicting new $Z^\prime$ bosons and theoretically motivated masses and couplings to quarks and leptons extend over a broad range [see, for example, the review by \citet{Langacker2009}]. $Z^\prime$ bosons that do not directly interact with standard model particles (and therefore reside in the so-called hidden sector) are commonly referred to as hidden photons \cite{Hol86}.

Hidden photons have some particularly notable features that deserve extra attention. As opposed to generic $Z^\prime$ bosons and some classes of dark photons, the only coupling of hidden photons to standard model fermions is through their mixing into a ``real'' electromagnetic field. The Lagrangian describing the hidden photon is of the form:
\begin{align}
\sL = J^{\mu} (A_{\mu}  + \kappa X_{\mu}) + m_{\gamma'}^2 X^2
\end{align}
where $J$ is the electromagnetic current, $A$ represents the photon field, $X$ represents the hidden-photon field, $\kappa$ is the mixing parameter, and $m_{\gamma'}$ is the hidden-photon mass. Notice that in the limit where $m_{\gamma'} \rightarrow 0$, there is no difference between the photon field and the hidden-photon field. In the $m_{\gamma'} = 0$ limit, one can redefine a linear combination $\sA = A + \kappa X$ which couples to electromagnetic current $J$ and a sterile component $\sX = X - \kappa A$ which does not interact at all electromagnetically. Essentially this means that all direct hidden-photon interactions are suppressed by powers of $m_{\gamma'}^2$ in the small mass limit. This argument reduces or eliminates many astrophysical bounds on hidden photons \cite{Pos08bosonic}. The nonrelativistic Hamiltonian for the spin-dependent hidden-photon interaction is
\begin{align}
H_{\gamma'} = \hbar g\mu_B \kappa \mathcal{B}' \cdot \hat{\bs{\sigma}}~,
\label{Eq:hidden-photon-spin-Hamiltonian}
\end{align}
which describes the interaction of spins $\hat{\bs{\sigma}}$ with a real magnetic field $\mathcal{B} = \kappa\mathcal{B}'$ always present wherever there is a hidden field $\mathcal{B}'$, $g$ is the Land\'e factor, and $\mu_B$ is the Bohr (or, if relevant, nuclear) magneton. Here the spin-coupling occurs via the usual magnetic-dipole interaction through the part of $\mathcal{B}'$ that is mixed into a ``real'' magnetic field. This means that observable effects of a hidden photon are suppressed within a shielded region \cite{Cha15,Dub15heating}: although the hidden-photon field is not blocked by the shield, it does affect charges and spins in the shield via the action of the ``real'' magnetic field $\mathcal{B} = \kappa\mathcal{B}'$. The effect of $\mathcal{B}$ in turn generates a ``compensating'' field $\mathcal{B}\ts{comp} \approx -\mathcal{B}$ within the shielded region, cancelling the observable effects of the hidden-photon field \cite{Cha15,Kim16}. This shielding suppression is on top of the small mixing parameter $\kappa$.

\subsubsection{Conclusions}

Even this brief survey portrays a compelling case for experimental searches for exotic spin-dependent interactions. Such interactions are a ubiquitous feature of theoretical extensions to the standard model and general relativity, and furthermore are intimately connected to the mysteries of dark energy, dark matter, the strong $CP$ problem, and even the hierarchy problem and grand unification.

\subsection{Parametrization}
\label{Sec:ExoticSpin:Parametrization}

\subsubsection{Introduction}
\label{Sec:ExoticSpin:Parametrization:intro}

Considering the vast theoretical jungle filled with hypothetical new particles (and even unparticles) possessing unknown properties outlined in Sec.~\ref{Sec:ExoticSpin:TheoreticalMotivation}, a reader may ask: `How are we to systematically search for their effect on atomic systems and quantify their existence or lack thereof?' To set up a general system enabling comparison between different experiments that search for the effects of such new particles and fields, let us consider the related question: if a heretofore undiscovered spin-dependent force exists, how might it affect atoms and their constituents: electrons, protons, and neutrons? It turns out that based on rather general principles, a framework to describe all possible types of interactions between electrons, protons, and neutrons can be quantified by ``exotic physics coupling constants'' for a range of length scales. Thus experimental goals are clarified: an experiment searches for an exotic interaction, and if nothing is found, a limit or constraint is established for coupling constants at the studied length scale for particular forms of interactions. Experimentalists seek to explore regions of parameter space that have not been previously studied to determine if as-yet-undiscovered physics exists with such properties. Then particle theorists can interpret the experimental results in terms of possible new bosons and derive limits on theories introduced in Sec.~\ref{Sec:ExoticSpin:TheoreticalMotivation}.

\subsubsection{Moody-Wilczek-Dobrescu-Mocioiu (MWDM) formalism}
\label{Sec:ExoticSpin:Parametrization:MWDM-formalism}

Generally speaking, the most commonly employed framework for the purpose of comparing different experimental searches for exotic spin-dependent interactions is that introduced by \citet{Moo84} to describe long-range spin-dependent potentials associated with the axion and extended by \citet{Dob06} to encompass long-range potentials associated with any generic spin-0 or spin-1 boson exchange; here we denote this framework the Moody-Wilczek-Dobrescu-Mocioiu (MWDM) formalism. Given basic assumptions within the context of quantum field theory (e.g., rotational invariance, energy-momentum conservation, locality), interactions mediated by new bosons can generate sixteen independent, long-range potentials between fermions. Most laboratory experiments search for interactions between electrons (e) and nucleons [either protons (p) or neutrons (n)]. In general, because of their different quark content, the couplings of protons and neutrons may be expected to differ [for example, in one of the most widely studied models of the QCD axion, the so-called Kim-Shifman-Vainshtein-Zakharov (KSVZ) model \cite{Kim79,Shi80}, the axion coupling to the proton is $\gtrsim 30$ times stronger than that of the neutron \cite{Raf99}]. Thus there are six fermion pairs (ee, ep, en, pp, nn, np) that can couple with sixteen different potentials. The potentials are ascribed dimensionless scalar coupling constants, $f^{XY}_i$, between different fermions (which in general are momentum-dependent, but can be approximated as momentum-independent in the nonrelativistic limit). Here $XY$ denotes the possible fermion pairs: $X,Y = e,n,p$ and $i = 1,2,\ldots,16$ labels the corresponding potential. The potentials can be written in terms of a dimensionless $r$-dependent function $y(r)$ that is determined by the exact nature of the propagator describing the exotic boson exchange. \citet{Dob06} originally derived the potentials in the so-called ``mixed representation'' of position $\mb{r}$ and velocity $\mb{v}$ of fermion $X$, which is useful for analysis of laboratory-scale experiments where $\mb{r}$ and $\mb{v}$ can be treated as classical variables. However, as noted by \citet{Fic17}, for calculations at the atomic scale it is useful to derive the potentials in position representation, keeping in mind that the momentum $\mb{p}$ should be treated as an operator; this derivation has been recently carried out in detail for all MWDM potentials by \citet{Fad18}. The potentials enumerated 1-8 by \citet{Dob06} encompass all possible $P$-even (scalar) rotational invariants, and can be written in the nonrelativistic limit (small fermion velocity and low momentum transfer) as:
\begin{widetext}
\begin{align}
\sV_1 & = f^{XY}_1 \hbar c \frac{y(r)}{r}~, \label{Eq:V1}\\
\sV_2 & = f^{XY}_2 \hbar c \prn{ \hat{\bs{\sigma}}_X \cdot \hat{\bs{\sigma}}_Y } \frac{y(r)}{r}~, \label{Eq:V2}\\
\sV_3 & = f^{XY}_3\frac{\hbar^3}{m^2c} \sbrk{ \hat{\bs{\sigma}}_X \cdot \hat{\bs{\sigma}}_Y \prn{ \frac{1}{r^3} -\frac{1}{r^2}\dbyd{}{r} } - 3 \prn{ \hat{\bs{\sigma}}_X \cdot \hat{\mb{r}} } \prn{ \hat{\bs{\sigma}}_Y \cdot \hat{\mb{r}} } \prn{ \frac{1}{r^3} - \frac{1}{r^2}\dbyd{}{r} + \frac{1}{3r} \dbyd{^2}{r^2}  }}y(r)~, \label{Eq:V3}\\
\sV_{4+5} & = -f^{XY}_{4+5}\frac{\hbar^2}{2m^2c} \hat{\bs{\sigma}}_X \cdot \cbrk{ \prn{ \mb{p} \times \hat{\mb{r}} }, \prn{ \frac{1}{r^2} - \frac{1}{r}\dbyd{}{r} } y(r) }~, \label{Eq:V4-5}\\
\sV_{6+7} & = -f^{XY}_{6+7}\frac{\hbar^2}{2m^2c} \cbrk{ \prn{\hat{\bs{\sigma}}_X\cdot\mb{p}}, \prn{\hat{\bs{\sigma}}_Y\cdot\hat{\mb{r}}} \prn{ \frac{1}{r^2} - \frac{1}{r}\dbyd{}{r} } y(r) }~, \label{Eq:V6-7}\\
\sV_{8} & = f^{XY}_8\frac{\hbar}{4m^2c} \cbrk{ \prn{\hat{\bs{\sigma}}_X\cdot\mb{p}}, \cbrk{ \prn{\hat{\bs{\sigma}}_Y\cdot\mb{p}}, \frac{y(r)}{r} } }~, \label{Eq:V8}
\end{align}
and those enumerated 9-16 encompass all possible $P$-odd (pseudoscalar) rotational invariants, given in the nonrelativistic limit by:
\begin{align}
\sV_{9+10} & = -f^{XY}_{9+10}\frac{\hbar^2}{m} \hat{\bs{\sigma}}_X \cdot \hat{\mb{r}} \prn{ \frac{1}{r^2} - \frac{1}{r}\dbyd{}{r} } y(r)~, \label{Eq:V9-10}\\
\sV_{11} & = -f^{XY}_{11}\frac{\hbar^2}{m} \prn{ \hat{\bs{\sigma}}_X \times \hat{\bs{\sigma}}_Y }\cdot \hat{\mb{r}} \prn{ \frac{1}{r^2} - \frac{1}{r}\dbyd{}{r} } y(r)~, \label{Eq:V11}\\
\sV_{12+13} & = f^{XY}_{12+13}\frac{\hbar}{2m} \hat{\bs{\sigma}}_X \cdot \cbrk{ \mb{p}, \frac{y(r)}{r} }~, \label{Eq:V12-13}\\
\sV_{15} & = f^{XY}_{15}\frac{3\hbar^3}{2m^3 c^2} \cbrk{ \prn{ \hat{\bs{\sigma}}_X \cdot \hat{\mb{r}} } \sbrk{ \prn{ \hat{\bs{\sigma}}_Y \times \hat{\mb{r}} } \cdot \mb{p}  } , \prn{ \frac{1}{r^3} - \frac{1}{r^2}\dbyd{}{r} + \frac{1}{3r} \dbyd{^2}{r^2} } y(r) }  \nonumber \\
& ~~~~ +f^{XY}_{15}\frac{3\hbar^3}{2m^3 c^2} \cbrk{ \prn{ \hat{\bs{\sigma}}_Y \cdot \hat{\mb{r}} } \sbrk{ \prn{ \hat{\bs{\sigma}}_X \times \hat{\mb{r}} } \cdot \mb{p}  } , \prn{ \frac{1}{r^3} - \frac{1}{r^2}\dbyd{}{r} + \frac{1}{3r} \dbyd{^2}{r^2} } y(r) }~, \label{Eq:V15}\\
\sV_{16} & = -f^{XY}_{16}\frac{\hbar^2}{8m^3c^2} \cbrk{ \hat{\bs{\sigma}}_Y \cdot \mb{p} , \cbrk{ \hat{\bs{\sigma}}_X \cdot \prn{ \mb{p} \times \hat{\mb{r}} }, \prn{ \frac{1}{r^2} - \frac{1}{r}\dbyd{}{r} } y(r) }} \nonumber \\
& ~~~~ -f^{XY}_{16}\frac{\hbar^2}{8m^3c^2} \cbrk{ \hat{\bs{\sigma}}_X \cdot \mb{p} , \cbrk{ \hat{\bs{\sigma}}_Y \cdot \prn{ \mb{p} \times \hat{\mb{r}} }, \prn{ \frac{1}{r^2} - \frac{1}{r}\dbyd{}{r} } y(r) }}~. \label{Eq:V16}
\end{align}
\end{widetext}
In Eqs.~\eqref{Eq:V1} - \eqref{Eq:V16}, $\hbar$ is Planck's constant, $c$ is the speed of light, $m$ is the mass of fermion $X$, $r$ is the distance between the fermions and $\hat{\mb{r}}$ is the unit vector along the line between them, $\hat{\bs{\sigma}}_i$ is a unit vector in the direction of the spin of fermion $i$, $\mb{p}$ is the momentum of particle $X$, and $\cbrk{ \Box,\Box }$ denotes the anticommutator. Where sums appear in the potential indices there is dependence on $\hat{\bs{\sigma}}_X$, the differences of the respective potentials depend on $\hat{\bs{\sigma}}_Y$. The potential $\sV_{14}$ (not listed above), proportional to $\prn{ \hat{\bs{\sigma}}_X \times \hat{\bs{\sigma}}_Y } \cdot \mb{p}$, turns out to vanish in the nonrelativistic limit \cite{Fad18}. It is interesting to note that the MWDM formalism applies whether or not the underlying theory obeys Lorentz invariance (so long as rotational invariance is preserved) and also applies in the case of multi-boson exchange between the fermions in question. Thus the MWDM formalism is quite general in nature and serves as a useful framework for comparing different experiments.

\subsubsection{MWDM formalism for Lorentz-invariant, single-boson exchange}
\label{Sec:ExoticSpin:Parametrization:MWDM-formalism:1-boson-exchange}

A specific form can be obtained for $y(r)$ if some assumptions are made about the propagator. Assuming one-boson exchange within a Lorentz-invariant quantum field theory, $y(r)$ takes on a Yukawa-like form:
\begin{align}
y(r) = \frac{1}{4\pi} e^{-r/\lambda}~,
\label{Eq:y-Yukawa-function}
\end{align}
where
\begin{align}
\lambda = \frac{\hbar}{Mc}
\label{Eq:Compton-wavelength}
\end{align}
is the reduced Compton wavelength of the new boson of mass $M$, which sets the scale of the new interaction. If there is multi-boson exchange or Lorentz invariance is violated, other forms of $y(r)$ can arise, but the spin dependence of the potential functions is preserved. Generally an experimental setup characterized by a distance scale $\ell$ is sensitive to new bosons of mass $M \lesssim \hbar/(c\ell)$. [Note that the derivative operators with respect to $r$ are understood to act only on $y(r)$ and not on wave functions.]

If particular spin and parity properties of the new boson are specified, correlations between the coupling strengths are found.  For example, if the new boson is a spin-0 particle such as an axion or ALP, $f^{XY}_3 = -{\rm g}_p^X{\rm g}_p^Y/\prn{4 \hbar c}$, where ${\rm g}_p^{X,Y}$ parameterizes the vertex-level pseudoscalar coupling (denoted by the subscript $p$) of the spin-0 field to the fermions. The quantity ${\rm g}_p^2/(\hbar c)$ is dimensionless. Under these assumptions, for example, the dipole-dipole potential of Eq.~\eqref{Eq:V3} can be written in the form most commonly encountered in the literature,
\begin{widetext}
\begin{align}
\sV_3(r) = -\frac{{\rm g}_p^X{\rm g}_p^Y\hbar^2}{16\pi m^2 c^2} \sbrk{ \hat{\bs{\sigma}}_X \cdot \hat{\bs{\sigma}}_Y \prn{ \frac{1}{\lambda r^2} + \frac{1}{r^3} } - \prn{ \hat{\bs{\sigma}}_X\cdot\hat{\mb{r}} } \prn{ \hat{\bs{\sigma}}_Y\cdot\hat{\mb{r}} } \prn{ \frac{1}{\lambda^2r} + \frac{3}{\lambda r^2} + \frac{3}{r^3}} } e^{-r/\lambda}~.
\label{Eq:V3-expanded-form}
\end{align}
\end{widetext}
If the new interaction possesses both scalar and pseudoscalar couplings, for example, $f^{XY}_{9+10} = {\rm g}_p^X{\rm g}_s^Y/\prn{\hbar c}$ (where the subscript $s$ denotes the scalar coupling) one obtains the following monopole-dipole potential for coupling of polarized fermions $X$ to a monopole source of fermions $Y$:
\begin{align}
\sV_{9+10}(r) = -\frac{{\rm g}_p^X{\rm g}_s^Y\hbar}{4\pi m c} \hat{\bs{\sigma}}_X \cdot \hat{\mb{r}} \prn{ \frac{1}{r\lambda} + \frac{1}{r^2} }  e^{-r/\lambda}~.
\label{Eq:V9-10-expanded-form}
\end{align}
Monopole-dipole and dipole-dipole potentials, and indeed the vast majority of the potentials enumerated Eqs.~\eqref{Eq:V1} - \eqref{Eq:V16}, can also be generated by exchange of spin-1 particles, such as a Z$^\prime$ boson \cite{Dob06,Gom15}. For example, if the new boson is a massive spin-1 boson, $f^{XY}_3 = \sbrk{\prn{1+m^2/m_Y^2}{\rm g}_A^X{\rm g}_A^Y + \prn{ 2m/m_Y }{\rm g}_V^X{\rm g}_V^Y }/\prn{8 \hbar c}$, where $m_Y$ is the mass of fermion $Y$ and the subscripts $A$ and $V$ refer to the axial vector and vector couplings, respectively. An axial vector coupling also generates the dipole-dipole potential
\begin{align}
\sV_{2}(r) = -\frac{{\rm g}_A^X{\rm g}_A^Y}{4\pi \hbar c}\frac{\hbar c}{r} \hat{\bs{\sigma}}_X \cdot \hat{\bs{\sigma}}_Y  e^{-r/\lambda}~,
\label{Eq:V2-expanded-form}
\end{align}
which has a different scaling with particle separation as compared to the $\sV_3(r)$ potential described by Eq.~\eqref{Eq:V3-expanded-form}. A complete enumeration of the coefficients $f_i^{XY}$ in terms of vertex-level couplings is given in the paper by \citet{Fad18}.

The relative signs of the potentials have recently been analyzed by \citet{daido2017sign} and some corrections to the work of \citet{Moo84} have been noted and incorporated into, for example, Eq.~\eqref{Eq:V3-expanded-form}. Although these sign errors have to some extent propagated throughout the literature, they do not affect existing constraints since experiments limit the absolute value of the coupling constants.

\subsubsection{Contact interactions}
\label{Sec:ExoticSpin:Parametrization:contact-interactions}

Another detail to be aware of is that the potentials described in Eqs.~\eqref{Eq:V1} - \eqref{Eq:V16} are long-range potentials that assume the fermions under investigation are separated by a finite distance. In searches for exotic spin-dependent interactions in atoms and molecules one must also take into account the possibility of wave function overlap and the contribution of terms in the potentials proportional to Dirac delta functions $\delta^3(\mb{r})$. For example, the term
\begin{align}
-\frac{{\rm g}_p^X{\rm g}_p^Y\hbar^2}{12 m_Xm_Y c^2} \hat{\bs{\sigma}}_X \cdot \hat{\bs{\sigma}}_Y \delta^3(\mb{r})
\end{align}
must be added to the expression for the dipole-dipole interaction generated by an ALP given in Eq.~\eqref{Eq:V3-expanded-form}. Additional contact terms appear in the potentials when higher-order terms in the particles' momenta are included, see \citet{Fad18}.

Of related interest is the fact that the Higgs boson \cite{ATL12,CMS12}, a spin-0 particle, is predicted to induce a Yukawa-like interaction between fermions \cite{Hab79}, leading to a delta-function-like potential which could be searched for in precision atomic physics experiments \cite{Del16,Ber17,DelFruFuc17}. The Higgs interaction can even produce a $P$-odd, $T$-odd electron-nucleon interaction generating EDMs of atoms and molecules \cite{Bar92a,Barr1992}. Because the mass of the Higgs boson is $\approx 125~{\rm GeV}$, the range of any force mediated by the Higgs is $\approx 10^{-17}~{\rm m}$ (the Higgs Compton wavelength), and thus meaningful constraints on Higgs-mediated interactions have not yet been experimentally obtained.

A closely related point is that measurements of permanent electric dipole moments (EDMs), discussed in Sec.~\ref{Sec:EDM}, also constrain some exotic spin-dependent forces. This is because a $P$- and $T$-violating interaction between particles will naturally induce a $P$- and $T$-violating atomic EDM, and indeed a number of the potentials $\sV_i$ violate $P$ and $T$ symmetries [$\sV_{9+10}$, $\sV_{14}$, and $\sV_{15}$ -- see Eqs.~\eqref{Eq:V9-10} and \eqref{Eq:V15}]. \citet{Gha15} have reinterpreted the results of the Hg EDM experiment \cite{Gri09} to constrain a $P$,$T$-odd interaction of electrons and nucleons through the exchange of a massive gauge boson, and have excluded vector bosons with masses $\gtrsim 1~{\rm MeV}$ with coupling strengths $\gtrsim 10^{-9}$. \citet{stadnik2017improved} have analyzed constraints from EDM experiments on spin-0-boson-mediated interactions.

\subsubsection{Position representation and permutation symmetry}
\label{Sec:ExoticSpin:Parametrization:position-representation}

As noted in Sec.~\ref{Sec:ExoticSpin:Parametrization:MWDM-formalism}, in atomic and molecular calculations for velocity-dependent potentials, it is often useful to convert the momentum $\mb{p}$ into the relevant operator in position space. Furthermore, for identical particles care must be taken to account for permutation symmetry. For example, the $\sV_8$ potential [Eq.~\eqref{Eq:V8}], which can arise from the exchange of axial-vector bosons, can be written for identical particles 1 and 2 as
\begin{widetext}
\begin{align}
\sV_{8}(r) = \frac{{\rm g}_A{\rm g}_A}{4\pi \hbar c} \frac{\hbar^3}{4 m^2 c} \cbrk{ \hat{\bs{\sigma}}_1 \cdot \prn{ \nabla_1 - \nabla_2 }, \cbrk{ \hat{\bs{\sigma}}_2 \cdot \prn{ \nabla_1 - \nabla_2 }, \frac{e^{-r/\lambda}}{r} } }~,
\label{Eq:V8-identical-particles-position-space}
\end{align}
\end{widetext}
where $\nabla_i$ is the vector differential operator in position space for particle $i$ in the center-of-mass frame. Details concerning this point are addressed by \citet{Fic17}.

\subsubsection{Quantum field theory details}
\label{Sec:ExoticSpin:Parametrization:QFT-details}

In order to check whether  different experiments are truly measuring the same quantity, it can sometimes be important to consider further specifics regarding the origin of a spin-dependent coupling within quantum field theory. For example, an ALP field $\varphi$ can generate the potential $\sV_3(r)$ described by Eq.~\eqref{Eq:V3-expanded-form} between fermions $\psi$ of mass $m$ in two different ways: either through a Yukawa-like coupling described by the Lagrangian \cite{Moo84,Vas09}
\begin{align}
\sL\ts{Yuk} = -i g_p \bar{\psi} \gamma^5 \psi \varphi~,
\label{Eq:V3-Yukawa-Lagrangian}
\end{align}
or through a derivative coupling described by the Lagrangian
\begin{align}
\sL\ts{Der} = \frac{g_p}{2m} \bar{\psi} \gamma_\mu\gamma^5 \psi \partial^\mu \varphi~,
\label{Eq:V3-derivative-Lagrangian}
\end{align}
where in Eqs.~\eqref{Eq:V3-Yukawa-Lagrangian} and \eqref{Eq:V3-derivative-Lagrangian} we have used the Dirac $\gamma$ matrices. Although experimental searches for dipole-dipole interactions are sensitive to both Yukawa-like and derivative couplings, various searches for spin-independent interactions and some astrophysical phenomena are sensitive only to one or the other type of coupling \cite{Fis99,Raf95,Raf12}. Similarly, \citet{Man14} have shown that by delving deeper into the quantum field theoretic origins of exotic spin-dependent interactions one can distinguish the effects of the QCD axion from generic ALPs by comparing the results of nuclear EDM searches with results of searches for new spin-dependent forces [see also the analysis of \citet{Gha15}]. It is also important to note that QCD axion models \cite{Din81,Shi80,Kim79,Zhi80} have a definite relationship between the interaction strength and the axion mass, whereas for a generic ALP the mass and the interaction strength are independent parameters.

\subsubsection{Connection between the MWDM formalism and various fundamental theories}

In most cases there is a clear one-to-one correspondence between potentials in the MWDM formalism and the fundamental theories predicting exotic spin-dependent interactions outlined in Sec.~\ref{Sec:ExoticSpin:TheoreticalMotivation}, although there are exceptions such as the predicted potentials generated by unparticles (Sec.~\ref{Sec:ExoticSpin:TheoreticalMotivation:unparticles}).

Consider, for example, the standard QCD axion discussed in Sec.~\ref{Sec:ExoticSpin:EarlyWork:axions-ALPs}. An axion (or ALP) is characterized by a symmetry breaking scale $f_a$ and an interaction scale $\Lambda$, which in the case of the QCD axion is the QCD confinement scale $\Lambda \approx 200~{\rm MeV}$ (ALPs may have different values for $\Lambda$). These scales determine, for example, the mass of the axion
\begin{align}
m_a c^2 = \frac{\Lambda^2}{f_a}~.
\end{align}
The interaction of an axion with a fermion $X$ is determined by a dimensionless coupling constant $C_X$ which can be predicted in the context of a specific theory, and related to the coupling constants in the MWDM formalism. For instance, the pseudoscalar coupling
\begin{align}
\frac{g_p^X}{\sqrt{ \hbar c }} = C_X \frac{m c^2}{f_a}~.
\end{align}
For a particular manifestation of the QCD axion referred to as the KSVZ axion \cite{Kim79,Shi80}, $C_p \approx -0.34$ for the proton, $C_n \approx 0.01$ for the neutron, and $C_e = 0$ for the electron \cite{Raf99}. Note that in this specific theoretical model a single parameter, the symmetry breaking scale $f_a$, determines both the axion mass and the coupling strength to particular fermions. This formalism connects searches for exotic spin-dependent interactions to the broader context of QCD axion searches: most QCD axion searches exploit the axion-photon coupling, also proportional to $1/f_a$, but with a different coupling constant. For example, the Axion Dark Matter eXperiment [ADMX, \cite{Asz10}] searches for axions converted into detectable microwave photons using the inverse Primakoff effect as first outlined by \citet{Sik83}. Since experiments such as ADMX probe a different coupling and generally speaking a different axion mass range as compared to searches for spin-dependent interactions, these experimental approaches are largely complementary [see Sec.~\ref{Sec:LightDarkMatter} and also the reviews by \citet{Kim10R} and \citet{Gra15review}].

As another example, a standard propagating gravitational torsion field (see Sec.~\ref{Sec:ExoticSpin:EarlyWork:torsion}) can generate a dipole-dipole interaction identical to the $\sV_3$ potential in the MWDM formalism \cite{Ham95,Ade09,Nev80,Nev82}, with the relationship
\begin{align}
\frac{g_p g_p}{\hbar c} = \beta^2 \frac{18 \pi G m^2}{\hbar c}~,
\end{align}
where $G$ is Newton's gravitational constant and the minimal torsion model predicts the torsion constant $\beta = 1$.

In general, there is similar one-to-one correspondence between the MWDM formalism and any model based on a quantum field theory with new force-carrying spin-0 and spin-1 bosons.

\subsubsection{Relationship between coupling constants for atoms and elementary particles}
\label{Sec:ExoticSpin:Parametrization:elementary-particles-to-atoms-nuclei}

Furthermore, theoretical knowledge of atomic, molecular, and nuclear structure is critical for interpretation of experiments. In order to meaningfully compare experimental results, the coupling of the exotic field to the atomic spin must be interpreted in terms of the coupling to electron, proton, and neutron spins. The basic scheme of such a parametrization of spin couplings to new physics can be cast in terms of an exotic atomic dipole moment $\bs{\chi}=\chi_a\mb{F}$ related to coupling constants $\chi_e$, $\chi_p$, and $\chi_n$ for the electron, proton, and neutron, respectively, where $\mb{F}$ is the total atomic angular momentum. It is generally assumed that such couplings do not follow the same scaling as magnetic moments. The coupling constants $f^{XY}_i$ describing the potentials enumerated in Eqs.~\eqref{Eq:V1} - \eqref{Eq:V16} can then be written in terms of $\chi_e$, $\chi_p$, and $\chi_n$ depending on the specific experiment, where for each different potential $\sV_i(r)$ the constants $\chi_e$, $\chi_p$, and $\chi_n$ may be different. The nucleon coupling constants $\chi_p$ and $\chi_n$ can in turn be related to quark and gluon couplings via measurements and calculations based on QCD \cite{Aid13,Fla04}.

It is generally assumed by most theories postulating new interactions \cite{Moo84,Dob06,Fla09,Gra13,Ark05,Geo07,Lia07} that there is no coupling of the exotic field to orbital angular momentum $\mb{L}$. In the context of quantum field theory, this theoretical bias can be understood as follows. If an exotic field couples to $\mb{L}$ then the field couples to particle current. However, the lowest-order coupling to particle current vanishes if the exotic interaction is mediated by a spin-0 particle such as an ALP \cite{Dob06}. On the other hand, a coupling of a generic massive spin-1 boson to particle current is forbidden by gauge invariance \cite{Dob05}, and constraints on couplings of massless spin-1 bosons are already quite stringent \cite{App03}. Thus, generally, couplings of exotic fields to particle current, and thus $\mb{L}$, are expected to be suppressed relative to spin couplings. Nonetheless, it should also be noted that there are theories that do postulate exotic couplings to $\mb{L}$. For example, hidden photons can mix with ordinary photons, and thus can produce real magnetic fields in magnetically shielded regions that would indeed couple to $\mb{L}$ \cite{Cha15}.

The relationship of the expectation value for total atomic angular momentum $\abrk{ \mb{F} }$ to electron spin $\abrk{ \mb{S} }$ and nuclear spin $\abrk{ \mb{I} }$ can be evaluated for the ground states of most low-to-intermediate mass atoms based on the Russell-Saunders {\it{LS}}-coupling scheme \cite{Bud08}:
\begin{align}
\abrk{ \mb{F} } &= \abrk{ \mb{S} } + \abrk{ \mb{L} } + \abrk{ \mb{I} }~, \nonumber \\
&= \frac{\abrk{ \mb{S}\cdot\mb{F} } }{ F(F+1) } \abrk{ \mb{F} } + \frac{\abrk{ \mb{L}\cdot\mb{F} }}{ F(F+1) } \abrk{ \mb{F} } + \frac{\abrk{ \mb{I}\cdot\mb{F} }}{ F(F+1) } \abrk{ \mb{F} }~,
\end{align}
where $\mb{L}$ is the orbital angular momentum. It follows that for the exotic atomic dipole moment coupling constant $\chi_a$,
\begin{align}
\chi_a = \chi_e\frac{\abrk{ \mb{S}\cdot\mb{F} }}{ F(F+1) } + \chi_N\frac{\abrk{ \mb{I}\cdot\mb{F} }}{ F(F+1) }~,
\label{Eq:exotic-atomic-dipole}
\end{align}
where $\chi_N$ is the exotic nuclear dipole coupling constant which can be expressed in terms of $\chi_p$ and $\chi_n$.

The projection of $\mb{S}$ on $\mb{F}$ can be calculated in terms of eigenvalues of the system according to:
\begin{widetext}
\begin{align}
\abrk{ \mb{S}\cdot\mb{F} } & = \frac{\abrk{ \mb{S}\cdot\mb{J} }}{J(J+1)} \abrk{ \mb{J}\cdot\mb{F} }~, \\
& = \frac{\sbrk{J(J+1)+S(S+1)-L(L+1)} \sbrk{F(F+1)+J(J+1)-I(I+1)}}{4J(J+1)}~,
\end{align}
\end{widetext}
where $\mb{J} = \mb{S} + \mb{L}$, and the projection of $\mb{I}$ on $\mb{F}$ is given by
\begin{align}
\abrk{ \mb{I}\cdot\mb{F} }  = \frac{1}{2}\sbrk{F(F+1) + I(I+1) - J(J+1)}~.
\end{align}

The next problem is a more difficult one: what is the relationship between $\chi_N$ and the nucleon coupling constants, $\chi_p$ and $\chi_n$? Traditionally constraints from atomic experiments on exotic couplings to neutron and proton spins have been derived using the single-particle Schmidt model for nuclear spin [see, for example, \citet{Ven92}]. In this model, particular atomic species are sensitive to either neutron or proton spin couplings, but not both. The single-particle Schmidt model assumes that the nuclear spin $\mb{I}$ is due to the orbital motion and intrinsic spin of one nucleon only and that the spin and orbital angular momenta of all other nucleons sum to zero \cite{Sch37,Kli52,Bla79}: in other words, the nuclear spin $\mb{I}$ is entirely generated by a combination of the valence nucleon spin ($\mb{S}_p$ or $\mb{S}_n$) and the valence nucleon orbital angular momentum $\bs{\ell}$, so that we have
\begin{align}
\chi_N & = \frac{\abrk{ \mb{S}_{p,n} \cdot \mb{I} }}{I(I+1)}\chi_{p,n}~, \\
& = \frac{S_{p,n}(S_{p,n}+1) + I(I+1) - \ell(\ell+1)}{2I(I+1)} \chi_{p,n}~, \label{Eq:Schmidt-model-chiN}
\end{align}
where it is assumed that the valence nucleon is in a well-defined state of $\ell$ and $S_{p,n}$.  However, it is well known that nuclear magnetic moments are not accurately predicted by the Schmidt model, since in most cases it is a considerable oversimplification of the nucleus. Thus, in general, the nuclear spin content and magnetic moment cannot be described by a single valence nucleon in a well-defined state of $\ell$ and $S_{p,n}$.  While there have been attempts to apply semi-empirical models employing nuclear magnetic moment data to derive new constraints for non-valence nucleons \cite{Eng89,Fla06,Fla09,Sta15}, \citet{Kim15} has shown that such models cannot reliably be used to predict the spin polarization of non-valence nucleons by analyzing known physical effects in nuclei and by comparisons with detailed large-scale nuclear shell model calculations [see, for example, \citet{Vie15,Bro16}]. Thus while the sensitivity of valence nucleons and electrons to exotic physics can be reliably estimated, evaluating the sensitivity of non-valence nucleons and electrons to exotic physics requires detailed theoretical calculations.

\subsubsection{Conclusions}

Keeping in mind the above caveats, the MWDM framework introduced by \citet{Moo84} and \citet{Dob06} [Eqs.~\eqref{Eq:V1} - \eqref{Eq:V16}], and analyzed in further detail by \citet{Fad18}, provides a useful tool to compare different experimental searches for exotic spin-dependent effects.

\subsection{Overview of experimental searches}
\label{Sec:ExoticSpin:ExptOverview}

A typical approach in experiments searching for exotic spin-dependent interactions is to develop a sensitive detector of torques or forces on particles (such as a torsion pendulum) and then bring the detector in close proximity to an object that acts as a local source of the exotic field (for example, a large mass or highly polarized spin sample). The object producing the exotic field acts analogously to a charged object
producing an electric field.   Usually the major difficulty in such measurements is understanding and eliminating systematic errors: in other words, distinguishing exotic torques and forces that would be evidence of new physics from prosaic effects such as magnetic interactions. For this reason, it is advantageous if the source can be manipulated in such a way as to modulate the exotic field in order to distinguish its effects from background processes. In lieu of this, possible sources of systematic errors can be constrained by independent measurements. Another approach, often used to probe exotic spin-dependent interactions at the atomic or molecular length scale, is to compare theory and measurement for some property of a system (such as the energy splitting between different hyperfine states in an atom) that would change if an exotic spin-dependent interaction existed.

As seen in Sec.~\ref{Sec:ExoticSpin:Parametrization}, the basic features of an experiment that characterize its particular sensitivity are the identities and properties of the particle constituents of the exotic field source and the detector (determining whether the experiment is searching for neutron-neutron interactions, electron-electron interactions, etc.) and the distance between the source and detector (which determines the range of the interaction to which the experiment is sensitive, or, alternatively, the mass of the exotic boson communicating the interaction). The precision of the experiment determines the strength of the interaction to which it is sensitive. Depending on whether one or both of the source and detector employ polarized particles and if the source and detector are in relative motion, the experiment can be sensitive to different potentials among those enumerated in the MWDM formalism (Sec.~\ref{Sec:ExoticSpin:Parametrization:MWDM-formalism}). Most experimental searches to date have been for velocity-independent interactions ($\sV_1$, $\sV_2$, $\sV_3$, $\sV_{9+10}$, and $\sV_{11}$, see Sec.~\ref{Sec:ExoticSpin:Parametrization:MWDM-formalism}).

While most experiments house a macroscopic source and detector in a single laboratory, thus allowing proximities between source and detector to range from slightly less than a millimeter to a few meters [see, for example, \citet{You96,Vas09,Tul13,Ter15}], the longest-range experiments use the Earth as a source mass [see, for example, \citet{Win91,Ven92,Kim17GDM}] or a source of polarized electrons \cite{Hun13}, and the shortest-range experiments probe atomic or molecular structure [see, for example, \citet{Ram79,Led13,Fic17}]. Experiments with the Earth as an exotic field source have the particular challenge of lacking a way to reverse or modulate the interaction. Atomic-range experiments suffer from a similar challenge insofar as they generally must rely on a comparison between calculations of energy levels and spectroscopic measurements.

Experiments searching for exotic spin-dependent interactions typically employ magnetic shielding between the source of the exotic field and the detector. Any such experiment must answer the basic question: what is the effect of the magnetic shield system on the signal detected by the spin-polarized ensemble? This question was considered by \citet{Kim16}, and the general conclusion is that for common experimental geometries and conditions, magnetic shields do not significantly reduce sensitivity to exotic spin-dependent interactions, especially when the technique of comagnetometry is used [where measurements are simultaneously performed on two or more atomic species, see \citet{Lam89}]. However, exotic fields that couple to electron spin can induce magnetic fields in the interior of shields made of a soft ferro- or ferrimagnetic material. This induced magnetic field must be taken into account in the interpretation of experiments searching for new spin-dependent interactions.

A particular case, discussed in detail by \citet{Cha15,Dub15heating} and also in Sec.~\ref{Sec:ExoticSpin:TheoreticalMotivation:hidden-photons-Z-bosons}, where careful consideration of electromagnetic shielding is crucial is that of hidden photons. The photon field and the hidden-photon field couple to standard model particles in the essentially the same way: observable effects of the hidden-photon field are nearly entirely through the effects of the mixing of the hidden field into a ``real'' electromagnetic field. Thus observable effects of hidden photons can be significantly reduced by electromagnetic shielding \cite{Kim16}. In contrast to hidden photons, generic spin-1 particles such as dark photons or $Z^\prime$ bosons may have no particular relationship with electromagnetism, and thus magnetic shielding generally does not suppress their effects \cite{Kim16}.

In the next sections we review the experiments establishing the best laboratory constraints on various exotic spin-dependent interactions.

\subsection{Experimental constraints on monopole-dipole interactions}
\label{Sec:ExoticSpin:MonopoleDipole}

%----------------------------------------------------------------
\begin{figure}
\center
\includegraphics[width=4.1 in]{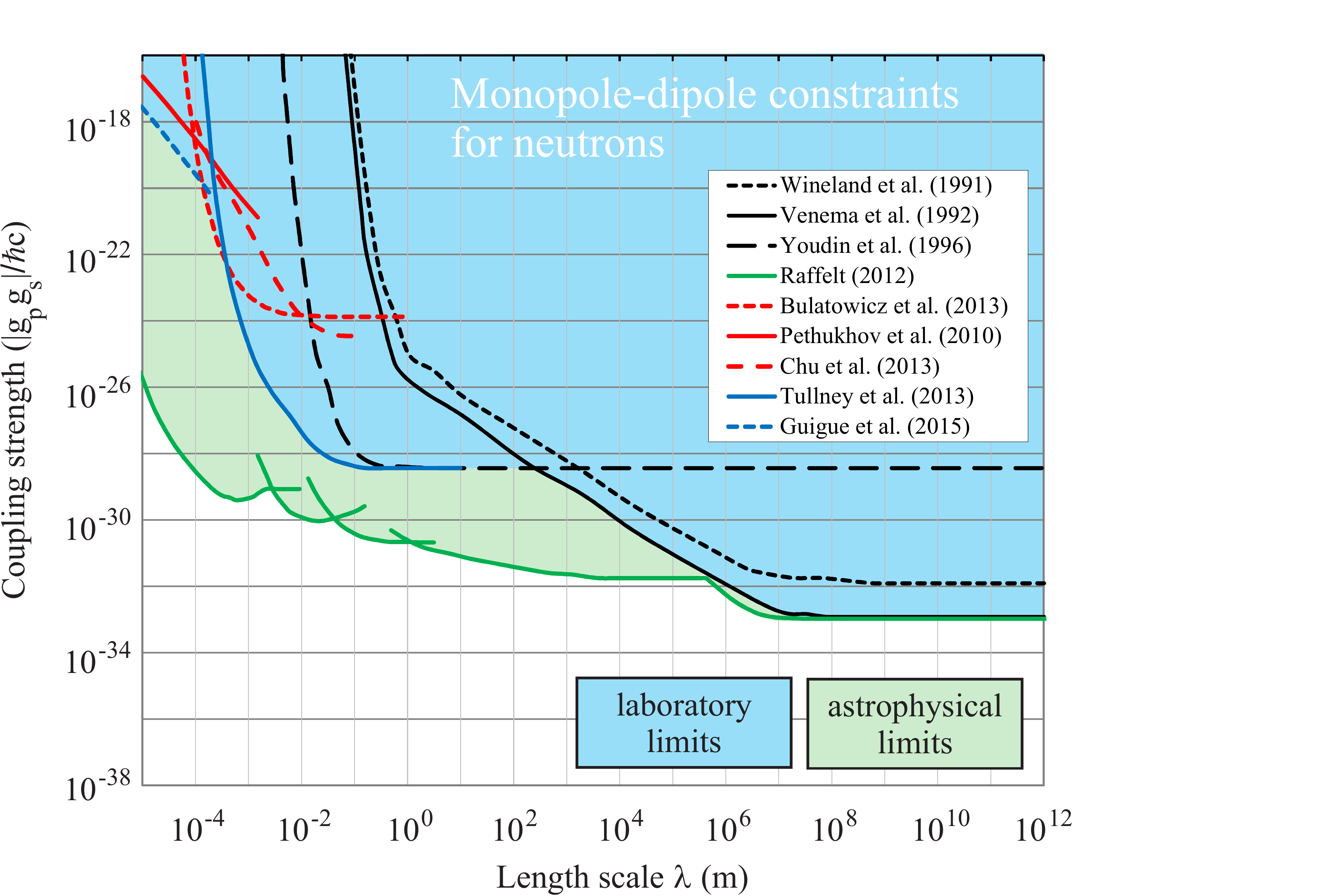}
\includegraphics[width=4.1 in]{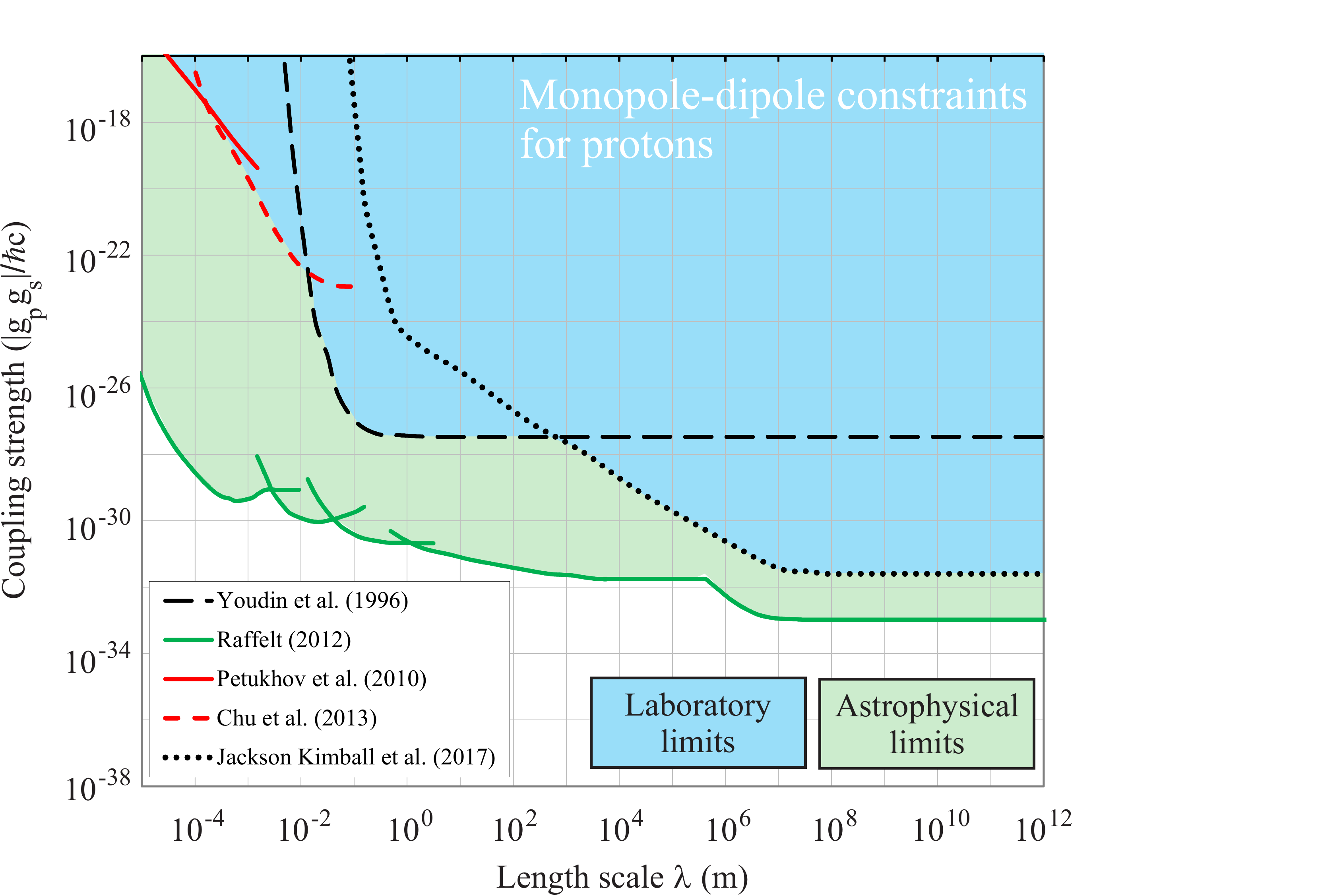}
\includegraphics[width=4.1 in]{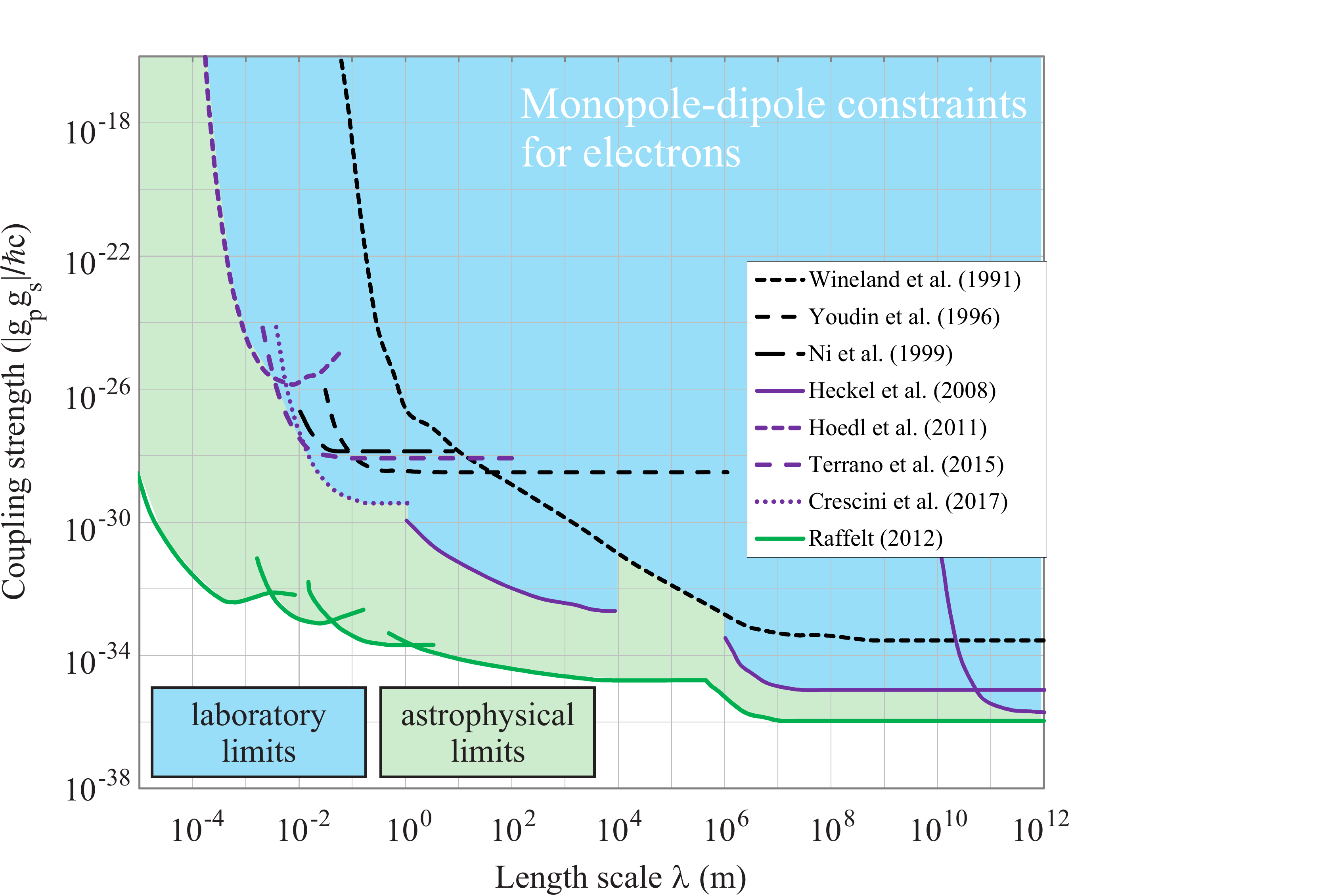}
\caption{Laboratory constraints (shaded light blue, see text for discussion of individual experiments) on monopole-dipole (scalar-pseudoscalar) couplings, $\left| {\rm g}_p {\rm g}_s \right|/\hbar c$ [the $\sV_{9+10}$ potentials as described in Eq.~\eqref{Eq:V9-10-expanded-form}], for neutrons, protons, and electrons as a function of the range $\lambda$ of the interaction (${\rm g}_p$ and ${\rm g}_s$ are the pseudoscalar and scalar coupling constants, respectively). Astrophysical constraints (excluded parameter space shaded light green) are from the analysis of \citet{Raf12}.}
\label{Fig:monopole-dipole-constraints}
\end{figure}
%----------------------------------------------------------------

Figure~\ref{Fig:monopole-dipole-constraints} shows the most stringent laboratory and astrophysical constraints on exotic monopole-dipole interactions, in particular the $\sV_{9+10}$ potentials as described by the MWDM formalism [Eq.~\eqref{Eq:V9-10-expanded-form}], which can be interpreted as a scalar-pseudoscalar coupling. The horizontal axes show the range of the interaction, inversely proportional to the mass of the boson communicating the interaction [Eq.~\eqref{Eq:Compton-wavelength}]. The vertical axes show the dimensionless coupling parameter $\left| {\rm g}_p {\rm g}_s \right|/\hbar c$ between the studied particles. Typically in experiments, the monopole (scalar) coupling is to an unpolarized sample with roughly equal numbers of protons, neutrons, and electrons, whereas the dipole (pseudoscalar) coupling is to a polarized sample of predominantly one species, so the upper, middle, and lower plots in Fig.~\ref{Fig:monopole-dipole-constraints} can be interpreted as constraints on $\left| {\rm g}^n_p {\rm g}^X_s \right|/\hbar c$, $\left| {\rm g}^p_p {\rm g}^X_s \right|/\hbar c$, and $\left| {\rm g}^e_p {\rm g}^X_s \right|/\hbar c$, respectively, where the superscripts $n$, $p$, and $e$ refer to neutrons, protons, and electrons, respectively, and $X=n,p,e$ for each case.

\subsubsection{Neutrons}

At the longest interaction ranges probed by experiments, the most stringent laboratory constraint on monopole-dipole interactions between spin-polarized neutrons and other particles is derived from the experiment of \citet{Ven92}, establishing the limit displayed on the upper plot of Fig.~\ref{Fig:monopole-dipole-constraints} with a solid black line. The experiment of \citet{Ven92} illustrates the principles involved in a broad class of experiments that rely on optical measurements of the spin precession of various atomic species in the gas phase [for reviews of these experimental techniques, see \citet{Bud13,Bud02,Bud07}]. \citet{Ven92} simultaneously measured the spin-precession frequencies of two isotopes of Hg (this exemplifies the technique of comagnetometry) as the orientation of a magnetic field $\mc{B}$ was changed relative to the Earth's gravitational field $\bs{g}$.  Since the ground electronic state of Hg is ${\rm ^1S_0}$, the ground-state polarization is entirely due to the nuclear spin $I$, with $^{199}$Hg having $I=1/2$ and $^{201}$Hg having $I=3/2$.  A heretofore undiscovered long-range, monopole-dipole interaction would generate spin precession about an axis directed along the local gravitational field $\bs{g}$.  In the presence of only $\mc{B}$ and $\bs{g}$, the spin precession frequencies for the two Hg isotopes are
\begin{align}
\Omega_{199} = \gamma_{199} B + \chi_{199} g \cos{\phi}~, \label{Eq:OmegaGDM-Hg199} \\
\Omega_{201} = \gamma_{201} B + \chi_{201} g \cos{\phi}~, \label{Eq:OmegaGDM-Hg201}
\end{align}
where $\gamma_i$ is the gyromagnetic ratio and $\chi_i$ is the so-called ``gyrogravitational ratio'' parameterizing the new interaction (where the subscripts $i$ denote the respective isotopes), and $\phi$ is the angle between $\mc{B}$ and $\bs{g}$.  As long as $\chi_{199}/\chi_{201} \neq \gamma_{199}/\gamma_{201}$ (as generally expected), the ratio $\sR = \Omega_{199}/\Omega_{201}$ acquires a dependence on $B$ and $\phi$ if the $\chi_i$'s are nonzero, enabling a search for the long-range monopole-dipole coupling.

%----------------------------------------------------------------
\begin{figure}
\includegraphics[width=3.25 in]{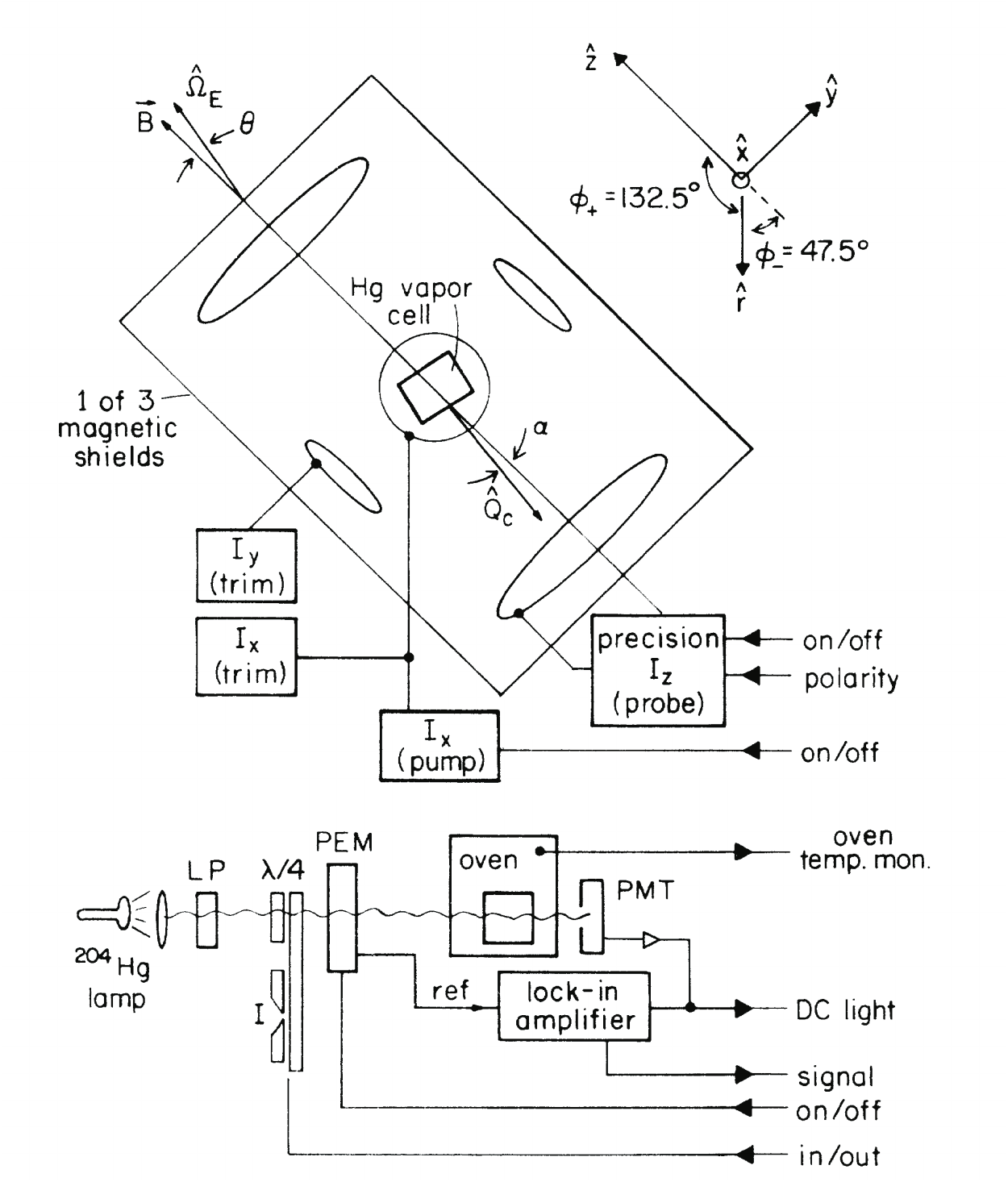}
\caption{Experimental setup from \citet{Ven92}.  LP = linear polarizer, $\lambda/4$ = quarter-wave plate, I = iris, PMT = photomultiplier tube, PEM = photoelastic modulator.  Arrows on right-hand side indicate computer control and data acquisition.  The angles $\phi_\pm$ indicate the projection of $\mb{\hat{r}}$ (parallel to $\bs{g}$) along $\mb{\hat{z}}$ (parallel to $\mc{B}$) for the two magnetic field orientations.}
\label{Fig:Hg-GDM-expt-setup}
\end{figure}
%----------------------------------------------------------------

The experimental setup is shown in Fig.~\ref{Fig:Hg-GDM-expt-setup}.  The Hg atoms were contained in a cylindrical vapor cell situated at the center of a three-layer cylindrical $\mu$-metal shield with internal coils to apply controlled magnetic fields to the atoms.  The axes of the concentric cylinders of the shield system (defined to be $z$) and quadrupole axis of the vapor cell ($\bs{\hat{Q}}_c$), as well as the magnetic field during spin precession, were oriented along the Earth's rotation axis ($\bs{\hat{\Omega}}_E$). This orientation is designed to make systematic errors related to the Earth's rotation quadratic in the misalignment angle of the apparatus, as discussed below. The experimental procedure consisted of a pump stage and a probe stage.  During the pump stage, the atoms were optically pumped
in the presence of a small magnetic field along $x$ ($B_x \lesssim 10~{\rm mG}$) by circularly polarized light propagating along $\mb{\hat{x}}$.
Optical pumping involves exciting atomic transitions with polarized light in order to generate spin polarization: the angular momentum of the light field is transferred to the atomic sample -- see, for example, reviews by \citet{Hap72} and \citet{Hap10}. In the probe stage, the magnetic field was re-directed along $\pm \mb{\hat{z}}$ in order to induce spin precession. The light intensity was reduced so as not to significantly perturb the atomic states, and a photoelastic modulator (PEM) rapidly alternated the light polarization between right- and left-circular in order to reduce vector light shifts and enable lock-in detection.  The detected signal was demodulated at the PEM frequency (42~kHz) and the free-precession-decay signal was analyzed to extract the precession frequencies.

Two important systematic errors required special consideration.  The first arose due to a collisional interaction of the $^{201}$Hg atoms with the walls of the cylindrical vapor cell, causing a $\approx 50~{\rm mHz}$ quadrupolar shift. The quadrupolar wall shift led to resolved splitting of the Zeeman frequencies for $^{201}$Hg. The quadrupolar wall shift, and an optical method to cancel it, has recently been studied in detail by \citet{Pec16} for Cs atoms -- although it should be noted that in this case the quadrupolar wall shift turns out to be of electronic origin rather than nuclear as is the case for Hg. The second systematic error arose because the experimental apparatus was attached to the Earth, while the Hg spins were effectively decoupled from the Earth's rotation during the probe stage (since the spins were freely precessing). Consequently, the Hg isotopes exhibited apparent precession at the rotation rate of the Earth, $\Omega_E \approx 2\pi \times 11.6~{\rm \mu Hz}$.  This effect, known as the gyro-compass effect \cite{Hec08}, can be understood as the result of viewing an inertial system, the Hg spins, from a noninertial frame, the surface of the rotating Earth. The gyro-compass effect was studied with even greater precision in the work of \citet{Bro10} and \citet{Gem10}.  Both systematic effects were constrained at or below the statistical sensitivity of the experiment by orienting the apparatus so that uncertainty in the effects were quadratic in the misalignment angles.

The experiment establishing the strongest laboratory-scale limit on monopole-dipole couplings of neutrons was that of \citet{Tul13}, shown by the solid blue curve in the upper plot of Fig.~\ref{Fig:monopole-dipole-constraints}. In the experiment of \citet{Tul13}, the spin-precession frequencies of co-located gaseous samples of $^{3}$He and $^{129}$Xe were measured using a multi-channel, low-$T_c$ Superconducting QUantum Interference Device (SQUID) to monitor the magnetization. This avoided issues related to light shifts that can be problematic in optical atomic magnetometry experiments \cite{Aco06,Kim13,Kim17GDM}. The source mass was a cylindrical unpolarized BGO crystal (Bi$_4$Ge$_3$O$_{12}$) whose position could be modulated using a compressed-air driven piston between $\approx 2~{\rm mm}$ and $\approx 200~{\rm mm}$ from a $^{3}$He/$^{129}$Xe cell in order to modulate the strength of the exotic interaction. The BGO crystal was chosen as the source mass based on its high nucleon number density, low conductivity (and thus low Johnson-Nyquist noise), and its vanishingly small low-field magnetic susceptibility.

At sub-mm distance scales, limits on monopole-dipole interactions of the neutron have been obtained by the experiments of \citet{Bul13}, \citet{Pet10}, and \citet{Gui15} shown by short-dashed red, solid red, and short-dashed blue curves, respectively, in the upper plot of Fig.~\ref{Fig:monopole-dipole-constraints}. \citet{Bul13} employed a dual species xenon nuclear magnetic resonance (NMR) gyroscope with polarized $^{129}$Xe and $^{131}$Xe to search for a monopole-dipole interaction when a zirconia rod was moved near the NMR cell. Again the technique of co-magnetometry was utilized: by simultaneously comparing the precession frequencies of the two Xe isotopes, magnetic field changes were distinguished from frequency shifts due to the monopole-dipole coupling between the polarized Xe nuclei and the zirconia rod source mass. The experiments of \citet{Pet10} and \citet{Gui15} used measurements of hyperpolarized $^3$He to constrain the contribution of short-range monopole-dipole interactions to relaxation rates. Although it is outside the range of the parameter space plotted in Fig.~\ref{Fig:monopole-dipole-constraints}, the work of \citet{Jen14} establishes the strongest bounds on $\left| {\rm g}_p {\rm g}_s \right|/\hbar c$ for distances between 1~$\mu$m and 100~$\mu$m. In the experiment of \citet{Jen14}, transitions between quantum states of ultracold neutrons confined vertically above a horizontal mirror by the Earth's gravity were driven by resonantly oscillating the mirror position. At even shorter distance ranges, $10^{-10}~{\rm m} \lesssim \lambda \lesssim 10^{-7}~{\rm m}$, the most stringent laboratory constraints on monopole-dipole interactions come from measurement of the diffraction of a cold neutron beam as it passed through a non-centrosymmetric quartz crystal \cite{Fed13}, setting the bound $\left| {\rm g}_p {\rm g}_s \right|/\hbar c \lesssim 10^{-12}$. The experiment of \citet{Afa15} used co-located samples of ultracold neutrons and $^{199}$Hg atoms to obtain constraints at a level similar to that of \citet{Pet10}.

\subsubsection{Electrons}

\begin{figure}
\includegraphics[width=3.25 in]{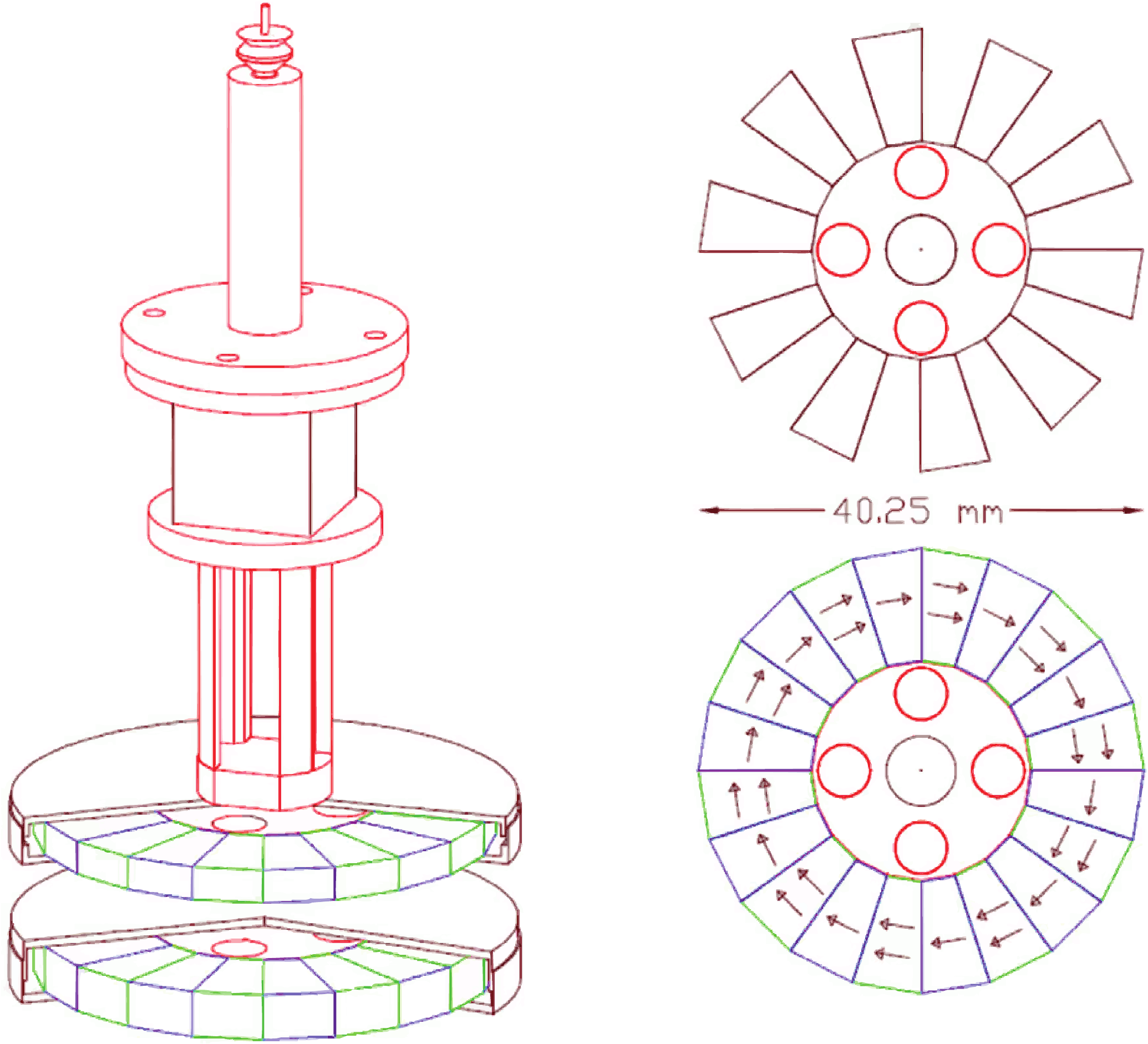}
\caption{The left panel shows the 20-pole spin-polarized torsion pendulum and the right panel shows the unpolarized and polarized sources (upper and lower figures, respectively) used to search for monopole-dipole and dipole-dipole interactions in the experiment of \citet{Ter15}. The $\mu$-metal shielding surrounding the spin-polarized pendulum and the sources is cut away to show the alnico (green) and SmCo$_5$ (blue) segments and one of the four pair of calibration cylinders (red). The mirror cube (in the middle of the pendulum's support structure) is used to monitor the pendulum twist angle. The entire apparatus is contained within a system of magnetic shields. The arrows on the spin attractor indicate net spin density and direction.}
\label{Fig:spin-polarized-torsion-pendulum}
\end{figure}

The lower plot of Fig.~\ref{Fig:monopole-dipole-constraints} shows constraints on monopole-dipole interactions of electrons. Most of the best limits for electrons come from a series of experiments using spin-polarized torsion pendulums carried out at the University of Washington \cite{Hec08,Hoe11,Ter15}, shown by the purple curves. A diagram of the spin-polarized torsion pendulum setup used by \citet{Ter15} is shown in Fig.~\ref{Fig:spin-polarized-torsion-pendulum}. The key piece of the experimental apparatus was a ring of 20 equally magnetized segments of alternating high- and low-spin density materials. The 20-pole spin ring was the active element of the torsion-pendulum detector. The high spin density material was alnico and the low spin density material was SmCo$_5$ -- a substantial degree of the magnetization of SmCo$_5$ comes from the orbital motion of electrons while the magnetization of alnico is almost entirely due to the electrons' spin. The magnetization of each alnico wedge is tuned by a localized external field so that the spin-polarized torsion pendulum has negligible variation in magnetization. Then either an unpolarized copper attractor or spin-polarized attractor (identical to the pendulum detector) was rotated below the torsion pendulum at a frequency $\omega$, producing a modulated torque at $10\omega$ as the source's high mass (or spin) density wedges passed below the high or low spin density segments of the pendulum. The pendulum's and both attractors' four cylinders (either tungsten or vacuum) provided gravitational calibration signals at $4\omega$. The twisting of the pendulum was measured optically using a reflector cube, and the torque was inferred from a harmonic analysis of the pendulum twist angle. The experimental setup allowed the attractors to be moved close to the pendulum, with a minimum separation of $\approx 4~{\rm mm}$. The experiment of \citet{Hec08} used a similar spin-polarized torsion pendulum but with the Earth and Sun as source masses. \citet{Hoe11} used a semiconductor-grade silicon single crystal attached to an ultrapure titanium bar as the torsion pendulum in order to have a highly non-magnetic detector, and then used a ferromagnet as a dipole source -- this setup enabled the spin source to be brought into close proximity of the detector, allowing sensitivity to monopole-dipole forces with ranges of fractions of a mm (i.e., boson masses $\gtrsim 1~{\rm meV}$).

Strong laboratory constraints on monopole-dipole couplings of electrons were also obtained by \citet{Ni99} by using a paramagnetic salt (TbF$_3$) and a dc SQUID to search for induced spin polarization in the TbF$_3$ sample caused by the proximity of a copper mass. This approach was recently improved upon by \citet{Crescini2017}, who used a dc SQUID to measure variation of the magnetization of a GSO crystal (${\rm Gd_2SiO_5}$) housed within a superconducting shield as a function of the distance to a lead mass under cryogenic conditions. Important constraints on both electron and neutron monopole-dipole interactions were also obtained in the experiment of \citet{Win91}. They carried out measurements on trapped $^9$Be$^+$ ions as an applied magnetic field was reversed relative to the local gravitation field $\bs{g}$: the resulting frequency shift between the $^9$Be$^+$ $^2 S_{1/2}~\ket{F=1,M=0}$ and $^2 S_{1/2}~\ket{F=1,M=-1}$ states was constrained to be $< 13.4~\mu{\rm Hz}$, leading to the limits shown in the lower plot of Fig.~\ref{Fig:monopole-dipole-constraints} with the dotted black line.

%----------------------------------------------------------------
%----------------------------------------------------------------

\subsubsection{Protons}

The middle plot of Fig.~\ref{Fig:monopole-dipole-constraints} shows constraints on monopole-dipole interactions of protons. Here experimental limits are somewhat sparser. Most measurements using atomic vapor comagnetometers to search for exotic spin-dependent interactions use noble gases with valence neutrons, and therefore, as discussed previously, they are insensitive to proton couplings. Experiments using spin-polarized torsion pendulums or solid-state systems are sensitive to electron couplings. The laboratory-range experiment of \citet{You96}, whose established constraints are shown by the long-dashed black curve, is an exception. \citet{You96} searched for monopole-dipole couplings between a 475-kg lead mass and the spins of $^{133}$Cs and $^{199}$Hg atoms using co-located atomic magnetometers (consisting of a Cs vapor cell sandwiched between a pair of $^{199}$Hg cells contained within a system of magnetic shields, with laser optical pumping and probing of the atomic spins). \citet{You96} originally interpreted the results of their experiment to constrain only electron and neutron spin couplings. However, because the $^{133}$Cs nucleus has a valence proton, \citet{Kim15} noted that in this case the single-particle Schmidt model, semi-empirical models, and large-scale nuclear shell model calculations are all in reasonable agreement concerning the contribution of the valence proton spin to the nuclear spin of $^{133}$Cs. Therefore the experiment of \citet{You96} reliably establishes laboratory constraints on exotic monopole-dipole couplings of the proton. Similarly, at short ranges, the experiments of \citet{Pet10} and \citet{Chu13} establish constraints for protons because of the well-understood contribution of proton spin to the nuclear spin of $^3$He \cite{Ant96,Kim15}. \citet{Chu13} search for a spin-precession frequency shift of polarized $^3$He when an unpolarized mass (either a ceramic block or a liquid mixture of $\approx 1~\%$ MnCl$_2$ in pure water) was moved between 5~cm and 10~$\mu$m of the $^3$He vapor cell. The particular source masses were chosen based on their nucleon densities, low magnetic impurities and magnetic susceptibilities, and minimal influence on the NMR measurement procedure. Although the work of \citet{Tul13} also uses polarized $^3$He, because the technique of comagnetometry with $^{129}$Xe is employed and there is considerable uncertainty regarding the contribution of the proton spin to the $^{129}$Xe nuclear spin \cite{Kim15}, we do not infer a limit on monopole-dipole interactions of the proton from this work. Recently \citet{Kim17GDM} completed a search for a long-range monopole-dipole coupling of the proton spin to the mass of the Earth using a $^{85}$Rb/$^{87}$Rb comagnetometer, improving on the long-range limits of \citet{You96} by over three orders-of-magnitude. The experiment of \citet{Kim17GDM} employed overlapping ensembles of $^{85}$Rb and $^{87}$Rb atoms contained within an evacuated, antirelaxation-coated vapor cell and simultaneously measured the spin precession frequencies using optical magnetometry techniques \cite{Bud13} as the magnetic field was reversed relative to the direction of the gravitational field, similar to the experiment of \citet{Ven92} discussed earlier. The measurement of \citet{Kim17GDM} establishes the best constraint on the proton GDM. The experiment was ultimately limited by systematic effects related to scattered light and magnetic field gradients.

\subsubsection{Astrophysical constraints}

The green curves and light green shading in Fig.~\ref{Fig:monopole-dipole-constraints} show the parameter space excluded by astrophysical considerations. \citet{Raf12} argues that the coupling constants $g_s$ and $g_p$ are individually constrained, and thus constraints on their product $g_sg_p$ can be derived. The scalar coupling constant $g_s$ is constrained by laboratory searches for monopole-monopole interactions [the potential $\sV_1(r)$, Eq.~\eqref{Eq:V1} -- see the review by \citet{Ade09} and also Sec.~\ref{Sec:5thForces}]. The pseudoscalar coupling constant $g_p$ for nucleons is constrained by the measured neutrino signal from supernova 1987A: the $\approx 10~{\rm s}$ duration of the signal excludes excessive new energy losses \cite{Raf88,Tur88}, although this constraint is based on the bremsstrahlung process in the collapsed supernova core and thus suffers from significant uncertainties related to dense nuclear matter effects \cite{Jan96}, and recent calculations \cite{Blu16SN1987a,Cha17SN1987a,Har17SN1987a,Mah17SN1987a} have suggested these limits may be weaker than first estimated and as displayed in Fig.~\ref{Fig:monopole-dipole-constraints}. The pseudoscalar coupling constant $g_p$ for electrons is constrained by star cooling rates \cite{Raf95}. Although the astrophysical constraints on $\left| {\rm g}_p {\rm g}_s \right|/\hbar c$ are more stringent than the laboratory limits in all cases, there is both a degree of model specificity \cite{Mas05} and some degree of uncertainty regarding the accuracy of stellar models \cite{Har17SN1987a}. Furthermore, it is possible that a so-called ``chameleon mechanism'' that screens interactions in regions of space with high mass density could invalidate astrophysical bounds on new interactions \cite{Jai06}.  Thus direct laboratory measurements play a crucial, comparatively less ambiguous role in determining the existence of exotic spin-dependent interactions even when they are somewhat less sensitive than astrophysical bounds.

\subsection{Experimental constraints on dipole-dipole interactions}
\label{Sec:ExoticSpin:DipoleDipole}

%----------------------------------------------------------------
\begin{figure}
\center
\includegraphics[width=3.5 in]{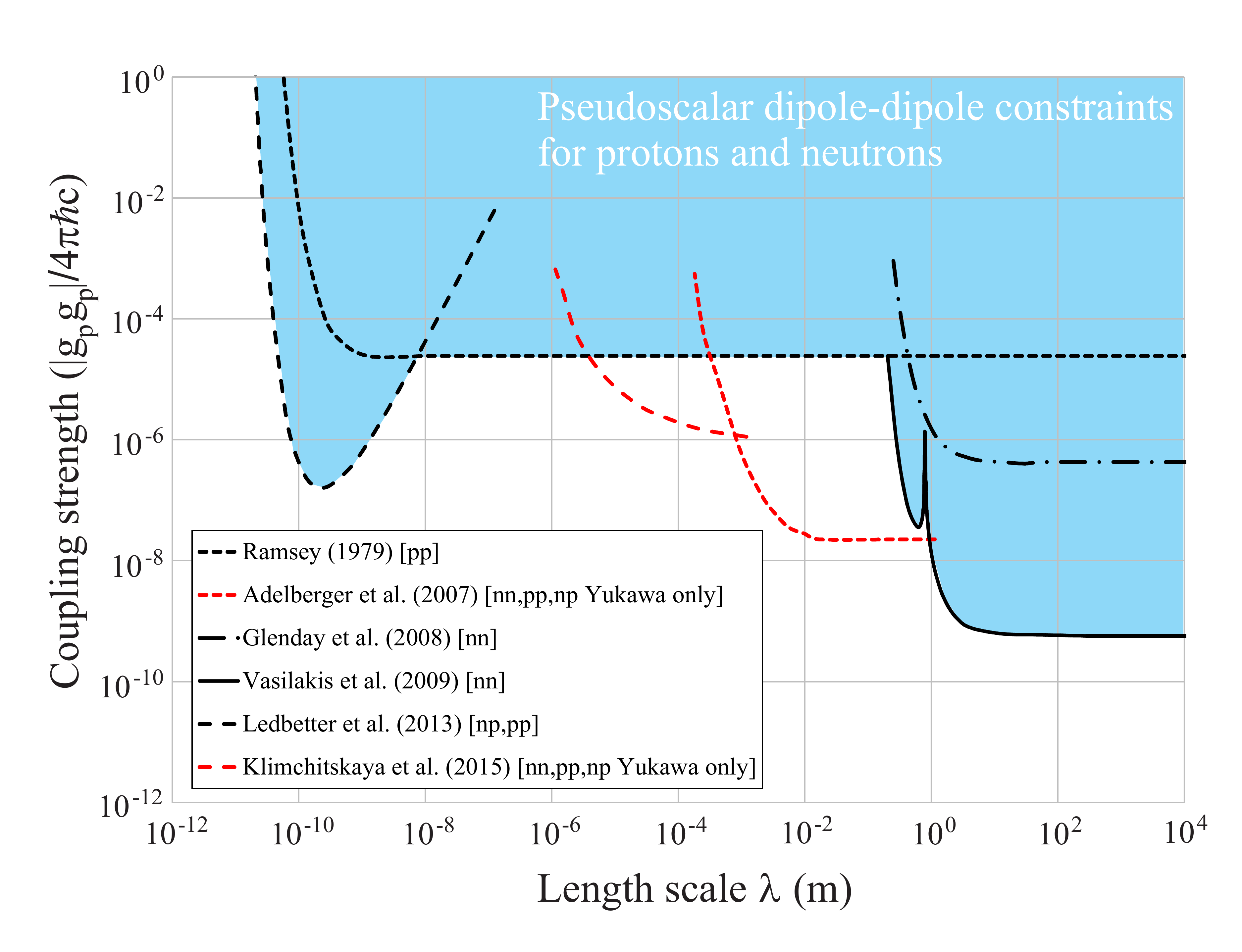}
\includegraphics[width=3.5 in]{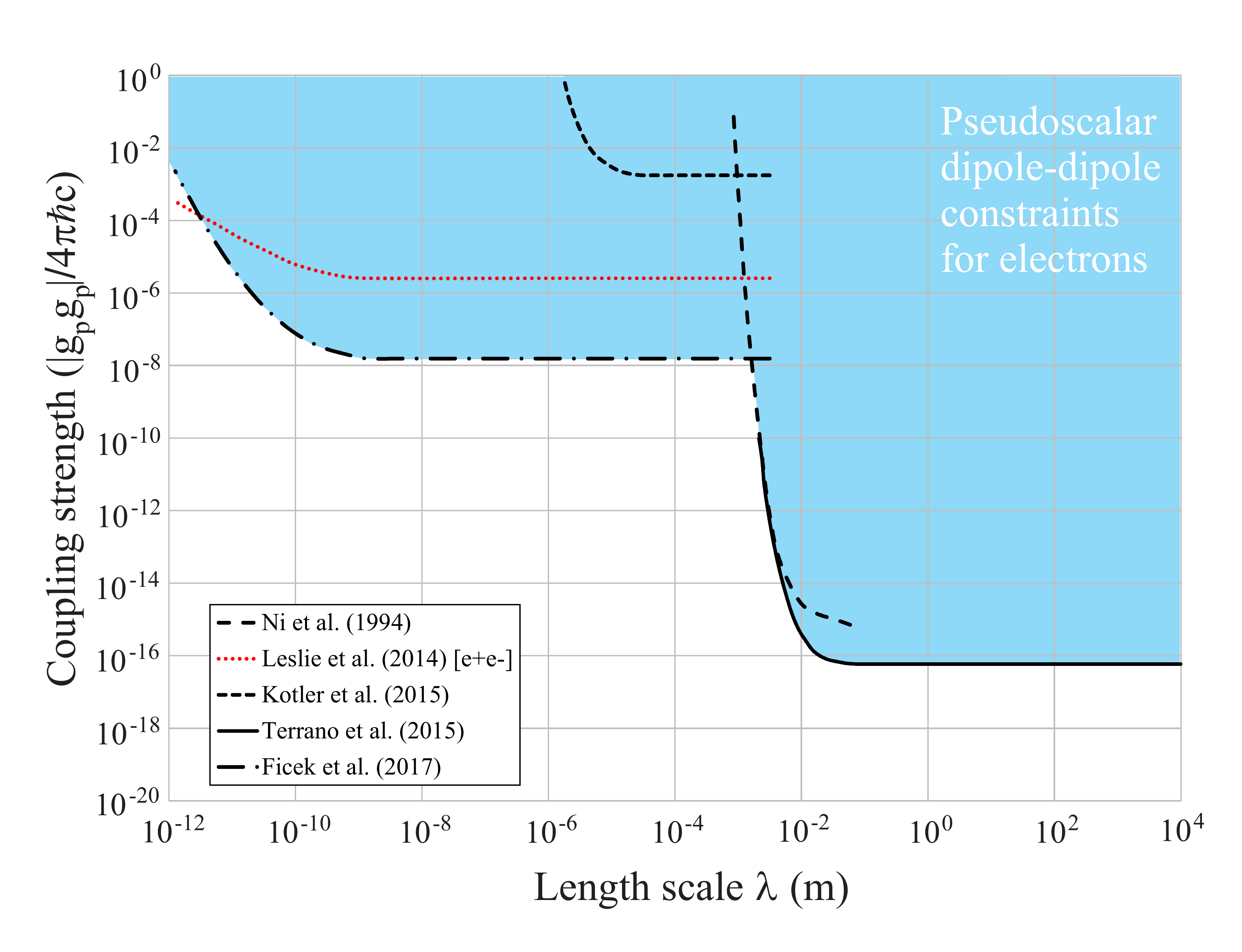}
\caption{Laboratory constraints (shaded light blue, see text for discussion of individual experiments) on pseudoscalar dipole-dipole couplings, $\left| {\rm g}_p {\rm g}_p \right|/(4 \pi\hbar c)$ [the $\sV_{3}$ potential as described in Eq.~\eqref{Eq:V3-expanded-form}], between nucleons and electrons as a function of the range $\lambda$ of the interaction. The short- and long-dashed red lines in the upper plot show constraints derived from spin-independent measurements that apply only to the Yukawa form of the pseudoscalar interaction [Eq.~\eqref{Eq:V3-Yukawa-Lagrangian}]. The long-dashed red line in the lower plot shows constraints based on positronium spectroscopy which in order to be compared with electron-electron constraints must assume $CPT$ invariance.}
\label{Fig:pseudoscalar-dipole-dipole-constraints}
\end{figure}
%----------------------------------------------------------------

Experimental searches for monopole-dipole interactions have certain appeal because such couplings violate invariance under both time reversal and spatial inversion, and hence one expects negligible background from standard-model physics. Dipole-dipole couplings, on the other hand, are even under both $T$ and $P$ and can arise from standard model physics. In this sense, dipole-dipole couplings may be problematic for exotic physics searches because one must carefully account for standard-model physics effects. Nonetheless, there has been impressive recent progress in laboratory searches for exotic dipole-dipole interactions.

\subsubsection{Constraints on $\sV_3(r)$}

\begin{figure}
\center
\includegraphics[width=3.25 in]{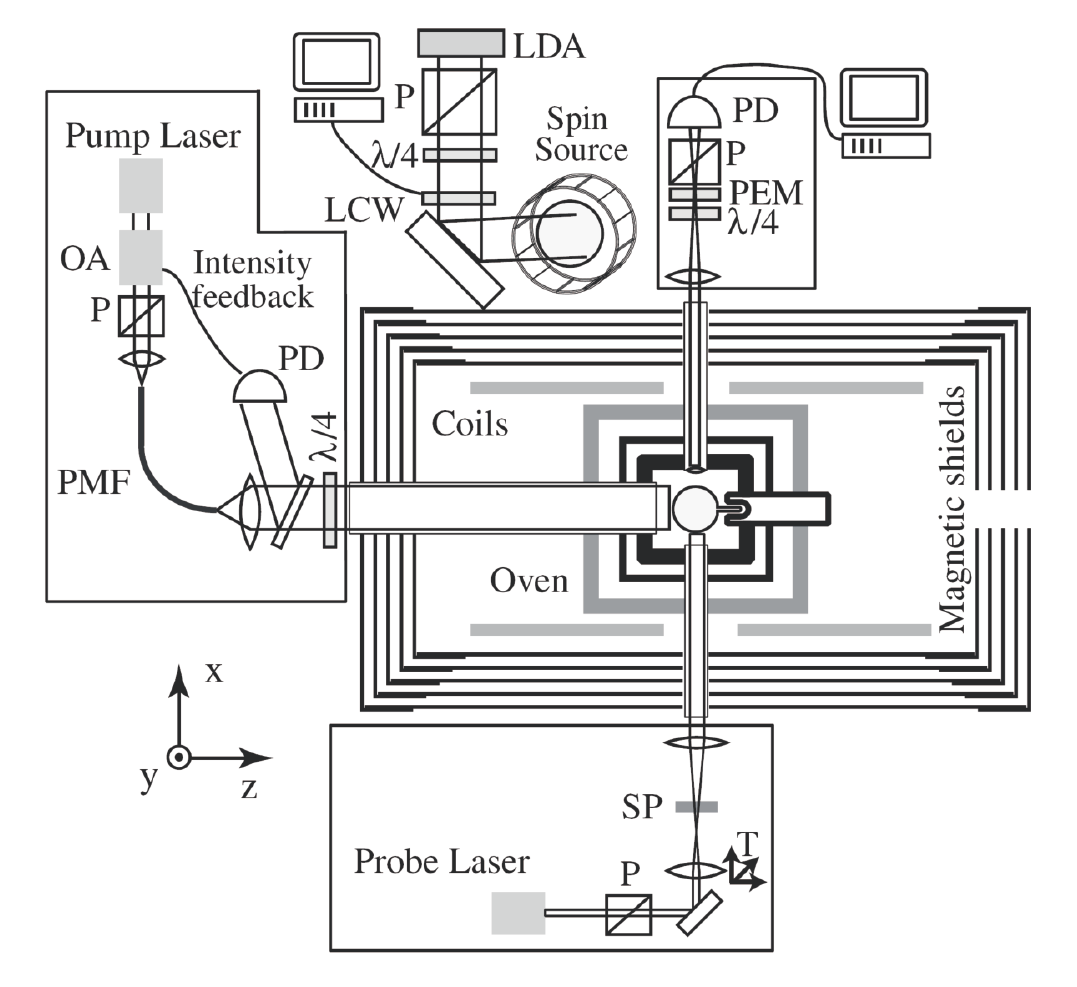}
\caption{Experimental setup of \citet{Vas09}.  PD = photodiode, SP = stress plate to control polarization of the probe beam, T = translation stage to shift the probe beam, P = polarizer, PMF = polarization maintaining fiber, OA = optical amplifier, LCW = liquid crystal wave plate,
PEM = photoelastic modulator, $\lambda/4$ = quarter-wave plate, LDA = laser diode array.}
\label{Fig:SERF-comagnetometer}
\end{figure}

The best limit on long-range pseudoscalar dipole-dipole interactions [of the form given by the $\sV_{3}$ potential described in Eq.~\eqref{Eq:V3-expanded-form}] between neutrons was achieved in the experiment of \citet{Vas09} (solid black curve in the upper plot of Fig.~\ref{Fig:pseudoscalar-dipole-dipole-constraints}) using the setup shown in Fig.~\ref{Fig:SERF-comagnetometer}.  The measurement technique is based on the principles of spin-exchange-relaxation-free (SERF) magnetometry \citet{All02,Kor02,Kom03,Kor05}.  The atomic sample consists of overlapping ensembles of potassium (K) and $^3$He at relatively high vapor densities ($^3$He density $\approx 10^{20}~{\rm atoms/cm^3}$ and K density $\approx 10^{14}~{\rm atoms/cm^3}$). The K sample is polarized through optical pumping and the $^3$He sample is polarized through spin-exchange collisions with K.  The vapor cell is located within a five-layer $\mu$-metal shield fitted with internal coils used to cancel residual magnetic fields and create a small field $\mc{B}$ parallel to the propagation direction of the pump beam.

%----------------------------------------------------------------
%----------------------------------------------------------------

Under these experimental conditions, for which the Larmor frequencies are comparable to or smaller than the frequency of spin-exchange collisions, the spin-exchange interaction between K atoms and the polarized $^3$He vapor strongly couples the two spin ensembles \cite{Kor02,Kor05}.  In a spherical cell this coupling can be represented as an effective magnetic field $\mc{B}\ts{eff}$ experienced by one spin species due to the average magnetization $\mb{M}$ of the other.  The applied field $\mc{B}$ is tuned so that it approximately cancels the $\mc{B}\ts{eff}$ experienced by the K atoms.  The K atoms are then effectively in a zero-field environment. Because the $^3$He magnetization $\mb{M}$ adiabatically follows $\mc{B}$, components of $\mc{B}$ transverse to $\mb{\hat{z}}$ are automatically compensated by $\mc{B}\ts{eff}$ to first order. Such cancellation only occurs for interactions that couple to spins in proportion to their magnetic moments, leaving the K-$^3$He comagnetometer sensitive to inertial rotation \cite{Kor05} and anomalous spin couplings \cite{Vas09}.  Thus the self-compensating K-$^3$He comagnetometer enables one to use high-sensitivity SERF magnetometry techniques for detection of anomalous spin-dependent interactions causing precession about axes transverse to $\mb{\hat{z}}$.

The K spin-polarization along $\mb{\hat{x}}$ is determined by measuring optical rotation of an off-resonant, linearly polarized probe light beam [see, for example, the review by \citet{Bud02} for a discussion of optical rotation].  After residual magnetic fields and light shifts are eliminated using zeroing routines [described in detail by \citet{Kor05}], the K spin-polarization along $\mb{\hat{x}}$ can only arise due to non-magnetic, spin-dependent interactions --- offering a highly sensitive probe of such anomalous interactions.  The spin source in the experiment of \citet{Vas09} consisted of a dense ($\approx 3 \times 10^{20}~{\rm cm^{-1}}$), highly polarized ($\approx 15~\%$ polarization) $^3$He gas located approximately 50~cm from the cell.  The nuclear spin direction of the $^3$He sample was reversed at a 0.18~Hz rate by adiabatic fast passage.  After approximately one month of data acquisition, no anomalous effect was detected at a level corresponding to a magnetic field value less than an attoTesla ($10^{-14}~{\rm G}$).

Constraints on pseudoscalar dipole-dipole couplings between protons at the molecular scale were deduced by \citet{Ram79} based on molecular-beam experiments with hydrogen (H$_2$). Comparing the measurements of \citet{Har53} to calculations of the magnetic dipole-dipole interaction between the protons in H$_2$ limited the possible contribution of an exotic dipole-dipole interaction to spin-dependent energy splittings, establishing the constraint shown by the short-dashed black curve in the upper plot of Fig.~\ref{Fig:pseudoscalar-dipole-dipole-constraints}. \citet{Led13} obtained the constraints on proton-proton and neutron-proton pseudoscalar dipole-dipole couplings shown by the long-dashed black curve in the upper plot of Fig.~\ref{Fig:pseudoscalar-dipole-dipole-constraints} by comparing NMR measurements to theoretical calculations of indirect nuclear dipole-dipole coupling (J-coupling) in deuterated molecular hydrogen (HD). The Hamiltonian describing J-coupling has the form $\sJ \mb{I}_1\cdot\mb{I}_2$ ($\mb{I}_{1,2}$ are the nuclear spins and $\sJ$ parameterizes the interaction strength) and arises due to a second-order hyperfine interaction where the interaction between the nuclear spins is mediated through the electron cloud. The measurements from which \citet{Led13} extracted constraints were performed with HD in the gas phase: thus the internuclear vector $\hat{\mb{r}}$ was randomly reoriented due to collisions. This effect leads to an averaging of Eq.~\eqref{Eq:V3-expanded-form}, so that its distance scaling becomes proportional to $e^{-r/\lambda}/(\lambda^2 r)$. The collisional averaging reduces sensitivity to exotic dipole-dipole forces for which $\lambda$ differs significantly from the mean internuclear separation, as seen in Fig.~\ref{Fig:pseudoscalar-dipole-dipole-constraints}. Of interest in regard to the constraints derived from J-coupling in HD are more recent measurements and calculations \cite{Gar14,Ner14}.

Other notable experiments searching for exotic dipole-dipole couplings of nucleons include the work of \citet{Gle08}, an experiment similar to that of \citet{Vas09} that employed a dual-species $^3$He-$^{129}$Xe maser as the detector, and constraints from \citet{Ade07} and \citet{Kli15} based on short-range tests of the gravitational inverse-square law and the Casimir effect. The work of \citet{Ade07} and \citet{Kli15}, which actually search for spin-independent interactions, constrain only the Yukawa form of the pseudoscalar coupling [Eq.~\eqref{Eq:V3-Yukawa-Lagrangian}] and are thus more model-specific than the other laboratory searches considered. The constraints from \citet{Ade07} and \citet{Kli15} do not apply to the derivative form [Eq.~\eqref{Eq:V3-derivative-Lagrangian}] that would be expected for Goldstone bosons such as the axion. Constraints on spin-dependent interactions derived from experimental searches for spin-independent interactions are also considered by \citet{Ald16}.

For electrons, the experiments of \citet{Ter15} and \citet{Ni94} establish the most stringent constraints on pseudoscalar dipole-dipole forces at interaction ranges $\gtrsim 1~{\rm mm}$ (solid and long-dashed black curves, respectively, in the lower plot of Fig.~\ref{Fig:pseudoscalar-dipole-dipole-constraints}). The experiment of \citet{Ter15} was addressed in the preceding section on monopole-dipole interactions (see Fig.~\ref{Fig:spin-polarized-torsion-pendulum} and surrounding discussion). \citet{Ni94} used a SQUID to measure the magnetization of a paramagnetic salt (TbF$_3$) induced by dipole-dipole interactions with rotating spin-polarized samples (Dy$_6$Fe$_{23}$ and HoFe$_3$). From atomic scales up to a mm, the agreement between energy structure calculations and spectroscopic measurements in He \cite{Fic17} provide the most stringent constraints, shown by the black dot-dashed line in the lower plot of Fig.~\ref{Fig:pseudoscalar-dipole-dipole-constraints}. Also of interest are electron spin resonance (ESR) measurements in iron using a scanning tunneling microscope (STM) by \citet{Luo17constraints}.

%----------------------------------------------------------------
\begin{figure}
\center
\includegraphics[width=3.5 in]{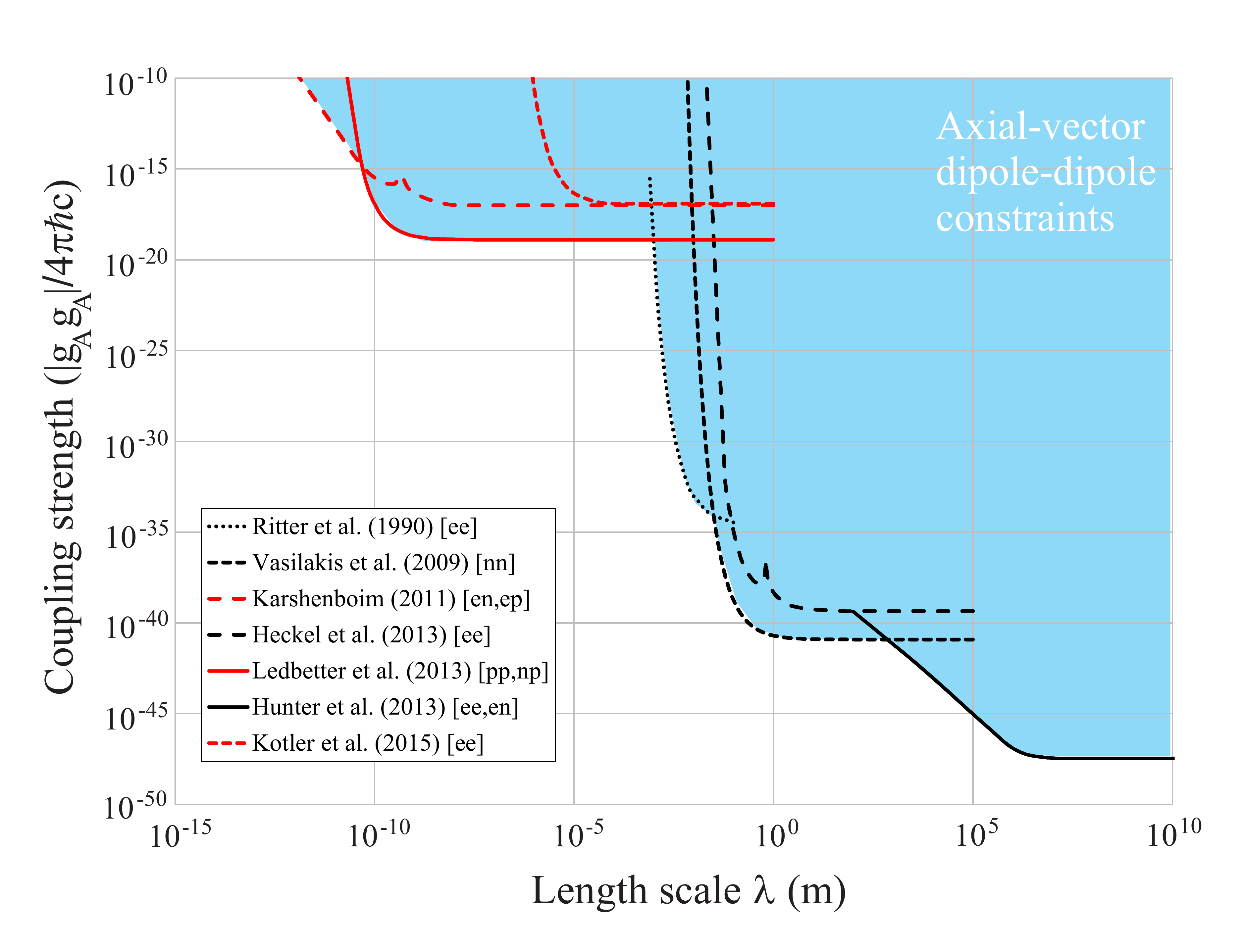}
\caption{Laboratory constraints (shaded light blue, see text for discussion of individual experiments) on axial-vector dipole-dipole couplings, $\left| {\rm g}_A {\rm g}_A \right|/(4 \pi \hbar c)$ [the $\sV_{2}$ potential as described in Eq.~\eqref{Eq:V2-expanded-form}], for various particles as a function of the range $\lambda$ of the interaction.}
\label{Fig:axial-vector-dipole-dipole-constraints}
\end{figure}
%----------------------------------------------------------------

\subsubsection{Constraints on $\sV_2(r)$}

Figure~\ref{Fig:axial-vector-dipole-dipole-constraints} shows the laboratory constraints on axial-vector dipole-dipole couplings, $\left| {\rm g}_A {\rm g}_A \right|/(4 \pi \hbar c)$, described by the $\sV_{2}$ potential in the MWDM formalism [Eq.~\eqref{Eq:V2-expanded-form}]. In terms of experiments, the critical difference between the $\sV_{2}$ and $\sV_{3}$ potentials is the scaling with particle separation: the $\sV_{2}$ potential scales as $1/r$ whereas the $\sV_{3}$ potential scales as $1/r^3$. Thus experiments searching for dipole-dipole interactions can have vastly different sensitivities to the two different potentials.

An excellent example illustrating the importance of the distance scaling is the work of \citet{Hun13} which established the long-range axial-vector constraints shown by the solid black curve in Fig.~\ref{Fig:axial-vector-dipole-dipole-constraints}. \citet{Hun13} took advantage of the large number of polarized electrons in the Earth: there are $\approx 10^{49}$ unpaired electron spins in the Earth, yielding $\approx 10^{42}$ polarized geoelectrons polarized by the Earth's magnetic field. Thus the number of polarized geoelectrons exceeds that of a typical laboratory source by a factor of $\gtrsim 10^{17}$. However, a typical laboratory source of polarized electrons can be placed closer than a meter away from a detector whereas the mean distance of a polarized geoelectron is $\gtrsim 10^5~{\rm m}$ from a detector on the surface of the Earth. For pseudoscalar dipole-dipole interactions, the $1/r^3$ distance scaling makes searches for exotic interactions with geoelectrons less competitive than searches employing polarized laboratory sources. On the other hand, the $1/r$ scaling of the axial-vector interaction makes the huge number of polarized geoelectrons a much stronger source with which to search for long-range interactions. \citet{Hun13} used data from optical atomic magnetometers \cite{Ven92,Pec12} and a spin-polarized torsion pendulum \cite{Hec08} to derive the limits shown in Fig.~\ref{Fig:axial-vector-dipole-dipole-constraints}. The experiments of \citet{Pec12} and \citet{Hec08} utilized rotatable mounts for their entire experimental apparatus in order to modulate the signal from the polarized geoelectrons, a technique also employed in the experiment of \citet{Bro10}.

Many of the experiments searching for exotic dipole-dipole interactions previously discussed also place strong constraints on axial-vector interactions between various particles \cite{Led13,Vas09,Hec08}. Between a $\mu$m and a mm, the best direct constraint on axial-vector dipole-dipole interactions between electrons comes from the measurement of the magnetic dipole-dipole interaction between two trapped $^{88}$Sr$^+$ ions \cite{Kot14,Kot15}. \citet{Kot14} trapped two $^{88}$Sr$^+$ ions using a linear radio-frequency Paul trap, and the ions were initialized in an entangled state that was insensitive to spatially homogeneous magnetic field noise. This technique enabled precise measurement of the magnetic dipole-dipole interaction between the ions, which when compared to a straightforward calculation gave good agreement at a level of $\approx 200~{\rm \mu Hz}$. \citet{Kot15} then used the agreement between experiment and theory to limit the strength of exotic dipole-dipole interactions as shown by the short-dashed red curve in  Fig.~\ref{Fig:axial-vector-dipole-dipole-constraints}. \citet{Rit90} (solid purple curve in Fig.~\ref{Fig:axial-vector-dipole-dipole-constraints}) carried out an experiment with a spin-polarized torsion pendulum made from Dy$_6$Fe$_{23}$, which had the characteristic that at a particular temperature (between 265~K to 280~K) the magnetization due to the orbital motion of the electrons approximately cancelled the magnetization from the electron spins, allowing a torsion pendulum with large net intrinsic spin but small magnetic moment, a similar idea to that behind the later work of the University of Washington group discussed previously \cite{Hec08,Ter15}. \citet{Kar10a,Kar10b,Kar10c} compared spectroscopic measurements of hyperfine structure to QED calculations for various atomic systems in order to derive constraints on axial-vector interactions, the strongest constraints coming from hydrogen, deuterium, and $^3$He$^+$.

\subsubsection{Astrophysical constraints}

It should be noted that laboratory limits on pseudoscalar interactions are weaker than relevant astrophysical constraints on $g_p$ from the neutrino signal from SN 1987A \cite{Eng90}, the metallicity of stars \cite{Hax91}, the maximum brightness of red giants \cite{Raf95}, and null searches for axion emission from the Sun \cite{Der09}. However, these astrophysical constraints do not necessarily apply to axial-vector or vector interactions \cite{Dob06}. Since both the $\sV_2$ and $\sV_3$ potentials can be generated by spin-1 bosons, astrophysical constraints -- specific to the vertex-level interactions for spin-0 bosons \cite{Raf99,Dob06} -- do not apply in general to the $\sV_2$ and $\sV_3$ potentials and are therefore not shown in Figs.~\ref{Fig:pseudoscalar-dipole-dipole-constraints} and \ref{Fig:axial-vector-dipole-dipole-constraints}.

\subsection{Experimental constraints on other forms of spin-dependent interactions}
\label{Sec:ExoticSpin:OtherSpinDependentPotentials}

A number of experiments have searched for some of the other forms of exotic spin-dependent potentials enumerated by \citet{Dob06} [Eqs.~\eqref{Eq:V1} - \eqref{Eq:V16}]. For example, \citet{Vas09} and \citet{Hun13} specifically searched for the $\sV_{11}$ potential [Eq.~\eqref{Eq:V11}]. \citet{Kim10} used measurements and calculated cross sections for spin exchange between alkali metal atoms and noble gases (specifically sodium and helium) to constrain anomalous spin-dependent forces between nuclei at the atomic scale, and established the first limits on the $\sV_8$ potential [Eq.~\eqref{Eq:V8}]. \citet{Hun14} used polarized geoelectrons to constrain many of the other velocity-dependent potentials: $\sV_{6,7}$, $\sV_8$, $\sV_{15}$, and $\sV_{16}$ [Eqs.~\eqref{Eq:V6-7}, \eqref{Eq:V8}, \eqref{Eq:V15}, and \eqref{Eq:V16}, respectively]. \citet{Yan13} used measurements of a $P$-odd spin rotation when a cold neutron beam passed through a liquid $^4$He target to set limits on $\sV_{12+13}$ at short ranges ($10^{-6}~{\rm m} \lesssim \lambda \lesssim 1~{\rm m}$), and \citet{Yan15} used $^{3}$He spin-relaxation rates with the Earth as an unpolarized source mass to constrain $\sV_{12+13}$ at long ranges ($\lambda \gtrsim 1~{\rm m}$). \citet{Pie12} were able to establish bounds on $\sV_{4+5}$ at the mm scale for neutrons using Ramsey's method of separated oscillatory fields with a cold neutron beam that travelled past a nearby copper plate.  \citet{Hec08} constrained long-range velocity-dependent potentials between their torsion pendulum and the Moon and Sun. \citet{ficek2018constraints} compared spectroscopic measurements and theoretical calculations for antiprotonic He to obtain the first constraints on exotic spin-dependent semileptonic interactions between matter and antimatter.

Measurements of atomic parity violation as described in Sec.~\ref{Sec:APV} can be used to search for interactions mediated by exotic bosons since several of the MWDM potentials violate $P$. Indeed, some of the best constraints on interactions mediated by new spin-1 ($Z^\prime$) bosons have been derived from atomic parity violation experiments \citet{BOUCHIAT2005,Dav12darkZ,Dzu17PNCbosons}.

It also bears mentioning that there have been several tests of the universality of free fall (UFF) performed with spin-polarized objects, in particular with cold atoms \cite{Fra04,Tar14,DuaDenZho16}. At present, such experiments are orders of magnitude less sensitive to the potentials described in Eqs.~\eqref{Eq:V1} - \eqref{Eq:V16} than the experiments described in Secs.~\ref{Sec:ExoticSpin:MonopoleDipole} and \ref{Sec:ExoticSpin:DipoleDipole}. The basic reason for this is that free-fall experiments essentially measure the spatial derivative of $\sV_i$ whereas the experiments using optical atomic magnetometers or torsion pendulums measure the energy shift due to $\sV_i$ directly. Sec.~\ref{Sec:GR} discusses UFF tests using both polarized and unpolarized test masses along with other experimental probes of the equivalence principle.

As noted in Sec.~\ref{Sec:ExoticSpin:TheoreticalMotivation}, there are a variety of other theories predicting spin-dependent interactions that are not well-described by the potentials outlined in Eqs.~\eqref{Eq:V1} - \eqref{Eq:V16}, and several experiments have specifically sought to measure such effects. \citet{Gle08}, \citet{Vas09}, and \citet{Hec08} searched for the hypothetical ghost condensate resulting from spontaneous breaking of Lorentz symmetry \cite{Ark04,Ark05}. \citet{Vas09} and \citet{Hun13} searched for the potentials arising from unparticles \cite{Geo07,Lia07}. Many experiments have analyzed their results in terms of gravitational torsion \cite{Hec08,Kim10,Kot15,Leh14,Leh15,Led13}. \citet{Iva17} have proposed that gravitational torsion generates a new type of $P$-even and $T$-odd potential that can be probed using spin-polarized particles moving through unpolarized matter that is rotating in the laboratory frame.  \citet{lehnert2017constraining} have experimentally investigated a different deviation from the predictions of general relativity known as nonmetricity by measuring the rotation of neutron spins as the neutrons propagate though liquid helium. Undoubtedly, the rich theoretical landscape of exotic spin-dependent interactions will continue to inspire a vibrant array of experiments as many possible interactions still remain unexplored.

\subsection{Emerging ideas}
\label{Sec:ExoticSpin:EmergingIdeas}

A major new direction in the search for exotic spin-dependent interactions is the push to study oscillating and transient signals from fields comprised of new bosons such as axions, ALPs, and hidden photons that may constitute dark matter or dark energy. These ideas are discussed in Sec.~\ref{Sec:LightDarkMatter}, and include global networks of optical atomic magnetometers \cite{Pos13,Pus13} and atomic clocks \cite{DerPos14} to search for correlated transient signals heralding new physics that might arise from topological defects \cite{Pos13,Sta14a} or clumps of virialized ultra-light fields \cite{Derevianko2016a}. There are also new experiments using NMR \cite{Bud14}, atomic spectroscopy \cite{Sta14b}, and resonant electromagnetic detectors \cite{Cha15} to search for coherently oscillating dark matter fields. A related proposal is that of \citet{Rom13}, who have noted that cosmological scalar fields, which may explain dark energy, have local spatial gradients that could have detectable electromagnetic couplings.

There are also new ideas being developed for novel sources and detectors that can be used to search for exotic spin-dependent interactions. \citet{Chu15} proposed the use of new paramagnetic insulators, in particular gadolinium gallium garnet (Gd$_3$Ga$_5$O$_{12}$, or GGG), to search for spin-dependent interactions. \citet{Led12} have proposed a new class of liquid state nuclear spin comagnetometers with potential sensitivities in the $10^{-11}~{\rm Hz}$ range for one day of measurement and \citet{limes2017} have demonstrated new techniques for He-Xe comagnetometry offering superior stability and accuracy.  Another new concept being developed by \citet{Arv14} combines the techniques used in short-distance tests of gravity employing torsion pendulums \cite{Kap07} and micro-cantilevers \cite{Ger08} with those used in NMR experiments in order to search for short-range monopole-dipole interactions.

\citet{Kim16-needle} have recently predicted that a ferromagnetic needle will precess about the axis of a magnetic field at a Larmor frequency $\Omega$ when $I\Omega \ll N\hbar$, where $I$ is the moment of inertia of the needle about the precession axis and $N$ is the number of polarized spins in the needle. In this regime the needle behaves as a gyroscope with spin $N\hbar$ maintained along the easy axis of the needle by the crystalline and shape anisotropy. Such a precessing ferromagnetic needle is a correlated system of $N$ spins which can be used to measure magnetic fields for long times. In principle, by taking advantage of rapid averaging of quantum uncertainty, the sensitivity of a precessing needle magnetometer can far surpass that of magnetometers based on spin precession of atoms in the gas phase. Under conditions where noise from coupling to the environment is subdominant, the scaling with measurement time $t$ of the quantum- and detection-limited magnetometric sensitivity is $t^{-3/2}$. If a magnetometer based on a precessing ferromagnetic needle can be experimentally realized, a measurement of needle precession averaged over $\approx 10^3~{\rm s}$ could reach a sensitivity to exotic electron-spin-dependent couplings at an energy scale of $\approx 10^{-26}~{\rm eV}$. If such an experimental sensitivity could be achieved in practice, it would probe exotic spin-dependent interactions more than five orders of magnitude weaker than present laboratory limits.

\section{Searches for exotic spin-independent interactions}
\label{Sec:5thForces}

\subsection{Introduction}
\label{Sec:5thForces:Intro}

One of the exotic potentials described by the MWDM formalism deserves special attention, namely $\sV_1$ [Eq.~\eqref{Eq:V1}] --- the sole potential among those discussed in Sec.~\ref{Sec:ExoticSpin} that has no dependence on the spins of the interacting fermions. Experimental searches for such exotic spin-independent interactions have a long history, mostly from the perspective of tests of the inverse-square law (ISL) of gravity. Originally the idea was to see if the gravitational force law followed the form \cite{Ade03}
\begin{align}
\mb{F}_G(r) = -G \frac{m_X m_Y}{r^{2+\epsilon}} \hat{\mb{r}}~,
\label{Eq:5thForces:ISL-violation}
\end{align}
where $\mb{F}_G$ is the gravitational force between test masses $m_X$ and $m_Y$ separated by a distance $r$, $G$ is Newton's gravitational constant, and $\epsilon$ is a parameter characterizing deviation from the ISL. Since the $r^{-2}$ scaling of the gravitational force law derives from the geometry of three-dimensional space, it turns out, generally, that a force law of the form given by Eq.~\eqref{Eq:5thForces:ISL-violation} is difficult to motivate from a theoretical perspective. Instead, the modern perspective follows the MWDM formulation, positing a Yukawa-like deviation from the ISL; the common $\alpha-\lambda$ parametrization \cite{Tal88} found in the literature proposes a modified form of the gravitational potential given by
\begin{align}
V'(r) = - \frac{G m_X m_Y}{r} \prn{ 1 + \alpha e^{-r/\lambda} }~,
\label{Eq:5thForces:alpha-lambda-parametrization}
\end{align}
where the parameter $\alpha$ characterizes the strength and $\lambda$ characterizes the range of the modified gravitational interaction. From the point-of-view of quantum field theory, such a modification of the gravitational interaction is equivalent to effects generated by the exchange of a new boson as in the MWDM formalism. Typically in the literature such a Yukawa-like, spin-independent interaction is referred to as a \emph{fifth force} \cite{Fuj71,Fis86}. Correspondence between the two viewpoints can be made explicit: exchange of scalar or vector bosons between fermions $X$ and $Y$ can be described with
\begin{align}
\alpha = \frac{\hbar c}{4\pi G m_X m_Y} \prn{ g_{s}^X g_{s}^Y - g_{v}^X g_{v}^Y }~,
\label{Eq:5thForces:alpha-in-MWDM-formalism}
\end{align}
where $g_{s,v}^{X,Y}$ characterizes the vertex-level scalar (subscript $s$) or vector (subscript $v$) coupling generating a long-range $\sV_1$ potential [Eq.~\eqref{Eq:V1}]. The range $\lambda$ is understood in this case to be the reduced Compton wavelength of the new scalar or vector boson. Although there have been numerous alternative theoretical proposals for specific forms of modified gravitational potentials, to a large degree these considerations are moot for experimental work since all searches for ISL violations have to date returned null results; Eq.~\eqref{Eq:5thForces:alpha-lambda-parametrization} is entirely adequate for phenomenological comparison of different experimental constraints. In the event a violation is detected, however, it will be necessary to pursue determination of the specific form of the new interaction.

There have been a number of recent comprehensive reviews on the topic of ISL tests and searches for exotic spin-independent interactions, we refer the reader to the works by \citet{Ade03,Gun05,Ono06,New09,Ant11,Lam12,Fis12,Mur15,brax2017bounding} for more details on this subject. In this section we offer a brief overview of the field and recent developments.

\subsection{Motivation and Theoretical Landscape}
\label{Sec:5thForces:motivation-theory}

Theories motivating searches for ISL violations and fifth forces are often inspired by the inherent conflict between general relativity and quantum field theory. One aspect of this conflict is the \emph{hierarchy problem}, the enormous gulf between the Higgs mass and the Planck mass (discussed in Sec.~\ref{Sec:ExoticSpin:TheoreticalMotivation}). An influential theoretical suggestion that inspired a new generation of short-range ISL tests was the proposal by \citet{Ark98,Ark99} that the hierarchy problem could be resolved if there existed relatively large (sub-mm scale) extra compact spatial dimensions in which gravitons could propagate but Standard Model particles could not. In this scenario, $n$ extra dimensions beyond the ordinary four are compactified with characteristic radius $R$ and the hierarchy problem is resolved by setting the ``true'' Planck mass $M\ts{Pl} \approx M\ts{EW}$, the electroweak scale. The observed long-range strength of gravity is a result of the dilution of the field through the extra dimensions, so from Gauss's Law the apparent ``four-dimensional'' Planck mass $M\ts{Pl}^* = \sqrt{ \hbar c / G }$ is given by
\begin{align}
\prn{ M\ts{Pl}^* }^2 \approx M\ts{Pl}^2 \prn{ \frac{R}{\ell\ts{Pl}} }^n \approx \frac{c^n R^n}{\hbar^n} M\ts{Pl}^{2+n}~,
\label{Eq:5thForces:Planck-mass-extra-dims}
\end{align}
where $\ell\ts{Pl} = \hbar/(M\ts{Pl}c)$ is the ``true'' Planck length. Setting $M\ts{Pl} \approx M\ts{EW}$, for $n=2$, $R \approx 100~{\rm \mu m}$. Although recent experiments \cite{Kap07,Bez11,Che16,Kam15,Sus11,Tan16,Yan12} and astrophysical constraints \cite{Ark99,Cul99,Bar99,Hal99} have excluded the $n=2$ possibility, scenarios with $n \geq 3$ and variations on the ideas of \citet{Ark98} involving, for example, extra dimensions with nonuniform compactification scales \cite{Lyk00} and alternative metrics for the extra dimensions \cite{Ran99}, including the possibility of infinite-sized extra dimensions \cite{Ran99b}, are still viable and provide motivation for continued tests of the ISL.

A second aspect of the conflict between general relativity and quantum field theory is the \emph{cosmological constant problem} or \emph{vacuum energy catastrophe} \cite{Wei89}. Observational evidence suggests that the accelerating expansion of the universe may be explained by a nonzero cosmological constant associated with a vacuum energy density $\rho\ts{vac} \approx 4 \times 10^3~{\rm eV/cm^3}$, the so-called \emph{dark energy}. However, rough estimates of $\rho\ts{vac}$ based on the Standard Model assuming no new physics up to the Planck scale suggest a vacuum energy density $\approx 10^{122}~{\rm eV/cm^3}$, a staggering discrepancy. The vacuum energy scale derived from cosmological observations corresponds to a length scale
\begin{align}
\ell\ts{vac} \approx \sqrt[4]{\frac{\hbar c}{\rho\ts{vac}}} \approx 100~{\rm \mu m}~.
\label{Eq:5thForces:vacuum-energy-length-scale}
\end{align}
A suggested theoretical path toward resolving the cosmological constant problem is the proposal that somehow the gravitational interaction with vacuum fluctuations ``cuts off'' at length scales $\lesssim \ell\ts{vac}$ \cite{Sun99}, indicating that one might generically expect a change in gravitational physics below $\approx 100~{\rm \mu m}$. It is suggestive that two of the most significant theoretical problems confronting quantum theories of gravity both indicate a benchmark scale of $\approx 100~{\rm \mu m}$ where a deviation from the ISL might be expected.

As noted in Sec.~\ref{Sec:5thForces:Intro}, the existence of new scalar or vector bosons could also give rise to apparent violations of the ISL due to the appearance of a new Yukawa potential between fermions. Such new bosons commonly appear in grand unification theories such as string theory \cite{Bai87} as well as in related theories involving extra dimensions such as those discussed above \cite{Ant98}, supersymmetric theories \cite{Tay90}, and many others \cite{Ade03,Ant11,Dob06}. Two specific examples from string theory are often cited as possible targets of searches: radions \cite{Bra61}, which are scalar bosons related to the radius of extra dimensions, and dilatons \cite{Arv15}, which are scalar bosons that determine the interactions between particles in string theory. Particles such as radions and dilatons are known collectively as moduli, scalar bosons whose expectation values determine key parameters in string theory \cite{Sch13}.

Another important theoretical motivation to search for new scalar bosons is the idea of \emph{quintessence}, the proposal that the accelerating expansion of the universe is a result of the potential energy of a scalar field; for reviews see \citet{Pee03,Pad03,Lin08,Fri08,Tsu13}. Furthermore, there have been a number of proposals that attempt to explain dark energy as a modification of gravity at cosmological distance scales; for a review, see \citet{Joy15}. To produce the observed accelerating expansion, the modification of gravity would correspond to a long-range scalar interaction. However, modified gravity at such large distance scales immediately confronts stringent observational tests at the solar system scale and shorter distances \cite{Wil14} and is ruled out. To avoid these observational constraints, there have been a number of proposals that the new scalar component of gravity is somehow screened within the solar system, for example via self-interactions \cite{Kho04,Oli08}, modified Newtonian dynamics [MOND; see, for example, the work of \citet{Mil83}], or other nonlinearities \cite{Vai72}. These screening mechanisms are, in turn, associated with new particles such as chameleons \cite{Kho13} and galileons \cite{Nic09} that can be searched for in laboratory experiments.

\subsection{Laboratory tests}
\label{Sec:5thForces:lab-tests}

Many experimental searches for fifth forces and tests of the ISL employ torsion pendulums, an experimental technique discussed in Sec.~\ref{Sec:ExoticSpin:MonopoleDipole} in the context of searches for spin-dependent interactions (see Fig.~\ref{Fig:spin-polarized-torsion-pendulum} and surrounding discussion). Torsion-pendulum tests of the ISL are reviewed by \citet{Ade09} and \citet{New09}; recent torsion-pendulum experiments by \citet{Kap07}, \citet{Yan12}, and \citet{Tan16} have established the most stringent constraints on $\alpha$ for $10^{-5}~{\rm m} \lesssim \lambda \lesssim 10^{-2}~{\rm m}$ [Eq.~\eqref{Eq:5thForces:alpha-lambda-parametrization}], probing the theoretically interesting region of parameter space covering up to three orders of magnitude below the nominal strength of gravity around the dark energy scale of $\ell\ts{vac} \approx 100~{\rm \mu m}$. Between 5 and 15 microns, the best constraint on a fifth force comes from measurements employing a $\approx 1~{\rm \mu g}$ test mass attached to cryogenic micro-cantilever and a source mass with alternating $100~{\rm \mu m}$-wide gold and silicon strips that is moved beneath the cantilever \cite{Ger08}.

A feature common to all recent torsion-pendulum tests of the ISL and micro-cantilever experiments is the use of a thin conducting membrane between the source and test masses that acts as an electrostatic shield. Because of the challenges related to manufacturing conducting membranes thinner than a few microns, experimental tests of the ISL below a few microns have generally had to contend with distant-dependent electromagnetic forces due to the Casimir effect \cite{Lam97} and electrostatic patch potentials \cite{Kim10casimir,Sus11a}. The Casimir effect [reviewed by \citet{Lam04}, for example] is the attraction or repulsion between objects due to modification of the electromagnetic vacuum modes in the space between the objects, which appears as an additional short-range force. Precise comparisons between Casimir effect measurements and calculations provide some of the best constraints on fifth forces for $10^{-7}~{\rm m} \lesssim \lambda \lesssim 10^{-5}~{\rm m}$ \cite{Bez11,Sus11,Mas09}. Experiments by \citet{Che16} employing a micromechanical torsional oscillator have recently probed the $4 \times 10^{-8}~{\rm m} \lesssim \lambda \lesssim 10^{-5}~{\rm m}$ range by coating the surface of an alternating density source mass with gold in order to keep the Casimir effect uniform as the position of the source mass is varied \cite{Dec05,Mat01}, improving on the Casimir effect measurement constraints on $\alpha$ by several orders of magnitude. Of note at this distance scale are continuing efforts to use ultracold atoms as force sensors near dielectric surfaces to probe short-range gravity \cite{ferrari06bloch,wolf07optical,sorrentino09quantum,pelle13state}.

At even smaller length scales, on the order of 0.01~nm to 10~nm, the best constraints on fifth forces come from experiments measuring the scattering of neutrons off of noble gas atoms \cite{Pok06,Nes08,Kam15}. Atomic and molecular spectroscopy can also produce meaningful constraints at this length scale. In particular, spectroscopy of atomic hydrogen \cite{Dah16,Wan15} and molecular hydrogen [H$_2$, HD, and D$_2$, \cite{Sal15,Gat15}] have been used in conjunction with theoretical calculations of atomic and molecular energy levels to constrain the models of gravity postulating extra dimensions discussed in Sec.~\ref{Sec:5thForces:motivation-theory}.

Another closely related class of experimental probes of gravity involve tests of the Einstein equivalence principle (EEP) that underpins general relativity. The EEP states that any local experiment (local in the sense that gravitational tidal effects may be neglected) cannot distinguish between a gravitational field and an acceleration of the laboratory. Tests of the EEP are discussed in Sec.~\ref{Sec:GR}, and include recent precise measurements of the gravitational redshift using atom interferometry by \citet{Mul10,Pol11,ZhoLonTan15} that verify the predictions of general relativity with an accuracy better than $10^{-8}$. An alternative approach to testing the EEP employing atomic spectroscopy has achieved a sensitivity matching that of atom interferometry: \citet{Hoh13} used measurements of the transition frequency between two nearly degenerate opposite-parity states of atomic dysprosium over the course of two years to constrain electron-related anomalies in gravitational redshifts at the $10^{-8}$ level.

As mentioned in Sec.~\ref{Sec:5thForces:motivation-theory}, theoretical attempts to ascribe the accelerating expansion of the universe to a long-range modification of gravity appear to require a screening mechanism in order to evade experimental limits on fifth forces. Experiments using atom interferometry have established the most stringent constraints on such theories \cite{Elder16chameleon,burrage2016constraining}. \citet{Ham15} used a Cs matter-wave interferometer near a spherical source mass in an ultra-high vacuum chamber, thereby reducing any screening mechanisms by searching for a fifth force with individual atoms rather than bulk matter (in contrast to the torsion pendulum, microcantilever, and Casimir-effect experiments discussed above).

It is notable that the types of scalar particles that would mediate fifth forces, such as dilatons \cite{Til15}, may also constitute the dark matter (in the same way that the axions and ALPs mediating spin-dependent interactions can be dark matter, as mentioned in Sec.~\ref{Sec:ExoticSpin:EmergingIdeas}). Consequently, atomic physics techniques can be employed to search for dark matter scalar bosons as discussed in detail in Sec.~\ref{Sec:LightDarkMatter}.  There are also a number of new proposals on the horizon that promise improved sensitivity to spin-independent interactions at various length scales and new ways to test the EEP and ISL: examples include experiments employing optically trapped microspheres and nanospheres \cite{Ger10,Ger15}, Bose-Einstein condensates \cite{Dim03}, novel atom interferometry experiments \cite{Hoh12}, and measurements employing trapped neutrons \cite{Abe10}. An alternative way to look for exotic interactions is to see if, for example, a mass can source a scalar field that changes fundamental constants; such an experiment can be competitive with those searching directly for new forces as surveyed in this section \cite{Lee16}.

\section{Searches for light dark matter}
\label{Sec:LightDarkMatter}

\subsection{Introduction}
\label{Sec:LightDarkMatter:Intro}

A variety of astrophysical and cosmological measurements \cite{Ber05,Fen10,Gor14} strongly suggest that over 80~\% of all matter in the Universe is invisible, nonluminous dark matter (DM).  Understanding the microscopic DM nature is one of the paramount goals of cosmology, astrophysics, and particle physics, since it will not only reveal the origins of the dominant constituent of matter in the Universe but also offer insights into the cosmology of the early Universe, uncover new physical laws, and potentially lead to the discovery of other fundamental forces.

The evidence for DM is derived from observations of DM's gravitational effects at the galactic scale and larger: galactic rotation curves \cite{Zwi33,Rub70,Rub80}, gravitational lensing \cite{Ref03,Tys98}, the Bullet Cluster \cite{Clo06} and other galactic cluster studies \cite{Lew03}, large-scale structure of the Universe \cite{All03}, supernovae surveys \cite{Rie98,Per99}, and the cosmic microwave background radiation \cite{Kom11}.  All these observations point toward the existence of DM. In order to fully elucidate the nature of DM, terrestrial experiments seek to measure non-gravitational interactions of DM with Standard Model particles and fields. However, the vast extrapolation from the $\gtrsim 10~{\rm kpc}$ distances associated with astrophysical observations to particle-physics phenomena accessible to laboratory-scale experiments leaves open a vast number of plausible theoretical possibilities worth exploring.

In order to develop an experimental strategy for terrestrial DM detection, it is useful to consider what can be surmised about the local DM environment of our solar system based on astrophysical observations. The local DM density is best estimated from the galactic rotation curve of the Milky Way, which, it turns out, is rather challenging to measure from the vantage point of our solar system. Furthermore, in the end, the galactic rotation curve only offers incomplete information on the local DM density since it is sensitive to the integrated mass density between our location and the center of the galaxy, and the mass density near the galactic center is notoriously difficult to determine. Nonetheless, based on numerical models \cite{Ber98} and observations of other similar spiral galaxies \cite{Sal03}, it is believed that the Milky Way is embedded within and rotates through a spherical DM halo.

The commonly used standard halo model predicts that the  DM energy density local to the Solar system is $\rho_\mathrm{DM} \approx
 (0.3 ~\rm{to}~ 0.4) \, \mathrm{GeV}/\mathrm{cm}^3$; this corresponds to a mass density equivalent to one hydrogen atom per a few cm$^3$. Further, in the galactic rest frame the velocity distribution of DM objects is isotropic and quasi-Maxwellian, with dispersion $v\approx 290\, {\rm km/s}$ and a cutoff above the galactic escape velocity of $v_{\rm esc} \approx 550\, {\rm km/s}$ \cite{Freese2013}.
The Milky Way rotates through the DM halo with the Sun moving at $\approx 220\, {\rm km/s}$ roughly towards the Cygnus constellation. Therefore one may think of a DM ``wind''  impinging upon the Earth, with typical relative  velocities $v_g \approx 300 \, \mathrm{km/s}\approx 10^{-3} c$. The speed of the Earth with respect to the DM halo is also seasonally modulated at a level of $\approx 10$~\% due to the Earth's orbit around the Sun. Furthermore, the prevailing view based on astrophysical observations is that the DM is cold, i.e., moving with velocities much smaller than the speed of light.

To date, experimental efforts to detect DM have largely focused on Weakly Interacting Massive Particles (WIMPs), with masses between 10 and 1000 GeV \cite{Ber05,Fen10}. Despite considerable effort, there are no conclusive signs of WIMP DM interactions, even as experimental sensitivities have improved rapidly in recent years. While the WIMP is theoretically well-motivated, it is by no means the only DM candidate. Observational limits permit the mass of DM constituents to be as low as $10^{-33}$~GeV or as high as $10^{48}$~GeV. A number of candidates inhabit this vast parameter space, ranging from ultra-light axions and axion-like particles (ALPs), which are discussed in relation to new interactions in Secs.~\ref{Sec:ExoticSpin} and \ref{Sec:5thForces}, to more complex dark sectors that lead to composite DM ``clumps.''

While particle detectors work by measuring energy deposition, precision measurement techniques are well suited for detecting candidates that act as coherent entities on the scale of individual detectors (or networks of detectors). Aided by recent advances in fields such as optical and atomic interferometry, magnetometry, and atomic clocks, several promising new experimental concepts have been recently proposed to employ these technologies to search for DM candidates with masses between $10^{-33}$~GeV and $10^{-12}$~GeV. Methods to probe ultra-heavy, composite DM candidates with astrophysical and terrestrial measurements have also emerged.

\begin{table*}[ht]
\begin{ruledtabular}
  \begin{tabular}{cccccc}
Spin & Type & Operator & Interaction & DM effects & Searches \\\hline
\rule{0ex}{3.2ex} \multirow{4}{*}{0} & \multirow{2}{*}{scalar} &  \multirow{1}{*}{$ \varphi \, h^\dagger h$, $\phi^n \, \mathcal O_{\rm SM}$} & \multirow{1}{*}{Higgs portal / dilaton} & fund.-constant variation& Atomic clocks, GPS.DM \\\cline{3-6}
%   &  &  &  & acceleration &  accelerometers, torsion balances \\\cline{5-6}
 %   &  &  &  & local gravity $g$ & gravimeters, atom interferometers  \\\cline{2-6}
\rule{0ex}{3.2ex}   & \multirow{5}{*}{pseudo-scalar} & $a \, G^{\mu\nu}\tilde{G}_{\mu\nu}$ & axion-QCD & nucleon EDM & CASPEr-Electric \\\cline{2-6}
\rule{0ex}{3.2ex}   &  & ${a}\, F^{\mu\nu}\tilde{F}_{\mu\nu}$ & axion-E\&M & EMF along $B$ field & ADMX, CULTASK, MADMAX \\\cline{3-6}
\rule{0ex}{3.2ex}   &  & $({\partial_\mu a}) \bar{\psi}\gamma^\mu \gamma_5 \psi$ & axion-fermion &  spin torque & CASPEr-Wind, GNOME, QUAX \\\hline
\rule{0ex}{3.2ex} \multirow{3}{*}{1} & \multirow{2}{*}{vector}
  % & $A_\mu' \bar{\psi} \gamma^\mu \psi$ & minimally coupled  & acceleration &accelerometers, torsion balances \\\cline{3-6}  %\: \slash\!\!\!\! A'
  & $F_{\mu\nu}' F^{\mu\nu}$ & vector--photon mixing & EMF in vacuum & DM Radio, ADMX \\\cline{3-6}
\rule{0ex}{3.2ex}   &  & $F_{\mu\nu}' \bar{\psi}\sigma^{\mu\nu} \psi $ & dipole operator & spin torque & CASPEr-Wind
   \\\cline{2-6}
\rule{0ex}{3.2ex}   & axial-vector & $A_\mu' \bar{\psi} \gamma^\mu \gamma^5 \psi$ & minimally coupled  & spin torque & CASPEr-Wind
%\\\hline
%2  & tensor & \,$h'_{\mu\nu} T^{\mu\nu} \, $ & gravity-like & gravitational wave-like &  \\
\end{tabular}
  \end{ruledtabular}
\caption{ Current experimental efforts in searches for bosonic ultralight dark matter candidates. The Table lists
illustrative couplings of bosonic ultralight dark matter candidates [scalar $\varphi$, axion $a$ and  dark photon $A_\mu '$] to SM fields, and their experimental effects.
$h$, $G^{\mu\nu}$, $F^{\mu\nu}$, and $\psi$ represent respectively SM Higgs, gluon, photon, and fermion fields, or operators of that form.
$\mathcal O_{\rm SM}$ stand for terms from the SM Lagrangian density. $n=1,2$ for linear/quadratic couplings. Free fields cause signal oscillations at the field Compton frequency and self-interacting fields forming DM ``clumps'' can cause transient effects. Specific experiments are discussed in Sec.~\ref{Sec:LightDarkMatter:Searches}. The table is not exhaustive, as for example, the GPS.DM and GNOME experiments could be sensitive to monopole topological defects which require vector fields.
Modified table from \citet{Gra16b}.
}
\label{Table:VULFcandidates}
\end{table*}

The key idea behind these concepts is that light DM particles have a large  mode occupation numbers and their phenomenology is described by a classical field. For this mass range the DM candidates are necessarily bosonic: non-interacting fermionic candidates would require larger masses to reproduce the standard halo model velocity distribution [if the mass of the DM particle is smaller than $\approx 10~{\rm eV}$ then the corresponding Fermi velocity exceeds the galactic escape velocity, see \citet{Derevianko2016a}]. The lower mass limit of $10^{-24}~\mathrm{eV}$ comes from requiring that the de Broglie wavelength is smaller than the size of galaxies where gravitational signatures of DM have been observed. While such classical fields may arise in a wide variety of DM models, their effects on Standard Model particles include a finite number of possibilities (see Table~\ref{Table:VULFcandidates}): the classical field can cause precession of nuclear/electron spins, drive currents in electromagnetic systems, and induce equivalence-principle-violating accelerations of matter~\cite{Gra13}. They could also modulate the values of the fundamental ``constants'' of nature, which can induce changes in atomic transition frequencies~\cite{DerPos14,Arv15} and local gravitational field~\cite{GeraciDerevianko2016-DM.AI}.
Some of these phenomena have been searched for in other contexts described throughout this review (see Secs.~\ref{Sec:FC},\ref{Sec:ExoticSpin},\ref{Sec:5thForces},\ref{Sec:LV}, and \ref{Sec:GR}). Here we examine the particular characteristics of effects induced by light DM fields and how precision measurement techniques such as nuclear magnetic resonance (NMR), atomic and SQUID (Superconducting QUantum Interference Device) magnetometry, electromagnetic resonators, atomic/optical interferometers, and atomic clocks can be used to search for these effects. When the mass of the particle constituting the DM is sufficiently light, the classical DM field leads to persistent time-varying signals that are localized in frequency at the DM mass, enabling rejection of technical noise while permitting signal amplification through resonant schemes. The classical fields sourced by ``clumpy'' DM could cause  transient signals that can be observed by correlating output from multiple synchronized detectors.

The entire field of laboratory cosmology, where table-top-scale precision measurement experiments search for terrestrial signatures of effects related to light DM, has emerged as a vibrant research area over the last few years with a number of promising new proposals joining several ongoing experiments. As noted above, based purely on the known properties of DM, the range of parameter space to be explored is  vast. However, experiments can be guided by clues from other fields of physics suggesting mysteries that can be solved by postulating, for example, new DM candidates with particular properties --- this is what distinguishes the most theoretically well-motivated light DM candidates (by the Occam's razor principle).

%\subsection{}
Among the most well-motivated light DM candidates is the quantum chromodynamic (QCD) axion,  discussed in Sec.~\ref{Sec:ExoticSpin:EarlyWork:axions-ALPs}; experimental axion searches were recently reviewed, for example, by~\citet{Gra15review}. Axions can naturally constitute a significant fraction of DM: for example, they can be produced in sufficient abundance in the early universe via the so-called vacuum realignment process \cite{Abb83,Pre83,Din83}. This process results from a misalignment between the axion field generated when axions are first produced via spontaneous symmetry breaking and the minimum of the potential generated by QCD interactions. Since the axion field is initially perturbed from the QCD potential minimum, it will oscillate; these oscillations are not significantly damped over the age of the universe and in fact in most models it is the energy stored in these coherent oscillations of the axion field that constitute the mass-energy ascribed to DM \cite{Pre83,Din83,Duf09}. Similar scenarios describe the production of most light bosons. Another axion production mechanism is through the decay of topological defects such as domain walls or strings \cite{Dav85,Cha98,Nag94}, where the topological defects interpolate between regions of space with different axion vacuum fields which can exist, for example, due to nontrivial vacuum structure (i.e., multiple equivalent local minima in the self-interaction potential).

QCD axions couple to photons, gluons, and fermion spins over a predictable range of axion mass-coupling strength parameter space~\cite{Pre83,Abb83,Din83}. There are three possible interactions of axions with Standard Model particles and fields: axions can couple to electromagnetism, induce electric dipole moments (EDMs) for nucleons (see Sec.~\ref{Sec:EDM}) via interaction with the gluon field, and can cause precession of electron and nucleon spins (see Table~\ref{Table:VULFcandidates}).

There are robust astrophysical constraints on QCD axions with masses $\gtrsim 10$~meV based on the observation of the neutrino signal from supernova 1987A and star cooling~\cite{Raf99}. Heavier axions would have produced observable effects in astrophysical objects, and much heavier axions would already have been seen in terrestrial experiments. Constraints have also been considered for QCD axions with masses $\lesssim 1~{\rm \mu eV}$ based on cosmology. However, these constraints depend upon assumptions about unknown initial conditions of the universe. Such lighter-mass QCD axions were never ruled out either by experimental or astrophysical observations, but theory prejudice held that they were less likely based on cosmology. It has now been realized that this was based on a particular scenario for the earliest epochs in the universe, a time about which we know little. Since the inception of this cosmological argument against lower-mass QCD axions, inflation has become the dominant paradigm for the birth of the universe. This (along with other factors) led to alternative possibilities for axion production in the early universe. As several authors have pointed out, these allow a much larger mass range for the QCD axion, and in fact bestows the lighter axions with a strong theoretical motivation \cite{Fre10,Lin88}.

Going beyond the QCD  axions, SM extensions offer a plenitude of  ultralight DM candidates.  We  collectively refer to such candidates  as  virialized ultralight fields (VULFs). Possibilities are compiled in Table~\ref{Table:VULFcandidates}, where the fields are characterized  by their spin and intrinsic parity. As noted above, in the considered mass range ($< 10\, \mathrm{eV}$) the DM candidates are necessarily bosonic.
 In particular, spin-1 particles, commonly referred to as dark or hidden photons (see Sec.~\ref{Sec:ExoticSpin:TheoreticalMotivation:hidden-photons-Z-bosons}) could  conceivably constitute a substantial fraction of the DM \cite{Nel11darklight,arias2012wispy,Foo15dissipative}.  Table~\ref{Table:VULFcandidates} also indicates various potential  DM couplings to SM fields. More broadly, the possible {\em non-gravitational} couplings can be classified according to  ``portals'' that correspond to different  gauge invariant operators of the
SM fields coupling to  operators that contain fields from the dark sector.  This phenomenological approach is widely used in particle physics for searches of DM and dark forces~\cite{Essig:2013lka}. For example for scalar DM fields $\varphi$, the following portals may arise~\cite{DerPos14}:
\begin{eqnarray*}
{\cal L}_1 = \frac{\partial_\mu \varphi}{\Lambda} \sum_{\rm SM~particles } c_\psi \bar\psi \gamma_\mu\gamma_5 \psi  ~~~~&&{\rm axionic~portal},
\\
{\cal L}_2 = \frac{ \varphi}{\Lambda} \sum_{\rm SM~particles } c^{(s)}_\psi m_\psi \bar\psi  \psi  ~~~~&&{\rm linear~scalar},
\\
{\cal L}_3 = \frac{ \varphi^2}{\Lambda^2} \sum_{\rm SM~particles }
 c^{(2s)}_\psi m_\psi \bar\psi  \psi  ~~~~&&{\rm quadratic ~ scalar},
\\
{\cal L}_4 = \frac{i\varphi^* \partial_\mu \varphi }{\Lambda^2 } \sum_{\rm SM~particles } g_\psi \bar\psi \gamma_\mu \psi
~~~~&&{\rm current-current} \,.
\end{eqnarray*}
Here $\psi$ are SM fermion fields with associated masses $m_\psi$, $\Lambda$ are the energy scales, and
$c_i$ are individual coefficients that can take on different values depending on type of $\psi$. This classification can be generalized to include the
SM gauge bosons, for example of the form $\varphi \,F_{\mu\nu}F^{\mu \nu}$ for electromagnetism and extended further to non-scalar DM fields.

While the phenomenological portal classification is broad, one should be aware of certain existing astrophysical and laboratory constraints on the coupling strengths or energy scales $\Lambda$. For example, the hypothesized DM fields can mediate forces and thus the limits from fifth-force experiments  (Sec.~\ref{Sec:5thForces}) immediately apply. Thereby an experiment searching for DM signatures through a specific portal must probe yet unexplored parameter space. In some cases, the broader search may soften such constraints. For example, the discussed bounds on the QCD axion are relaxed for ALPs \cite{Mas05}. ALPs are pseudo-scalar particles similar in nature to the QCD axion  that do not solve the strong $CP$ problem, but rather emerge naturally from other frameworks such as string theory. ALPs may also have the properties necessary to solve the hierarchy problem, as discussed in Sec.~\ref{Sec:ExoticSpin:TheoreticalMotivation}.

%ALPs in particular can have a wide range of coupling strengths to Standard Model particles and fields, and thus it is important to search for such particles over a wide range of mass/frequency parameter space using detection schemes based on all the different possible couplings.

As mentioned earlier, a distinct theoretical possibility is that  DM is not distributed uniformly but rather occurs in the form of ``clumps.'' Even the ever-present gravitational interaction leads to instabilities and clumping.  Examples of  ``clumpy'' objects include ``dark stars'' \cite{Kol93,Bar11,Bra16,Iwa15},  $Q$-balls~\cite{Coleman1985,Kusenko2001}, solitons, and clumps formed due to dissipative interactions in the DM sector.  Alternatively, a significant fraction of the DM mass-energy could be stored in ``topological defects'' manifesting as monopoles, strings, or domain walls \cite{Vil85}. If DM takes such a form, terrestrial detectors would not register a continuous oscillating  signal associated with VULFs, but rather would observe  transient events associated with the passage of the detector through such a DM object~\cite{Pos13,Bud15,DerPos14}. Self-interacting fields  can include bosonic and fermionic DM candidates.
The characteristic spatial extent of topological defects is determined by the Compton wavelength of the underlying DM field. For an Earth-sized object, the characteristic mass is $\approx 10^{-14} \, \mathrm{eV}$ which places such DM fields in the category of ultralight candidates.

\subsection{Experimental Searches}
\label{Sec:LightDarkMatter:Searches}

Axion and ALP searches can be classified in different categories depending on where the detected particles are produced. For example, in ``light shining through walls'' (LSW) experiments \cite{Rob07,Red11}, axions are created in the experiment from an intense laser light field and a static field of a strong magnet which facilitates mixing between photons and axions. These axions are then detected by converting them back to photons after they cross a wall that is transparent to them but completely blocks the light. In ``helioscope''  experiments~\cite{Raf99,Gra15review}, the task of producing axions or ALPs is ``subcontracted'' to the Sun (hence the name), but detection is accomplished as in LSW experiments. Finally, ``haloscopes'' {\emph{directly}} detect the DM from the galactic halo. In a somewhat complementary approach, {\emph{indirect}} experiments search for modifications of the known interactions via exchange of virtual exotic particles. Such experiments include the ``fifth-force'' searches and experiments looking for exotic spin-dependent interactions or modification of fundamental constants in the presence of massive and/or spin-polarized objects that presumably act as sources of the virtual exotic particles.
We discuss some examples of direct experimental searches of different kinds below, while indirect searches are discussed in Secs. \ref{Sec:ExoticSpin} and \ref{Sec:5thForces}.
The direct detection of ``clumpy'' DM objects requires networks of precision measurement tools, and we discuss here two ongoing searches with  magnetometers and atomic clocks.

\subsubsection{Microwave cavity axion experiments}
\label{Sec:LightDarkMatter:Searches:cavities}

\begin{figure}
\includegraphics[width=3.35 in]{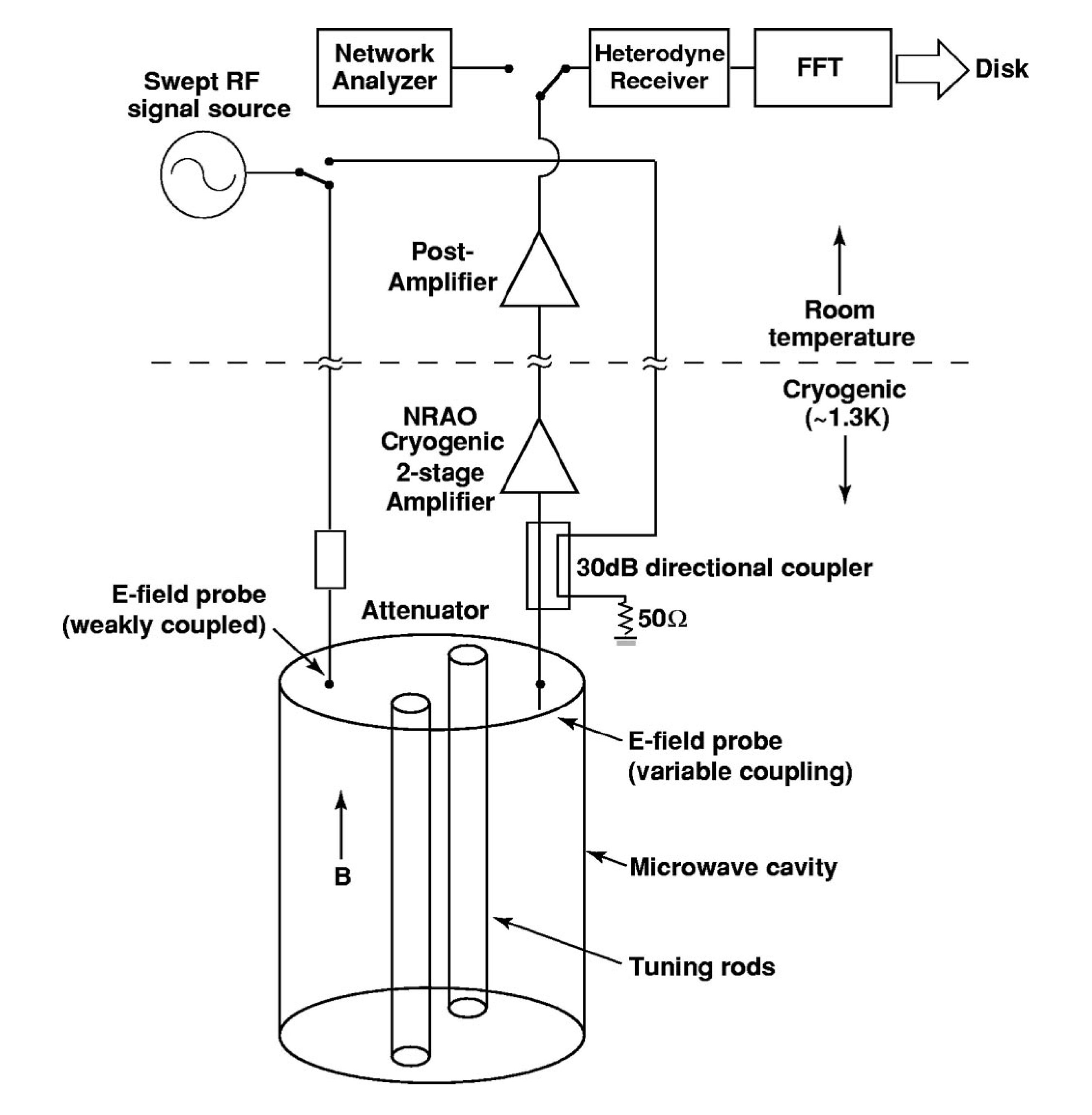}
\caption{Schematic diagram of the ADMX experiment from \citet{Asz01}. Photons produced in the microwave cavity by the interaction of an axion field $a({\mb{r},t})$ with the magnetic field $\mc{B}$ [Eq.~\eqref{Eq:axion-photon-Lagrangian}] are detected by the electric-field probes. Tuning rods enable the resonant frequency of the cavity to be scanned to search for axions of different masses. (Fields from the RF signal source can be sent through the setup for calibration purposes.) The signals are recorded after multiple amplification stages and heterodyning. The 2001 experiment employed cryogenic heterojunction field-effect transistors built by the National Radio Astronomy Observatory (NRAO), while new versions of ADMX employ SQUID amplifiers \cite{Asz10}.}
\label{Fig:ADMX-expt}
\end{figure}

Microwave cavity searches for dark-matter axions were reviewed by \citet{Bra03}. The first experiment to search for light DM composed of QCD axions was the Axion DM eXperiment (ADMX), which began its work in the 1990s \cite{Asz01,Asz10}. This experiment exploits the coupling of the QCD axion to the electromagnetic field to convert axions into microwave photons in a strong magnetic field $\mc{B}$ (Fig.~\ref{Fig:ADMX-expt}). In general, pseudoscalar particles such as axions and ALPs can be produced by the interaction of two photons via a process known as the Primakoff effect \cite{Pri51}, and consequently an axion/ALP interacting with an electromagnetic field can produce a photon via the inverse Primakoff effect \cite{Raf88}. This process was proposed by \citet{Sik83,Sik85} as a method to search for DM axions (haloscope experiment) as well as axions emitted by the Sun (helioscope experiment). The nonrelativistic Lagrangian describing this interaction is
\begin{align}
\sL_{a\gamma\gamma} = g_\gamma \frac{\alpha}{\pi} \frac{a({\mb{r},t})}{f_a} \mc{E}\cdot\mc{B}~,
\label{Eq:axion-photon-Lagrangian}
\end{align}
where $g_\gamma$ is the coupling constant describing the strength of the axion-photon interaction, $\alpha$ is the fine structure constant, $a({\mb{r},t})$ is the axion field, $f_a$ is the symmetry breaking scale associated with the axion (see Sec.~\ref{Sec:ExoticSpin:Parametrization}), and $\mc{E}$ and $\mc{B}$ are the electric and magnetic fields. This interaction corresponds to the axion-E\&M entry in Table~\ref{Table:VULFcandidates}.
In the ADMX experiment, $\mc{B}$ is generated with a superconducting solenoid and $\mc{E}$ is the electric field of the resultant microwave photon produced by the inverse Primakoff effect. The resonant frequency of the cavity can be tuned so that it matches the frequency of the microwave photons produced by the interaction of $a({\mb{r},t})$ with $\mc{B}$, which have energy corresponding to
\begin{align}
E_\gamma \approx m_a c^2 + \frac{1}{2} m_a c^2\beta^2~,
\end{align}
where $m_a$ is the axion mass and $\beta = v/c \approx 10^{-3}$ is the relative velocity of the laboratory with respect to the rest frame of the axion field. As noted in the introduction to this section, the dispersion of the axion velocities is roughly on the same order as $\beta$, i.e., $\Delta \beta \approx 10^{-3}$, so the axionic DM would produce a relatively narrow microwave resonance:
\begin{align}
\frac{\Delta \omega}{\omega} \sim \prn{\Delta \beta}^2 \sim 10^{-6}~.
\end{align}

%----------------------------------------------------------------
%----------------------------------------------------------------

ADMX is to date the first and only experiment to probe the particularly interesting region of parameter space corresponding to standard QCD axion models, namely the Kim-Shifman-Vainshtein-Zakharov (KSVZ) and Dine-Fischler-Srednicki-Zhitnitskii (DFSZ) family of models \cite{Kim79,Shi80,Din81,Zhi80}, in a band of axion masses near $\approx 2\times 10^{-6}~{\rm eV}$ to
 $\approx 4 \times 10^{-6}~{\rm eV}$. A new effort to extend the ADMX experiment to search for higher mass axions using correspondingly higher frequency microwave cavities, known as HAYSTAC -- Haloscope At Yale Sensitive To Axion Cold dark matter \cite{Bib13}, has recently produced its first results \cite{Bru16}. HAYSTAC was able to probe higher axion masses with improved sensitivity by pushing to lower temperatures and leveraging recent progress in quantum electronics; HAYSTAC has probed the KSVZ parameter space in a band of axion masses near $\approx 24 \times 10^{-6}~{\rm eV}$ \cite{Bru16}.  The ADMX and HAYSTAC collaborations plan a systematic search for QCD axions with masses between $m_ac^2 \approx 10^{-6}~{\rm eV}$ and $\approx 10^{-4}~{\rm eV}$ by 2021.

Another significant microwave cavity experimental program is underway at the Center for Axion and Precision Physics Research (CAPP) at KAIST in South Korea \cite{Sem17,You16}, where researchers are developing stronger magnets, new low-noise amplifiers (e.g., based on Josephson parametric amplifiers), and superconducting cavities with novel designs to increase their Q and expand their volume. The CAPP haloscope, known as CULTASK (CAPP's Ultra Low Temperature Axion Search in Korea), aims to target an axion mass range near $\approx 10^{-5}~{\rm eV}$.

A new broadband axion DM haloscope experiment aimed at detecting axions with $m_a \approx 10^{-4}~{\rm eV}$ proposed by \citet{Jae13}  is under development at the Max Planck Institute for Physics \cite{Maj16}. This project, named MADMAX, is based on axion-photon conversion at the transition between different dielectric media. By using $\approx 80$ dielectric discs immersed in a $\approx 10~{\rm T}$ magnetic field, the emitted power is enhanced by a factor of $\approx 10^{5}$ over that from a single mirror (flat dish antenna).

\subsubsection{Spin-precession axion experiments}
\label{Sec:LightDarkMatter:Searches:cavities1}

A new experiment recently proposed by \citet{Bud14} to search for lighter QCD axions and ALPs using different couplings from those exploited in ADMX and similar microwave cavity experiments is the Cosmic Axion Spin Precession Experiment (CASPEr). CASPEr exploits both the axion-gluon coupling, which generates a time-varying electric dipole moment (EDM) of nuclei\footnote{As noted by \citet{Sch63}, the interaction of an EDM of a point-like particle with an applied electrostatic field is screened in atomic systems, since the constituent charged particles redistribute themselves to cancel the field. However, the screening is incomplete because of finite nuclear size and relativistic effects, which can even enhance the atomic EDM relative to the electron or nuclear EDM in heavy atoms (see Sec.~\ref{Sec:EDM}).} (CASPEr Electric), and the coupling of the axion to nuclear spins (CASPEr Wind), see \citet{Gra13}. CASPEr uses nuclear magnetic resonance (NMR) techniques for detecting spin precession caused by background axion DM. This approach complements ADMX,  HAYSTAC, and CULTASK which are sensitive to higher axion masses and a different coupling.

%----------------------------------------------------------------
\begin{figure}
\includegraphics[width=2.8 in]{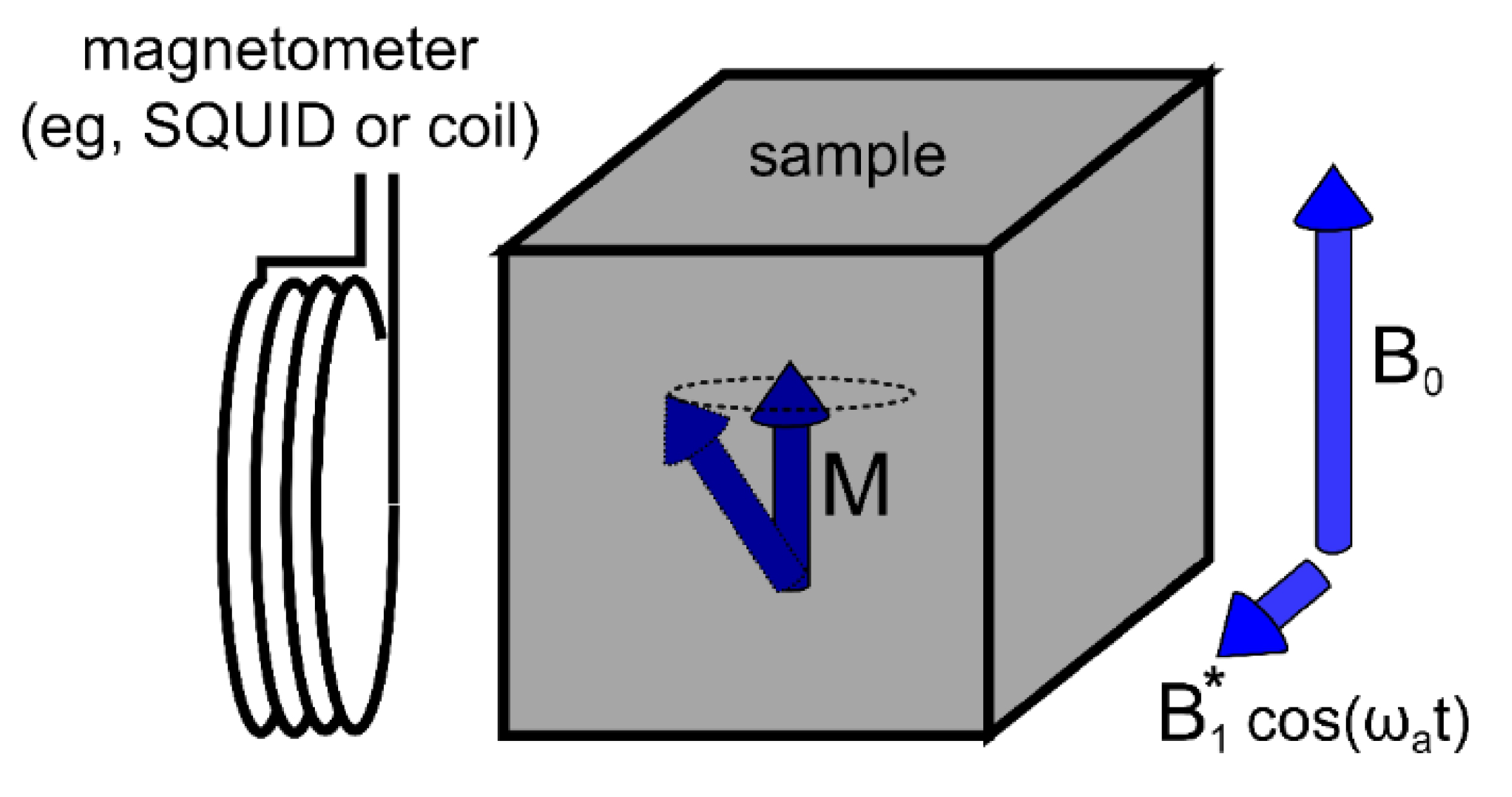}
\caption{Experimental concept of CASPEr \cite{Bud14}. The oscillating axion field $a({\mb{r},t})$ acts as a pseudo-magnetic field $\mc{B}_1^*$, either by inducing an oscillating electric dipole moment (EDM) via the axion-gluon interaction that couples to an electric field (CASPEr Electric), or via the interaction of spins with the gradient of $a({\mb{r},t})$ arising from the motion of the sample through the axion field (CASPEr Wind). The oscillating $\mc{B}_1^*$ causes polarized nuclear spins to tip away from the leading field $\mc{B}_0$ and precess at the Larmor frequency. The approach is based on the principles of NMR experiments.}
\label{Fig:CASPEr-expt-concept}
\end{figure}
%----------------------------------------------------------------

The key idea underlying the CASPEr concept is that axion DM can cause the precession of nuclear spins in a manner similar to that discussed for exotic spin-dependent interactions and EDMs (Secs.~\ref{Sec:ExoticSpin} and \ref{Sec:EDM}, respectively). Nuclear spins in a non-centrosymmetric crystal, such as a ferroelectric, experience a large effective electric field \cite{Leg78,Muk05}, a phenomenon analogous to the large internal effective electric fields experienced by electrons in polar molecules \cite{Gra11}. The coupling of the axion DM field to nuclear spins (via the generation of electric dipole moments through the axion-gluon coupling) in such a material has the form of a pseudo-magnetic field $\mc{B}_1^*$ oscillating at the axion Compton frequency. If the external bias magnetic field $\mc{B}_0$ is set to a value such that the nuclear spin splitting matches this frequency, a resonance condition is achieved, and the nuclear spins and the corresponding magnetization $\mb{M}$ tilt and undergo Larmor precession (see Fig.~\ref{Fig:CASPEr-expt-concept}). A sensitive magnetometer, such as a Superconducting Quantum Interference Device (SQUID), placed next to the sample, detects the oscillating transverse magnetization. The experimental protocol of CASPEr-Electric is to sweep the externally-applied bias magnetic field and search for a non-zero magnetometer response, which is a signature of spin coupling to the axion DM field. CASPEr Electric has the potential to reach sensitivity to QCD axions over a mass range of $10^{-14}~{\rm eV} \lesssim m_a c^2 \lesssim 10^{-9}~{\rm eV}$ and search a significant fraction of unexplored parameter space for ALPs up to masses of $\approx 10^{-7}~{\rm eV}$ \cite{Bud14,kimball2017overview}.

CASPEr Wind is an example of an experiment specifically sensitive to ALP DM (at least in its present form, it will not have sufficient sensitivity to reach parameter space corresponding to the QCD axion). CASPEr Wind is analogous to CASPEr Electric, except that the pseudo-magnetic field $\mc{B}_1^*$ is generated by a different mechanism: the coupling of nuclear spins to the spatial gradient of the ALP DM field (the so-called ``ALP wind''). This enables the use of materials such as liquid xenon without electric fields. Xenon can be efficiently spin-polarized to enhance the signal. A variety of experimental approaches \cite{kimball2017overview}, including the use of atomic magnetometers \cite{graham2017spin} and zero-to-ultralow-field (ZULF) NMR \cite{garcon2018CASPEr}, have been proposed as methods to search for the ALP wind. \citet{abel2017search} have used the CASPEr Wind approach to analyze data from a search for the neutron EDM to constrain ALP DM with $10^{-24}~{\rm eV} \lesssim m_a c^2 \lesssim 10^{-17}~{\rm eV}$.

In the KSVZ family of QCD axion models \cite{Kim79,Shi80}, the coupling of the axion to electron spins is nominally zero, whereas in the DFSZ family of models \cite{Din81,Zhi80}, the axion is predicted to couple to the electron spin. Thus, in addition to searches for axion couplings to nuclear spins as searched for in CASPEr, it is of interest to search for axion-electron couplings: this is the target of the QUAX (QUaerere AXion) experiment \cite{Ruo16}. The essence of the experiment, originally outlined by \citet{Kra85,Bar89,Kak91,Tur90}, is quite similar to that of CASPEr, with the important difference that in the QUAX experiment a Yttrium Iron Garnet (YIG) sphere is used as the sample of polarized electron spins as opposed to the polarized nuclear samples studied in CASPEr.

\subsubsection{Radio axion searches}
\label{Sec:LightDarkMatter:Searches:radio}

ADMX, CASPEr, and related experiments are also sensitive to another class of particles known as dark or hidden photons \cite{Wag10}, discussed in Sec.~\ref{Sec:ExoticSpin:TheoreticalMotivation:hidden-photons-Z-bosons}. Like ordinary photons, hidden photons are vector particles with spin 1. However, hidden photons have mass and could constitute the DM in a manner similar to axions and ALPs \cite{Ark09}. Hidden-photon DM can be described as a weakly coupled ``hidden electric field,'' oscillating at the hidden-photon Compton frequency, and able to penetrate shielding \cite{Kim16}. At low frequencies (where the wavelength is long compared to the size of the shielding), the interaction of electrons in the shielding material with the hidden-photon field generates a real, oscillating magnetic field. It has recently been proposed that such hidden-photon DM can be searched for using a tunable, resonant LC circuit designed to couple to this magnetic field, a ``dark matter  radio'' \cite{Cha15}. Hidden-photon DM has an enormous range of possible Compton frequencies, but current experiments (such as ADMX, which is also sensitive to hidden photons) search only over a few narrow parts of that range \cite{Wag10}. In contrast, DM Radio has potential sensitivity many orders of magnitude beyond current limits over an extensive range of hidden photon masses, from $10^{-12}~{\rm eV} \lesssim m_{\gamma'}c^2 \lesssim 10^{-3}~{\rm eV}$, where $m_{\gamma'}$ is the hidden photon mass.

Related proposals for broadband axion/ALP detection with LC circuits were developed by \citet{Sik14} and \citet{Kah16}. The concept of these experiments can be understood by noting that the axion-photon coupling effectively modifies Maxwell's equations \cite{Sik83,Sik85} such that dark-matter axions/ALPs generate an oscillating current density in the presence of a magnetic field. These ideas also apply to the high-frequency (10 to 100)~GHz axion search proposed by the MADMAX collaboration~\cite{Jae13}.

\subsubsection{Atomic clocks and accelerometers, and spectroscopy}
\label{Sec:LightDarkMatter:Searches:atomic-clocks-spectroscopy}

As noted in the introduction to this section, one of the  generic signals VULFs can produce are time-oscillating interactions. An example is DM consisting of dilatons, ultralight scalar particles arising in string theories \cite{Arv15}. Like axions and ALPs, dilatons form a  gas described as a scalar field oscillating at the Compton frequency of the dilaton. This field feebly interacts with normal matter leading to temporal variation of fundamental ``constants'' which in turn affects the ``ticking'' rates of atomic clocks. Since clocks based on distinct atomic transitions have different sensitivities to a change of constants such as the fine structure constant $\alpha$, comparisons between such clocks is a sensitive way to detect the variation of the constants (see Sec.~\ref{Sec:FC}), including those caused by a time-varying DM field. A ``differential atomic clock'' based on microwave transitions between nearly-degenerate metastable states in dysprosium was used by \citet{Til15} to search for dilatons in the mass range of 10$^{-24}$ to 10$^{-16}\ $eV, improving existing constraint on the electron coupling of a DM dilaton by up to four orders of magnitude. These limits were further improved by \citet{Hee16}. Modern atomic clocks based on single trapped ions \cite{HunSanLip16} and ensembles of neutral atoms in optical lattices \cite{Nem16} are reaching into relative frequency instability levels of a part in $10^{18}$, promising a boost in the sensitivity of dilaton searches by about two orders of magnitude in the near future.

Also of note is that recently \citet{Gra16b} proposed using using accelerometers (e.g., torsion balances and atom interferometers) to search for  DM-induced forces, $-\nabla[M_a(\mathbf{r},t) c^2]$, where $M_a$ is the DM-renormalized atomic mass.
For atomic interferometers, the effects of  atomic mass variation during the interferometric sequence and also DM-induced renormalization of the local gravity $g$ dominate over the direct DM-induced forces~\cite{GeraciDerevianko2016-DM.AI}.
Accelerometers are particularly sensitive to vector and scalar VULFs.

\citet{arvanitaki2017resonant} have proposed an entirely new spectroscopic scheme for detection of bosonic DM with boson masses between 0.2~eV and 20 eV: a search for resonant transitions between states in polyatomic molecules driven by the oscillating DM field.

\begin{figure}[h]
\begin{center}
\includegraphics*[width=3.2in]{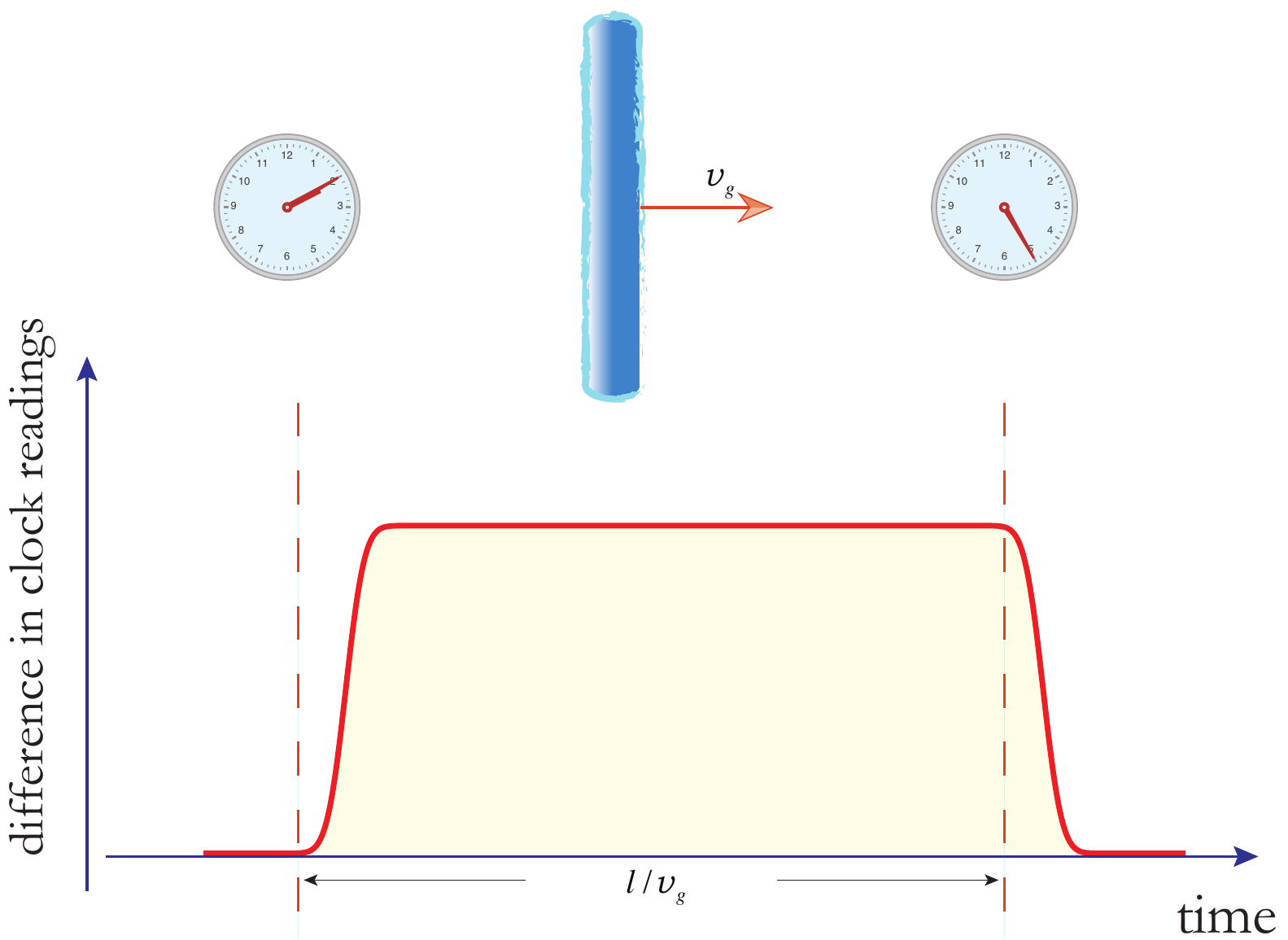}
\end{center}
\caption
{Spatially-separated and initially-synchronized identical clocks are expected to exhibit a distinct de-synchronization and re-synchronization pattern due to an encounter with a DM object. Two  clocks are separated by a distance $\ell$, and because the wall propagates through the network with a speed $v_g\approx 300~{\rm km/s}$,
the characteristic ``hump'' persists  over time $\ell/v_g$. Adopted from~\citet{DerPos14}.
\label{Fig:Clock-wall}}
\end{figure}

\subsubsection{Exotic spin-dependent forces due to axions/ALPs}
\label{Sec:LightDarkMatter:Searches:exotic-forces}

Section~\ref{Sec:ExoticSpin} explores the exotic spin-dependent interactions generated by axions, ALPs, and dark/hidden photons. A recent proposal by \citet{Arv14} to search for short-range monopole-dipole interactions between nuclei using NMR techniques, the Axion Resonant InterAction Detection Experiment (ARIADNE), has particular relevance to axion DM searches. The aim of ARIADNE is to detect monopole-dipole interactions between the spins of $^3$He nuclei and a rotating unpolarized tungsten attractor. The geometry of the experiment is specially designed to be sensitive to QCD axions in the range $10^{-6}~{\rm eV} \lesssim m_a c^2 \lesssim 10^{-3}~{\rm eV}$ \cite{geraci2017progress}. The upper end of the axion mass range to be explored by ARIADNE, well within the astrophysically and cosmologically allowed region, is particularly difficult for DM detection experiments such as ADMX and CASPEr to access, and so ARIADNE has the potential to fill in an important gap in the explored axion parameter space.

\subsubsection{Magnetometer and clock networks for detection of transient dark matter signals}
\label{Sec:LightDarkMatter:Searches:transients}

%A detection of the QCD axion or other VULF candidates
%in an experiment such as ADMX, CASPEr, or DM Radio would not only constitute the discovery of the nature of DM but would also provide insights into the high-energy scales $f_a$ from which the axion/ALP arises, possibly near the fundamental scales of particle physics such as the scale of grand unification and the Planck scale.
If a detection of the QCD axion or other VULF candidates  is made, a network of such experiments can be used to verify it, since the signal in all of them should be centered at the axion/ALP Compton frequency, a fundamental constant. A network would also enable the study of spatial coherence  of the  DM field~\cite{Derevianko2016a} and search for deviations from the standard halo model predictions due to non-uniform/non-isotropic DM flows~\cite{Duffy2008}. Networks of sensors are crucial in order to search for ``clumpy'' DM. Here the searches rely on the characteristic time delay of DM-induced signals between the nodes (see Fig.~\ref{Fig:Clock-wall}), as on average the ``clumps'' would sweep the network at galactic velocities.
%Moreover, as long as the nodes are within the VULF correlation length, SNR is proportional to the  number of nodes, $N_d$, while for uncorrelated devices SNR $\propto \sqrt{N_d}$. The correlation length is in the order of the de Broglie wavelength for DM fields~\cite{Derevianko2016a}.

%----------------------------------------------------------------
\begin{figure}
\includegraphics[width=2.8 in]{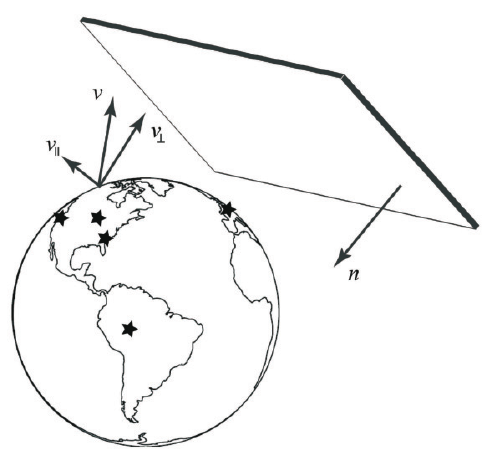}
\caption{Schematic figure of an axion/ALP domain-wall crossing event as searched for by the GNOME; figure from \citet{Pos13}. The crossings recorded in four distinct locations (marked with stars) allow determination of the normal velocity $v_\perp$ to the wall and prediction of the timing of the 5$^{th}$ event.}
\label{Fig:domain-wall-crossing}
\end{figure}
%----------------------------------------------------------------

The Global Network of Optical Magnetometers to search for Exotic physics (GNOME) collaboration \cite{Pos13,Pus13} is searching for such transient signals due to passage of the Earth through compact DM objects, such as DM domain walls \cite{Pos13} or DM ``stars'' \cite{kimball2017axionstars}, that couple to atomic spins (similar to the ALP wind coupling searched for by CASPEr). While a single magnetometer system could detect such transient events, it would be exceedingly difficult to confidently distinguish a true signal generated by light DM from ``false positives'' induced by occasional abrupt changes of magnetometer operational conditions (e.g., magnetic-field spikes, laser-light-mode jumps, electronic noise, etc.). Effective vetoing of false positive events requires an array of magnetometers. Furthermore, there are key benefits in terms of noise suppression and event characterization to widely distributing the magnetometers geographically; see Fig.~\ref{Fig:domain-wall-crossing}. The Laser Interferometer Gravitational Wave Observatory (LIGO) collaboration has developed sophisticated data analysis techniques to search for similar correlated ``burst'' signals from a worldwide network of gravitational wave detectors \cite{And01,All12}, and the GNOME collaboration has demonstrated that these data analysis techniques can be adapted to analyze data from the GNOME \cite{Pus13}. Presently the GNOME consists of over 10 dedicated atomic magnetometers located at stations throughout the world.

%\begin{figure*}
%\includegraphics[width=6in]{clock-glitch.pdf}
%\caption{Spatially-separated and initially-synchronized identical clocks are expected to exhibit a distinct de-synchronization and re-synchronization pattern due to an encounter with a DM object.}
%\label{Fig:clock-glitch}
%\end{figure*}

 If DM leads to variation of fundamental constants, DM ``clumps'' can also manifest themselves as ``glitches'' of atomic clocks, for example those onboard  satellites of Global Positioning System \cite{DerPos14}: if particular interactions exist, the clocks would become desynchronized as they are swept by a DM object (Fig.~\ref{Fig:Clock-wall}). The glitches would propagate through the GPS constellation at galactic velocities, $\approx 300~{\rm km/s}$, characteristic of DM halo. The GPS.DM collaboration is mining over a decade of archival GPS data to hunt for such DM objects, effectively using the GPS constellation as a 50\,000 km-aperture DM detector. While the initial search has not found evidence for such DM objects at the current search sensitivity, it improves the current constraints on certain DM couplings by several orders of magnitude~\cite{roberts2017first}. Recently, it has  been also shown that a single optical atomic clock (composed of two independent clock ensembles probed by the same laser) can be sensitive to transient DM interactions  and constraints on scalar quadratic DM-clock couplings have been obtained by \citet{Wci16}. While this co-located clock technique can be used to place limits on DM-clock couplings, the positive DM detection still requires a geographically-distributed network. Transient variations of fundamental constants can be also searched for with a global network of laser interferometers~\cite{StaFla2015-LaserInterferometry,Stadnik2016-laserinterferometersPRA}.

\section{General relativity and gravitation}
\label{Sec:GR}

\subsection{Tests of the Einstein equivalence principle}

The equivalence principle can be traced back to the sixteenth-century observation that all bodies fall to Earth at the same rate of acceleration \cite{Wil14}.  This was a remarkable discovery, for it leads to the conclusion that a body's mass is proportional to its weight.  The constant of proportionality seems to be independent of material composition or any other detail of the body. That is the basic principle of  the equivalence of gravitational mass and inertial mass.

Within the framework of Einstein's theory of general relativity (GR), there is the Einstein equivalence principle (EEP), which includes the following postulates  \cite{Wil14}:

\begin{enumerate}
\item The weak equivalence principle (WEP): the trajectory of a freely
falling ``test'' body  is independent of its internal structure and composition. All bodies in a common gravitational field
fall with the same acceleration according to WEP. This is also called the universality of free fall (UFF).
\item Local Lorentz invariance (LLI):  the outcome of any local non-gravitational experiment conducted in free fall is independent of the velocity  and the orientation
of the apparatus.
\item Local position invariance (LPI): the outcome of any local non-gravitational experiment  is independent of where and when in
the universe it is performed.
\end{enumerate}
Different versions of the EEP appear in the literature; precise formulations of its variants are discussed by \citet{CasLiSon15}.
AMO tests of  LLI are discussed in Sec.~\ref{Sec:LV}.
%\ref{LV}.
%Gravity tests with antimatter are discussed in Sec.~[CPT].
%\ref{CPT}
Tests of LPI include searches of the temporal and spatial variation of fundamental constants, as discussed in detail in Sec.~\ref{Sec:FC}.
%\ref{FC}.

Both GR and the SM are assumed to be
low-energy limits of a more complete
theory at the high-energy scale.
The EEP implies  a universal coupling between matter and gravity, i.e.  all forms of matter-energy respond to gravity in the same way.
However, this may not be the case for most theories aimed at unifying all four fundamental interactions, such as string theories.
Any theories in which the coupling constants are spatially dependent violate the WEP, as discussed in Sec.~\ref{Sec:FC}.
%\ref{FC}.
On the other hand, the WEP can be tested with experiments complementary to those used to test fundamental constant variation. Thus WEP  tests provide additional opportunities to open a low-energy window into the nature of unification  theories.
Various theoretical arguments that the EEP is violated at  small but  measurable  levels are discussed in detail by
\citet{Dam12}. ``Runaway dilaton'' models \cite{DamPol94,DamPiaVen02,Dam12}, estimate that onset of  WEP violation may start just beyond the sensitivity of current experiments.

Tests of the equivalence principle as well as more general tests of gravity are reviewed by \citet{Wil14}.
A review of the past WEP tests and future proposals is given by \citet{SonDit16}. Modern tests of the WEP include torsion-balance experiments, free-fall experiments and measurement of relative motions of celestial bodies  (for example, lunar laser ranging).
To quantify violations of  WEP, we suppose that the gravitational mass of a body, $m_{\rm g}$, is not equal to its inertial mass, $m_{\rm I}$. Then, the acceleration $a$ of a body  in a gravitational field $g$ is given by $\mb{a}=\frac{m_{\rm g}}{m_{\rm I}}\mb{g}$. To test WEP, one compares the accelerations $a_1$, $a_2$ of two falling bodies which differ in their composition, and measures the ``E\"otv\"os'' ratio $\eta$
\begin{equation}
\eta=2\left|\frac{a_1-a_2}{a_1+a_2}\right|.
\end{equation}
The ``E\"ot-Wash'' torsion-balance  experiments tested WEP to $10^{-13}$ by comparing differential accelerations of beryllium-aluminum and beryllium-titanium test-body pairs \cite{SchChoWag08,WagSchGun12}. Lunar laser-ranging experiments, which measure   the  differential accelerations of the Earth and Moon towards the Sun,
 provided similarly stringent limits to  the violation of the  equivalence principle \cite{WilTurBog04,WilTurBog12}. Both the torsion-balance and lunar-laser ranging WEP tests are close to their fundamental limits of accuracy.
 Macroscopic (i.e., classical)  Earth-based free fall WEP tests are less accurate, reaching the $10^{-10}$ level \cite{KurMio89}.

 Significant improvement in probing WEP is expected to come from future space-based missions. MicroSCOPE is a Centre National d'\'{E}tudes Spatiales (CNES)/ European Space Agency (ESA) gravity-research minisatellite mission \cite{BerTouRod15} that aims to test the WEP in space to $10^{-15}$  by comparing the acceleration experienced by two free-falling test masses in the Earth's gravity field. The satellite was launched in April 2016 and the mission is planned to last two years. In 2017, the first results were reported \cite{TouMetRod17}.
 Other macroscopic  proposals to test the WEP include the Sounding Rocket Principle of Equivalence
 Measurement \cite{ReaPatPhi12} [free-fall stratosphere experiment], Galileo Galilei \cite{NobShaPeg12} [space-based torsion-balance experiment], the
 Satellite Test of the Equivalence principle  \cite{OveEveWor12} [free-fall space based experiment], and others.

% Tim Kovachi
% Den Shlipper

The WEP test using both quantum and  classical  objects was reported by \citet{PetChuChu99}.
This experiment compared the values for the local acceleration due to the Earth's gravity, $g$,  obtained using an atom interferometer
based on a fountain of $^{133}$Cs laser-cooled atoms  and a Michelson-interferometer classical gravimeter which used macroscopic
glass object, demonstrating  agreement to 7 parts in 10$^9$. In subsequent related work, \citet{MerBodMal10} compared the performance of optical and atom-interferometric gravimeters. \citet{FraBauUll15} reported an international comparison of 25 different gravimeters.
\citet{FreHauSch16} reported absolute measurements with a mobile atom-interferometric gravimeter, demonstrating  an accuracy of 39~nm/s$^2$ and long-term stability of 0.5~nm/s$^2$.

The theoretical framework for WEP tests in the quantum domain is
discussed by \citet{HerDitLam12}.
Weak-equivalence tests using quantum matter were made possible by techniques for production and control of ultracold atoms and by the attainment
of dilute atomic Bose-Einstein condensates (BECs) \cite{CorWie02,Ket02}.
Atom interferometers measuring the difference in phase between matter waves traveling along different paths can be used as accelerometers,
offering potential precision tests of GR with quantum rather than classical matter.
A good general review of atom interferometry is given by
 \citet{CroSchPri09,TinKas14}. \citet{HogJohKas09,KleKajRou15} and several of the chapters in \citet{TinKas14} review the light-pulse
atom interferometry which is employed in most high-precision measurements mentioned in the
review and its applications to tests of fundamental physics.

Fully  quantum WEP tests with atomic interferometry directly compare the phase shifts of two different types of matter waves without the use of classical gravimeters.
The potential of matter-wave  interferometers using  quantum gases for probing  fundamental concepts of quantum mechanics and GR has been discussed by \citet{DimGraHog07,HerDitLam12,MunAhlKru13,Bie15}. Testing the limits of quantum mechanical
superpositions with different systems  has been discussed by \citet{ArnHor14}.

The sensitivity of atom interferometers to WEP violations increases linearly with the momentum difference between
the two matter waves emerging from a beam splitter
and quadratically with the time of free fall. Sensitivity can be increased by increasing  the momentum difference or the time in free fall (or both). Therefore, the space-based experiments promise a breakthrough in sensitivity because of long free-fall times.
Current and proposed tests of gravity with atom interferometry include splitting free-falling BECs
in atomic fountains \cite{SchHarAlb14,HarAbeSch15},  drop towers \cite{MunAhlKru13}, parabolic flights \cite{GeiMenSte11,BarAntChi16}, sounding rocket missions \cite{SeiLacBec15}, and outer space \cite{WilChiYu16,TinSorAgu13,AguAhlBat14}.

A number of  experiments, implementing the longest atom-interferometry
times to date, have been performed in Stanford's 10-meter atomic fountain \cite{DicHogSug13}, where superpositions of atomic wave packets with spatial separations of up to 50 cm were created \cite{KovAseOe15}. An
important step was the preparation through atomic lensing of narrow momentum distributions
corresponding to effective temperatures in the picokelvin regime \cite{KovHogSug15}.

Gravity gradients pose a major challenge for high-precision tests of the WEP with atom interferometry.
They degrade
 contrast in the interference signal and
 impose
severe requirements on the level at which the relative positions and velocities of the initial wave packets for the two atomic species must be controlled. Otherwise, gravity gradients could mimic spurious violations of the WEP.
These difficulties can be mitigated by employing wave packets with narrower position and momentum widths until reaching the Heisenberg uncertainty limit.
A scheme to compensate the effects of gravity
gradients and overcome these difficulties has recently been proposed \cite{Rou17}, and has been
demonstrated experimentally \cite{AmiRosZha17,OveAseKov17}. \citet{OveAseKov17} have shown that one can bring down
the systematic effects associated with gravity gradients in WEP tests to one part in $10^{14}$. This makes atom interferometry  competitive with tests employing macroscopic masses.

In 2010, a preparation and observation of a Bose-Einstein
condensate during free fall in a 146-meter-tall evacuated drop tower of the Center of Applied Space Technology and Microgravity (ZARM) in Bremen, Germany was reported \cite{vanGaaSin10}.
  The realization
of an asymmetric Mach-Zehnder interferometer operated with a Bose-Einstein condensate  in extended free fall
at ZARM was reported by \citet{MunAhlKru13}. These proof-of-principle
experiments
  demonstrated a feasibility of coherent
matter-wave experiments in microgravity paving the way toward for matter-wave experiments in space.
In 2017, the QUANTUS collaboration \cite{SeiLacBec15} conducted
successful  MAIUS~1 (Matter-Wave Interferometry in Microgravity) experiment aboard a sounding rocket at altitude up to 243 km above the Earth's surface, well above the K\'{a}rm\'{a}n line that marks the boundary of outer space (the International Space Station's orbit is about 400 km above the surface of the Earth).  About 100 discrete matter-wave experiments were conducted during the six-minute experimental phase of this flight.
% \cite{SoundingRocket2017}.  A schematic of those experiments is shown in Fig.~\ref{SoundingRocketFig}.

 Atom-interferometry quantum tests of the universality of free fall with cold rubidium $^{85}$Rb and $^{87}$Rb atoms were performed
 by \citet{Fra04} and \citet{BonZahBid13} at the $10^{-7}$ level.
 A scheme to suppress common-mode noise in lasers
used for atom interferometry was demonstrated by \citet{ZhoLonTan15}, resulting in measurement of $\eta(^{87}\mr{Rb},^{85}\mr{Rb})
= (2.8\pm 3.0) \times 10^{-8}$.

 One of the advantages for using cold atom clouds for gravity tests is the opportunity to perform qualitatively
different WEP tests  with well-characterised ``test masses'' with a definite spin
for a search of the spin-gravity coupling effects.
\citet{Tar14} reported such an experimental comparison
of the gravitational interaction for a  $^{88}$Sr (boson, $I = 0$) with that of a
$^{87}$Sr (fermion, $I = 9/2$). The E\"otv\"os ratio and possible spin-gravity coupling were constrained at the 10$^{-7}$ level. Note that such a test is completely insensitive to the types of spin-gravity interactions probed in spin-precession experiments such as those of \citet{Ven92} and \citet{Kim17GDM}; see the discussion in Sec.~\ref{Sec:ExoticSpin:OtherSpinDependentPotentials}.

\citet{DuaDenZho16} reported  a test of the universality of free fall   with $^{87}$Rb  atoms in different spin orientations.
 They used a Mach-Zehnder-type atom interferometer  to alternately measure the free fall acceleration of the atoms
in $m_F = +1$ and  $m_F = -1$ magnetic sublevels, with the resultant E\"otv\"os ratio of $\eta =(0.2 \pm 1.2) \times 10^{-7}$.

\citet{RosAmiCac17} reported a novel WEP test performed on rubidium atoms prepared in coherent superpositions of different energy eigenstates.
A Bragg atom interferometer in a
gravity gradiometer configuration was used to compare their free fall.  This experiment tested quantum aspects of EEP than have no classical counterpart. \citet{RosAmiCac17}  also measured the E\"otv\"os  ratio of atoms in two
hyperfine levels with relative uncertainty in the range of 10$^{-9}$.

\citet{GeiTru18} proposed an experiment to study the possible influence of entanglement between two test masses on the universality of free fall. They devised a test of the weak equivalence principle with $^{85}$Rb and $^{87}$Rb atoms entangled by a vacuum stimulated rapid adiabatic passage protocol implemented in a high-finesse optical cavity.

The first quantum test of the UFF  with matter waves of two different atomic species  was reported by \citet{SchHarAlb14}. This experiment compared the free-fall accelerations of laser-cooled ensembles of $^{87}$Rb and $^{39}$K  atoms by measuring the gravitationally induced shift in two Mach-Zehnder-type interferometers. \citet{SchHarAlb14} measured the E\"otv\"os ratio
\begin{equation}
\eta_{\rm Rb, K}=2\frac{g_{\rm Rb}-g_{\rm K}}{g_{\rm Rb}+g_{\rm K}}
= (0.3\pm 5.4)\times 10^{-7},
\end{equation}
where $g_{\rm Rb}$ and $g_{\rm K}$ are free-fall accelerations of the $^{87}$Rb and $^{39}$K atoms, respectively.
A non-zero value of the E\"otv\"os ratio would indicate a UFF violation resulting from either a difference of the inertial and gravitational masses or an additional (unknown) force which depends on the composition of the atomic species and differs for $^{87}$Rb and $^{39}$K atoms.
The same apparatus may be used to improve the precision by two orders of magnitudes, to the ppb level.

\citet{HarAbeSch15} proposed a long baseline atom interferometer test of EEP
with Rb and Yb.
With over 10 meters of free fall, their experiment
is estimated to reach $7 \times 10^{-13}$ accuracy in the E\"otv\"os ratio.
Use of the heavy alkaline earth Yb will broaden the scope of atom interferometric EEP
tests in view of  EEP violation parametrization based on the dilaton model described by \citet{Dam12}.

A number of  quantum WEP tests in microgravity are being pursued, with the promise of greatly increased precision
over current quantum tests. This is a goal of the QUANTUS collaboration mentioned above.
The I.C.E. (Interf\'{e}rometrie atomique \'{a} sources Coh\'{e}rentes pour l'Espace - Coherent atom interferometry for space applications) experiment \cite{GeiMenSte11} is a compact and transportable atom interferometer, designed to test WEP
by comparing the accelerations of free-falling  clouds of ultracold  Rb and K atoms  inside an airplane in free fall.
Searching for  WEP violation at high-precision  is the primary science objectives of the Space–Time Explorer and QUantum Equivalence Space
Test (STE -- QUEST) space mission \cite{AguAhlBat14,AguAhlBat14a,AltBaiBla15} designed to measure the E\"otv\"os ratio between matter waves of two Rb isotopes in a
differential atom interferometer at the
$2 \times 10^{-15} $ uncertainty level. Although QUEST was not selected for the European Space Agency M3 Cosmic Vision Programme, it demonstrated
the potential of future  space-based quantum WEP tests.
Other proposals for quantum atom interferometry space-based WEP tests include
Quantum Test of the Equivalence principle and Space Time (QTEST) \cite{WilChiYu16} and a Quantum WEP test (Q-WEP) \cite{TinSorAgu13} on the International Space Station.

\subsection{Determination of the Newtonian gravitational constant}
The Newtonian gravitational constant, $G$, was the second  fundamental constant subject to an absolute measurement, which was first conducted by Cavendish in 1797-98 \cite{Cavendish1798}. The 2014 CODATA  recommended value \cite{CODATA2014} is $G = 6.674 08(31) \times 10^{-11}$ m$^3$ kg$^{-1}$ s$^{-2}$. The relative standard uncertainty in $G$ of $4.7 \times 10^{-5}$ is by far the largest of any of the primary fundamental constants, exceeding that of the Boltzmann constant, $k$, by two orders of magnitude. The slow rate of progress in reducing the uncertainty in $G$ and large unresolved disagreements
between the most precise measurements is a matter of concern in the precision-measurement community \cite{Schlammi2014,2015Anderson,PhysRevD.91.121101}.
Reflecting the isolation of gravity and GR from the Standard Model, $G$ is also unique among the fundamental constants in having no dependence upon any of the other constants included in the CODATA least-squares fit \cite{CODATA2014}.

The first experimental observation of a gravitational shift of a de Broglie wave was made in 1974 using neutron interferometry in the Earth's gravitational field \cite{PhysRevLett.34.1472}. Until 2014, all experimental determinations  that contributed to the CODATA recommended value of $G$ involved measurement of classical forces. The first  atom-interferometric measurements of $G$  were performed by \citet{BerLamCac06} and \citet{Fixler74}.  These experiments demonstrated the gravitational action of a laboratory source mass upon an atomic de Broglie wavelength, an intrinsically quantum-mechanical effect. They yielded values of $G$ consistent with the CODATA recommended value then, but with much larger uncertainties. Atom-interferometric measurements of $G$ certainly offer the prospect of having systematic uncertainties qualitatively different from those of classical experiments, and they may eventually provide a link between $G$ and other fundamental constants, as in the example of the dependence of the fine structure constant, $\alpha$, on the ratio $\hbar/M(^{87}\mathrm{Rb})$. \citet{RosSorCac14} reported an atom-interferometric determination of $G$ was
with a relative standard uncertainty of 0.015~\%. Their measurement was included in the least-squares fit that determined 2014 CODATA recommended value.

The coupling of the initial position and velocity of the
atomic wave packets to gravity gradients has so far been the main source of systematics in the gradiometry measurements for the determination of $G$ with atom interferometry. However, it has been argued that the compensation technique of \citet{Rou17} can also be exploited in this context to achieve accuracies competitive with those of measurements employing macroscopic masses or even better \cite{AmiRosZha17,Ros17}. Past and ongoing $G$
determinations based on atom interferometry are reviewed by \citet{Ros16}.

Use of multiple atomic samples in an interferometer also enables measurements of higher-order spatial derivatives of the
gravity field. \citet{RosCacSor15} reported the first direct measurement of the gravity-field curvature based on three conjugated atom
interferometers. The gravity curvature was produced by nearby source masses along one axis.
 In the experimental set up designed by \citet{RosCacSor15},
 three atomic clouds  launched in the vertical direction are simultaneously interrogated by
the same atom interferometry sequence probing the gravity field at three equally spaced positions.
Such atomic sensor is capable to measure
gravity, gravity gradient, and curvature along the vertical
direction at the same time, important for
geodesy studies and  and Earth monitoring applications. The same scheme may be used for a novel approach to G measurement.

\citet{AseOveKov17} used a dual light-pulse atom interferometer to
measure a phase shift associated with  spacetime-curvature induced tidal forces on the wave function of a single quantum system. A macroscopic spatial superposition state in each interferometer, extending over 16 cm, acted as a nonlocal probe of the spacetime manifold.

\subsection{Detection of gravitational waves}

The detection of the gravitational waves (GW) by the Advanced LIGO  in 2015 \cite{AbbAbbAbb16,AbbAbbAbb16b} initiated the field of gravitational-wave astronomy. This opens a new window on the universe since many of the GW cosmic sources do not
have detectable electromagnetic emissions. Theoretical physics implications of the observed  binary black-hole mergers
and probes of new physics and cosmology enabled by the detection of the
gravitational waves are described by \citet{YunYagPre16}.
Once at full sensitivity, the Advanced LIGO detectors will be able to see inspiralling binaries made up of two 1.4 solar-mass neutron stars to a distance of 300 megaparsecs (Mpc, 1 parsec = 3.3 light years) and coalescing black-hole systems at the  cosmological distance, to the red shifts $z=0.4$, significantly increasing the number of potentially detectable events. Advanced LIGO at full capacity will be  essentially operating at the quantum noise limit. With the Advanced Virgo GW detector in Italy along with future detectors,  the GW signals may be triangulated.
 There are already proposals for 10~km and 40~km laser interferometers, the Einstein telescope \cite{SatAbeAce12} and the Cosmic explorer \cite{AbbAbbAbb16a}, significantly longer than the Advanced LIGO 4~km arms, and thus able to measure lower frequencies at smaller fractional sensitivity.

The detection capability of the laser interferometry terrestrial detectors is limited to GWs with frequencies above ${\approx 10}$ Hz by the
seismic noise and Newtonian noise (fluctuations of the terrestrial gravity field which creates a tidal effect on separated test masses) \cite{Sau84}.
The ability to detect gravitational waves of lower frequencies  will significantly increase the number of binary star mergers
 from which the gravitational waves may be detected and allow for longer observation of the inspiralling binary stars before the merger.
Stochastic gravitational
waves, i.e. relic GWs from the early evolution of the universe, from cosmological (and possibly unforeseen) sources, such as inflation
and reheating, a network of cosmic strings, or phase
transitions in the early Universe, etc., can also be easier to
detect at these low frequencies, see \citet{DimGraHog08}.
 Proposals for the detection of the gravitational waves at lower frequencies include space-based laser interferometry detector [Laser Interferometer Space Antenna (LISA) \cite{DanPriBin11} and Evolved-LISA (eLISA) \cite{AmaAouBab12}] and both terrestrial and space based matter-wave detectors, using either atom interferometers or atomic clocks \cite{TinVet07,DimGraHog08,YuTin11,HogJohDic11,GraHogKas13,ChiWilYu15,ChaGeiCan16,HogKas16,GraHogKas16,KolPikLan16}.

Terrestrial atom interferometers have been proposed  \cite{DimGraHog08,ChaGeiCan16} for GW detection in the 0.3~Hz to 3~Hz frequency band.
\citet{ChaGeiCan16} proposed a new detection strategy based on a
correlated array of atom interferometers which allows reduction of the Newtonian noise
limiting all ground based GW detectors below $\approx 10$~Hz.
The matter-wave laser interferometer gravitation antenna
(MIGA) is a hybrid detector that couples laser and
matter-wave interferometry aimed at both the geophysical studies and sub-hertz GWs detection
in a
low-noise underground laboratory to minimise effects of laboratory vibrations
\cite{GeiAmaBer14}.

LISA/eLISA are space-based
laser  interferometric detectors analogous to LIGO, to be composed of three spacecraft forming either a two- or
three-arm Michelson interferometer \cite{DanPriBin11,AmaAouBab12}, with the GW frequency-detection range from 0.1 mHz to 1 Hz.

Both atomic clocks and atom interferometry technologies improved tremendously over the past decade, also leading to fast development
of the gravitational wave detection schemes \cite{DimGraHog08,HogJohDic11,GraHogKas13,ChiWilYu15,ChaGeiCan16,HogKas16,GraHogKas16,KolPikLan16} with improved sensitivities, realistic requirements for the underlying technologies, and
addressing  some problems \cite{Ben11,Ben14} of earlier proposals.
The intrinsic noise sources and sensitivity limits of atom-based vs. light-based interferometers for GW detection are being clarified \cite{Ben11,BakTho12,Ben14}.
 Most importantly, the restriction of a space-based atom interferometry
GW detector
to a relatively short baseline, $\approx 1000$~km, in comparison with LISA,
has been lifted in the recent 2016 proposal \cite{HogKas16}.
Due to considerable evolution of the AMO GW detector proposals, we very briefly describe only the most recent proposals based on atom interferometers and atomic clocks.

\citet{HogKas16}  proposed a space-based GW detector based on two satellites with light-pulse atom interferometers separated
 by a long baseline (over 100\,000~km), capable of detecting GWs in the 0.1-mHz to 1-Hz frequency band. The light
pulses are sent back and forth across the baseline from
alternating directions, driving atomic single-photon transitions. Use of  single-photon
transitions in alkaline-earth atoms (Sr) with long lifetimes of the excited state
significantly reduces laser frequency noise \cite{GraHogKas13} in comparison to the GW detector proposals based on alkali-metal atoms requiring implementation of  two-photon transitions \cite{DimGraHog08}.
The GWs are detected
by monitoring the phase difference between the two interferometers caused by the variation of the light travel time across
the baseline due to a passing GW.
 As described by \citet{GraHogKas16}, the atom interferometric   GW detector essentially compares time kept by the laser and atom ``clocks.``
A gravitational wave affects the flat-space relation
between these clocks by a factor proportional to the
distance between them, and such change oscillates in time with
the frequency of the gravitational wave resulting in a detectable signal.

 In previous proposals \cite{DimGraHog08,GraHogKas13}, the same laser would drive atom
transitions at both ends, requiring the laser to remain collimated over the optical path between two satellites, significantly restricting
the maximum baseline length.
 In the  \citet{HogKas16} scheme, both satellites house a master laser and a
  local-oscillator laser that have sufficient
intensity to drive transitions in the local atom interferometer. The master laser beam interacts with its satellite's atomic cloud,
and then propagates to the second satellite acting now as a reference beam which does not have to be
collimated as it reaches the opposite satellite. A  local oscillator in the other  satellite is phase referenced
or phase locked to the incoming reference laser beam and drives the transitions in this satellite's atomic cloud. An identical scheme is implemented
in the reverse direction. Since much less intensity of the reference beam is required for phase reference than for atomic excitations,
this scheme allows for much larger baseline leading to enhanced sensitivity, simplified atom optics, and reduced atomic-source flux
requirements.
 Such a GW detector scheme with 12 photon recoil atom optics and $6\times10^8$~m baseline is evaluated to exceed the sensitivity of proposed the LISA detector by a factor of ten \cite{HogKas16}.

 \citet{GraHogKas16} proposed an atom interferometric  GW  detector that can operate in a resonant
detection mode and can switch between the broadband and narrowband  detection
modes to increase sensitivity.

\citet{Gei17} reviewed the perspective of using atom interferometry for GW detection in the mHz to about 10 Hz frequency band.

\citet{KolPikLan16} proposed to use a two-satellite scheme sharing ultrastable optical laser light over a single baseline, with
atomic optical lattice clocks (rather than atom interferometers)
as sensitive, narrowband detectors of the local frequency of the shared
laser light.
A passing GW induces effective
Doppler shifts and the GW signal is detected as a differential frequency shift of
the shared laser light due to the relative velocity of the satellites. Such a scheme can detect GWs with frequencies ranging from 3 mHz to 10 Hz without loss of sensitivity. The clock scheme may be integrated with an optical interferometric detector.
The next stage of matter-wave gravitational  detector development is a   demonstration of ground-based prototype
systems and characterization of the noise sources.

\subsection{Gravity experiments with antimatter}
\label{antigrav}
One of the recent foci of the experimental efforts on matter-antimatter comparisons is testing whether antimatter is affected by gravity in the same way as matter. For example, the CERN based GBAR collaboration is developing an ingenious technique \cite{Perez2015} where they will first  create a positive ion consisting of an antiproton and two positrons that will be sympathetically cooled with Be$^+$ ions and then ``gently'' photoionize to produce a cold neutral antihydrogen atom that will fall under gravity over a known distance before being detected.
The  AEgIS experiment at CERN is also aimed at the direct measurement of the Earth's gravitational acceleration on antihydrogen but has a
 completely different design \cite{Aegis15}. In the AEgIS  expariment, a cold, pulsed beam of antihydrogen will pass through  a moir{\'e} deflectometer \cite{Aegis14}, coupled to a position-sensitive detector to measure the strength of the gravitational interaction between matter and antimatter to a precision of 1~\%.
  A second goal of the AEgIS experiment is to carry out spectroscopic measurements on antihydrogen atoms. There is a possibility of laser cooling of the negative ion of lanthanum (La, $Z=57$) \cite{JorCerFri15}. Laser-cooled La$^-$ might be used  for sympathetic cooling of antiprotons
for subsequent  antihydrogen formation \cite{KelCanFis09}.

A method to directly
measure the ratio of the gravitational to inertial mass
of antimatter accomplished by searching for the free
fall (or rise) of ground-state antihydrogen atoms
was proposed by ALPHA collaboration \cite{Alpha13}. The antihydrogen atoms are
released from the  trap; the escaping anti-atoms are then detected
when they annihilate on the trap wall.

 One should emphasize that the possibility of a difference in gravity between matter and antimatter is already constrained, under some assumptions,  at the 1 ppm level by experiments of \citet{Gabrielse1999} which found no differences in gravitational red shift of matter and antimatter clocks.
\citet{BASE15} interpreted their result for the sidereal variation of the $q/m$ ratios given by Eq.~(\ref{eq1-CPT}) as a test of the weak equivalence
 principle for baryonic antimatter.
Following \citet{HugHol91}, they
expressed a possible gravitational anomaly acting on
antimatter with  a parameter $a_g$, which modifies the
effective newtonian gravitational potential $U$ to give $a_gU$, setting an upper limit of
  $|a_g-1| < 8.7 \times 10^{-7}$.
 The role of the internal kinetic energy of bound systems of matter in tests of the Einstein
equivalence principle was considered by \citet{HohMulWir13},
letting the limits on equivalence principle violations in antimatter from tests
using bound systems of normal matter.
We emphasize that any difference between matter and antimatter gravity would run into theoretical conceptual troubles \cite{Karshenboim2016}.

Furthermore, as discussed in previous sections, there have been numerous and stringent matter-based tests of the equivalence principle and $CPT$ invariance, and these must have a direct bearing on the proposed tests with antimatter, especially considering that most of the mass of the antiproton comes from the quark binding energy (i.e., the gluon field).  The ``true'' antimatter mass/energy content of the antiproton in the form of antiquarks can reasonably be assumed to be less than or about 1~\%, while the antimatter content of ordinary matter due to virtual particles is non-negligible.  This implies that there is a connection between matter-based equivalence principle tests and proposed antihydrogen experiments. Another compelling, albeit model-dependent, argument \cite{Ade91} limiting the difference between gravitational acceleration for matter and antimatter to less than a part in $10^5$ or perhaps much better (considering the considerably more sensitive updated versions of the torsion pendulum experiment) is based on reasonable assumptions of equivalence principle violations arising from a scalar/vector gravitational coupling and combining data from the exquisitely precise measurements using torsion pendulums with stringent limits on $CPT$ invariance.

\subsection{Other AMO tests of gravity}

A test of the local Lorentz invariance of post-Newtonian gravity was performed by \citet{MulChiHer08}  by monitoring Earth's
gravity with a Mach-Zehnder atom interferometer (see Sec.\ref{Sec:LV}).
 \citet{Hoh12} proposed an experimental realization of the
 gravitational Aharonov-Bohm effect: measurement of phase shifts with an atom interferometer  due to a gravitational
potential $U$ in the absence of a gravitational force.
A pair of laboratory masses will be used as a source of the gravitational potential.
A matter-wave interferometry experiment  to measure such phase shifts
in the absence of a classical force is
currently under  construction at the University of California, Berkeley.

Testing sub-gravitational forces on atoms from a miniature, in-vacuum source mass has been reported by \citet{JafHasXu16}.

Tests of gravity are interconnected with the searches for exotic forces.
\citet{Lee16} have suggested and implemented the use of atomic spectroscopy to search for Yukawa-type fifth-forces. By studying the behaviour of atomic transition frequencies at varying distances away from massive bodies (e.g., the Sun, Moon, heavy masses in the laboratory), \citet{Lee16}  have placed constraints on possible non-gravitational interactions of a scalar field with the photon, electron and nucleons. This work also placed constraints on combinations of interaction parameters that cannot otherwise be probed with traditional anomalous-force measurements. \citet{Lee16}  suggested further measurements to improve on the current level of sensitivity. Such measurements include the use of more precise atomic clocks and other systems (molecular, highly-charged ionic and nuclear transitions), and implementing different experimental geometries (e.g., the size of the effect can be increased by up to four orders of magnitude by measuring atomic transition frequencies first on Earth, then on a space probe headed towards the Sun).

%Antimatter interferometry for gravity measurements
%have been proposed in \cite{HamZhmRob14}.

\section{Lorentz symmetry tests}
\label{Sec:LV}
Local Lorentz invariance (LLI) is one of the foundations of the current laws of physics: the outcome of any local non-gravitational experiment is independent of the velocity  and the orientation of the (freely-falling) apparatus. The first test of Lorentz invariance was Michelson's 1881 experiment \cite{Mic81} aimed at detecting the ether (erroneously assumed to be the medium for electromagnetic
wave propagation). This experiment was further improved by \citet{MicMor87}. Michelson and Morley's apparatus measured the interference between two beams of light travelling back and forth along two perpendicular paths. This light interferometer was rotated relative to the Earth to test the isotropy of the speed of light.

In the 1960s, the first spectroscopic tests of Lorentz symmetry were performed by \citet{HugRobBel60} and \citet{Dre61} where they searched for sidereal variation of nuclear magnetic resonance (NMR) lines in $^7$Li. The Hughes-Drever tests were inspired by the suggestion of \citet{Coc58search} that it might be possible, based on Mach's principle, for inertial mass to acquire a tensor character due to anisotropic distribution of matter in the universe.  This would cause a particle's inertial mass to depend on the orientation of its orbit with respect to the matter anisotropy, which in turn would generate energy shifts in atoms and nuclei. Experiments similar to the Hughes-Drever test have come to be known as a ``clock-comparison tests'' in which the frequencies of different atomic ``clock'' transitions are compared as the clocks rotate with the Earth. Since these early tests, the field of Lorentz symmetry tests has flourished, encompassing almost all fields of physics \cite{Mat05,LibMac09,KosRus11}. The \textit{Data Tables for Lorentz and $CPT$ Violation}, an extraordinary effort by \citet{KosRus17}, provides
yearly updates of experimental progress of the last decade and gives tables of the measured and derived values of coefficients for Lorentz and $CPT$ violation in the Standard Model Extension discussed below. The listed experiments include searches for Lorentz violation (LV) in the matter, photon, neutrino, and gravity sectors.
The \textit{Data Tables for Lorentz and $CPT$ Violation} has grown in length by 50~\% in the past three years demonstrating large number of new experiments in many sectors.

This recent interest in tests of Lorentz symmetries is motivated by theoretical developments in quantum gravity  suggesting that Lorentz symmetry may be violated at some energies,  tremendous progress in experimental precision, and development of a theoretical framework to analyze different classes of experiments. A particular attraction of the LLI tests is a tantalizing possibility of a positive result: a confirmed measurement of Lorentz violation would be an unambiguous signal of new physics.  The natural energy scale for strong LV induced by quantum gravity is the Planck scale ($M_{\mathrm{Pl}} \approx 10^{19}$ GeV/$c^2$), which is far beyond the reach of existing observations: even ultra-high energy cosmic rays still fall eight orders-of-magnitude short of the Planck scale. The good news is that strong LV at the Planck scale may also lead to tiny but potentially observable low-energy LV. Therefore, high-precision tests of LLI with matter, gravity, or
light may provide insight into possible new physics and set limits on various theories
such as quantum gravity. The bad news is that there are no predictions of the magnitude of LV violation at low energies.
Lorentz-violating effects may be suppressed
by some power of the ratio $R$   between the electroweak scale and the natural (Planck) energy
scale for strings: $R=m_{ew}/M_{Pl}=2\times10^{-17}$ \cite{KosPot95} or electron mass to Plank scale $4\times 10^{-23}$ \cite{LibMac09}.

Lorentz violation tests are analyzed in the context of an effective field theory  known as the Standard Model extension (SME).
Two approaches are used when constructing such an effective field theory to describe Lorentz violations: (1) add renormalizable Lorentz-violating terms to the Standard Model
Lagrangian \cite{ColKos98} and (2) explicitly break Lorentz invariance by
introducing nonrenormalizable operators  \cite{MyePos03}.

In minimal SME, corresponding to the first approach, the  Standard Model Lagrangian is augmented with every possible combination of the SM fields that are not term-by-term Lorentz invariant, while maintaining gauge invariance, energy--momentum conservation, and Lorentz invariance of the total action \cite{ColKos98}. Separate  violations of LLI are possible for each
type of particle, making it essential to verify LLI in different systems at a high level of precision.
\citet{LibMac09} reviewed non-minimal SME experimental tests, and all current limits are given in 2017 edition of the \textit{Data Tables for Lorentz and $CPT$ Violation} \citet{KosRus17}. We limit this review to recent AMO tests and  proposals. The diverse set of  AMO Lorentz symmetry tests involves experiments with atomic clocks \cite{WolChaBiz06}, other precision spectroscopy measurements \cite{Hoh13}, magnetometers \cite{SmiBroChe11,AllHeiKar14}, electromagnetic  cavities \cite{EisNevSch09}, and
quantum-information-trapped-ion technologies \cite{PruRamPor15}.

In minimal SME, a general expression for the quadratic Hermitian Lagrangian density
describing a single spin-$1/2$ Dirac fermion of mass
$m$  (electron, proton, or neutron) in the presence of Lorentz violation is given by \cite{KosLan99}
\begin{equation}
{\cal{L}}=\frac{1}{2} i c \overline{\psi} \Gamma_\nu \overleftrightarrow{\partial^\nu} \psi - M c^2 \overline{\psi} \psi,
\end{equation}
where
$\psi$ is a four-component Dirac spinor,
$$f \overleftrightarrow{\partial^\nu}g = f \partial^\nu g - g \partial^\nu f $$,
\begin{equation}
\label{LV1}
M =  m + a_{\mu}\gamma^{\mu} + b_{\mu} \gamma_5 \gamma^{\mu} + \frac{1}{2} H_{\mu \nu} \sigma^{\mu \nu}
\end{equation}
and
\begin{equation}
\label{LV2}
\Gamma_{\nu} = \gamma_{\nu} + c_{\mu \nu} \gamma_{\nu}+d_{\mu \nu} \gamma_5 \gamma_{\nu} + e_\nu +  i \gamma_5 f_\nu +
\frac{1}{2} g_{\lambda \mu \nu} \sigma_{\lambda \mu}.
\end{equation}
The first terms in the expressions for $M$ and $\Gamma_{\nu}$ give the usual SM Lagrangian. Lorentz
violation is quantified by the parameters $a_{\mu}$,  $b_{\mu}$, $c_{\mu \nu}$, $d_{\mu \nu}$, $e_\mu$, $f_\mu$, $ g_{\lambda \mu \nu}$, and $H_{\mu \nu}$. The coefficients in Eq.~(\ref{LV1}) have dimensions of mass; the coefficients in Eq.~(\ref{LV2}) are
dimensionless.  The field operators in Eqs.~(\ref{LV1},\ref{LV2}) containing the coefficients  $c_{\mu \nu}$, $d_{\mu \nu}$, and
$H_{\mu \nu}$
are even under $CPT$ and the remaining ones are odd under $CPT$.
The framework of interpreting the laboratory experiments involving monitoring atomic or nuclear frequencies in terms of the
SME coefficients is described in detail by \citet{KosLan99,KosMew02}.
Such  atomic experiments may be interpreted as Lorentz-invariance tests for the photon, electron,
and nuclear constituents, such as proton and neutron, with varying sensitivities to different combinations of LLI effects.
A number of experiments are sensitive to either electron or nucleon sectors, with photon contributions appearing in all atomic
experiments. AMO tests of LLI also include testing isotropy of gravity,
a test of the LLI of post-Newtonian gravity was performed by \citet{MulChiHer08}  by monitoring Earth’s
gravity with a Mach-Zehnder atom interferometer.
 Expressed within the standard model extension, the analysis limits four coefficients describing anisotropic gravity at the
ppb level and three others at the 10 ppm level. Using the SME, \citet{MulChiHer08} explicitly demonstrated
how their experiment actually compares the isotropy of gravity and electromagnetism.

\subsection{Electron sector of the SME}

Testing LLI of the electron motion in an atom has an advantage of
testing for new physics in a well understood system. In atomic experiments aimed at the LLI tests in the electron-photon sector
\cite{Hoh13,PruRamPor15}, one searches for variations of the atomic energy
levels when the orientation of the electronic wave function is rotated with respect to a standard
reference frame. Generally, one uses the Sun centered celestial-equatorial frame (SCCEF) for the analysis of the
experiments \cite{KosMew02}, indicated by the coordinate indexes $T$, $X$, $Y$, and $Z$.
For example,  the $c_{\mu \nu}$ tensor has 9 components that need to be experimentally determined: parity-even $c_{TT}$ and $c_{JK}$ and parity-odd $c_{TJ}$, where $J, K={X,Y,Z}$.
 The elements $c_{JK}$ which describe the dependence of the kinetic energy on the direction of the
momentum have a leading order time-modulation period related to the sidereal day (12-hr and
24-hr modulation) in the laboratory experiments described below. The $c_{TJ}$  and $c_{TT}$  describe the
dependence of the kinetic energy on the boost of the laboratory frame and have a leading
order time-modulation period related to the sidereal year.
The terms $c_{TJ}$ are proportional to the ratio of the Earth's orbital velocity to the speed of
light $\beta_{\oplus} \approx 10^{-4}$; and the $c_{TT}$ term is suppressed by $\beta^2_{\oplus} \approx 10^{-8}$, resulting in
weaker bound on these components of the $c_{\mu \nu}$ tensor.
The indexes (0,1, 2, 3) are used for the laboratory frame.
The most sensitive LLI tests for electrons have been conducted with neutral Dy atoms
\cite{Hoh13} and Ca$^+$ ions \cite{PruRamPor15} as described below.

Violations of Lorentz invariance  in bound electronic states
result in a perturbation of the Hamiltonian that can be described by \citet{KosLan99,Hoh13}
\begin{equation}
\delta H=-\left(  C_{0}^{(0)}-\frac{2U}{3c^{2}}c_{00}\right)
\frac{\mathbf{p}^{2}}{2m_e}-\frac{1}{6m_e}C_{0}^{(2)}T^{(2)}_{0},\label{LV3}%
\end{equation}
where $\mathbf{p}$ is the momentum of a bound electron. The second term in the parentheses  gives the leading order
gravitational redshift anomaly in terms of the
Newtonian potential $U$. We refer the reader to \citet{KosTas11}  for the study of the gravitational couplings
of matter  in the presence of Lorentz and $CPT$ violation and the derivation of the
relativistic quantum Hamiltonian  from the gravitationally coupled minimal SME.
The parameters $C_0^{(0)}$ and $C_{0}^{(2)}$
 are elements of the $c_{\mu \nu}$ tensor in the laboratory frame introduced by Eq.~(\ref{LV2}):
 \begin{eqnarray}
 C^{(0)}_0 &=& c_{00}+(2/3)c_{jj},\\
 C^{(2)}_0 &=& c_{jj}+(2/3)c_{33},
 \end{eqnarray}
where $j={1,2,3}$. The $C^{(2)}_{\pm 1}$ and $C^{(2)}_{\pm 2}$ do not contribute to the energy shift of bound states.
The values of the $C^{(0)}_0$ and $C^{(2)}_0$ in the laboratory frame are the functions of the $c_{\mu \nu}$ tensor in SCCEF
frame and the velocity and orientation of the lab.

The nonrelativistic form of the $T^{(2)}_{0}$ operator is
$T^{(2)}_{0}=\mathbf{p}^{2}-3p_{z}^{2}$.
Predicting the energy shift due to LV involves  calculating of the expectation value
 of the above Hamiltonian for the atomic states of interest. The larger the matrix elements, the more sensitive
 is this atomic state. One has to take into account that only a transition-energy shift
  can be measured, so the difference of the sensitivities of the upper and lower states
  is important for the final experimental analysis in terms of the $c_{\mu \nu}$ tensor.
    The most accurate tests in the electron-photon sector
 can be conducted in atoms or ions  with highest possible sensitivities which are amenable to
 high-precision measurement techniques.
   Since the operators in Eq.~(\ref{LV3}) contain the second power of the
momentum operator $p$, the corresponding matrix elements are expected to be large  for orbitals with large
kinetic energy.
This happens for atomic $4f$-electrons localized deep inside the atom in the area of large (negative)
potential and kinetic energy in some atomic systems.

We note that the formalism is the same for the LV violation in the nuclei, and the
 expectation values  of the same operators (but for the nuclear states) determine the sensitivity.

\subsubsection{LLI tests with dysprosium}

A joint test of local Lorentz invariance and the Einstein equivalence principle for electrons was reported by
\citet{Hoh13}
using long-term measurements of the transition frequency between two nearly degenerate states of atomic
dysprosium.

Dy, a lanthanide element with partially
filled electronic $4f$ shell, has two near-degenerate, low-lying
excited states with significant momentum quadrupole
moments, opposite parity, and leading configurations:
[Xe]$4f^{10}5d6s$, $J = 10$ (state A) and [Xe]$4f^{9}5d^26s$, $J = 10$
(state B). The energy difference
between  states A and B can be measured directly by driving
an electric-dipole transition  with a radio-frequency
(rf) field. The average shift in the the $B \rightarrow A$
transition frequency $\omega_{\mathrm{rf}}$, properly weighted for transition frequencies for different magnetic sublevels, is given by
\begin{equation}
\frac{\delta \omega_{\mathrm{rf}}}{2\pi}= \left(10^{14} \mathrm{Hz} \right)\left[
 500\left(C^{(0)}_0  -\frac{2U_{\odot}}{3c^2}c_{00} \right)
 +9.1 C^{(2)}_0 \right],
\end{equation}
where $U_{\odot}$ is the Sun's gravitational potential. The
sign of the frequency  shift is opposite for $^{162}$Dy and $^{164}$Dy.
 The uncertainty in the numerical coefficient
in front of the first term in the square brackets may be large due to the compilations in the evaluation of the
matrix elements of the $p^2$ operator.
There is no LV contribution from the nucleus since both Dy
isotopes used in the experiment, $^{162}$Dy and $^{164}$Dy, have nuclear spin $I=0$.

 The Dy experiment used repeated measurements acquired over
nearly two years to
  obtain constraints on eight of the
nine elements of the $c_{\mu \nu}$ tensor. \citet{Hoh13} tightened the previous  limits  \cite{Alt06,MulStaTob07,Alt10} on four of the
six parity-even components  by factors
ranging from 2 to 10,
  limiting Lorentz
violation for electrons at the level of $10^{-17}$ for the $c_{JK}$ components.
Previous studies used  rotating optical Fabry-Perot resonators and  microwave whispering-gallery sapphire resonators \cite{MulStaTob07} and high-energy astrophysical sources, synchrotron and inverse Compton data  \cite{Alt06,MulStaTob07,Alt10} to  constrain $c_{\mu \nu}$ coefficients for electrons.

\citet{Hoh13} also improved bounds on gravitational redshift anomalies for electrons \cite{VesLevMat80,Hoh11} by 2
orders of magnitude, to $10^{-8}$.

\subsubsection{LLI test with calcium ion}
\begin{figure}[t]
            \includegraphics[scale=0.43]{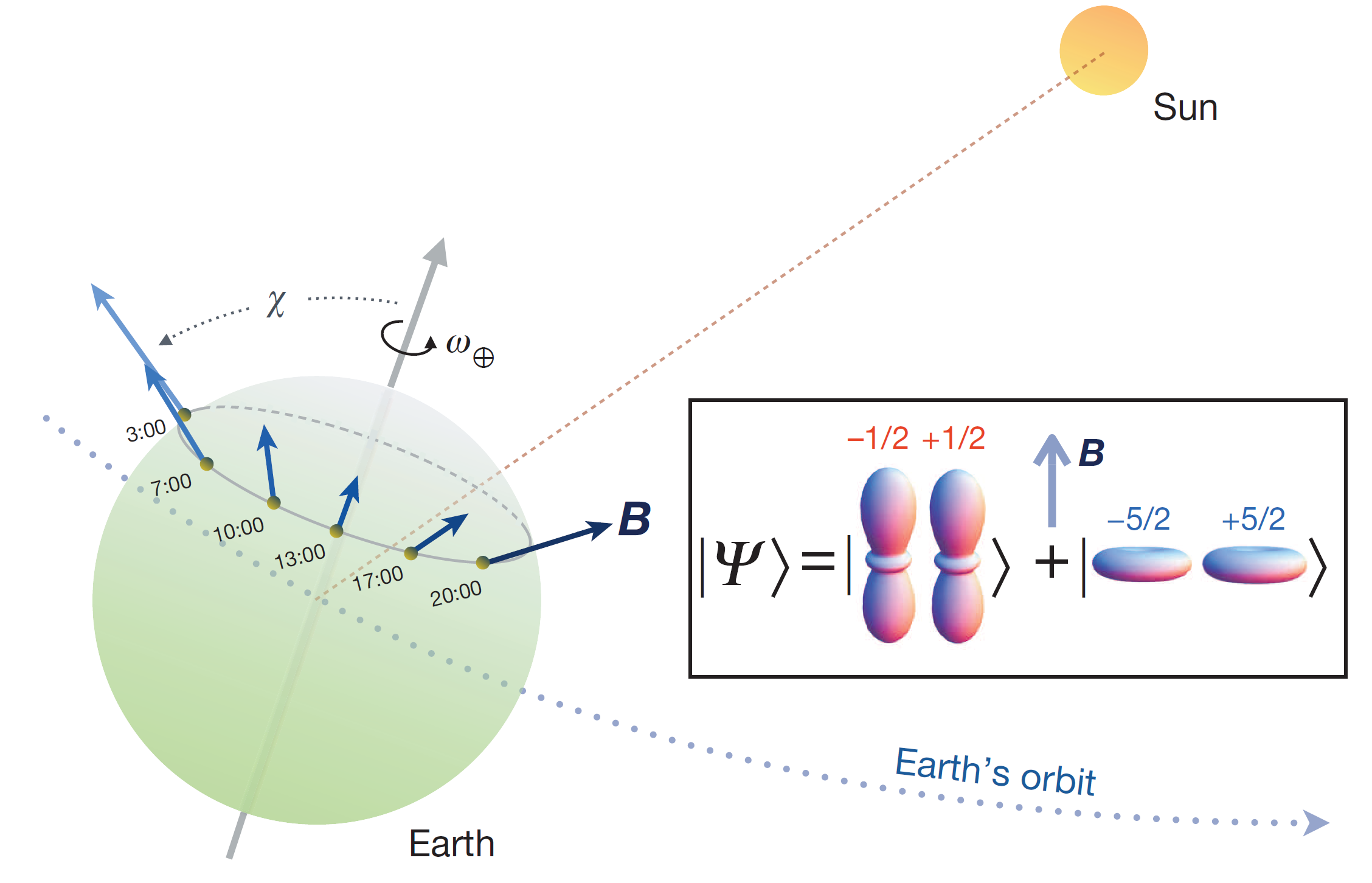}
            \caption{(color online).
         Rotation of the quantization axis of the experiment with respect to the Sun as the Earth
rotates. A magnetic field (B) is applied vertically in the laboratory frame to define the
eigenstates of the system. As the Earth rotates with an angular frequency given by
$\omega_{\oplus} =2\pi/(23.93 ~\mathrm{h})$, the orientation of the magnetic field  and, consequently, that of the electron
wave packet (as shown in the inset in terms of probability envelopes) changes with respect to the
Sun's rest frame (positions at
various times UTC are illustrated). The angle $\chi$ is the colatitude of
the experiment.  From \citet{PruRamPor15}.}
\label{fig1-LV}
\end{figure}

 \citet{PruRamPor15}  performed a test of Lorentz symmetry using an electronic analogue of a Michelson-
Morley experiment using the $^2$D$_{5/2}$  atomic states of
$^{40}$Ca$^+$  ion with anisotropic electron momentum distributions. The experiment involved interfering such states
aligned along different directions.
 A pair of
$^{40}$Ca$^+$ ions was trapped in a linear Paul trap, with a static magnetic field applied defining the eigenstates of the system. The direction of this magnetic field changes with
respect to the Sun as the Earth rotates, resulting in a rotation of the interferometer
as illustrated in Fig.~\ref{fig1-LV}.

In the magnetic field,  the $3d~^2D_{5/2}$ atomic state splits into six states with the
magnetic quantum numbers  $m_J=\pm1/2, \pm 3/2$, and $\pm 5/2$. Using the Hamiltonian
given by Eq.~(\ref{LV3}), and calculating the corresponding matrix elements of the $T^{(2)}_0$ operator,  the energy shift of these $^2D_{5/2}$ atomic states  induced by the Lorenz violation in the electron-photon sector
is given by
\begin{equation}
\frac{\delta E}{h}=\left[ \left( 2.16\times 10^{15} \right) - \left( 7.42\times 10^{14} \right) m^2_J  \right] C^{(2)}_0.
\end{equation}
Since the LV energy shift depends on the magnetic quantum number, monitoring the energy difference between the  $m_J=\pm 1/2$ and the $m_J=\pm 5/2$   Zeeman substates during the Earth rotation probes the $c_{JK}$ components of the LV tensor.
The frequency difference (in Hz) between the LV shifts of the $m_J=5/2$ and $m_J=1/2$ substates of the $3d~^2D_{5/2}$ manifold is given by
\begin{equation}
\frac{1}{h}\left( E_{m_J=5/2} -E_{m_J=1/2}\right) = \left[-4.45(9)\times10^{15}~ \textrm{Hz} \right]\times \,C_{0}^{(2)}.
\end{equation}
 This experiment is not sensitive to the scalar
$C^{(0)}_0$ coefficient of Eq.~(\ref{LV3}).

The main source of decoherence in this experiment is magnetic field noise, since it also shifts the
energies of the Zeeman substates. This problem is resolved by applying quantum-information inspired
techniques and creating a two-ion product state that is insensitive to magnetic field fluctuation to first order. The energy difference between
the two-ion states $|\pm5/2,\mp5/2 \rangle$ and $|\pm1/2,\mp1/2 \rangle$ was measured for 23 hours,
resulting in the limit of $h \times 11$ mHz.
\citet{PruRamPor15} pointed out that the experimental results may be interpreted in terms of
either photon or electron LV violation described via $c_{\mu \nu}^{\prime}=c_{\mu \nu}+k_{\mu \nu}/2$, where the first term refers to the electron LV and the second term to the photon LV. The Ca$^{+}$
experiment improved the limits to the
$c_{JK}^{\prime}$ coefficients of the LV-violation in the electron-photon sector to the 10$^{-18}$ level.
Because $^{40}$Ca$^+$ nucleus has nuclear spin $I=0$, there is no nuclear LV contribution, just as in the case of the Dy experiment.
The same experiment can be interpreted as testing
anisotropy in the speed of light with the
sensitivity similar to that of more recent work reported by \citet{NagParKov15}.

\subsubsection{Future prospects and other experiments}

With
optimization, both Dy and Ca$^+$ experiments could yield significantly
improved constraints. An optimized Dy experiment may reach sensitivities
on the order of $9\times 10^{-20}$ in one year for the $c_{JK}$ components \cite{Hoh13}.

Further significant improvement of LV constraints calls for another system with a long-lived or ground state that has a large $\langle j|T^{(2)}_0|j \rangle$ matrix element.
\citet{DzuFlaSaf16} carried out a systematic  study of this quantity for various systems and
identified general rules for the enhancement of the reduced matrix elements of the $T^{(2)}$ operator.
The authors identified  the ytterbium ion Yb$^+$ to be an ideal system for future LV tests with high
sensitivity, as well as excellent experimental controllability. The   sensitivity of the $4f^{13} 6s^2$~$^2F_{7/2}$ state of Yb$^+$ to LV is over an order of magnitude higher than that of the Ca$^+$ $^2D_{5/2}$ state.
This state also has an exceptionally long lifetime on the order of several years \cite{HunOkhLip12}, so the proposed experiment is not limited by spontaneous decay during a measurement in contrast to the Ca$^+$ case.

 Experimental techniques for precision control and manipulation of Yb$^+$ atomic states are particulary well developed owing to atomic clock \cite{HunSanLip16} and quantum information \cite{IslSenCam13} applications making it an excellent candidate for searches of the Lorentz-violation signature.

\citet{DzuFlaSaf16} estimated that experiments with the metastable $4f^{13} 6s^2$~$^2\text{F}_{7/2}$  state of Yb$^+$  can reach sensitivities of $1.5\times10^{-23}$ for the $c_{JK}$ coefficients, over $10^5$ times more stringent than current best limits. Moreover, the projected sensitivity to the $c_{TJ}$ coefficients will be at the level of
$1.5\times 10^{-19}$, below the ratio between the electroweak and Planck energy scales.
Similar sensitivities may potentially be reached for LV tests with highly charged ions \cite{DzuFlaSaf16,ShaOzeSaf17}, given future development of experimental techniques
for these systems \cite{SchVerSch15}.

\citet{ShaOzeSaf17} proposed a broadly applicable experimental scheme to search for the LLI violation with atomic systems using dynamic decoupling which can be implemented
in current atomic clocks experiments, both with single ions and arrays of neutral atoms.
Moreover, this scheme can be performed on systems with no optical transitions, and therefore it is
also applicable to highly charged ions which exhibit particularly high sensitivity to Lorentz invariance
violation.

Another interesting future possibility is measuring transition energies of rare-earth ions doped in crystalline
lattices, which can be highly sensitive to the electron SME parameters \cite{harabati2015effects}.

Also of  note is the work of \citet{botermann2014test}, testing the time dilation predictions of special relativity, using  a
clock-comparison test  performed at relativistic speeds using Li$^+$ ions in a storage ring. Another test of time dilation has been reported by \citet{DelLodBil17} who searched for variations of the frequency differences between four strontium optical lattice clocks
in different locations in Europe, connected by fiber optic  links.

\subsection{Proton and neutron sectors of the SME}

\subsubsection{Cs clock experiment}

Another example of a clock comparison test is the work of \citet{WolChaBiz06} who used a cold Cs atomic clock to test LLI in the matter sector, setting limits on the tensor Lorentz-violating coefficients for the proton.
The Cs clock, which is also the primary frequency standard defining the second, operates on the
$|F=3\rangle \longleftrightarrow |F=4\rangle $ hyperfine transition of the $^{133}$Cs $6S_{1/2}$
ground state, where $\mb{F}=\mb{J}+\mb{I}$ is the total angular momentum and Cs nuclear spin is $I=7/2$. In the magnetic field, $F=3$ and $F=4$ clock states split into 7 and 9 Zeeman
substates with $m_{F}=[-3,3]$ and $m_{F}=[-4,4]$, respectively. The atomic clock operates
on the
\begin{eqnarray}
|F=3, m_F=0\rangle \leftrightarrow |F=4, m_F=0 \rangle
\label{Cs0}
\end{eqnarray}
hyperfine transition at 9.2~GHz, which is insensitive to either Lorentz violation or first-order magnetic field effects, but the other transitions with $\delta m_{F}\neq 0$ are used for magnetic field
  characterization. To test Lorentz symmetry, \citet{WolChaBiz06} monitored a combination of clock
\begin{eqnarray}
|F=3, m_F=3\rangle \longleftrightarrow |F=4, m_F=3 \rangle,
\label{Csp}
\end{eqnarray}
   and
   \begin{eqnarray}
   |F=3, m_F=-3\rangle \longleftrightarrow |F=4, m_F=-3 \rangle
   \label{Csm}
  \end{eqnarray}
    transitions to form a combined observable
 \begin{equation}
 \nu_c=\nu_{+3} + \nu_{-3} -2 \nu_0.
 \end{equation}
 The $\nu_0$, $\nu_{+3}$, and $\nu_{-3}$ are frequencies of
 (\ref{Cs0}), (\ref{Csp}), and (\ref{Csm}) transitions above.
 The combined observable is used to avoid the dominant noise source -
 the first order Zeeman shift due to the magnetic field fluctuations which strongly affect the states with $m_F\neq0$, but cancels for the $\pm m_F$ combination. Since the $m_F$ is the same for upper and lower states of all transitions, there is no Lorentz-violating tensor
 component from the electron sector. The $^{133}$Cs nucleus has one unpaired
 proton, the experiment is interpreted in terms of the proton LV parameters of the
 $c_{\mu \nu}$ tensor, using the Schmidt nuclear model.
 The Cs clock experiment set the limits for parameters for the proton at the $10^{-21}-10^{-25}$ level.
  A reanalysis of this experiment using an improved model linking the frequency shift of the
$^{133}$Cs hyperfine Zeeman transitions and SME coefficients of proton and neutron carried out by  \citet{PihGueBai17}
placed improved bounds on the LV for the proton and neutron.

\subsubsection{Comagnetometer experiments}

Some of the most stringent clock-comparison tests of LLI \cite{Bro10,SmiBroChe11} have been carried out using the self-compensating spin-exchange relaxation-free (SERF) comagnetometry scheme \cite{Kor02,Kor05}, which was also used for constraining anomalous dipole-dipole interactions \cite{Vas09} as discussed in detail in Sec.~\ref{Sec:ExoticSpin}.

%----------------------------------------------------------------
\begin{figure}
\includegraphics[width=3.5 in]{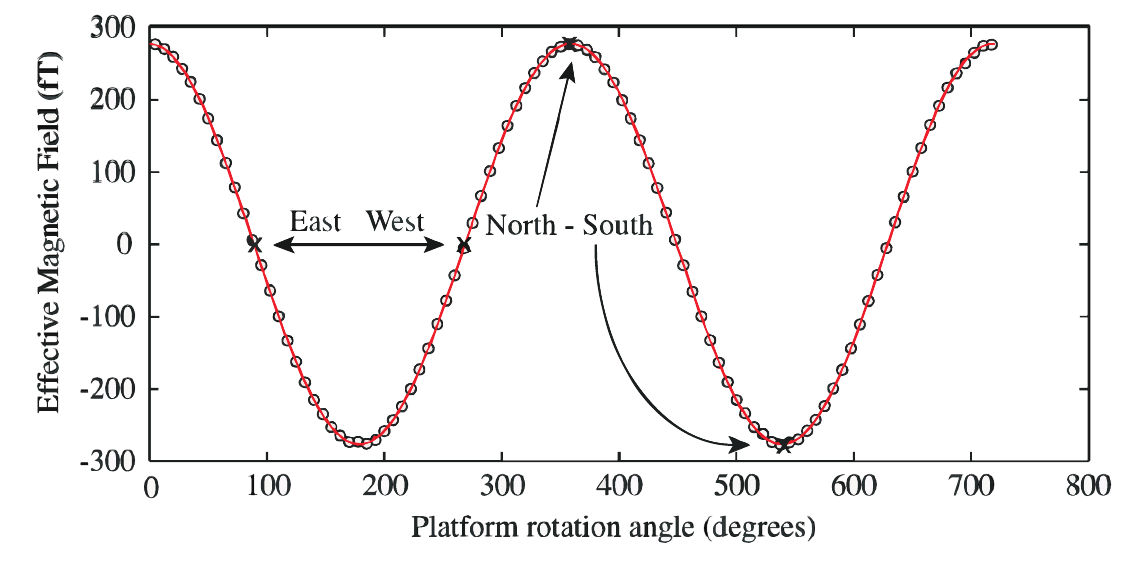}
\caption{Change of K-$^3$He self-compensating SERF comagnetometer signal for 180$^\circ$ platform rotation as a function of the initial platform angle. Figure from \citet{Bro10}. }
\label{Fig:K-He-comagnetometer-gyro-compass}
\end{figure}
%----------------------------------------------------------------

The experiment of \citet{Bro10} employs overlapping ensembles of K and $^3$He coupled via spin-exchange collisions.  The atoms are in the low-magnetic-field SERF regime where broadening of the Zeeman resonances due to spin-exchange collisions is eliminated.  The magnetic field along the $z$-direction is tuned to the compensation point where the K-$^3$He SERF comagnetometer is insensitive to magnetic fields but highly sensitive to anomalous interactions that do not scale with the magnetic moments.  The spin polarization of the K atoms along the $x$-direction, probed via optical rotation with a linearly polarized laser beam propagating along $x$, is given to leading order by:
\begin{align}
P_x^e = P_z^e \frac{\gamma_e}{\Gamma\ts{rel}}\prn{ \beta_y^N - \beta_y^e + \frac{\Omega_y}{\gamma_N} }~,
\label{Eq:K-He-comag-signal}
\end{align}
where $\beta_y^N$ and $\beta_y^e$ describe the phenomenological SME background fields along the $y$-direction coupling to the $^3$He nucleus ($N$) and valence electron ($e$) of K, respectively, $P_z^e$ is the K electron spin polarization along $z$, $\Gamma\ts{rel}$ is the relaxation rate for K polarization, $\gamma_e$ and $\gamma_N$ are the gyromagnetic ratios for electrons and $^3$He nuclei, respectively, and $\Omega_y$ is the rotation rate of the apparatus.  A number of steps are taken to eliminate various sources of noise and systematic error.  For example, the signal described by Eq.~\eqref{Eq:K-He-comag-signal} depends explicitly on the rotation rate of the apparatus, so a nonzero $P_x^e$ is generated by the gyro-compass effect  due to the
Earth's rotation \cite{Ven92,Hec08}.  To compensate for this effect, the experiment is mounted on a rotary platform.  Figure~\ref{Fig:K-He-comagnetometer-gyro-compass} shows the change in the comagnetometer signal for a 180$^\circ$ rotation as a function of the initial platform angle, demonstrating a significant effect of Earth's rotation on $P_x^e$.  In order to test LLI, the orientation of the apparatus is alternated between North-South or East-West (marked by crosses on the plot of Fig.~\ref{Fig:K-He-comagnetometer-gyro-compass}) every 22~s over the course of many days. Nonzero values of $\beta_y^N$ or $\beta_y^e$ would lead to sidereal oscillation of the amplitude of the difference between the North-South or East-West signals.  The results of the measurements, carried out over 143 days, are consistent with no LLI violation. Combined with the constraints on electron couplings to SME background fields from \citet{Hec08}, the results of this experiment probe neutron couplings to SME background fields at energy scales $\approx 10^{-25}~{\rm eV}$ \cite{Bro10}.

The closely related experiment of \citet{SmiBroChe11} uses a $^{21}$Ne-Rb-K SERF comagnetometer with a shot-noise-limited sensitivity to LLI violations that is an order of magnitude better than the K-$^3$He comagetometer.  Additionally, because the nuclear spin of $^{21}$Ne  is $I=3/2$, the $^{21}$Ne-Rb-K comagnetometer is sensitive to tensor anisotropies as well as the vector anisotropies probed by the K-$^3$He comagetometer ($I=1/2$ for $^3$He). A new version of this experiment, performed at the South Pole to better control for the gyro-compass effect \cite{hedges2015south}, is expected to improve on these constraints by yet another order-of-magnitude.

A different scheme was used by \citet{AllHeiKar14} to test LV in the neutron sector at a similar level of accuracy by measuring precession of overlapping ensembles of $^3$He and $^{129}$Xe atoms [although note the discussion between \citet{Rom14comment} and \citet{All14reply} regarding these results].

\subsection{Quartz oscillators}

\citet{lo2016acoustic}  proposed and demonstrated a novel approach to LLI tests in the matter sector taking advantage of new, compact, and reliable quartz oscillator technology. Violations of LLI in the matter and photon sector of the SME generate anisotropies in particles' inertial masses and the elastic constants of solids,
giving rise to anisotropies in the resonance frequencies of acoustic modes in solids. Thus the spatial-orientation-dependence of acoustic resonances can be used to constrain LV: the initial experiment of \citet{lo2016acoustic} set constraints on certain SME parameters some 3 orders-of-magnitude more stringent than other laboratory tests and ten times more stringent than astrophysical limits.

\subsection{Photon sector of the SME}

At the close of this section, we return to the experimental setup that was the basis of the first tests of Lorentz invariance, the rotating interferometer. Recent experiments with rotating optical and microwave resonators establish some of the most stringent constraints on LV in the photon sector \cite{MulStaTob07,Her09rotating,EisNevSch09,Hoh10improved,NagParKov15,Che16LLIcrossed-resonator}. For example, \citet{EisNevSch09} searched for a spatial anisotropy of the speed of light using two orthogonal standing-wave optical cavities contained in a single block of glass with ultralow thermal expansion coefficient. The orthogonal cavities were probed with a laser and rotated nearly 200,000 times over the course of 13 months using an air cushion rotation table with low axis wobble, low vibration level, and active stabilization of optical elements. The quantity of interest in the experiment was the spatial-orientation-dependence of the beat frequency between the light from the two cavities, which was found to be invariant at a level below a part in $10^{17}$. The experiment of \citet{NagParKov15} improved upon this result by a factor of 10 through the use of two cryogenic cylindrical copper cavities loaded with identical sapphire dielectric crystals whose axes were oriented orthogonally to one another. Whispering gallery mode resonances near 13 GHz were excited and the apparatus was rotated with a period of $\approx 100~{\rm s}$. Again the observable was the spatial-orientation-dependence of the beat frequency between the signals from the two microwave cavities. Compared to the original experiments of \citet{MicMor87}, this is an improvement of 17 orders of magnitude. \citet{kostelecky2016searching} also point out that gravitational wave detectors, km-scale laser interferometers with exquisite sensitivity, establish constraints on certain LV parameters of the SME that are several orders-of-magnitude more stringent than previous limits.  Even more stringent constraints come from re-interpretation of existing data: \citet{Fla17LLIcoulomb} have noted that by analyzing the Coulomb interactions between the constituent particles of atoms and nuclei, comagnetometer experiments testing LLI \cite{SmiBroChe11} establish that the speed of light is isotropic to a part in $10^{28}$.

\section{Search for violations of quantum statistics, spin-statistics theorem}
\label{SPINSTATISTICS}

The concept of \emph{identical particles} is unique to quantum physics. In contrast to, for example, identical twins or  so-called ``standard-candle'' supernovae, all electrons, helium atoms, $^{85}$Rb nuclei, etc., are, as far as we can tell, \emph{truly} identical to each other. This means that if we have a wavefunction representing a system containing identical particles, particle densities should not change upon interchange of two identical particles. As a consequence, the wavefunction should either remain invariant or change sign under permutation of identical particles. This is the essence of the permutation-symmetry postulate (PSP). The spin-statistics theorem (SST) dictates which of the two options is realized given the particular intrinsic spin of the particles. (This connection is nontrivial and, one might argue, a-priori unexpected.)  The resulting division of particles into fermions and bosons is one of the cornerstones of modern physics.

The SST is proved in the framework of relativistic field theory using the assumptions of causality and Lorentz invariance in 3 + 1 spacetime dimensions, along with a number of more subtle implicit assumptions enumerated  by \citet{Wic01}.

While it is notoriously difficult to build a consistent relativistic theory incorporating SST and PSP violations (a feat that has not as yet been accomplished, to the best of our knowledge), it is important to put these properties to rigorous tests given their fundamental importance in our understanding of Nature. One may think of such tests as probing all the assumptions in the SST proof, as well as providing a possible experimental window into theories that go beyond  conventional field theory, for instance, string theory. For example, plausible theoretical scenarios for small spin-statistics violations include excitations of higher dimensions allowing particles to possess wrong-symmetry states in the usual 3-dimensional space while maintaining the correct symmetry in an $N$-dimensional space \cite{Gre89}.

Since all our observations so far are consistent with PSP and SST, the experiments should search for \emph{small} violations of PSP and SST, the effects sometimes referred to as ``violations of quantum statistics.''

Comprehensive reviews of the literature on the spin-statistics connection and related issues such as the Pauli
exclusion principle and particle indistinguishably, including theoretical background and experimental searches, can be found in a book edited by \citet{HilTin00}  and a paper
by \citet{CurGilHil12}. Here we limit our discussion to examples of recent experiments to give the reader a  flavor of atomic, molecular, and optical techniques that are used in this field.

The strongest limit on a possible violation of the Pauli exclusion principle for electrons currently comes from the VIP experiment at Gran Sasso \cite{Marton2013}. Here strong electric current is flown through a copper sample and Pauli-forbidden atomic transitions involving occupied atomic orbitals are searched for by measuring x-rays at the anticipated transition energy. The limit on the probability for two electrons to be in a symmetry-forbidden state is currently $<4.7\cdot 10^{-29}$ with expected improvement in the upgraded VIP2 experiment by further two orders of magnitude \cite{Shi2016,Marton2017}.

Molecular spectroscopy has played an important historical role in establishing the experimental basis for the PSP and the SST \cite{CurGilHil12}. The general idea is that in a molecule containing two identical nuclei, rotational states corresponding to the overall molecular wavefunction being symmetric (in the case of half-integer-spin nuclei) or antisymmetric (in the case of integer-spin nuclei) are forbidden by quantum statistics, and so the spectral lines involving these molecular states are absent from the molecular spectrum. A powerful experimental methodology for testing for statistics violations is to look for such forbidden lines \cite{Tin01}.

Recent experiments \cite{Pas12} using saturated-absorption cavity ring-down spectroscopy searched for forbidden rovibrational lines at a 4.25 $\mu$m wavelength in the spectra of the $^{12}$C$^{16}$O$_2$ molecule containing two bosonic oxygen nuclei. They limited the relative probability for the molecule to be in a wrong-symmetry state at $< 3.8\times 10^{-12}$ level, significantly improving on earlier results. An interesting extension is to molecules containing more than two identical nuclei that would allow to probe for more complex permutation symmetries than are allowed for just two identical particles \cite{Tin00,Tin01}.

An experimental test of Bose-Einstein (BE) statistics and, consequently, the SST as
it applies to photons interacting with atoms was carried out by \citet{EngYasBud10}. The experiment, extending earlier work \cite{DeM99}, used a selection rule for atomic transitions that is closely related to the Landau-Yang theorem \cite{Lan48,Yan50} in high-energy physics. The selection rule states that two collinear, equal-frequency, photons cannot be in a state of total angular momentum one. An example in high-energy physics is that the neutral spin-one $Z_0$ boson cannot decay to two photons. (According to the Particle Data Group, the branching ratio for this process is limited to $< 5.2\times 10 ^{-5}$, although there are additional reasons that suppress such decay.)
For atoms, the selection rule means that two collinear equal-frequency photons cannot stimulate a transition between
atomic states of total angular momentum zero and one. The experiment employed an atomic beam of barium optically excited in a power-buildup cavity and resulted in a limit for two photons to be in wrong (i.e., fermionic) symmetry state of $< 4.0\times 10 ^{-11}$. Further improvements by several orders of magnitude are expected in ongoing experiments using ultra-cold Sr atoms \cite{GuzInaPen15}.\\

As mentioned above, quantum-statistics violation would be an effect outside of the framework of conventional field theory, in contrast to most other ``exotic'' effects discussed in this review. Combined with the absence of a consistent alternative framework, discussion of such effects often leads to conceptual difficulties, including questions like: \textit{What is the experiment really testing?}, \textit{How can we compare results from different experiments?}, etc.

The results of some early experiments were dismissed as they did not take into account a so-called \textit{superselection rule} stating that the permutation symmetry of a system of identical particles cannot change in the course of the system's evolution. Being truly identical implies that the particles cannot be distinguished by any measurement. In particular, this means that all operators corresponding to physical observables must commute with all exchange operators $\xi$, for example $\sbrk{H,\xi}=0$ for any Hamiltonian $H$. This fact is used by \citet{Ama80} to derive the aforementioned superselection rule, which implies:
%--------------------------------------------------------------------
\begin{align}
    \bra{A}H\ket{S} = 0~,
\end{align}
%--------------------------------------------------------------------
where $\ket{S}$ and $\ket{A}$ are exchange symmetric and antisymmetric states, respectively. The superselection rule prevents, for example, the transition between a symmetric and antisymmetric state (as was searched for in some early experiments purporting to test the SST) based purely on the fact that the particles are identical and not on the PSP or SST. However, it is important here to note that the superselection rule does not prevent creation of particles with mixed statistics. The quon algebra \cite{Gre91,Gre99}, for example, takes advantage of this exception by postulating creation and annihilation operators which do not obey the usual commutation relations, leading to the creation of particle states which are neither symmetric nor antisymmetric.  Another immediate consequence of the superselection rule is that a description in terms of a wavefunction with a mixed permutation symmetry is not acceptable and a density matrix should be used instead. A further discussion of these points and related references can be found in the paper by \citet{EllLarGeh12}.

\section{Conclusion}

AMO physics has been crucially important in laying the foundations of our understanding of the fundamental laws of nature ever since the advent of precision spectroscopy in the 19th century. The most remarkable success is the discovery of the inevitability of quantum theory and its subsequent spectacular development, including firming up such fundamental concepts as indistinguishability of identical particles, the spin-statistics connection,
the role of discrete symmetries such as parity and time-reversal,
 entanglement, relativistic quantum mechanics and quantum field theory, and many others. From the early days, AMO physics has been closely connected to astronomy and astrophysics, from the discovery of new elements in the solar spectrum to determining the velocities of stars and measuring the expansion of the  Universe via red shifts of spectral lines. The list of seminal fundamental physics discoveries using AMO techniques can, of course, be made almost arbitrarily long.

Remarkably, two centuries after its birth, the field of precision AMO tests of fundamental physics continues to be at the forefront of discovery, showing no signs of slowing down! Conversely, with collider physics becoming more and more expensive and potentially reaching saturation in terms of accessible particle energies and intensities, AMO physics beautifully complements high-energy physics and, in some cases, provides powerful ways to indirectly explore potential new phenomena at energy scales reaching orders of magnitude beyond what can be expected to be directly accessible with accelerators in any foreseeable future.

Having powerful AMO tools for fundamental-physics inquiry is especially important because there are many basic properties of the Universe that we do not understand:
\begin{itemize}
\item What are dark matter and dark energy?
\item Why is there so much more matter in the Universe than antimatter?
\item Why are the masses of all known particles so much smaller than the fundamental energy scales such as the grand-unification and the Planck scales?
\item Why do strong interactions appear to respect the $CP$ symmetry?
\item What lies beyond the Standard Model of particles and interactions?
\item How can general relativity be unified with quantum theory? ...
\end{itemize}
 These questions are, in a sense, ``urgent.'' For instance, dark matter constitutes most of the mass in galaxies including our own, and so it is likely that a discovery of the dark-matter composition is ``around the corner.''

We hope that with this review, we have succeeded in conveying to the reader our own excitement and anticipation of forthcoming paradigm-shifting discoveries in fundamental physics with atoms, molecules, and light.

\section*{Acknowledgements}
  We are grateful to Eric Benck, Klaus Blaum, Sid Cahn, David Cassidy, Catalina Curceanu, Vladimir Dzuba, Pavel Fadeev, Takeshi Fukuyama, Victor Flambaum, Remi Geiger, Peter Graham, David Hanneke, Larry Hunter, Kent Irwin, Masatoshi Kajita, Mikhail Kozlov, Ivan Kozyrev, Konrad Lehnert, John McFerran, Holger M\"uller, Sergey Porsev, Maxim Pospelov, Surjeet Rajendran, Eric Shirley, Yannis Semertzidis, Michael Snow, Yevgeny Stadnik, Alexander Sushkov, Guglielmo Tino, Sunny Vagnozzi, Antoine Weis, Xing Wu, Nodoka Yamanaka, Z.-C. Yan, Max Zolotorev  for helpful discussions and comments on the manuscript. AD acknowledges the support of the National Science Foundation under grants PHY-1506424 and PHY-1607396.
DB acknowledges the support of the DFG Koselleck program, the Heising-Simons and Simons Foundations, and the National Science Foundation under grant PHY-1507160, as well as the European Research Council (ERC) under the European Union’s Horizon 2020 research and innovation program (grant agreement No. 695405).
DD acknowledges the support of the National Science Foundation under grant PHY-1404146, the Templeton Foundation, and the Heising-Simons Foundation.
DFJK acknowledges the support of the National Science Foundation under grant PHY-1707875 and the Heising-Simons and Simons Foundations.
MSS acknowledges the support of the National Science Foundation under grants PHY-1404156 and PHY-1620687.
This research was performed in part under the sponsorship of the
U.S. Department of Commerce, National Institute of Standards
and Technology and the National Science
Foundation via the Physics Frontiers Center at the Joint
Quantum Institute.
\appendix
\section{Notations, units, and abbreviations} %==================================

\subsection{Atomic and molecular properties as encoded in spectroscopic notation}

Atoms and molecules make wonderful clocks and precision measurement instruments because their electronic states offer read, write and storage capabilities extending across many decades of bandwidth. Key properties of a given state can be understood by symmetry considerations, for which an understanding of conventional spectroscopic notation is a useful aid. We present a brief summary of notation that is germane to most of the specific examples discussed in this paper. The concepts can be found on display in the periodic table of the elements constructed for use in atomic spectroscopy \cite{NISTSP699}.  More comprehensive accounts can be found in \citet{MW2002} and \citet{IUPAC1,IUPAC2,IUPAC3}.

\subsection{Atomic symmetries}
\label{Atomicsymmetries}

The conventional periodic table of the elements \cite{NISTSP699}  is laid out in a way that displays the {\em Aufbau} principle. As the atomic number $Z$ increases, electrons are added one by one to atomic electron shells, $n,l$.  These are labeled  by  the integer principal quantum number $n \ge 1$ and orbital angular momentum quantum number $l \, , \left(0 \leq l < n \right)$.  These two quantum numbers are encountered in the nonrelativistic quantum theory of the hydrogen atom.  The beginning of the $n^\mathrm{th}$ row of the periodic table marks the start of filling the electron shell $n,0$, and the end marks the filling of an electron shell.

This representation of atomic structure is only an approximate model, but it also defines a zeroth-order basis of many-electron wavefunctions that can be consistently improved upon and enlarged by techniques of quantum many-body theory.  Good guidance for rough estimates of energies and transition probabilities is communicated in a standard notation.  This notation expresses, in order of descending magnitude of energy: the atomic mean field ({\em configuration}), electron-electron interaction ({\em term}), spin-orbit interaction ({\em electronic level}), possible electron-nucleus interactions ({\em hyperfine level}), and projection of the total angular momentum ({\em Zeeman sublevel}).

We illustrate this using the example of the ground state of cerium (Ce, Z = 58). Its electron {\em configuration} is conventionally described as
\begin{equation}
1s^2 2s^2 2p^6 3s^2 3p^6 3d^{10} 4s^2 4p^6 4d^{10} 5s^2 5p^6 4f^1 5d^1 6s^2.
\label{Ceconfig}
\end{equation}
\noindent Here $s,p,d,f$ designate $l = 0,1,2,3$, and the superscripts designate occupation numbers. Thus, starting from the left, the expression (\ref{Ceconfig}) indicates that there are two electrons in the $n=1, l=0$ shell, two more in $2,0$ subshell of the $n=2$ shell, six more in $2,1$ subshell and so on up to the last closed subshell, $5p$. That consolidated list of subshells is the same as that for the ground state of Xe, so it is convenient to rewrite the expression (\ref{Ceconfig}) as
\begin{equation}
\left[\mathrm{Xe}\right] 4f \,  5d \, 6s^2 ,
\label{Ceconfig1}
\end{equation}
where no superscript indicates single occupancy of the electron shell.

This shows that Ce has four electrons outside an isotropic closed-shell Xe-like core. These electrons determine the symmetries of the electronic wavefunction and have predominant influence on the atom's chemical and physical properties.

There are 140 independent electronic states that are members of the (\ref{Ceconfig1}) configuration. These are differentiated by the term and level hierarchies.
 The ground state of Ce has the term and level designation
\begin{equation}
\left[\mathrm{Xe}\right] 4f \,  5d \, 6s^2 \, ^1\mathrm{G}^\circ_4 .
\label{lev}
\end{equation}
Four properties are encoded in the rightmost expression,  $^1\mathrm{G}^\circ_4$,  of  expression (\ref{lev}):

\begin{itemize}
\item The state's total electronic spin angular momentum $S$, (in units of $\hbar$), which is encoded as $2S+1$ in the $^1\mathrm{G}$ superscript. Here the state is a ``spin singlet'' with $S = 0$.

\item The state's total electronic orbital angular momentum $L$, (in units of $\hbar$),  which is encoded as a capital letter. The string SPDFGHIK expresses the character values for $0 \leq L \leq 7$ . Here the state has $L = 4$.

\item  The state's total electronic angular momentum $J$, (in units of $\hbar$), which is shown in the $^1\mathrm{G}^\circ_4$ subscript. Here $J=4$, consistent with $L=4$ and $S=0$.

\item  The state's parity under inversion of spatial coordinates.  This is sometimes shown by $^\circ$ if the parity is odd, the superscript is omitted for even-parity states. Indeed it is redundant, because the parity is given by $(-1)$ to the power of  $\sum_k l_k$, where the sum runs over all atomic electrons. Here it is odd, which is readily verified since $4f$ is the only electron shell that makes an odd contribution to the sum.
\end{itemize}

Of these four indices, only two are exact: parity (to the extent that electroweak interactions are negligible) and total angular momentum $J$ (in cases where there is no nuclear angular momentum or neglecting hyperfine structure).  Concerning $J$, when there is no hyperfine structure, the application of a weak magnetic field reveals that the ground level has $2J+1$ distinct Zeeman sublevels. As for $S$ and $L$, there are cases in which it is meaningful to consider them as good quantum numbers.  For example, helium was once considered to consist of two elements, ortho- and para-helium, because it evinced distinctive singlet and triplet  spectra, between which there seemed to be no connection \cite{Keesom}. Now we understand those to be spectra associated with states that are (predominantly) $S=0$ and $S=1$, respectively.
In many cases of atoms with several valence electrons, different configurations and terms are strongly mixed, and the dominant
configuration and $LS$ term are listed.  Indeed, our example expression, (\ref{lev}), is a case in point!
The Ce ground state approximately described by (\ref{lev}) is, in fact, a mixture of different configuration and terms, where the weight of (\ref{lev}) is about 60\% \cite{NIST}.

When the atomic nucleus has no angular momentum, as is the case for all even-even isotopes in their nuclear ground state, then a level designation such as expression~(\ref{lev}) identifies $2J + 1$ degenerate states, corresponding to the distinct values of $M_J$, the projection of {\bf J} upon some arbitrary quantization axis.
When the nucleus has spin  $\mb{I} \neq 0$, then the total atomic angular momentum is designated $\mb{F}=\mb{I}+\mb{J}$. As above, the corresponding quantum numbers are $I$, $J$, and $F$. The magnetic quantum number of an atomic state, $M_F$, takes one of $2F + 1$ discrete values. The separate values of $F$ correspond to different relative arrangements  of electronic and nuclear magnetic and electric moments, whereby they have slightly different energies. These energy differences were called ``hyperfine structure'' when they were first interpreted by  \citet{Pauli}, because they were a minute detail of atomic spectra.
\subsection{Molecular symmetries}
\label{Molecularsymmetries}

The molecular term symbols that designate the electronic states of a diatomic
molecule take the form

\begin{equation}
^{2S+1}\Lambda^{(+/-)}_{\Omega,(g/u)}.
\label{molecularterm}
\end{equation}

\noindent The symbols $\Lambda, S,\Omega$ are analogous to their atomic counterparts $L,S,J$. Indeed, $S$ designates the same net electronic spin in both cases. In the {\em body frame} of the molecule, {\em i.e.} a frame in which the internuclear axis is fixed in space, rotations of all electrons about the internuclear axis commute with the Hamiltonian, so projections of electronic angular momentum upon that axis can be taken to be good quantum numbers. The absolute value of the projection upon this axis of electronic orbital angular momentum, $L$, is designated $\Lambda$
(in units of $\hbar$). Thus, $\Lambda = 0, 1, 2, 3, \dots $, designated respectively by uppercase Greek letters,
$\Sigma$, $\Pi$, $\Delta$, $\Phi$, $\dots$ in analogy with the atomic $\rm{S, P, D, F, }\dots$.  As atomic $L$ is to
$\Lambda$, so is atomic $J$ to $\Omega$, which is the magnitude of the projection of electronic total angular momentum upon the internuclear axis, again in units of $\hbar$. As for $J$ in atoms, $\Omega$ is an  integer or half integer. As an example, we consider a state of the thorium oxide molecule, ThO, that is mentioned in
Sec. \ref{Sec:EDM:Future:Diamagnetic}. Its electronic state is labeled there as $^3\Delta_1$, thus $S = 1$,
$\Lambda = 2$ and $\Omega = 1$.
\begin{table}
\caption{Mathematical symbols used and their meanings.}
\medskip \begin{tabular}{ll} \hline \hline
Symbol~~~~~~ & Meaning \\
\hline
\rule{0ex}{2.6ex} $c$ & speed of light\\
\rule{0ex}{2.6ex} $\epsilon_0$ & electric constant \\
\rule{0ex}{2.6ex} $G$ & Newtonian constant of gravitation\\
\rule{0ex}{2.6ex} $h$& Planck constant, $\hbar=h/2\pi$ \\
\rule{0ex}{2.6ex} {\textit{e}} & elementary charge \\
\rule{0ex}{2.6ex}  $\alpha$ & fine structure constant \\
\rule{0ex}{2.6ex} $R_{\infty}$ & Rydberg constant \\
\rule{0ex}{2.6ex}  $a_0$ & Bohr radius \\
\rule{0ex}{2.6ex} $\mu_\mr{N}$ & nuclear magneton \\
\rule{0ex}{2.6ex}  $G_\mr{F}$ & Fermi constant \\
\rule{0ex}{2.6ex} $g$  & local acceleration due to the Earth's gravity\\
\rule{0ex}{2.6ex}  $m_\mb{e}$ & electron mass \\
\rule{0ex}{2.6ex}  $m_\mb{p}$ & proton mass \\
\rule{0ex}{2.6ex}  $M_Z$ & Z-boson mass \\
\rule{0ex}{2.6ex} $\theta_\mr{W}$ & weak mixing angle\\
\rule{0ex}{2.6ex}  $\sigma_{i}$ & Pauli matrices, $i=1,2,3$ \\
\rule{0ex}{2.6ex} $\gamma _\mu$ & Dirac matrices, $\mu=0,1,2,3$ \\
\rule{0ex}{2.6ex} $\gamma _5$ & Dirac matrix associated with pseudoscalars\\
\rule{0ex}{2.6ex} $\sigma^{\mu \nu}$ & $\sigma^{\mu \nu}=\frac{i}{2}\left( \gamma^\mu \gamma^\nu - \gamma^\nu \gamma^\mu \right)$\\
\rule{0ex}{2.6ex} $\mathbf{s}$ & single electron spin \\
\rule{0ex}{2.6ex} $\mathbf{S}$ & multi-electron atom total spin\\
\rule{0ex}{2.6ex} $\mb{p}$ & linear momentum \\
\rule{0ex}{2.6ex} $\Evec$ & electric field (vector)\\
\rule{0ex}{2.6ex} $\Bvec$ & magnetic field (vector)\\
\rule{0ex}{2.6ex} $C, P, T$ & charge conjugation, parity, and time-reversal \\
\rule{0ex}{2.6ex} &  transformations\\
\rule{0ex}{2.6ex}  $\mu=m_\mb{p}/m_\mb{e}$ & proton to electron mass ratio, $\overline{\mu}=1/\mu$ (Sec.~\ref{Sec:FC})\\
\rule{0ex}{2.6ex}  $K$& dimensionless sensitivity factor of an energy level\\
                      & to $\alpha$-variation (Sec.~\ref{Sec:FC})\\
\rule{0ex}{2.6ex}  $K_\mu$& dimensionless sensitivity factor of an energy level\\
                          & to $\mu$-variation (Sec.~\ref{Sec:FC})\\
    \rule{0ex}{2.6ex}  $m_{q}$ & average mass of light quarks (Sec.~\ref{Sec:FC})\\
\rule{0ex}{2.6ex}  $\Lambda_{\mr{QCD}}$& QCD energy scale (Sec.~\ref{Sec:FC})\\
\rule{0ex}{2.6ex}  $\kappa$& dimensionless sensitivity factor of an energy level\\
            & to a variation of $X_q=m_q/\Lambda_{\mr{QCD}}$ (Sec.~\ref{Sec:FC})\\
 \rule{0ex}{2.6ex}           $k_X$& dimensionless factor quantifying the spatial\\
  \rule{0ex}{2.6ex}    &  variation of the fundamental constant $X$ (Sec.~\ref{Sec:FC})\\
\rule{0ex}{2.6ex} $Q_\mr{W}$ & nuclear weak charge (Sec.~\ref{Sec:APV})\\
\rule{0ex}{2.6ex}  $\mathbf{d}$ & electric dipole moment (Sec.~\ref{Sec:EDM})\\
\rule{0ex}{2.6ex}  $\mathcal{P}$ & dimensionless electrical polarization (Sec.~\ref{Sec:EDM}) \\
\rule{0ex}{2.6ex}  \hbox{\boldmath{$\mathcal{S}$}} & Schiff moment (Sec.~\ref{Sec:EDM})\\
\rule{0ex}{2.6ex}  $\tilde{d}$ & chromo-EDM (Sec.~\ref{Sec:EDM}) \\
\rule{0ex}{2.6ex} $\hat{\bs{\sigma}}$ & unit vector along spin (Sec.~\ref{Sec:ExoticSpin})\\
\hline \hline
\end{tabular}
\label{Table:symbols}
\end{table}

\begin{table}
\caption{Abbreviations and their meanings.}
\medskip \begin{tabular}{ll} \hline \hline
Abbreviation~~ & Meaning \\
\hline
\rule{0ex}{2.6ex} AMO & atomic, molecular and optical physics \\
\rule{0ex}{2.6ex} ALPs & axion-like particles \\
\rule{0ex}{2.6ex} APV & atomic parity violation \\
\rule{0ex}{2.6ex} CPT  & combined operation $CPT$ \\
\rule{0ex}{2.6ex} CPV & $CP$-violation \\
\rule{0ex}{2.6ex} cEDM & chromo-EDM \\
\rule{0ex}{2.6ex} DFSZ & Dine-Fischler-Srednicki-Zhitnitskii \\
\rule{0ex}{2.6ex} DM & dark matter \\
\rule{0ex}{2.6ex} EDM & electric dipole moment \\
\rule{0ex}{2.6ex} eEDM & electron EDM \\
\rule{0ex}{2.6ex} EEP & Einstein equivalence principle \\
\rule{0ex}{2.6ex} GDM & gravitational dipole moment \\
\rule{0ex}{2.6ex} GR & general relativity \\
\rule{0ex}{2.6ex} GPS & Global Positioning System \\
\rule{0ex}{2.6ex} GW & gravitational wave \\
\rule{0ex}{2.6ex} HCI & highly charged ion \\
\rule{0ex}{2.6ex} ISL & inverse-square law \\
\rule{0ex}{2.6ex} KSVZ & Kim-Shifman-Vainshtein-Zakharov \\
\rule{0ex}{2.6ex} LHC & Large Hadron Collider \\
\rule{0ex}{2.6ex} LLI& local Lorentz invariance\\
\rule{0ex}{2.6ex} LPI& local position invariance\\
\rule{0ex}{2.6ex} LV & Lorentz symmetry violation \\
\rule{0ex}{2.6ex} MWDM & Moody-Wilczek-Dobrescu-Mocioiu \\
\rule{0ex}{2.6ex} MQM & magnetic quadrupole moment \\
\rule{0ex}{2.6ex} NAM & nuclear anapole moment \\
\rule{0ex}{2.6ex} NIST & National Institute of Standards and Technology \\
\rule{0ex}{2.6ex} NMR & nuclear magnetic resonance \\
\rule{0ex}{2.6ex} n.r. &  nonrelativistic \\
\rule{0ex}{2.6ex}  PSP & permutation-symmetry postulate\\
\rule{0ex}{2.6ex} QCD & quantum chromodynamics \\
\rule{0ex}{2.6ex} SLI & semileptonic interaction \\
\rule{0ex}{2.6ex} SM & Standard Model \\
\rule{0ex}{2.6ex} SME & Standard Model extention\\
\rule{0ex}{2.6ex} SMt & Schiff moment \\
\rule{0ex}{2.6ex} SQUID & Superconducting QUantum Interference Device \\
\rule{0ex}{2.6ex} SCCEF & Sun centered celestial-equatorial frame \\
\rule{0ex}{2.6ex} SST & spin-statistics theorem \\
\rule{0ex}{2.6ex} SUSY & supersymmetry \\
\rule{0ex}{2.6ex} T,PV  & simultaneous $T$- and $P$-violation \\
\rule{0ex}{2.6ex} TV & $T$-violating but $P$-conserving  \\
\rule{0ex}{2.6ex} VULF & virialized ultralight field \\
\rule{0ex}{2.6ex} WEP & weak equivalence principle\\
\rule{0ex}{2.6ex} WIMP & weakly-interacting massive particle \\
\rule{0ex}{2.6ex} UFF & universality of free fall\\
\hline \hline
\end{tabular}
\label{Table:abbreviations}
\end{table}

For isolated molecules, only total angular momentum is rigorously conserved. Total angular momentum and parity also depend on  the rotational motion of the molecule. The rotational quantum number is designated $J = 0, 1, 2 \dots$; the corresponding inversion symmetry is $(-1)^J$. For homonuclear diatomic molecules, there is an additional
 quantum number associated with inversion with respect to the symmetry plane bisecting the line connecting the two nuclei. It can be even (German: ``gerade'') or odd (``ungerade'') under this transformation, which is represented in the term symbol as $g$ or $u$.

Finally, there is a symmetry of the molecular Hamiltonian under reflection in any plane that contains the internuclear axis. The electronic wavefunction may be even or odd under this transformation, which accounts for the $+$ or $-$ superscript that is an option in the expression (\ref{molecularterm}). It is used only for $\Sigma$ states, the best-known example being the $^3\Sigma_g^-$ ground state of molecular oxygen, O$_2$.

\subsection{Units}
The International System of Units (SI) is used throughout this paper, unless noted otherwise. Atomic units are often used in the source literature.
In atomic units, the values
of elementary charge  $e$, the electron mass $m_{\rm e}$, and the reduced
Planck constant $\hbar$ have  numerical value 1, and the electric constant $\epsilon_0$ has numerical value $1/(4\pi)$.
The conversion between SI and atomic units for commonly used quantities, including formulas and numerical values, is given, for example, in Table~XXXVII of \citet{CODATA2014}, p.~62.
For example, atomic unit of electric field is $\mathcal{E}_{\rm at} = e/(4\pi\epsilon_0 a_0^2)$.
\subsection{Symbols and abbreviations}
The common mathematical symbols and abbreviations which appear throughout the review are listed in Tables \ref{Table:symbols} and \ref{Table:abbreviations} for convenience. Chapter-specific notations are given under the chapter headings. The designations specific to a single subtopic and used only briefly are not tabulated below, but are defined
the first time they are introduced. CODATA and Particle Data Group designations are adopted in the review for common quantities.
Every effort is made to use notations and abbreviations which most commonly appear in the literature.

% \bibliography{master}
%\end{document} 
\bibliography{RMPref}
\end{document}